\documentclass[10pt,pra,aps,notitlepage,superscriptaddress,reprint]{revtex4-2}

\usepackage[hyperindex,breaklinks,colorlinks=true]{hyperref}

\usepackage{bm}

\usepackage[utf8]{inputenc}
\usepackage[T1]{fontenc}
\usepackage[english]{babel}  

\usepackage{tikz}

\usepackage{colortbl}
\usepackage{tabu,diagbox}

\usepackage{pgfplots}
\usepackage{lipsum}
\usepackage{youngtab}
\usepackage{subfigure}

\usepackage{times}
\usepackage{dsfont}

\usepackage{amssymb,amsmath,amsfonts,amsthm}
\usepackage{mathtools}
\usepackage{verbatim}
\usepackage{enumerate}
\usepackage{graphicx}

\usepackage[capitalise]{cleveref}
\usepackage{bbm}
\usepackage{booktabs}

\newcommand{\appsection}[1]{\let\oldthesection\thesection
  \renewcommand{\thesection}{Appendix \oldthesection}
  \section{#1}\let\thesection\oldthesection}

\def\Dbar{\leavevmode\lower.6ex\hbox to 0pt
{\hskip-.23ex\accent"16\hss}D}

\newcommand{\nc}{\newcommand}

\nc{\cA}{{\cal A}} \nc{\cB}{{\cal B}} \nc{\cC}{{\cal C}}
\nc{\cD}{{\cal D}} \nc{\cE}{{\cal E}} \nc{\cF}{{\cal F}}
\nc{\cG}{{\cal G}} \nc{\cH}{{\cal H}} \nc{\cI}{{\cal I}}
\nc{\cJ}{{\cal J}} \nc{\cK}{{\cal K}} \nc{\cL}{{\cal L}}
\nc{\cM}{{\cal M}} \nc{\cN}{{\cal N}} \nc{\cO}{{\cal O}}
\nc{\cP}{{\cal P}} \nc{\cQ}{{\cal Q}} \nc{\cR}{{\cal R}}
\nc{\cS}{{\cal S}} \nc{\cT}{{\cal T}} \nc{\cU}{{\cal U}}
\nc{\cV}{{\cal V}} \nc{\cW}{{\cal W}} \nc{\cX}{{\cal X}}
\nc{\cZ}{{\cal Z}}

\def\affa{\affiliation{Shenzhen SpinQ Technology Co., Ltd., Shenzhen, China}}
\def\affb{\affiliation{Shenzhen Institute for Quantum Science and Engineering and Department of Physics, Southern University of Science and Technology, Shenzhen, China}}
\def\affc{\affiliation{Guangdong Provincial Key Laboratory of Quantum Science and Engineering, Southern University of Science and Technology, Shenzhen, China}}
\def\affd{\affiliation{Department of Physics, The Hong Kong University of Science and Technology, Clear Water Bay, Kowloon, Hong Kong}}

\begin{document}

\title{QUANTUM COMPUTING: PRINCIPLES AND APPLICATIONS}

\author{Guanru Feng}
\email{gfeng@spinq.cn}
\affa

\author{Dawei Lu}
\affb
\affc

\author{Jun Li}
\affb
\affc

\author{Tao Xin}
\affb
\affc

\author{Bei Zeng} 
\email{zengb@ust.hk}
\affd

\begin{abstract}
People are witnessing quantum computing revolutions nowadays. Progress in the number of qubits, coherence times and gate fidelities are happening. Although quantum error correction era has not arrived, the research and development of quantum computing have inspired insights and breakthroughs in quantum technologies, both in theories and in experiments. In this review, we introduce the basic principles of quantum computing and the multilayer architecture for a quantum computer. There are different experimental platforms for implementing quantum computing. In this review, based on a mature experimental platform, the Nuclear Magnetic Resonance (NMR) platform, we introduce the basic steps to experimentally implement quantum computing, as well as common challenges and techniques.  
\end{abstract}

\date{\today}

\maketitle

\section{Introduction}

Quantum computing is a revolutionary approach to computation that leverages the principles of quantum mechanics to process and store data \cite{1,2,3,4,5}.  Quantum computers are believed to be exponentially faster than classical computers in certain problems, such as prime number factorization, data search and sorting, and simulation of complex quantum systems, etc. 

Nowadays, quantum computing is in the Noisy Intermediate-Scale Quantum (NISQ) era. In this stage, researchers and engineers are working to build and understand quantum devices subject to noise and other constraints. Despite these imperfections, NISQ devices offer opportunities to test and refine quantum algorithms, error correction techniques, and hardware development. The insights gained from NISQ devices will be invaluable in realizing scalable, fault-tolerant universal quantum computers in the future. In this stage, researchers are developing a range of quantum computing platforms. These include superconducting qubits, trapped ions, topological qubits, and photonic quantum computers, etc. Each platform presents unique advantages and challenges, and the scientific community has not yet reached a consensus on the most promising approach.

Although Nuclear Magnetic Resonance (NMR) \cite{6,7,8,9,10} may not rank among the most viable platforms for scalable quantum computing, it has significantly contributed to the field's experimental development. In the early stages of quantum computing research, NMR was broadly employed for the demonstration of quantum algorithms. Presently, the field of NMR quantum computing continues to be an active area of research. Meanwhile, the maturity of quantum control techniques in NMR, such as precise pulse shaping and qubit manipulation, has been instrumental in facilitating the development of other quantum computing platforms. Additionally, NMR serves as an excellent template system for introducing fundamental quantum computing concepts, such as superposition and quantum gates. It also offers a robust foundation for explaining the underlying theory of quantum control and demonstrating the implementation of essential quantum algorithms like Deutsch-Josza and Grover's algorithms. Consequently, NMR is an outstanding educational tool for understanding the principles of quantum computing and quantum control techniques.

In this review, we first introduce the basic principles of quantum computing and the multi-layer architecture for a quantum computer in Sections 2 and 3. Then, in Sections 4 and 5, based on desktop NMR quantum computing platforms, we present the fundamental steps for experimentally implementing quantum computing, along with common challenges and techniques.

\section{Basic principles of quantum computing}

\subsection{Quantum bit}

\subsubsection{Quantum states and quantum superposition}
The state of a quantum system is referred to as a "quantum state". It can be expressed by a vector of Hilbert space, which is usually denoted as $|\psi \rangle$ ($| ~ \rangle$ is the Dirac notation). A quantum state of a $N$-dimensional Hilbert space is expressed as  $\lvert \psi \rangle = (c_0 , c_1, \cdots, c_{N-1} )^{T}$, which is a linear combination of the basis vector $|0\rangle$, $|1\rangle$, $|2\rangle$, $\cdots$, $|N-1\rangle$ in this $N$-dimensional space:
\begin{align}
\lvert \psi \rangle = \begin{pmatrix} c_0 \\ c_1 \\ \vdots \\ c_{N-1} \end{pmatrix} = c_0 \lvert 0 \rangle + c_1 \lvert 1 \rangle + \cdots + c_{N-1} \lvert N-1 \rangle. \label{4.1}
\end{align}
The basis vectors $|0\rangle$, $|1\rangle$, $|2\rangle$, $\cdots$ , $|N-1\rangle$ in Eq. (\ref{4.1}) are quantum states, e.g. the corresponding states of electrons at different energy levels, i.e. the ground state, the first excited state and the second excited state, etc. These basis states are 1 in length and mutually orthogonal, that is, the basis states are orthonormal:
\begin{align}
\langle i \rvert j \rangle = 
\begin{cases} 
0, & \text{if } i \neq j \\
1, & \text{if } i = j 
\end{cases} \label{4.2}
\end{align}
In fact, the vectors of any state are normalized:
\begin{align}
\langle \psi | \psi \rangle = &\begin{pmatrix} c_0^* & c_1^* & \cdots & c_{N-1}^* \end{pmatrix} \begin{pmatrix} c_0 \\ c_1 \\ \vdots \\ c_{N-1} \end{pmatrix}\nonumber\\ =& (c_0^* \langle 0 | + c_1^* \langle 1 | + \cdots + c_{N-1}^* \langle N-1 |)\nonumber\\\cdot &(c_0 | 0 \rangle + c_1 | 1 \rangle + \cdots + c_{N-1} | N-1 \rangle)
\nonumber\\=& |c_0|^2 + |c_1|^2 + \cdots + |c_{N-1}|^2 = 1, \label{4.3}
\end{align}
where, $c_i^*$ refers to the complex conjugate of $c_i$. 

\begin{figure*}
\centerline{
\includegraphics[width=3.5in]{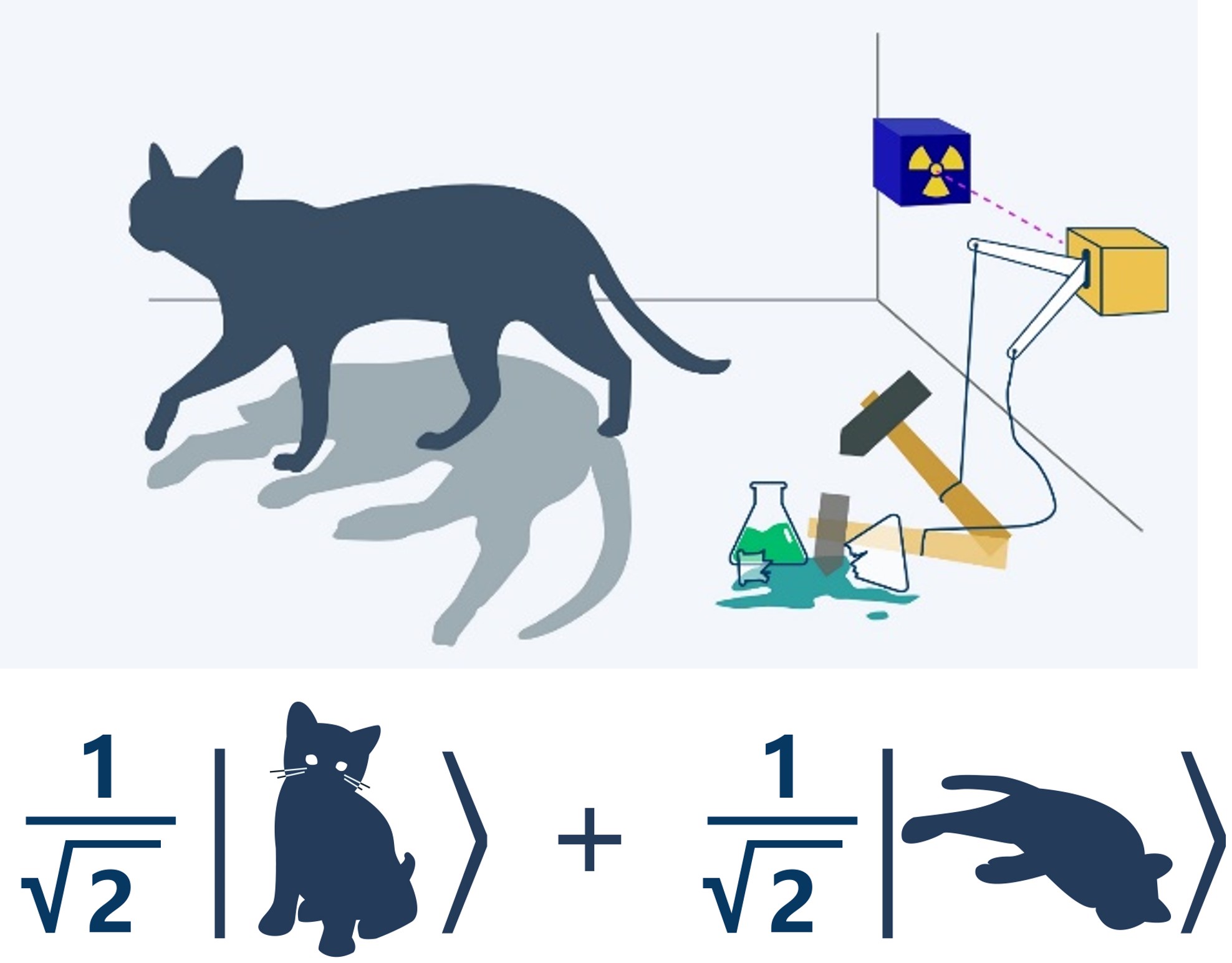}
}
\caption{Schrödinger's cat thought experiment: A cat is put in a closed box that releases poison gas, and there is also a radioactive source in the box. If the radioactive source decays and the emitted radiation is detected by the internal detector, the poison gas switch is then turned on, and the cat will be poisoned to death; if the radioactive source does not decay, the poison gas switch will not be turned on, and the cat will not be poisoned to death. The radioactive source is in the superposition of decay and non-decay, and the cat is in the superposition of life and death. }
\label{s1f2}
\end{figure*}

Equation (\ref{4.1}) has its special physical meaning, i.e. quantum superposition. Different from the classical world where the states are deterministic, $|\psi \rangle$ is in a superposition state of $N$ basis states at the same time, and the probability of being in each state is expressed as $|c_i|^2,\,i=0 \cdots N-1$. As the total probability should be 1, the sum of all $|c_i|^2$ is required to be 1, which is the physical meaning of Eq. (\ref{4.3}).   Quantum superposition is a key property of quantum states and is illustrated in the famous Schrödinger's cat thought experiment (see Fig. \ref{s1f2}).  In this thought experiment, a device is set so that if the atomic nucleus in the device decays, the mechanism will be triggered to give off poisonous gas to poison the cat, while if the atomic nucleus does not decay, the cat will not be dead.  The atomic nucleus is in the superpostion of decay and non-decay, and the cat is in a superposition of being alive and dead.

\subsubsection{Two-level quantum system and qubit}
\begin{figure*}
\begin{center}
\includegraphics[width=5.5in]{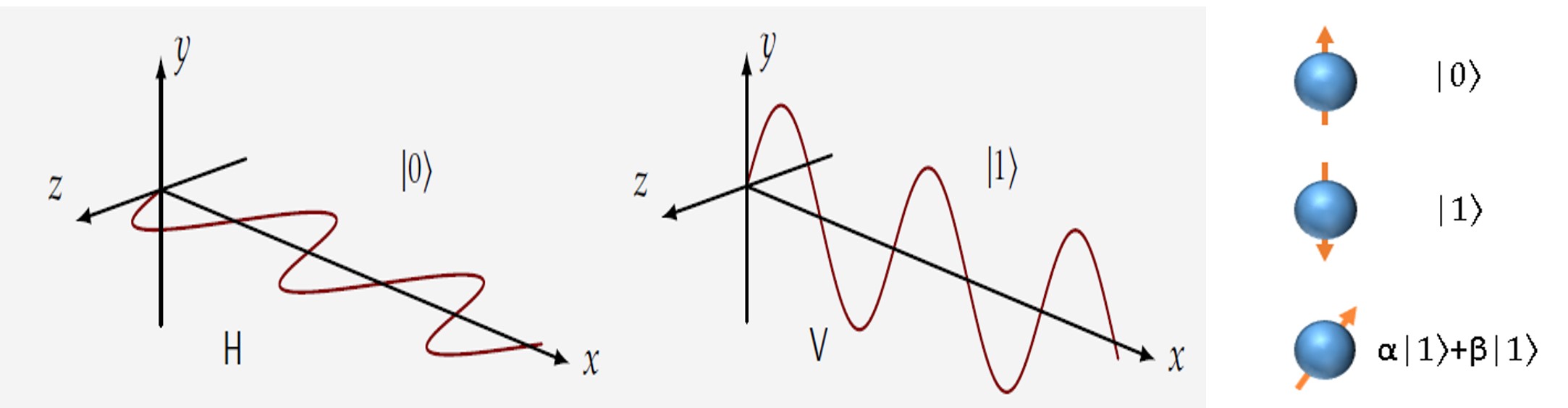}
\end{center}
\caption{A qubit can be either a nucleus with two spin states (left), or a photon with two polarization states (right).}
\label{s1f1}
\end{figure*}

In classical computing, a bit is the basic unit of information and can represent either a 0 or a 1. In quantum computing, the basic unit of information is known as a quantum bit, or qubit. In classical computing, different physical states of a system are used to represent different logical values of the classical bits. For example, a voltage of zero might represent a 0, while a positive voltage represents a 1. These 0s and 1s are known as the computational basis vectors used in classical computation. In quantum computing,  a qubit is represented by a two-level system, such as a photon with two polarization states, an atom with two spin states (see Fig. \ref{s1f1}), or a ground state and an excited state. These two different physical states of the two-level system  form the fundamental basis units of quantum computation,  which are usually denoted as $|0 \rangle$  and $|1 \rangle$. Unlike classical bits, where a state can only be either 0 or 1, qubits can be in a superposition of $|0 \rangle$  and $|1 \rangle$, i.e., $a|0\rangle + b|1\rangle$ (Eq. (\ref{4.1})). When measuring in the 0,1 basis, $|a|^2$ is the probability of obtaining the state 0, and $|b|^2$ is the probability of obtaining the state 1.

For a single qubit, it has two baisc states, in other words, the dimension of its Hilbert space is 2. For a multi-qubit system, the Hilbert space is the tensor product of the state space of each qubit. For example, the base for the two-qubit vector space could be $|0\rangle \otimes |0\rangle$, $|0\rangle \otimes |1\rangle$, $|1\rangle \otimes |0\rangle$, $|1\rangle \otimes |1\rangle$, denoted as $|00\rangle$, $|01\rangle$, $|10\rangle$, $|11\rangle$. A quantum state of two qubits is then a superposition of these basis states.

\subsubsection{Quantum entanglement}

Quantum entanglement is a concept of quantum correlation, which is significantly different from classical correlations. In addition to quantum superposition, quantum entanglement is another exotic property of the quantum system predicted by quantum mechanics. For a multi-qubit system, if the quantum state of this composite system cannot be expressed in a product form of single-qubit states, then the system is in an entangled state. Here we use a two-qubit system as an example. As mentioned in the previous section, its state can be written as
\begin{align}
|\psi\rangle=c_0|00\rangle+c_1|01\rangle+c_2|10\rangle+c_3|11\rangle.\label{2qubit}
\end{align}
If $|\psi\rangle$ can be written as the product form of two single-qubit states
\begin{align}
|\psi\rangle&=|\psi_1\rangle\otimes|\psi_2\rangle=(a_1 |0\rangle+b_1|1\rangle)\otimes(a_2 |0\rangle+b_2|1\rangle)\nonumber\\&=a_1a_2|00\rangle+a_1b_2|01\rangle+b_1a_2|10\rangle+b_1b_2|11\rangle,\label{2qubitproduct}
\end{align}
then, it is not an entangled states. Otherwise, $|\psi\rangle$ is an entangled state. Some examples of unentangled states of $|\psi\rangle$ are $|00\rangle$, $1/\sqrt{2}(|00\rangle+|01\rangle)$, etc. Some examples of entangled states of $|\psi\rangle$ are $1/\sqrt{2}(|01\rangle+|10\rangle)$, $1/\sqrt{2}(|00\rangle+|11\rangle)$, etc.

Quantum entanglement can also be illustrated in the famous Schrödinger's cat thought experiment (Fig. \ref{s1f2}). We can consider this cat and atomic nucleus each as a qubit, respectively. The life or death of the cat is correlated to the decay or not of atomic nucleus. In other words, the cat's state is entangled with the  atomic nucleus's state.
However, why such cat that is dead and alive simultaneously is not seen in our daily life? This is because what we see in daily life are generally macroscopic objects, whose superposition and entanglement states are very easy to be destroyed. Therefore, microscopic objects were used at the very beginning to study quantum entanglement, such as photons and electrons, etc.

\subsubsection{Bloch sphere}
Any quantum state of a qubit can be represented as
\begin{equation}
|\psi\rangle = e^{i\gamma} \left(\cos \frac{\theta}{2} |0\rangle + e^{i\varphi} \sin \frac{\theta}{2} |1\rangle\right), \label{1.1}
\end{equation}
The global phase $e^{i\gamma}$ has no effect on any observation of the system. So we rewrite Eq. (\ref{1.1}) as:
\begin{equation}
|\psi\rangle = \cos \frac{\theta}{2} |0\rangle + e^{i\varphi} \sin \frac{\theta}{2} |1\rangle = \begin{pmatrix} \cos \frac{\theta}{2} \\\ e^{i\varphi} \sin \frac{\theta}{2} \end{pmatrix}. \label{1.2}
\end{equation}
Although the global phase can be neglected, the relative phase $\varphi$  is a very important quantity of the qubit state. $|\psi\rangle$ can be represented by a unit vector on the Bloch sphere (see below), with coordinates $(\sin \theta \cos \varphi, \sin \theta \sin \varphi, \cos \theta)$. Note that $\theta$ and $\varphi$ are two angles that determine the direction of the unit vector.  It can be easily recognized that the unit vector that points in positive z direction represents $ |0\rangle$ state, that in negative z direction represents $ |1\rangle$ state, that in positive x direction represents $ (|0\rangle+ |1\rangle)/\sqrt{2}$, and that in positive y direction represents $ (|0\rangle+ i|1\rangle)/\sqrt{2}$. The difference between $ (|0\rangle+ |1\rangle)/\sqrt{2}$ and $ (|0\rangle+ i|1\rangle)/\sqrt{2}$ is in the relative phase ($\varphi$) and the difference is 90°.

\begin{figure*}
\centerline{
\includegraphics[width=2in]{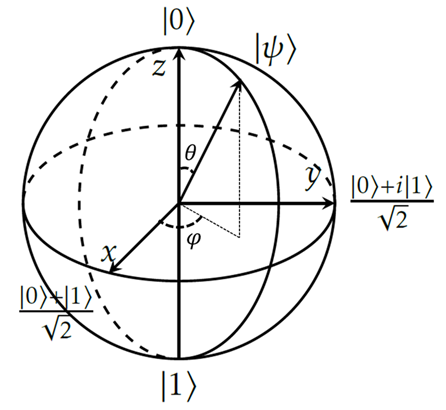}
}
\caption{ Bloch Sphere. The two poles of the Bloch sphere represent $ |0\rangle$ and $ |1\rangle$ states. Any unit vector on the x-y plane is an equal-probability superposition of $ |0\rangle$ and $ |1\rangle$. Unit vectors along any other direction represents superposition of $ |0\rangle$ and $ |1\rangle$ with unequal probabilities.}
\label{s1f3}
\end{figure*}

Any vector inside the Bloch sphere represents a mixed quantum state, which is a mixture of certain quantum states with some probability distribution (discussed in the next section). The Bloch sphere is a powerful tool used to analyze and visualize single qubit operations and relaxation processes.

\subsubsection{Density matrix, pure and mixed states}
Previously, we discussed the concept of quantum states, which are vectors in some complex vector space. These vectors are also referred to as "pure quantum states." In contrast, there is the concept of "mixed quantum states." To introduce the concept of mixed quantum states, let's first discuss the density matrix.

First, let us denote a (pure) quantum state by $|\psi\rangle = \begin{pmatrix} a \ b \end{pmatrix}^{T}$. The density matrix of $|\psi\rangle$ is written as $|\psi\rangle\langle\psi|$, where $\langle\psi| = (a^*, b^*)$ is the conjugate transpose of $|\psi\rangle$. In matrix form, the density matrix is given by:
\begin{align}
\rho =& |\psi\rangle\langle\psi|\nonumber\\ = &|a|^2 |0\rangle\langle 0| + |b|^2 |1\rangle\langle1| + ab^* |0\rangle\langle1| + a^*b |1\rangle\langle 0| \nonumber\\   =& \begin{pmatrix} |a|^2 & ab^* \\ a^*b & |b|^2 \end{pmatrix}, \label{1.3}
\end{align}
The diagonal elements of the density matrix represent the probabilities of $|0\rangle$ and $|1\rangle$ when measuring the system in the $|0\rangle$, $|1\rangle$ basis. The off-diagonal elements of the density matrix are called the coherent elements.

The density matrix can also be used to represent a mixed quantum state, which is a probabilistic mixture of pure quantum states. It can be used to describe the statistical properties of a quantum system, even when the exact state of the system is not known. 
In general, a density matrix is defined as
\begin{equation}
\rho = \sum_{i} p_i |\varphi_i\rangle\langle\varphi_i|, \label{1.4}
\end{equation}
\begin{equation}
p_i \geq 0, \quad \sum_{i} p_i = 1 \label{1.5}
\end{equation}
Note that $\rho$ is Hermitian, meaning that its conjugate transpose is equal to itself. The trace of $\rho$ (the sum of its diagonal elements) is 1, which corresponds to the fact that the probability must add up to 1 when measuring the quantum state in the diagonal basis. If the trace of the square of $\rho$ is also 1, then $\rho$ must be a pure state, which has one non-zero $p_i$ and can be represented by a state vector of the form $|\psi\rangle\langle\psi|$ for some $|\psi\rangle$. If the trace of the square of $\rho$ is less than 1, then $\rho$ represents a mixed state, which cannot be represented by a state vector. In the physical world, mixed states are more common because qubits interact with their environment, causing the off-diagonal elements of the density matrix to decay to zero (a process called decoherence). After decoherence, a pure quantum state becomes a mixed state. For example, in Eq. (\ref{1.3}), after decoherence the state becomes $|a|^2 |0\rangle\langle0| + |b|^2 |1\rangle\langle1|$, with its square as $|a|^4 |0\rangle\langle0| + |b|^4 |1\rangle\langle1|$ and the corresponding trace as $|a|^4 + |b|^4 \leq |a|^2 + |b|^2 = 1$ (the equality holds if and only if one of $a$ and $b$ is 0). That is, unless one of $a$ and $b$ is 0, the state in Eq. (\ref{1.3}) after decoherence is always a mixed state.

\subsection{Quantum measurement}

\begin{figure*}
\centerline{
\includegraphics[width=3.5in]{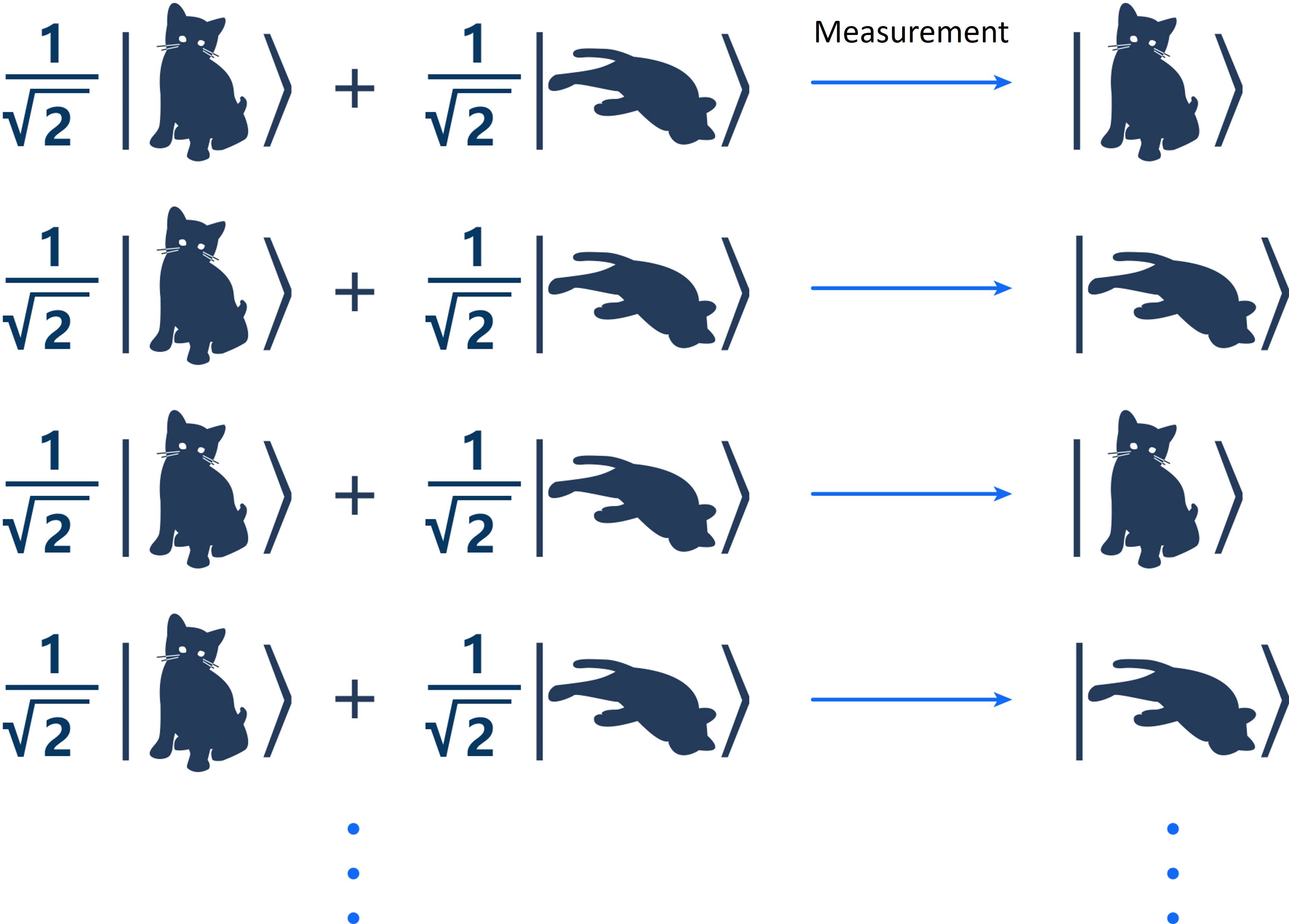}
}
\caption{ Consider the situation that there are many cats in the same state, that is, a superposition of dead and alive, then if any cat is observed, the result is either dead or alive. If all cats are observed, the probability of observing dead and alive is determined by the probability amplitude in the superposition state in which the cat was originally located. As shown in the superposition state, the probability of observing dead and alive is 1/2.}
\label{s1f4}
\end{figure*}

Quantum measurements differ from classical measurements in a number of ways. In quantum mechanics, any measurement corresponds to a Hermitian operator, and the eigenstates of this operator can be used as a set of basis vectors in the quantum state space. When a measurement is made on a quantum state, the state of the system will randomly collapse to one of the eigenstates of the measurement operator, and the measurement result will be the corresponding eigenvalue. If the quantum state has multiple copies, different eigenstates will be obtained upon measurement, with the probability of obtaining each eigenstate determined by the original quantum state (see Fig. \ref{s1f4}). This probabilistic nature of quantum measurements is a key feature of quantum mechanics. 

As an example, consider the spin angular momentum. Quantum systems also have angular momentums, which are quantized and can be some discrete values only. Spin is a type of the angular momentums, and nuclear spin is usually expressed by the symbol $\boldsymbol{I}$. Here we consider the nuclei with 1/2 spin (spin number I=1/2). It can have (2I+1)=2 discrete values along a specified axis. These two discrete values correspond to two states of a spin, i.e. parallel or anti-parallel along a specified axis. Without loss of generality, we denote the two states with spin components parallel and antiparallel to z axis as $|0\rangle$ and $|1\rangle$. Any pure state of the qubit can be expressed as a vector in the state space span by $|0\rangle$ and $|1\rangle$. The matrix expressions of spin operators $I_x$, $I_y$ and $I_z$, which are the x, y and z components of the spin angular momentum of the spin, are as follows:
\begin{equation}
\sigma_x = \begin{pmatrix} 0 & 1 \\ 1 & 0 \end{pmatrix}, \quad \sigma_y = \begin{pmatrix} 0 & -i \\ i & 0 \end{pmatrix}, \quad
\sigma_z = \begin{pmatrix} 1 & 0 \\ 0 & -1 \end{pmatrix} \label{1.6}
\end{equation}
\begin{equation}
I_i = \frac{\hbar}{2} \sigma_i, \quad i = x, y, z \label{1.7}
\end{equation}
The three operators in Eq. (\ref{1.6}) are known as Pauli matrices. For the state $a|0\rangle + b|1\rangle$, if we measure the $z$ component of the spin operators, i.e., $I_z$, with probability $|a|^2$ we will obtain the eigenstate $|0\rangle$ with a measurement result of $\frac{\hbar}{2}$, and with probability $|b|^2$ we will obtain the eigenstate $|1\rangle$ with a measurement result of $-\frac{\hbar}{2}$. Note that both $|0\rangle$ and $|1\rangle$ are eigenstates of $I_z$, with corresponding eigenvalues of $\frac{\hbar}{2}$ and $-\frac{\hbar}{2}$. If we have many copies of the state $a|0\rangle + b|1\rangle$, then measuring $I_z$ many times will return the expectation value:
\begin{equation}
\langle I_z \rangle = |a|^2 \frac{\hbar}{2} - |b|^2 \frac{\hbar}{2} = \mathrm{Tr}(I_z \rho) \label{1.8}
\end{equation}
Equation (\ref{1.8}) shows that the expectation value of a measurement (denoted by $\langle \rangle$) can be obtained as the trace of the product operator that is obtained by multiplying the density matrix and the measurement operator. This relationship between the expectation value of a measurement and the trace of the product of the density matrix and the measurement operator is known as the quantum mechanical trace rule. It is a fundamental principle of quantum mechanics that allows us to predict the statistical properties of quantum systems, even when the exact state of the system is not known.

\subsubsection{Quantum tomography }
Earlier we mentioned the Pauli matrices, which are the matrix representations of the $x$, $y$, and $z$ components of the spin operator. It can be shown that any density matrix of a single qubit can be written as a superposition of the Pauli matrices and the identity matrix:
\begin{equation}
\rho = \frac{1}{2} I + \frac{1}{2} \langle \sigma_x \rangle \sigma_x + \frac{1}{2} \langle \sigma_y \rangle \sigma_y + \frac{1}{2} \langle \sigma_z \rangle \sigma_z, \label{1.9}
\end{equation}
\begin{equation}
I = \begin{pmatrix} 1 & 0 \\ 0 & 1 \end{pmatrix} \label{1.10}
\end{equation}
\begin{align}
\langle \sigma_x \rangle = \mathrm{Tr}(\rho \sigma_x), &\quad \langle \sigma_y \rangle = \mathrm{Tr}(\rho \sigma_y), \nonumber\\\quad \langle \sigma_z \rangle =& \mathrm{Tr}(\rho \sigma_z). \label{1.11}
\end{align}
Here $I$ is the identity matrix. $\langle \sigma_x \rangle$, $\langle \sigma_y \rangle$, and $\langle \sigma_z \rangle$ are the expectations values of the $x$, $y$, and $z$ components of the Pauli matrices, respectively.

For any single-qubit quantum state, if we measure all three Pauli operators and obtain the corresponding expectation value as given in Eq. (\ref{1.11}), then we can reconstruct the quantum state by Eq. (\ref{1.9}). This process is called "quantum state tomography", that is, to reconstruct the density matrix of the quantum system by measuring a set of operators, to obtain complete information of the quantum states.

The above-mentioned method of quantum tomography for a single qubit can be generalized to a multi-qubit system. For example, for a density matrix of a two-qubit system, we can write the density matrix as:
\begin{equation}
\rho = \frac{1}{4}  I^{\otimes 2} + \frac{1}{4} \sum_{i,j} c_{i,j} \sigma_i^1 \sigma_j^2, \label{1.12}
\end{equation}
\begin{equation}
c_{i,j} = \mathrm{Tr}(\rho \sigma_i^1 \sigma_j^2) \label{1.13}
\end{equation}
Here, $i(j) = 0, x, y, z$ and $(i, j) \neq (0, 0)$, and the upper indices of the Pauli matrices represent the qubit that the Pauli matrix is acting on. $\sigma_0 = I$ is the unit matrix. By measuring all possible combinations of these operators and obtaining the corresponding expectation values, the density matrix can be reconstructed using Eq. (\ref{1.12}).

We remark that for a single-qubit pure state, such as the one given in Eq. (\ref{1.2}), measuring the Pauli operators on the corresponding density matrix will return the following expectation values:
\begin{align}
\langle \sigma_x \rangle = \sin \theta \cos \varphi, &\quad \langle \sigma_y \rangle = \sin \theta \sin \varphi,\nonumber\\ \quad \langle \sigma_z \rangle =& \cos \theta. \label{1.14}
\end{align}
Therefore, for any single-qubit pure state, the corresponding direction on the Bloch sphere, i.e. $(\sin \theta \cos \varphi, \sin \theta \sin \varphi, \cos \theta)$, gives the direction of the spin angular momentum, i.e. $(\langle \sigma_x \rangle, \langle \sigma_y \rangle, \langle \sigma_z \rangle)$. This result also applies to mixed states, where the direction of the spin angular momentum is given by $(\langle \sigma_x \rangle, \langle \sigma_y \rangle, \langle \sigma_z \rangle)$. However, for mixed states, the length of this vector is less than 1, resulting in a vector inside the Bloch sphere.

\subsection{Quantum state initialization}
In classical computing, the starting state of the bits is usually known. Similarly, quantum computing also begins with a known initial state of qubits. Without loss of generality, it is common to start quantum computing from the all-0 state. For example, in the case of a single qubit, the initial state is $|0\rangle$ and the corresponding density matrix is:
\begin{align}
\rho = |0\rangle \langle 0| = \frac{1}{2} I + \frac{1}{2} \sigma_z.\label{1.15}
\end{align}

The polarization, or $\langle \sigma_z \rangle$, of this initial state is 1. From Eq. (\ref{1.5})  we know that $|\langle \sigma_z \rangle| \leq 1$. The goal of initialization in quantum computing is to increase the purity or polarization of the system, ideally to a value of 1.

Different physical systems use different methods for initialization. For instance, in a liquid-state NMR system, the thermal equilibrium state at room temperature has a very low polarization, making it difficult to achieve "real" initialization. As a result, the initial state used is often a pseudo-pure state \cite{8,9,10}, which will be discussed in later sections. In the diamond color center system, initialization is often achieved through fluorescence, resulting in the bit system being initialized in the ground state $|0\rangle$. In the superconducting qubit system, one method of initialization is to perform a projective measurement on the system. After the measurement, the system is in the $|0\rangle$ or $|1\rangle$ state, and the $|0\rangle$ system is then further processed for quantum computing.

\subsection{Quantum gates}

A quantum gate is an operation that transforms one quantum state into another. Quantum computing performs certain transformations by performing a series of quantum gate operations on quantum states. The evolution of quantum states follows the Schrödinger equation \cite{11},
\begin{equation}
ih \frac{d}{dt} | \psi \rangle = \mathcal{H} | \psi \rangle, \label{1.16}
\end{equation}
which leads to the evolution of the density matrix
\begin{equation}
ih \frac{d}{dt} \rho = [\mathcal{H}, \rho]. \label{1.17}
\end{equation}
By controlling the "parameter" $\mathcal{H}$, the system Hamiltonian, and the evolution time in Eqs. (\ref{1.16}) and (\ref{1.17}), we can implement transformations that map a quantum state $1$ to another quantum state $2$. This is how quantum gates are implemented.

A quantum gate can be represented by a unitary matrix $U$:
\begin{equation}
U = \mathcal{T}\exp \left( \int_{t_1}^{t_2} -\frac{i\mathcal{H}}{h} dt \right) \label{1.18}
\end{equation}
\begin{equation}
U | \psi_1 \rangle = | \psi_2 \rangle \label{1.19}
\end{equation}
\begin{equation}
U \rho_1 U^\dagger = \rho_2. \label{1.20}
\end{equation}
where Eq. (\ref{1.18}) is derived from Eqs. (\ref{1.16}) and (\ref{1.17}). It is worth noting that $U$, also known as the evolution operator, is determined by the Hamiltonian $\mathcal{H}$ and the evolution time $t$. Equation (\ref{1.18}) is a time-ordered exponential, which is a general solution to Eqs. (\ref{1.16}) and (\ref{1.17}) for time-varying Hamiltonians. It works for both cases of $[\mathcal{H}(t_1),\mathcal{H}(t_2)] = 0$ and $[\mathcal{H}(t_1),\mathcal{H}(t_2)] \neq 0$. In the case of $[\mathcal{H}(t_1),\mathcal{H}(t_2)] = 0$, the time-ordered exponential symbol "$\mathcal{T}$" can be omitted. Eqs. (\ref{1.19}) and (\ref{1.20}) show the results of evolution after applying $U$ to the state vector and the density matrix, respectively. For an $n$-qubit system, the evolution operator $U$ is a $2^n \times 2^n$ matrix. As an example, in NMR system, controlling the Hamiltonian involves adjusting the pulse intensity and frequency, and by controlling the pulse length, various quantum gates can be realized.

Some commonly used single-qubit quantum gates include the Hadamard gate (denoted by H), the P($\pi/4$) gate, and the rotation gate $R$. Typical two-qubit gates include the controlled-NOT gate (denoted by $\text{CNOT}$). These gates are represented by the following matrices:
\begin{equation}
\text{H} = \frac{1}{\sqrt{2}} \begin{pmatrix} 1 & 1 \\ 1 & -1 \end{pmatrix}, \quad \text{P}(\frac{\pi}{4}) = \begin{pmatrix} 1 & 0 \\ 0 & e^{i\pi/4} \end{pmatrix}, \label{1.21}
\end{equation}
\begin{equation}
R = \exp \left[-i\frac{\theta}{2}(n_x \sigma_x + n_y \sigma_y + n_z \sigma_z) \right], \label{1.22}
\end{equation}
\begin{equation}
\text{CNOT} = \begin{pmatrix} 1 & 0 & 0 & 0\\0&1&0&0\\0&0&0&1\\0&0&1&0\end{pmatrix}. \label{1.23}
\end{equation}

The H gate transforms the basis states $|0\rangle$ and $|1\rangle$ into an equal probability superposition of the two basis states. The P($\pi/4$) gate will not change the distribution probability of the quantum state on $|0\rangle$ and $|1\rangle$. It only changes the phase of the state $|1\rangle$ in a superposition of $|0\rangle$ and $|1\rangle$, and increase the phase by $\pi$/4. The rotation gate $R$ performs a rotation of the vector on the Bloch sphere, with $(n_x, n_y, n_z)$ as the rotation axis and $\theta$ as the rotation angle.

\begin{figure*}
\centerline{
\includegraphics[width=3.5in]{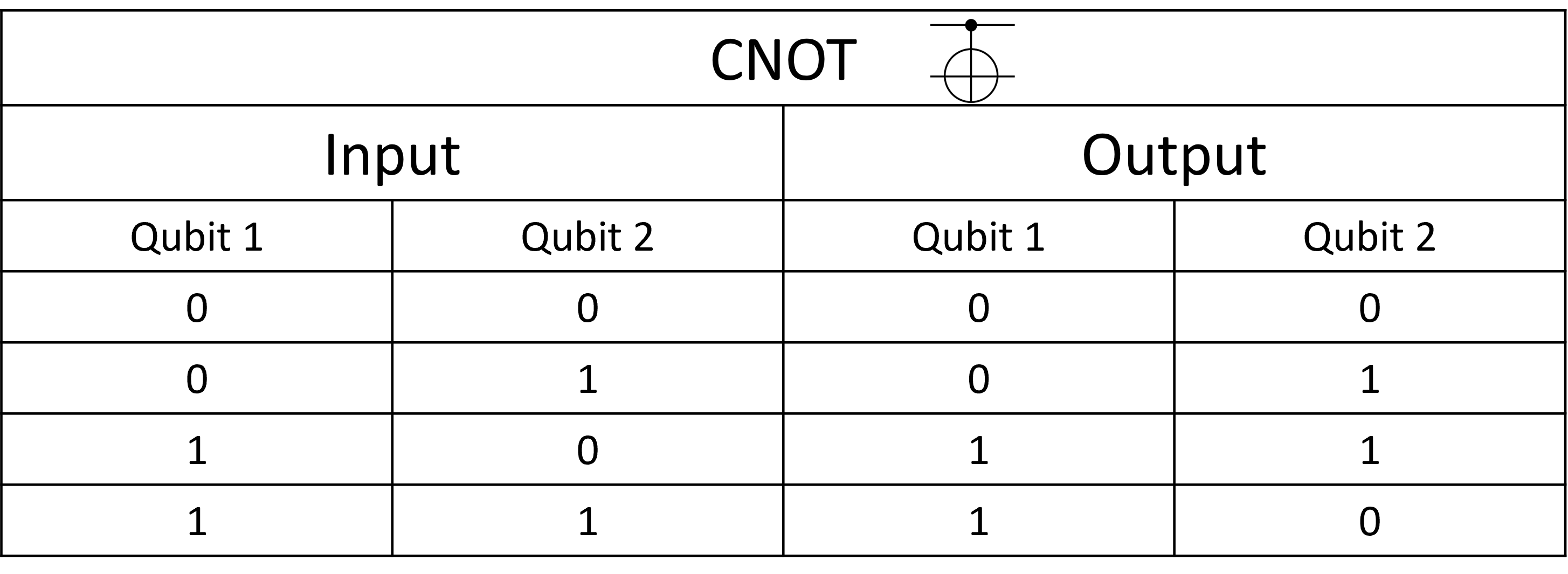}
}
\caption{The output states of CNOT gate, for  $|00\rangle$, $|01\rangle$, $|10\rangle$, and $|11\rangle$ as the initial state. The representative symbol of CNOT gates in the quantum circuit diagram is also given.}
\label{s1t1}
\end{figure*}

The $\text{CNOT}$ gate, with the first qubit as the control bit, transforms the four two-qubit basis states $|00\rangle$, $|01\rangle$, $|10\rangle$, and $|11\rangle$ as follows:
\begin{align}
\begin{pmatrix} \begin{matrix} 1 & 0 \\ 0 & 1 \end{matrix} & \begin{matrix} 0 & 0 \\ 0 & 0 \end{matrix} \\ \begin{matrix} 0 & 0 \\ 0 & 0 \end{matrix} & \begin{matrix} 0 & 1 \\ 1 & 0 \end{matrix} \end{pmatrix} \begin{pmatrix}1\\0\\0\\0\end{pmatrix}=\begin{pmatrix}1\\0\\0\\0\end{pmatrix},
\begin{pmatrix} \begin{matrix} 1 & 0 \\ 0 & 1 \end{matrix} & \begin{matrix} 0 & 0 \\ 0 & 0 \end{matrix} \\ \begin{matrix} 0 & 0 \\ 0 & 0 \end{matrix} & \begin{matrix} 0 & 1 \\ 1 & 0 \end{matrix} \end{pmatrix} \begin{pmatrix}0\\1\\0\\0\end{pmatrix}=\begin{pmatrix}0\\1\\0\\0\end{pmatrix}\label{1.24}\\
\begin{pmatrix} \begin{matrix} 1 & 0 \\ 0 & 1 \end{matrix} & \begin{matrix} 0 & 0 \\ 0 & 0 \end{matrix} \\ \begin{matrix} 0 & 0 \\ 0 & 0 \end{matrix} & \begin{matrix} 0 & 1 \\ 1 & 0 \end{matrix} \end{pmatrix} \begin{pmatrix}0\\0\\1\\0\end{pmatrix}=\begin{pmatrix}0\\0\\0\\1\end{pmatrix},
\begin{pmatrix} \begin{matrix} 1 & 0 \\ 0 & 1 \end{matrix} & \begin{matrix} 0 & 0 \\ 0 & 0 \end{matrix} \\ \begin{matrix} 0 & 0 \\ 0 & 0 \end{matrix} & \begin{matrix} 0 & 1 \\ 1 & 0 \end{matrix} \end{pmatrix} \begin{pmatrix}0\\0\\0\\1\end{pmatrix}=\begin{pmatrix}0\\0\\1\\0\end{pmatrix}\label{1.25}
\end{align}
In other words, the $\text{CNOT}$ gate does nothing if the first qubit is in the $0$ state and flips the second qubit if the first qubit is in the $1$ state. This can be summarized in Fig. \ref{s1t1}.

\begin{figure*}
\centerline{
\includegraphics[width=3.5in]{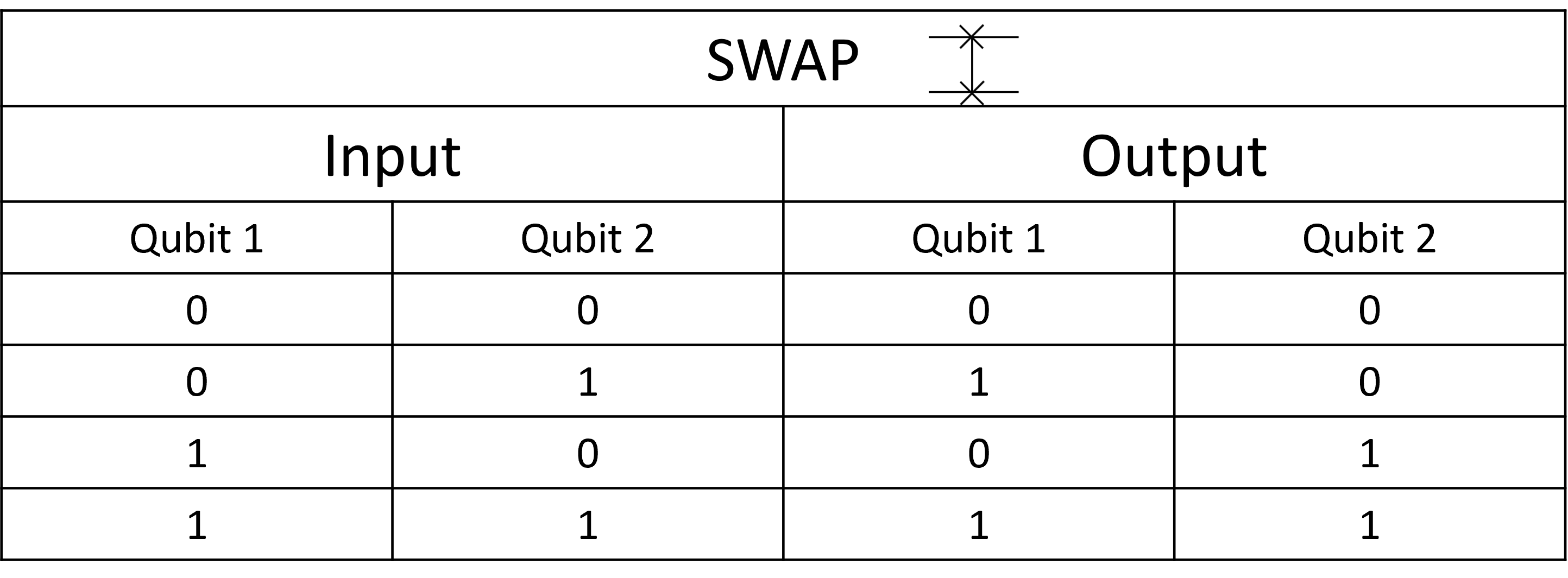}
}
\caption{The output states of SWAP gate, for  $|00\rangle$, $|01\rangle$, $|10\rangle$, and $|11\rangle$ as the initial state. The representative symbol of SWAP gates in the quantum circuit diagram is also given. }
\label{s1t2}
\end{figure*}

The controlled-NOT gate (CNOT) is important because any multi-qubit gate can be decomposed into a combination of CNOT gates and single-qubit gates. \cite{12} As an example, consider the quantum state swap gate (SWAP gate), which swaps the states of two bits. Figure \ref{s1t2} shows the relationship between the input and output states of the SWAP gate.

\begin{figure*}
\centerline{
\includegraphics[width=3in]{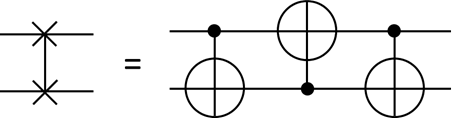}
}
\caption{Implementing SWAP gate using three CNOT gates, the 2nd CNOT gate has a different controlled bit from the 1st and the 3rd CNOT gates.}
\label{s1f5}
\end{figure*}

We can implement the SWAP gate using three CNOT gates, with the second CNOT gate having a different controlled bit from the first and third CNOT gates, as shown in Fig. \ref{s1f5}
.

\begin{figure*}
\centerline{
\includegraphics[width=5.5in]{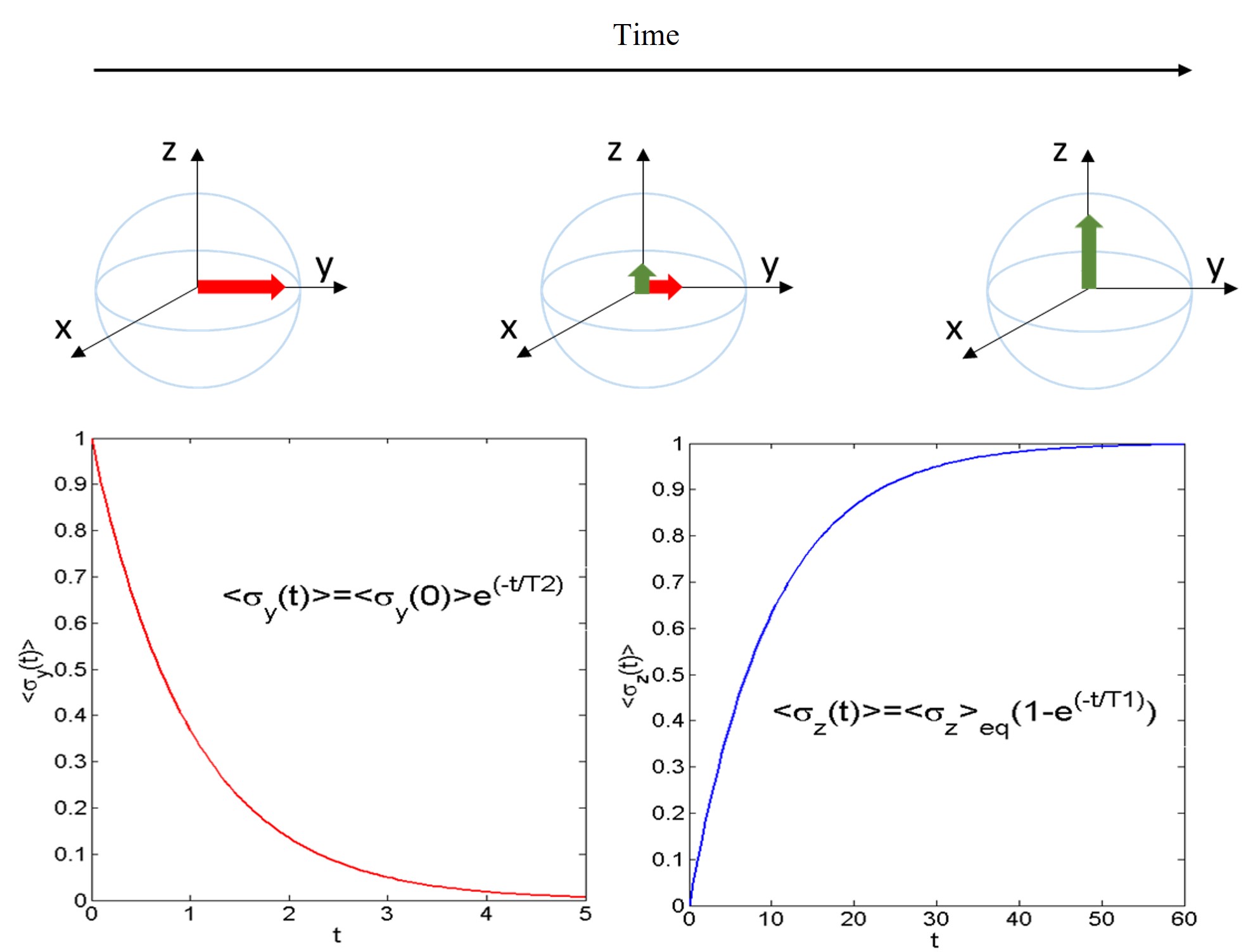}
}
\caption{Transverse and longitudinal relaxation processes of a single-spin system. The red vector in the Bloch sphere represents the transverse polarization of the spin, and the green vector represents the longitudinal polarization of the spin. As shown in the figure, assuming that only the transverse polarization exists at the initial moment, as time goes by, the transverse relaxation causes the transverse polarization gradually to become 0, and the longitudinal relaxation causes the longitudinal polarization gradually to recover from 0 to the thermal equilibrium state value. The two graphs here show the variation of transverse and longitudinal polarization. It is assumed here that the system is at the absolute zero temperature, then the longitudinal polarization can finally achieve a maximum value of 1, which is to say, the system is finally in the ground state $|0\rangle$. In this case, the longitudinal relaxation process can be used to initialize the system.}
\label{s1f6}
\end{figure*}

\subsection{Lifetime of quantum bits}

In an ideal scenario where a qubit is completely isolated from the environment, if spontaneous emission is not considered, its state is completely determined by the initial state, Hamiltonian, and evolution time. If the Hamiltonian is the identity matrix, the state of the qubit will remain unchanged. However, in practice, qubits cannot be completely isolated from the environment, so the state of a qubit will always change due to its interaction with the environment and any quantum state of the qubit must have a lifetime. This change is usually a relaxation process determined by the environment. There are two common types of relaxation: transverse and longitudinal.

As an example, consider a single-qubit NMR system. Suppose the static magnetic field is along the z direction, and in the thermal equilibrium state, the nuclear spins are arranged along the direction of the magnetic field due to the interaction with the static magnetic field. The probability distribution of the different energy levels of the qubit follows the Boltzmann distribution. In this case, the expectation value of the z component of the spin operator is nonzero, but the x and y components are zero, i.e., $\langle \sigma_z \rangle \neq 0$, $\langle \sigma_x \rangle = 0$, $\langle \sigma_y \rangle = 0$. If we apply pulses to the system and rotate the spin operator to the x-y plane, we have $\langle \sigma_z \rangle = 0$, $\langle \sigma_x \rangle \neq 0$, $\langle \sigma_y \rangle \neq 0$. In this situation, transverse relaxation will result in both $\langle \sigma_x \rangle$ and $\langle \sigma_y \rangle$ approaching 0, while longitudinal relaxation will result in a change in $\langle \sigma_z \rangle$ such that the system approaches the thermal equilibrium state determined by the temperature of the environment. Both relaxations have their characteristic times, denoted by $T_2$ and $T_1$, respectively. $T_2$ is the time it takes for $\langle \sigma_x \rangle$ and $\langle \sigma_y \rangle$ to reduce to $1/e$ of their initial values, and $T_1$ is the time it takes for $\langle \sigma_z \rangle = 0$ to restore to $(1-1/e)$ of its expectation value in the thermal equilibrium state.

Transverse relaxation, also known as decoherence, can be harmful to quantum computing as it often transforms a pure state into a mixed state, leading to the loss of information and calculation errors. On the other hand, longitudinal relaxation is often used to initialize qubits by increasing their polarization to the maximum allowed by the environment, that is, by maximizing $|\langle \sigma_z \rangle|$ to prepare for subsequent quantum computing.

\subsection{Fidelity}
The concept of distance in state space can be useful for expressing the degree of similarity between states. For instance, the closeness between the final state obtained after applying a quantum gate and the theoretically expected target state can be measured to evaluate the actual effect of the quantum gate. One commonly used metric for this purpose is quantum state fidelity \cite{4,13}, which reflects the degree of overlap between two quantum states. If the two states are exactly the same, the degree of overlap is at a maximum value of 1, while if the two states are completely different, such as $|0\rangle$ and $|1\rangle$, the fidelity is at its minimum value of 0. The fidelity between two pure states $|\varphi\rangle$ and $|\phi\rangle$, the fidelity between a pure state $|\varphi\rangle$ and a mixed state $\rho$, and the fidelity between two mixed states $\rho$ and $\sigma$ are given by:
\begin{align}
\mathrm{F}(|\varphi\rangle,|\phi\rangle) = |\langle \varphi | \phi \rangle|^2, \label{1.26} \\
\mathrm{F}(|\varphi\rangle,\rho) = \langle \varphi | \rho | \varphi \rangle, \label{1.27} \\
\mathrm{F}(\sigma,\rho) = (\operatorname{Tr}\sqrt{\sqrt{\sigma}\rho\sqrt{\sigma}})^2. \label{1.28}
\end{align}

Note that Eq. (\ref{1.28}) is the most general definition, and it can be reduced to Eqs. (\ref{1.26}) and (\ref{1.27}) in special cases.

\begin{figure*}
\centerline{
\includegraphics[width=5.5in]{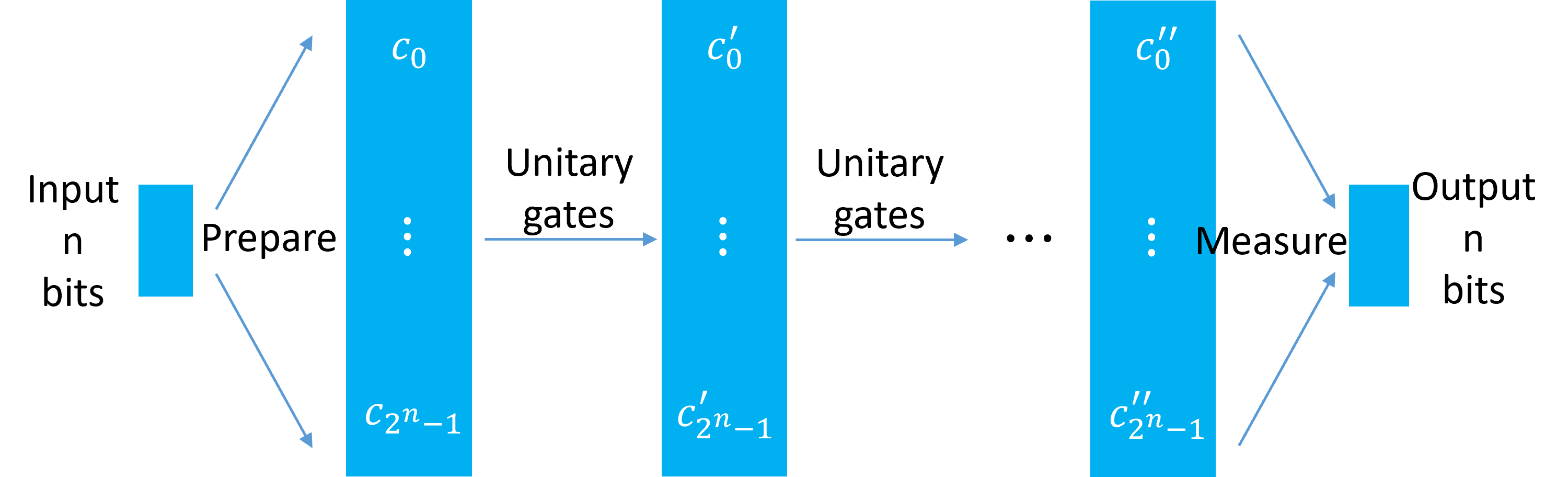}
}
\caption{The general flow of implementing quantum algorithms using a quantum computer. The system is generally initialized in a certain basis vector state. After a series of unitary operations, a superposition state is obtained. Then the system is measured and collapses to a certain basis vector state for the readout results.}
\label{s1f8}
\end{figure*}

\begin{figure*}
\centerline{
\includegraphics[width=5.5in]{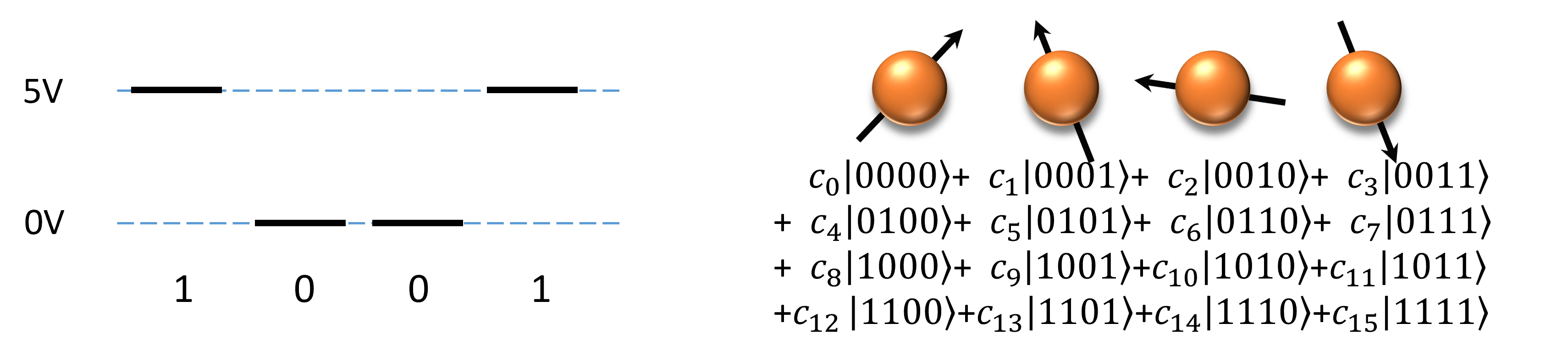}
}
\caption{ State comparison of four classical bits and four qubits. The four classical bits can only be in a specific state, such as the 1001 state shown in the figure. And four qubits can be any superposition of 16 states. }
\label{s1f7}
\end{figure*}

\subsection{Quantum computer}

Quantum computers operate by encoding information in qubits and encoding algorithms into quantum gates. The final state of the quantum computer is measured to obtain the solution to a problem (Fig. \ref{s1f8}). One of the key features of quantum computers is their ability to perform calculations in parallel, which is made possible by the superposition property of qubits. A single qubit can exist in a superposition of two states, while $n$ qubits can exist in a superposition of $2^n$ states. In contrast, classical bits can only be in one of the $2^n$ computational basis vectors at any given time (Fig. \ref{s1f7}). This allows quantum computers to process multiple pieces of information simultaneously, which can lead to significant speedups on certain types of problems, such as factoring large numbers into primes and searching disordered databases.

Quantum computers can also achieve significant speedups over classical computers when it comes to quantum simulation tasks \cite{1,14,15,16}. Systems that are used for quantum simulation are often referred to as quantum simulators. The state space of a microscopic system that follows the laws of quantum mechanics grows exponentially with the size of the system, which means that simulating such a system becomes increasingly difficult as the system size increases. Even with today's most powerful supercomputers, it is not possible to efficiently simulate large-scale quantum systems. In 1982, Feynman \cite{1} proposed using a quantum computer, which is based on the principles of quantum mechanics, to simulate quantum systems in order to overcome the challenge of storing and processing the large amounts of data required for simulating such systems.

\begin{figure*}
\centerline{
\includegraphics[width=3in]{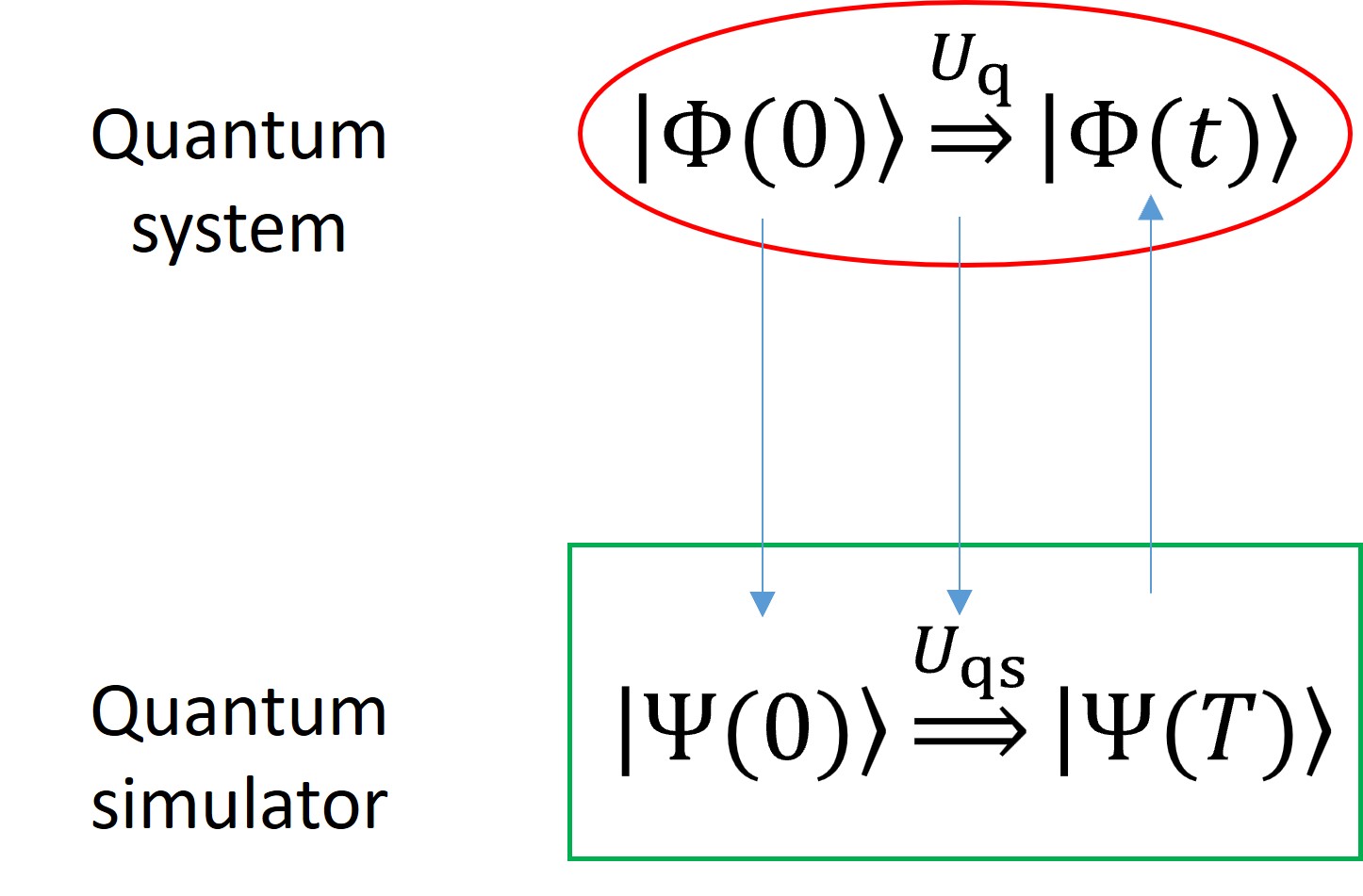}
}
\caption{Schematic diagram of quantum simulation. Prepare an initial state on the quantum simulator, which corresponds to the initial state of the real quantum system, and then use a controllable method to realize the evolution of the quantum simulator to simulate the evolution of the real quantum system, and finally measure the state of the quantum simulator. By doing a quantum simulation like this, useful information about the final state of the simulated system can be obtained. }
\label{s1f9}
\end{figure*}

\subsection{Quantum circuit model}

The quantum circuit model is one of the earliest models of quantum computation \cite{17} and is equivalent to the quantum Turing machine model \cite{18}. It is also the most widely used model for quantum computation. 

\begin{figure*}
\centerline{
\includegraphics[width=4in]{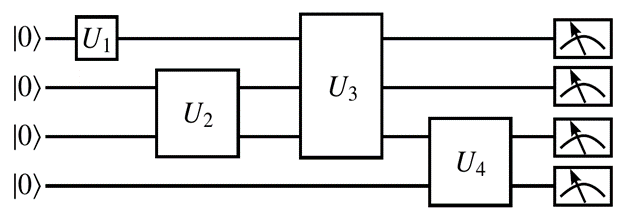}
}
\caption{An example of a four-qubit quantum circuit. Each qubit is represented by a horizontal line. The initial state is $|0000\rangle$. After four quantum gate operations $U_1$, $U_2$, $U_3$  and $U_4$, four single-qubit measurements are then performed to readout the computation result. }
\label{s1f10}
\end{figure*}

For an $n$-qubit system ($n \ge 1$), we can choose the basis of the state vector space as $\{ |i_1 i_2 \cdots i_n \rangle \} (i_k = 0, 1, 1 \le k \le n)$, which is also called the "computational basis". To implement a certain computation, a series of specially designed quantum gates are applied to the $n$-qubit system in sequence. A quantum circuit can be represented visually by a circuit diagram, where qubits are represented by horizontal lines and the evolution of the qubits over time proceeds from left to right. Quantum gate operations are represented by rectangles. In the circuit diagram in Fig. \ref{s1f10}, a four-qubit system is initialized to the state $|\psi_\text{initial}\rangle = |0000\rangle$, which enters the circuit from the far left. It is then followed by the quantum gate operations $U_1$, $U_2$, $U_3$, and $U_4$, leading to the final state $|\psi_\text{final}\rangle$ as the output state. Single-qubit measurements are then performed to read out the computation result.

Recall that any quantum unitary operation can be decomposed into a combination of single-bit gates and CNOT gates. Therefore, in principle, any complex quantum circuit diagram can be realized by single-bit gates and CNOT gates, which can then be implemented experimentally.

\subsection{Quantum algorithm -- an example}

Human beings have a long history of using algorithms to solve problems, such as the Euclidean algorithm in ancient Greece and the $\pi$ algorithm developed by Liu Hui in the Wei$/$Jin dynasty. In the Middle Ages, tools like the abacus and arithmetic were used for calculations. In the early 20th century, mathematical logicians such as Hilbert, Turing, and Gödel helped to formalize the concept of algorithms. In the modern era, the development of electronic computers has led to numerous technological innovations and achievements that rely on the efficient design of algorithms. Examples include the use of genetic algorithms to optimize ammunition loading, the application of information encryption algorithms in network transmission, and the use of parallel algorithms for network transmission and data mining. 

An algorithm is a step-by-step process for completing a task within a finite amount of time. To run an algorithm, a computing machine begins with an initial input and follows a series of well-defined steps. One should be clear about the set of states and rules allowed by the machine in the design and analysis of the algorithm. For example, in ruler and compass drawing, only compasses and rulers are used, and geometric problems must be solved using a limited number of operations. The use of certain tools, such as compasses and rulers, imposes constraints on the rules that can be followed. Both mechanical and electronic computing are based on the laws of classical physics, which dictate the set of allowed states and transformations. However, as we move into the microscopic world, the laws of nature become very different, following the principles of quantum mechanics, allowing a set of state representation and transformation laws that are different from classical physics. Quantum computing, which is based on the laws of quantum mechanics, has the potential to fundamentally change the way we process information and open up new algorithmic possibilities.
\begin{figure*}
\centerline{
\includegraphics[width=5.5in]{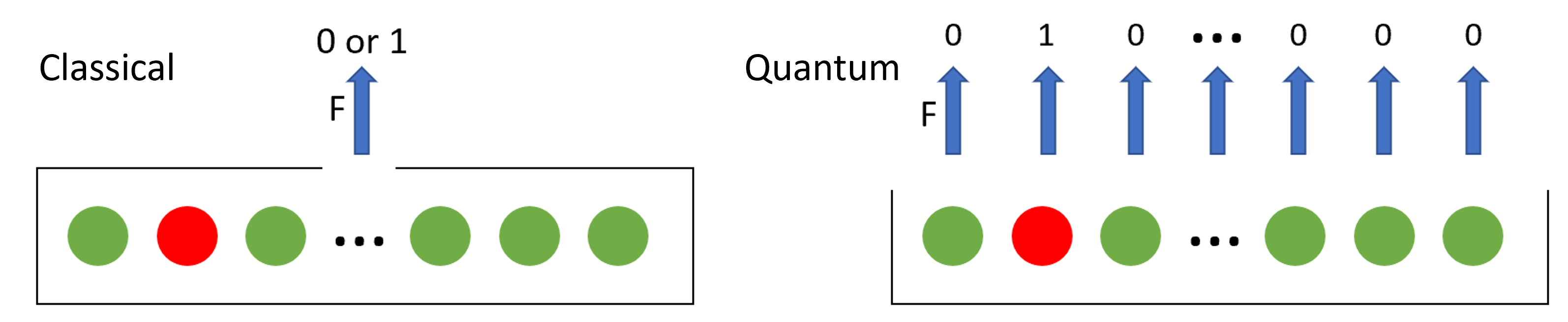}
}
\caption{ When calling the function F$(x)$ to check the color of the balls stored in the database, if the database is unsorted, the classical algorithm needs to check the balls one by one. Using the quantum algorithm, due to superposition of quantum states, F$(x)$ can be called for all $x$  at once to get the information of all the ball colors.}
\label{s1f11}
\end{figure*}
\begin{figure*}
\centerline{
\includegraphics[width=5in]{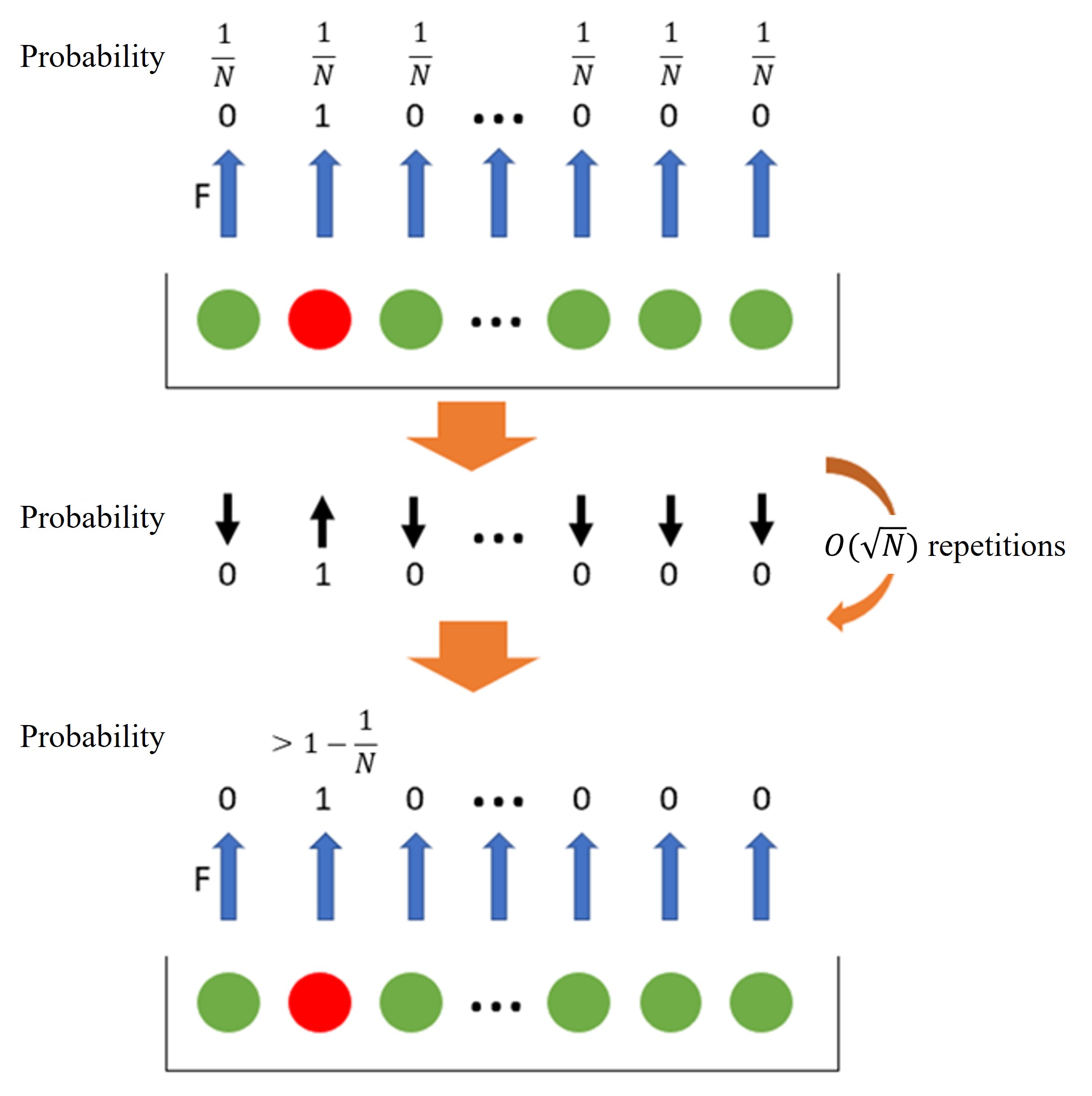}
}
\caption{The color information of all balls is stored in the quantum state with the same probability amplitude. If the quantum state is directly measured, the correct result can only be obtained with probability $1/N$. Therefore, it is necessary to use quantum operations to gradually increase the probability of the quantum state for obtaining the correct result, before measuring the quantum system. After a certain number of repetitions of this quantum operation, the system can be measured to return a correct answer with a probability very close to one.}
\label{s1f12}
\end{figure*}

\subsubsection{Unsorted data base search}
One of the most well-known quantum algorithms is Grover's algorithm, which was proposed in 1995 for the task of searching an unsorted database \cite{19,20}. This is a common problem in data processing, for example, when a teacher wants to retrieve the personal information of a student named Lee from a database containing the personal information of multiple students. On a classical computer, the process would involve comparing the names in the database with "Lee" one by one until the information is found.

To illustrate the search problem in a concrete example, consider a database containing the colors of $N$ balls (labeled 1 to $N$) with different colors. Let a function F be defined such that F($x$) is 0 if the ball is green and 1 if the ball is red. Suppose there is only one ball that is red, so F($x$)= 1 for one of the balls among 1 to $N$. Our goal is to find the ball number $x_0$ such that F($x_0$)= 1. On a classical computer, we would need to look at F($x$) one-by-one for each of the balls from 1 to $N$ (see Fig. \ref{s1f11} Left). In the worst case scenario, we would need to calculate F($x$) $N$ times, leading to a computational complexity of $\mathcal{O}$($N$).

Grover's quantum algorithm can significantly speed up this search process. Due to quantum superposition, data of $N$ elements can be stored in log$_2$$N$ qubits at the same time, and the value of the function F($x$) corresponding to these $N$ elements can be calculated simultaneously (as shown in Figure \ref{s1f11} Right). In other words, in a single round of quantum computing, the color information of all the balls can be obtained at once. This allows for a much faster search process compared to the classical method, which involves looking at each element one by one. 

However, the information about the colors of the balls is stored in the quantum state with equal probability. If the quantum algorithm stops at this point and a measurement is made, the quantum state will randomly collapse into an eigenstate, and there is only a 1$/N$ probability of obtaining the index of the red ball. To increase the probability of obtaining the index of the red ball, we need to perform additional quantum operations on the quantum states containing the color information of all the balls. After $\mathcal{O}$($\sqrt{N}$) operations, we can achieve a correct result with a probability very close to 1 (as shown in Fig. \ref{s1f12}). This allows Grover's quantum algorithm to significantly speed up the search process compared to classical methods, which have a linear complexity of $\mathcal{O}$($N$).

\begin{figure*}
\centerline{
\includegraphics[width=4in]{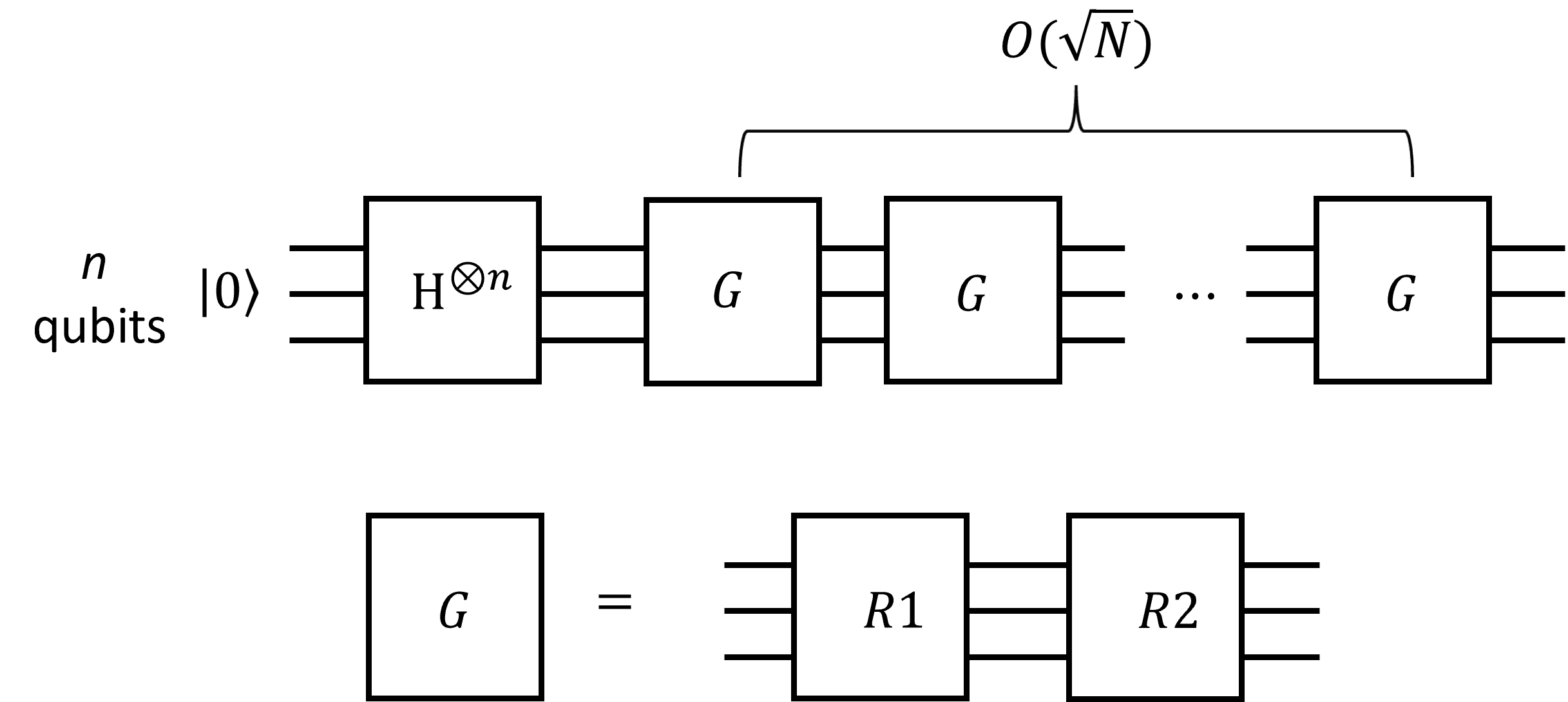}
}
\caption{ Quantum circuit diagram for Grover’s search algorithm. }
\label{s1f13}
\end{figure*}

Figure \ref{s1f13} shows the quantum circuit diagram of Grover's algorithm, which we will explain in more detail in Section 5.

Grover's algorithm reduces the complexity of the classical search from  $\mathcal{O}$($N$) to $\mathcal{O}$($\sqrt{N}$). There are also examples of other quantum algorithms that can exponentially speed up classical algorithms, such as the Shor prime factorization algorithm. Grover's algorithm is significant not only because the unsorted database search problem is an important problem in its own right, but also because it demonstrates the potential for quantum algorithms to solve classically hard problems. In addition, a series of algorithms that utilize the concept of probability amplitude amplification have been derived from Grover's algorithm, which remains an active research topic in quantum algorithm research today.

Grover's search algorithm was first implemented on an NMR system and is the first quantum algorithm to be fully experimentally implemented. It continues to be a benchmark experiment for various quantum computing platforms. 

\section{Quantum computer architecture and physical platforms}
\subsection{Quantum computer architecture}

\begin{figure*}
\centerline{
\includegraphics[width=4in]{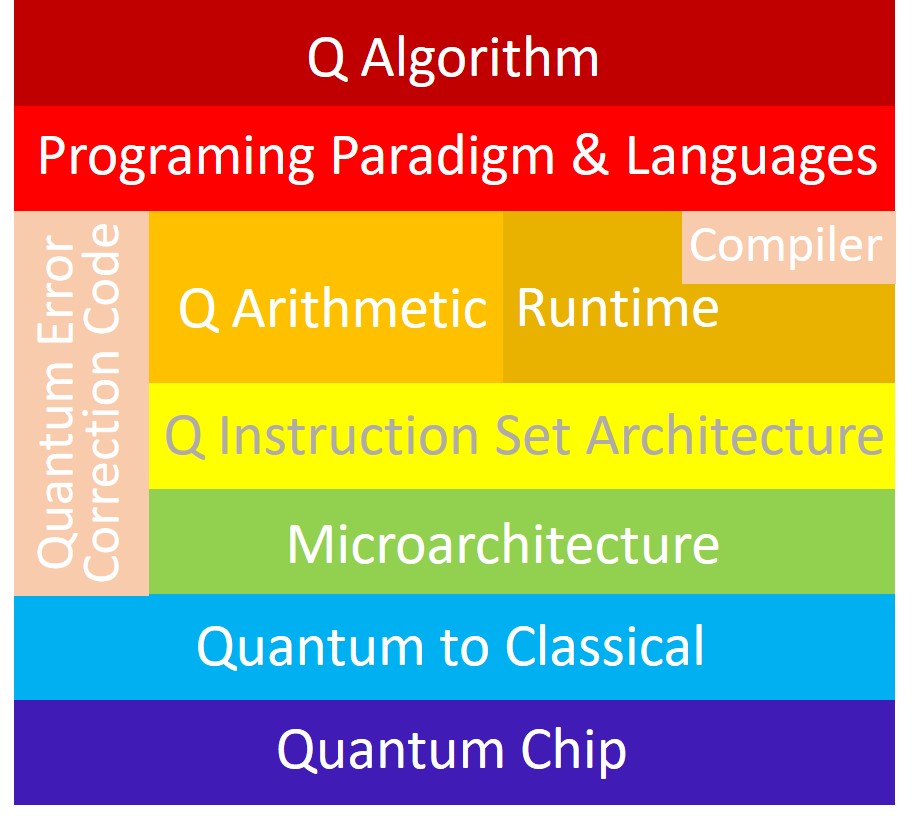}
}
\caption{Quantum Computer Architecture. Reproduced from An Experimental Microarchitecture for a Superconducting Quantum Processor [21]. }
\label{s2f1}
\end{figure*}
In order to develop a fully programmable quantum computer, a multi-layered architecture is necessary, as shown in Fig. \ref{s2f1} \cite{21}. This architecture includes several layers: quantum algorithms; quantum software and programming; quantum compilation and circuit optimization; quantum instruction set and microarchitecture; and quantum computing physical platforms. When a user wants to execute a quantum algorithm, they first describe it using a quantum programming language or software. This description is then passed to a quantum compiler, which optimizes the circuit based on the chosen quantum error-correcting code. The optimized, fault-tolerant quantum circuit is then compiled into a set of instructions in the quantum instruction set language. The microarchitecture system then translates these instructions into control and measurement signals that can be implemented on the quantum chip. This process involves precise timing control, real-time feedback data processing, and optimal quantum control algorithms. The control and measurement signals may be further translated into specific pulses, such as microwave pulses for superconducting systems, before being applied to the quantum chip. This process allows for a complete control chain from the user to the quantum chip, and from classical to quantum.

\subsubsection{Quantum software and quantum compilers}

The rapid advancement in quantum hardware design and manufacturing technology has led to optimistic predictions that a special-purpose quantum computer with hundreds of qubits will be developed within the next 5 years. However, the experience of traditional computer science research and development has shown that quantum software is a critical factor in unlocking the supercomputing power of quantum computers. Quantum software refers to software that can be executed on quantum computers or software that can program quantum algorithms. 

Designing quantum algorithms is a crucial task in harnessing the power of quantum computers. In particular, certain problems that are difficult to solve with classical computing can be significantly faster and more efficient when solved with quantum algorithms, due to their improved time or space complexity. Some notable early quantum algorithms include the Deutsch-Jozsa algorithm \cite{22}, the Bernstein-Vazirani algorithms \cite{23}, and Simon’s algorithm \cite{24}. The discovery of Shor's polynomial-time quantum algorithm for factoring large numbers \cite{25} was a significant milestone, and Grover's algorithm \cite{26} provided a quadratic speedup for the unsorted database search problem. More recently, Harrow, Hassidim, and Lloyd \cite{27} developed a quantum algorithm for solving systems of linear equations in logarithmic time, which represents an exponential improvement over classical algorithms. In the past two decades, many researchers have focused on issues related to the design of quantum algorithms, including quantum walks \cite{28}, element distinctness \cite{29,30}, quantum adiabatic algorithms \cite{31,32}, and quantum algorithms for solving Pell's equations \cite{33}.

Quantum algorithms are fundamentally different from classical algorithms, so quantum programming languages are also distinct from their classical counterparts. Quantum programming languages are essential for implementing quantum algorithms and fully leveraging the benefits of quantum computing. Currently, several quantum programming languages have been developed, including QCL \cite{34}, Q$|$SI$\rangle$ \cite{35}, Q language \cite{36}, Quipper \cite{37}, LIQUi$|\rangle$ \cite{38}, and Q$\#$ \cite{39}.

\begin{figure*}
\centerline{
\includegraphics[width=5.5in]{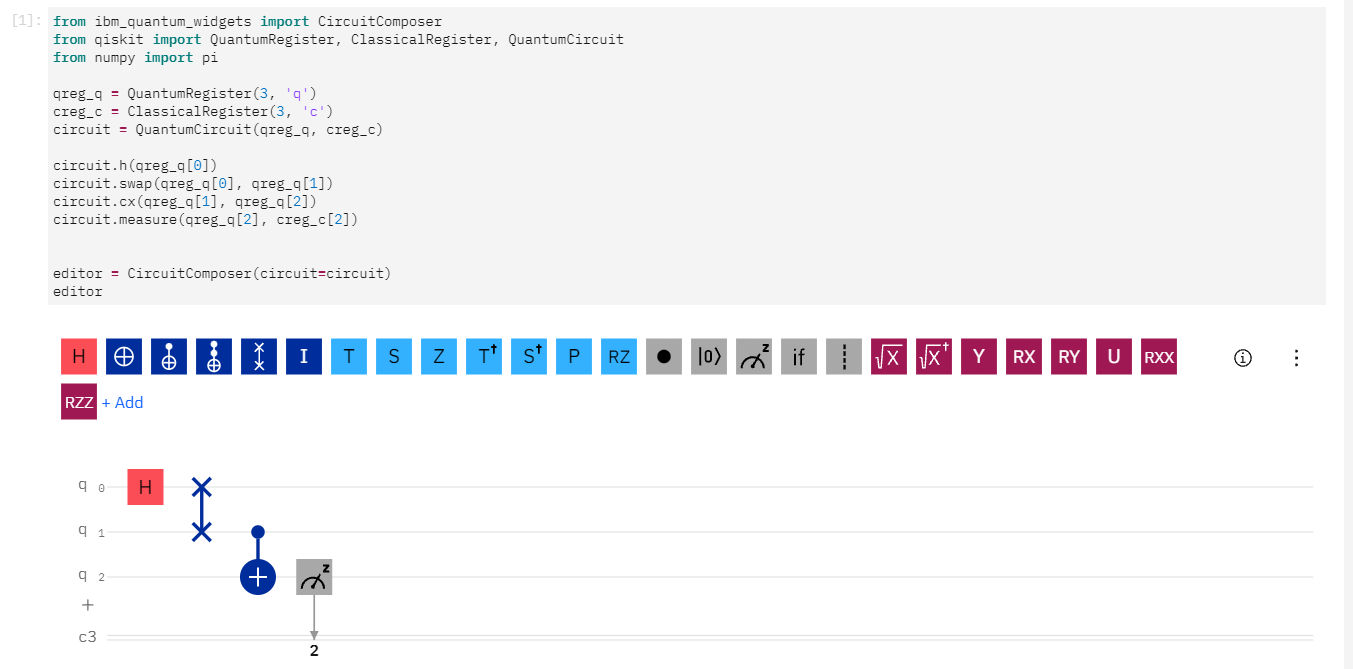}
}
\caption{An example case of using Qiskit to create a quantum circuit. }
\label{s2f2}
\end{figure*}

Several research institutions and companies have also developed quantum programming software, such as Qiskit \cite{40} by IBM, ProjectQ \cite{41} by ETH Zurich, and Forest \cite{42} by Rigetti. Among these different quantum programming softwares, Python is often used to create and edit quantum circuits. Figure \ref{s2f2} shows an example using Qiskit to create a quantum circuit. 

In the early days of quantum software research, much attention was paid to the development of quantum programming languages, while other software aspects of the quantum architecture received less attention. However, in recent years, there has been increasing research on other areas such as quantum software debugging, reuse, and other related fields \cite{43}.

\subsubsection{Quantum compiler and circuit optimization}
Like in classical computing, we need to design both low-level quantum assembly languages that can be run on quantum computers and high-level quantum languages that are suitable for programming and analysis. We also need a quantum compiler that can convert high-level languages into low-level languages \cite{44}. Most of the quantum software mentioned earlier also includes quantum compilation and quantum circuit optimization functions. One goal of quantum circuit optimization is to reduce the number of multi-bit gates. In real quantum systems, not all qubits can interact directly with each other, so implementing an arbitrary two-bit gate often requires using SWAP gates between different qubits. Therefore, the quantum compiler should optimize the circuit according to the topology of the quantum system to minimize the number of SWAP gates and improve the overall performance of quantum computing.

\subsubsection{Quantum instruction set and microarchitecture}

In the development of traditional electronic computers, architecture plays a crucial role. For example, the von Neumann architecture enables the principle of stored programs, which allows for the generalization and automation of computers through the digitization of control functions and the use of instructions and programs. In contrast, the early electronic computer ENIAC did not use the principle of stored programs, requiring manual changes to the circuit connections in the system every time the program was modified. On average, it took about 2 weeks to modify a program on ENIAC. Current research in computer architecture covers all aspects of computer system design and implementation. The 2017 ACM Turing Award was awarded to John L. Hennessy and David A. Patterson, two leading researchers in the field of computer architecture.

Like classical computers, architecture is also crucial for building quantum computers. Quantum instruction sets and microarchitectures \cite{45,46,47} serve as a bridge between quantum software and hardware. The microarchitecture acts as a link between the two, providing an instruction set for the top-level software system and control signals for the underlying physical system. It also runs quantum control algorithms (which vary depending on the physical system being used), generates the required control signals with precise timing, and performs real-time quantum error detection and correction. The instruction set and the control microarchitecture that implements it are responsible for organizing, manipulating, and managing the physical system, hiding the details and differences of the underlying physical system, and enabling the software system to control various physical systems. At the same time, the step-by-step compilation of the control sequence must also be designed and optimized to reduce the memory requirements of the hardware and increase control flexibility. This includes decisions such as how to implement the sequence cycle, how to manage storage, and how to call pulse waveforms. Quantum optimal control algorithms are also incorporated into the microarchitecture to improve performance. For example, an optimization algorithm might be used to design pulses that are robust against environmental noise, or to design pulses that take hardware transfer functions into account. Several institutions and companies have developed quantum instruction sets, such as Quil \cite{42} by Rigetti, Blackbird \cite{48} by Xanadu, OpenQASM \cite{49} by IBM, and eQASM \cite{47} by TUDelft. 

\subsection{Physical platforms for quantum computing}

Before discussing the various platforms for quantum computing, it is useful to review the DiVincenzo criteria \cite{50}, which are considered necessary conditions for constructing a quantum computer.

1. A scalable physical system with well-characterized qubits. These qubits can be two-level systems such as spin-half nuclear spins or photons with two polarizations, or they can be two-level subspaces of high-dimensional systems such as the ground state and the first excited state of an atom. The system must be scalable, meaning that it can support an arbitrary number of qubits.

2. The ability to initialize the state of the qubits to a simple fiducial state. Initialization is necessary for both quantum and classical computers, and is typically the first step in a computation procedure.

3. Long decoherence times that are relevant to the computation. Decoherence times should be much longer than the duration of quantum gates to minimize errors caused by decoherence, or they should be able to be corrected using quantum error correction.

4. A "universal" set of quantum gates. Quantum gates are unitary operations, and a universal gate set should be able to implement any quantum gate.

5. A qubit-specific measurement capability. Measurement is used to obtain the results of the computation and is typically the final step in a computation procedure.

\subsubsection{Superconducting qubit system}

It is clear that superconducting circuits \cite{51,52,53} are the dominant platform for quantum computers. In addition to widespread use in academia, tech giants such as Google, IBM, and Alibaba, as well as emerging companies such as Rigetti, Origin Quantum, and SpinQ Technology, have all invested in developing superconducting quantum computers. This indicates that almost everyone is focusing on superconducting quantum computers. In 2019, Google's superconducting quantum computing team published an article in Nature announcing that "Quantum Supremacy" had been achieved for the first time, which sparked widespread excitement in the quantum computing community \cite{54}. The goal of the "Quantum Advantage" experiment was to identify and implement a computing task that would be beyond the capabilities of the most powerful supercomputers, but could be solved quickly by a quantum computer. The Google team demonstrated an algorithm of a random quantum circuit implemented on a superconducting chip containing 53 qubits, which took only 3 minutes and 20 seconds to run. By comparison, the world's top supercomputer would take years to perform the same calculations. This marked the first time that quantum advantage had been achieved on a fully programmable universal quantum computer, which is a promising development for the future of fault-tolerant quantum computers.

\begin{figure*}
\centerline{
\includegraphics[width=3in]{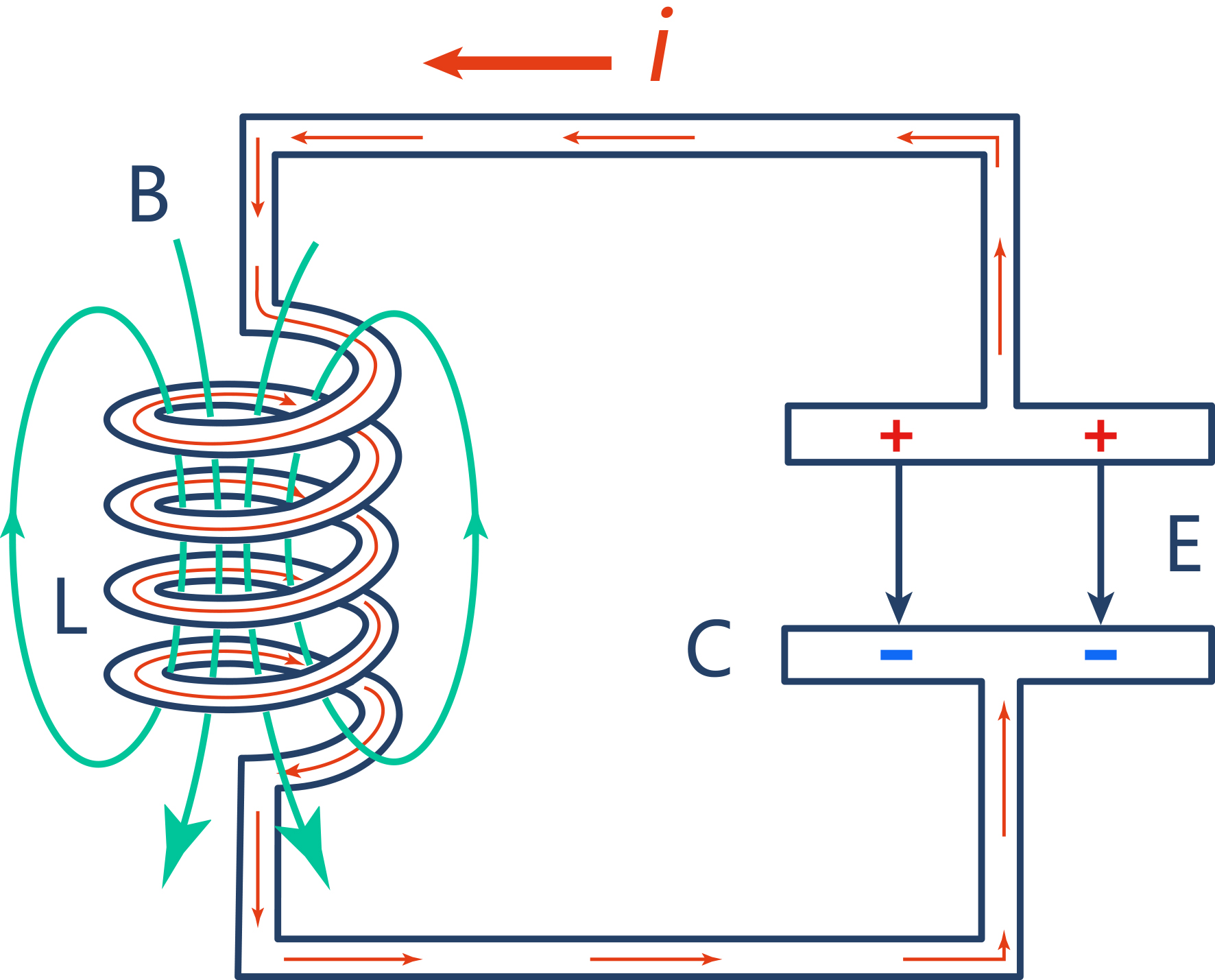}
}
\caption{ Current, electric field and magnetic field in a classical LC circuit.}
\label{s2f3}
\end{figure*}

So, what are the fundamental principles behind superconducting quantum computers? As mentioned earlier, the basic unit of a quantum computer is a quantum bit, which requires a two-level system that can be controlled in some superposition state. Now consider the LC oscillator circuit that you learned about in middle school. In an LC circuit, we know that the direction of the magnetic flux produced in the inductor can be determined using the right-hand rule, as shown in Fig. \ref{s2f3}. If the current is reversed, the direction of the magnetic flux will also reverse. In a classical LC circuit, although the direction of the magnetic flux oscillates periodically, there can only be one definite orientation at any given time.

\begin{figure*}
\centerline{
\includegraphics[width=3in]{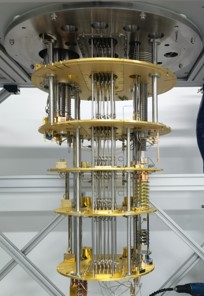}
}
\caption{The ultra-low temperature environment of the superconducting quantum computing system is provided by a dilution refrigerator. The picture shows the inside of the dilution refrigerator.}
\label{s2f4}
\end{figure*}

In the microscopic quantum world, the state of matter is described by a wave function and can be in a superposition state. If we make the LC circuit device small enough and place it in an ultra-low temperature environment to suppress classical thermal noise, it will exhibit some quantum superposition phenomena. For example, the counter-clockwise and clockwise directions of current can exist simultaneously, meaning that the magnetic flux can have two orientations at the same time. LC circuit devices entering the realm of quantum mechanics provide a possibility for realizing qubits that is different from systems such as atoms and photons. Because the LC circuit device is placed in an ultra-low temperature environment  (see Fig. \ref{s2f4}), the entire circuit becomes superconducting, resulting in a resistance of 0 and no heat generation. Therefore, this physical realization of quantum computing is called superconducting circuit quantum computing, or superconducting quantum computing for short. In low-temperature superconductors, electrons can combine to form Cooper pairs. The potential energy of Cooper-pair-formed condensates is a variable with quantum properties that can be changed by macroscopically controlling inductance and capacitance, providing a method for designing and constructing qubits. Likewise, this potential energy can also be rapidly changed by electrical signals, providing a well-established method of quantum control. These devices are similar to classical integrated circuits and can be readily fabricated using existing technologies.

\begin{figure*}
\centerline{
\includegraphics[width=5.5in]{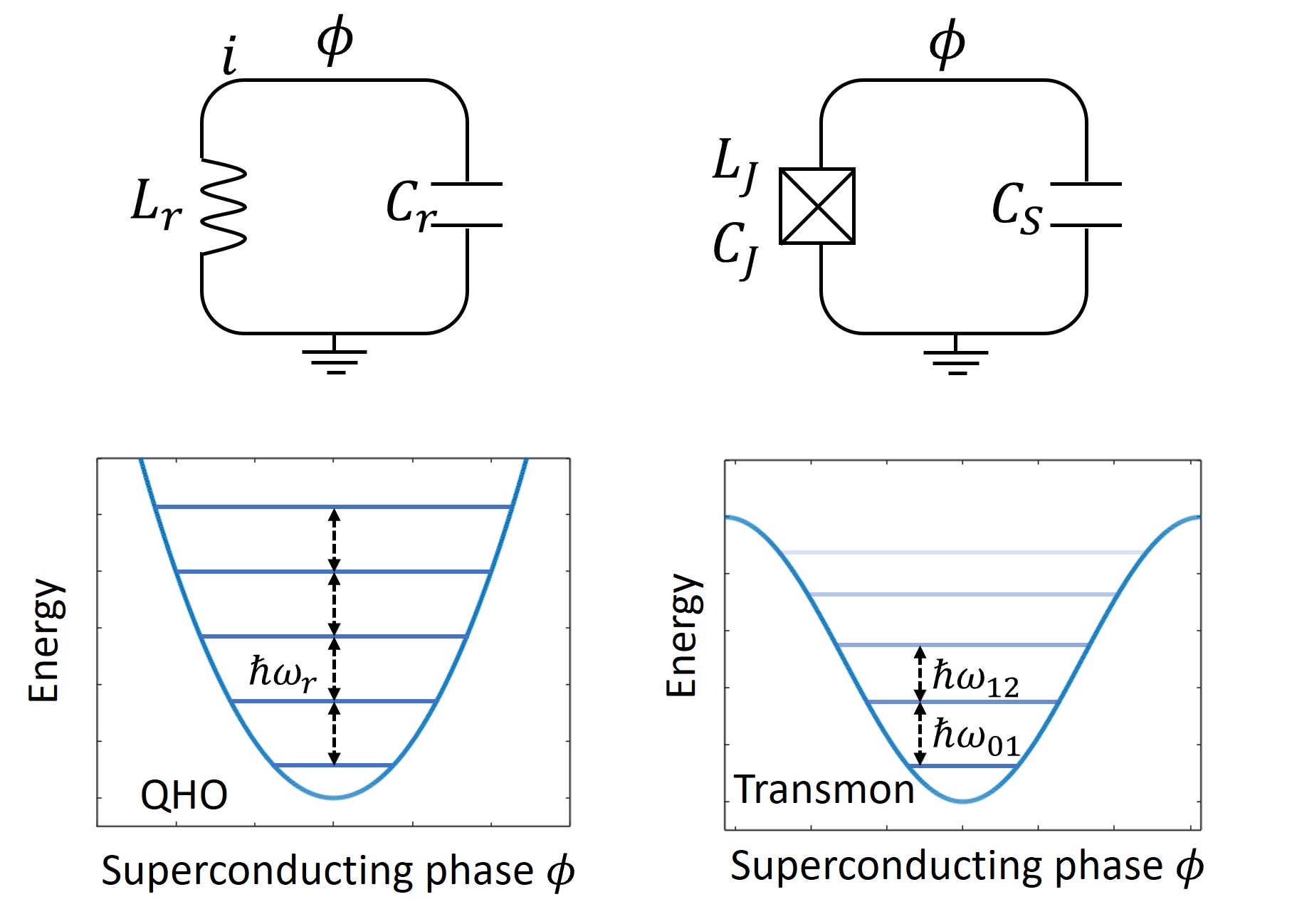}
}
\caption{The figures on the left show the energy level structure of the LC circuit and the simple harmonic quantum oscillator. The figures on the right show the addition of a Josephson junction in the LC circuit, and the system energy level structure after adding this device. It can be seen that the energy level interval is no longer the same, and the lowest two energy levels can be used to make a qubit. Reproduced from Ref. \cite{54ad}.}
\label{s2f5}
\end{figure*}

The physics behind superconducting qubits can be explained through analogy with single-particle quantum mechanics in a potential field. First, a conventional LC oscillator circuit provides a quantum harmonic oscillator. The magnetic flux $\Phi$ passing through the inductor and the charge $Q$ on the capacitive plate satisfy the commutation relation $[ \Phi, Q ] = i \hbar $, which means that $\Phi$ and $Q$ can be analogous to position and momentum in a single-particle quantum system, respectively. The dynamics of the system are determined by the "potential" energy $\Phi^2/2L$ and the "kinetic" energy $Q^2/2C$, leading to the well-known equidistant energy levels of the quantized harmonic oscillator. In other words, the system has an infinite number of energy levels with equal spacing. However, we need only two energy levels for a qubit, which requires anharmonicity. The anharmonicity can be obtained from the Josephson junction, a key component of superconducting qubits. A Josephson junction is a thin insulating layer that separates two superconductors. The quantization of the tunneling charge through the junction brings a cosine function term, with the magnitude of the Josephson energy $E_J$, to the potential energy of a parabolic function such as $\Phi^2/2L$. Ultimately, the two lowest-energy levels of the anharmonic potential form a qubit (Fig. \ref{s2f5}).

In general, the excitation frequency of qubits is designed to be in the range of 5-10 GHz, which is high enough to suppress thermal effects in cryogenic dilution refrigerators (temperature $T\sim10$mK; $k_B T/h\approx0.2$GHz), and microwave technology at this frequency range is also relatively mature. Single-bit gates can be implemented with resonant microwave pulses of 1-10 ns duration, which are delivered to local qubits through microwave transmission lines on the chip. Adjacent qubits are naturally coupled together directly via capacitance or inductance, providing a simple two-qubit quantum logic gate. However, for large-scale quantum computer architectures, we need more tunable coupling schemes. To turn on and off the interaction between qubits, indirect coupling mediated by tunable couplers was developed. At the same time, the application of tunable coupled qubits in adiabatic quantum computing is also being researched.

High-fidelity readout schemes are also being developed. The transition behavior of Josephson junctions at critical currents has been widely used for threshold discrimination of two states \cite{55}. Another promising direction is the demonstration of quantum non-destructive measurements, in which a qubit provides a state-based phase shift for electromagnetic waves in a transmission line. High-fidelity readout of around 92\% as well as quantum non-destructive readout within a dozen nanoseconds was achieved \cite{56}. Currently, readout as fast as around 48 ns with a fidelity as high as 98.25\% has already been realized \cite{57}.

Superconducting quantum computers have made great progress in recent years, with several companies and academic institutions working on their development. Companies that develop superconducting quantum computers mainly include Google, IBM, Alibaba, and startup companies Rigetti, Origin Quantum and SpinQ Technology, etc.; academic institutions include Yale University, University of California Santa Barbara, MIT, University of Science and Technology of China, Tsinghua University, etc.

\subsubsection{Ion trap system}

\begin{figure*}
\centerline{
\includegraphics[width=4in]{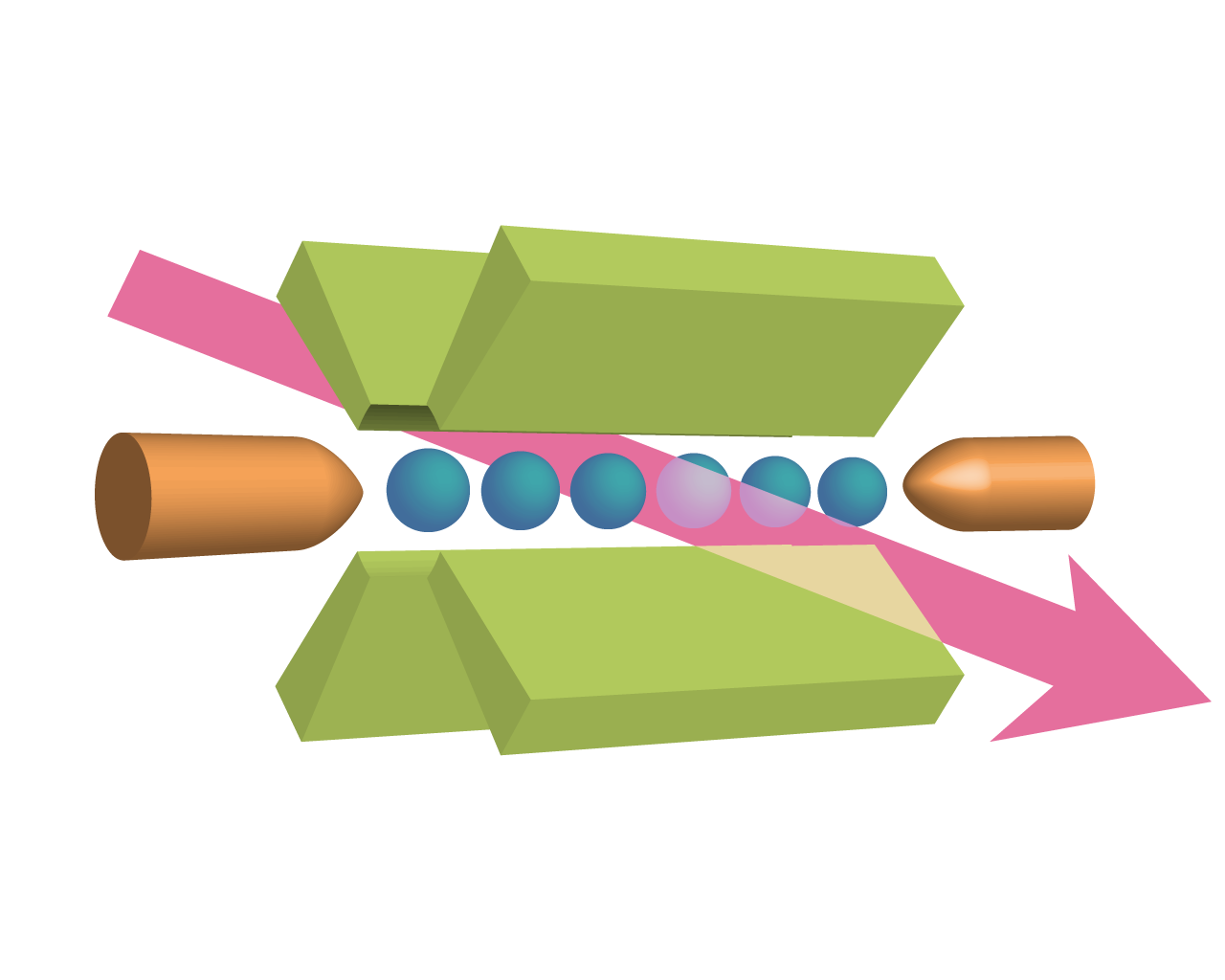}
}
\caption{Ion trap quantum computer. Reproduced from https://www.laserfocusworld.com/test-measurement/test-measurement/article/16550291/photonic-frontiers-quantum-computing-in-pursuit-of-quantum-computing. }
\label{s2f6}
\end{figure*}

Ion trap quantum computing is another promising platform for quantum computing. It involves trapping ions in a potential well created by an electromagnetic field (see Fig. \ref{s2f6}) \cite{58}. For instance, a single atomic ion can be confined in free space with nanometer-scale precision using an electric field and nearby electrodes. Qubits are realized by the different electronic states of ions. The qubits are initialized through optical pumping and laser cooling, and measurements are made using laser-induced fluorescence \cite{59,60}. Single-bit operations are performed using lasers, while two-bit control gates are realized through spin coupling, which is achieved through the collective vibrational modes of the ions \cite{58,61}. One simple realization of an entangling quantum gate was first proposed by Cirac and Zoller in 1995 \cite{58} and experimentally verified later that same year \cite{62}.

As the size of an ion trap system increases, it becomes more difficult to realize entanglement gates with a large number of ions. This is due to a number of challenges, including reduced laser cooling efficiency, increased sensitivity to decoherence caused by noisy electric fields and motion modes, and the potential for dense, stacked motion modes to disrupt quantum gates through mode crosstalk and nonlinearity. One potential solution to these problems is the use of ion trap electrodes to apply tunable electric power and control the movement of individual ions within complex ion trap structures \cite{63,64}. By doing so, it is possible to manipulate a smaller number of ions, potentially enabling the realization of entanglement gates in larger systems.

Another way to increase the number of qubits in ion trap systems is to couple multiple small-scale ion traps through optical interactions \cite{64,65}. This approach has the advantage of providing a long-distance communication channel, and has already been used to achieve macroscopic distance entanglement with atomic ions. This method is similar to probabilistic linear optical quantum computing, but the ion trap system can act as a quantum relay, enabling more efficient long-distance communication. In addition, such a system could scale the protocol of distributed probabilistic quantum computing to larger numbers of qubits.

The ion trap system has the advantage of a qubit coherence time that is much longer than the time required for initialization, multi-qubit control, and measurement. However, the biggest challenge facing ion trap quantum computers in the future is how to extend the high-fidelity operation achieved in few-qubit systems to more complex, multi-qubit systems \cite{64}.

At present, companies building ion trap quantum computers are mainly IonQ, Quantinuum, etc, and academic institutions include the University of Maryland, Duke University, the University of Innsbruck, Tsinghua University, etc.

\subsubsection{Dimond NV color center system}

The diamond Nitrogen-Vacancy (NV) center system is a promising platform for quantum computing and other applications due to its ability to leverage both nuclear and electron spins for quantum operations \cite{66,67}. This system can be controlled using lasers or microwaves and can be initialized by laser pumping. One of the advantages of the diamond NV center system is its ability to operate at room temperature, making it a potentially useful tool in a variety of fields including quantum computing, quantum communication, and quantum precision measurement.

\begin{figure*}
\centerline{
\includegraphics[width=5.5in]{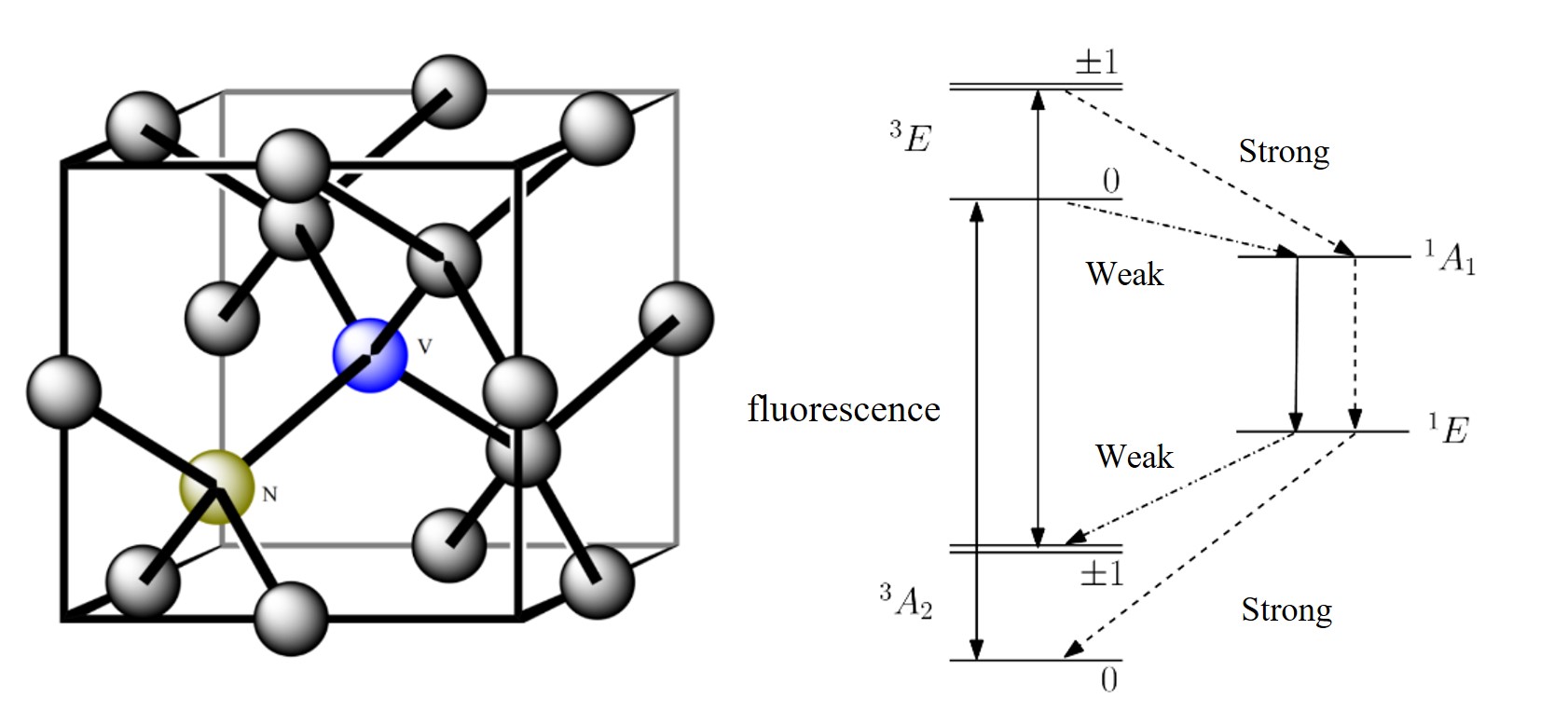}
}
\caption{Left: Structure of NV color center in diamond. Right: NV color center energylevels and fluorescence intensities.}
\label{s2f8}
\end{figure*}

The NV centers found in diamond consist of a vacancy created by replacing a carbon atom with a nitrogen atom and the absence of a carbon atom in an adjacent position (see Fig. \ref{s2f8}) \cite{68}. These NV centers have two charge states: neutral and negatively charged. Much of the current research on NV centers has focused on the negatively charged state, known as the NV$^-$ system, due to its ease of preparation, manipulation, and readout.
For the NV$^-$ center in diamond, the nitrogen atom and three carbon atoms surrounding a vacancy have a total of 5 unbonded electrons. Together with the one captured electron, there are a total of 6 electrons, resulting in an equivalent electron spin of 1. The electronic states of the ground state and excited state of the NV$^-$ center are both spin triplet states. By selecting two of these states, such as $|m_S = 0\rangle$ and $|m_S = -1\rangle$, a qubit can be formed. Even in the absence of an external magnetic field, the triplet states of the ground state of the NV$^-$ center will split into two energy levels, $|m_S = 0\rangle$ and $|m_S = \pm1\rangle$, due to the interaction between the spins. This zero-field splitting has a frequency of 2.87 GHz \cite{68}, which allows the system to be manipulated using microwaves at resonance.
In addition to the electron spin qubits, the $^{14}$N ($^{15}$N) atom of the color center can be regarded as a nuclear spin qutrit (qubit) with a spin of 1 (1/2). Similarly, if there are $^{13}$C atoms present near the NV center, they can also be used as qubits with a spin of 1/2 \cite{66}. This ability to leverage nuclear spin qubits is a key method for expanding the number of qubits available on the NV center experimental platform.

To perform quantum computing, we need not only well-controlled qubits, but also the ability to initialize and read out the qubit state. For NV color centers, the electron spins are initialized and read out by laser \cite{68}. At room temperature, a 532 nm laser is often used to excite the NV$^{-}$ center to its excited state. When in this excited state, the NV center has two paths to return to the ground state. The first path is a direct transition from the excited state back to the ground state, which results in fluorescence in the 637 nm - 750 nm range. The second path involves returning to the ground state through an intermediate state. This second path does not conserve the electron spin, and does not result in 637nm-750nm fluorescence. If the electron of the NV center is in the $|m_S = \pm1\rangle$ spin state before being excited, it is more likely to return to the ground  $|m_S = 0\rangle$ spin state through the intermediate state after being excited to the first excited state. On the other hand, if the electron of the NV center is in the ground state of  $|m_S = 0\rangle$ before being excited, it is more likely to follow the path of radiative transition and directly return to the ground state with $|m_S = 0\rangle$. As a result, the population of electrons in the $|m_S = \pm1\rangle$ state decreases, while the population in the $|m_S = 0\rangle$ state increases. This process can be used to initialize the electron spins of the NV center to the $|m_S = 0\rangle$ state. Similarly, the fluorescence intensity of the NV center can be used to determine the path the electron spins take when returning to the ground state. The tendency for electron spin transition paths to be different for different spin states results in different radiation fluorescence intensities for NV centers in different spin states. This allows for the readout of the NV center's spin state.

The NV center's ground state spin has the longest single-spin decoherence time ($T_2$) among all electron spins in a solid at room temperature, with some samples exhibiting $T_2$ values greater than 1.8 ms \cite{69,70,71}. Therefore, due to its relatively mature control technology and long decoherence time, the NV color center system has become the choice of many scientific research groups to implement quantum algorithms in experimental systems, and it is also one of the most promising platforms for realizing quantum sensing at room temperature. 

Currently, institutions that carry out research on diamond color center quantum computing include Delft University of Technology in the Netherlands, the University of Chicago, the University of Science and Technology of China, etc. 

\subsubsection{NMR systm}

In NMR quantum computers, qubits are implemented using nuclear spin-1/2s in a static magnetic field. The two states of spin up and down represent the 0 and 1 of the qubit, respectively. Single-qubit quantum gate operations can be performed using radio-frequency electromagnetic waves, which are tuned to the Larmor frequency of the nuclear spin in the static magnetic field. This allows for the manipulation of the nuclear spin between the 0 and 1 states. Two-qubit quantum gate operations can be achieved by using the coupling between different nuclear spins and radio-frequency electromagnetic waves.
 
The NMR system is one of the earliest quantum computing platforms developed \cite{7,8,13,72,73,74,75,76,77,78,79}. Its origins can be traced back to the discovery of the famous Rabi oscillation phenomenon in 1938, which demonstrated that nuclei in a magnetic field can be aligned parallel or antiparallel to the field and have their spin direction reversed by applying a radio frequency field. In 1946, Bloch and Purcell discovered that specific nuclear spins in an external magnetic field absorb radio-frequency field energy of specific frequencies, which laid the foundation for the study of NMR. Over the years, NMR has been widely used in fields such as chemistry and medicine, and its mature manipulation techniques has allowed for the precise manipulation of coupled two-level quantum systems in NMR. After the concept of quantum computing was proposed, NMR has also been extensively studied as a platform with a large number of controllable qubits and high precision qubit control \cite{80,81,82,83}. Despite its early beginnings, NMR remains a promising platform for quantum computing due to its versatility and robustness.

NMR quantum computing techniques is highly developed compared to other platforms, making it a good demonstration platform for more complex quantum algorithms \cite{84,85,86,87,88,89,90,91,92,93,94,95,96,97,98,99,100,101}. Several research institutions around the world are currently conducting research on NMR quantum computing, including the University of Waterloo in Canada, the University of Dortmund in Germany, the University of Science and Technology of China, and Tsinghua University. One company that specializes in the development of NMR quantum computing products is SpinQ Technology, which produces portable NMR quantum computers for educational use.

\subsubsection{Silicon quantum dot systems}

Silicon-based quantum dot systems have received significant attention in recent years as a potential platform for quantum computing. There are several types of silicon-based quantum dot systems, which can be broadly classified into two categories: those that use electron spins in electrostatic quantum dots as qubits, and those that use the impurity nuclear spin implanted on the silicon base and the electron spin in its vicinity as qubits \cite{102,103,104}. Single-qubit gates can be implemented using a variety of methods, including the control of spin qubits using microwaves and the control of spin qubits using spin-orbit coupling and electric fields (EDSR) \cite{104}. Two-qubit gates can be realized using exchange interactions, Coulomb interactions, or coupling through resonators \cite{104}. The readout of the spin quantum state typically involves converting the spin state information into an electrical signal that can be measured using a charge sensor \cite{104}.

Silicon-based quantum dot qubits have the advantage of being scalable and customizable due to the availability of advanced micro-nano processing technologies in the semiconductor industry \cite{105}. However, one limitation of this approach is that the large number of degrees of freedom that couple the quantum dot qubits with the environment can lead to a short coherence time. In recent years, researchers have made significant progress in increasing the coherence time of quantum dot qubits to the order of milliseconds \cite{106} by reducing the number of nuclear spins in the underlying silicon material, providing a cleaner environment for the qubits \cite{105}. Control fidelities of over 99.9\% and 99\% for single-qubit gates and two-qubit gates have also been reported \cite{107,108}. While the number of qubits in this system is currently lower than that of superconducting qubit systems and ion trap systems, researchers are optimistic about the potential for rapid development of the silicon-based quantum dot system in the near future.

\subsection{Quantum computing cloud platform}

As quantum information technology continues to advance, it is expected that quantum computing will become more widely available in the near future. However, it is unlikely that quantum computers will be in the hands of individual consumers anytime soon. Instead, quantum computers are expected to work alongside classical computers to solve computational problems. To access quantum computers, it is more likely that users will rely on specific software through cloud services. Quantum cloud computing, which provides a range of services including access to quantum computing hardware and software, has emerged as a key form of quantum computing service and development. The quantum computing cloud platform is expected to be a crucial enabler for the use of quantum computing in the future.

In 2016, IBM announced the launch of its quantum cloud platform, IBM Quantum Experience. The company has stated its commitment to building a general-purpose quantum computing system that is commercially viable. IBM Quantum Experience provides users with access to quantum computing systems and related services through the IBM cloud platform, which is designed to handle complex scientific computing tasks that traditional computers cannot handle. The first field that is expected to benefit from the use of this platform is chemistry, where it may be used to develop new drugs and materials. Currently, users can access superconducting qubit systems with up to 127 qubits through IBM Quantum Experience. In addition to building its quantum cloud platform, IBM has also developed quantum software, including its own quantum programming language and the quantum software development platform, Qiskit, which allows users to easily create quantum programs and access IBM quantum cloud services.

In addition to IBM, there are several other leading quantum computing cloud platforms in the world.
One of the most notable is Amazon Web Services' (AWS) Amazon Braket, which was launched in December 2019. As the world's largest cloud computing provider, Amazon's entry into the quantum cloud computing field was highly anticipated. Amazon Braket is a fully managed solution that allows users to connect to a variety of third-party quantum hardware devices, including IonQ's ion trap quantum device, Rigetti's superconducting quantum device, and D-Wave's quantum annealing device. The platform provides researchers and developers with a development environment for designing quantum algorithms, a simulation environment for testing algorithms, and a platform for running quantum algorithms on three types of quantum computing devices. One advantage of Amazon Braket is that it allows researchers and developers to fully explore the design of complex tasks in quantum computing.
Microsoft has also made significant investments in quantum software development and community operations. In November 2019, the company launched the open-source quantum cloud ecosystem Azure Quantum, which allows users to access ion trap quantum computing systems from Honeywell (now Quantinuum) and IonQ, as well as QCI's superconducting quantum computing system.
Several Chinese universities and companies have also launched their own quantum computing cloud platforms, including NMRCloudQ \cite{109} from Tsinghua University, "Taurus" from Shenzhen Spinq Technology Co., Ltd., a quantum computing cloud platform jointly developed by the Chinese Academy of Sciences and Alibaba Cloud, and the quantum computing cloud platform from Hefei Origin Quantum Computing Company. Tsinghua University's NMRCloudQ is the world's first quantum cloud computing platform based on the NMR platform, and it includes four qubits with a fidelity of over 98\%. "Taurus" from Shenzhen Spinq Technology Co., Ltd. is a cloud service platform that can connect multiple quantum computing systems. It currently includes NMR systems that can perform computing tasks with up to 6 qubits, making it well-suited for research and education. "Taurus" also includes a two-qubit desktop NMR quantum computer (Gemini) that provides hobbyists in the field of quantum computing with ample time to explore. In the future, "Taurus" plans to offer access to superconducting quantum computing systems to provide higher-performance quantum computing power.

\section{NMR quantum computing basics}
\subsection{Fundamentals of NMR}
NMR is a phenomenon that was first observed in 1946 by Purcell and Bloch. It involves the absorption and emission of electromagnetic waves by magnetic nuclei, which results in transitions between different energy levels. NMR has a variety of applications, including the study of the dynamic properties of molecules in liquids and solids, the determination of molecular structures, and the use of magnetic resonance imaging for medical diagnosis.

\begin{figure*}
\centerline{
\includegraphics[width=3in]{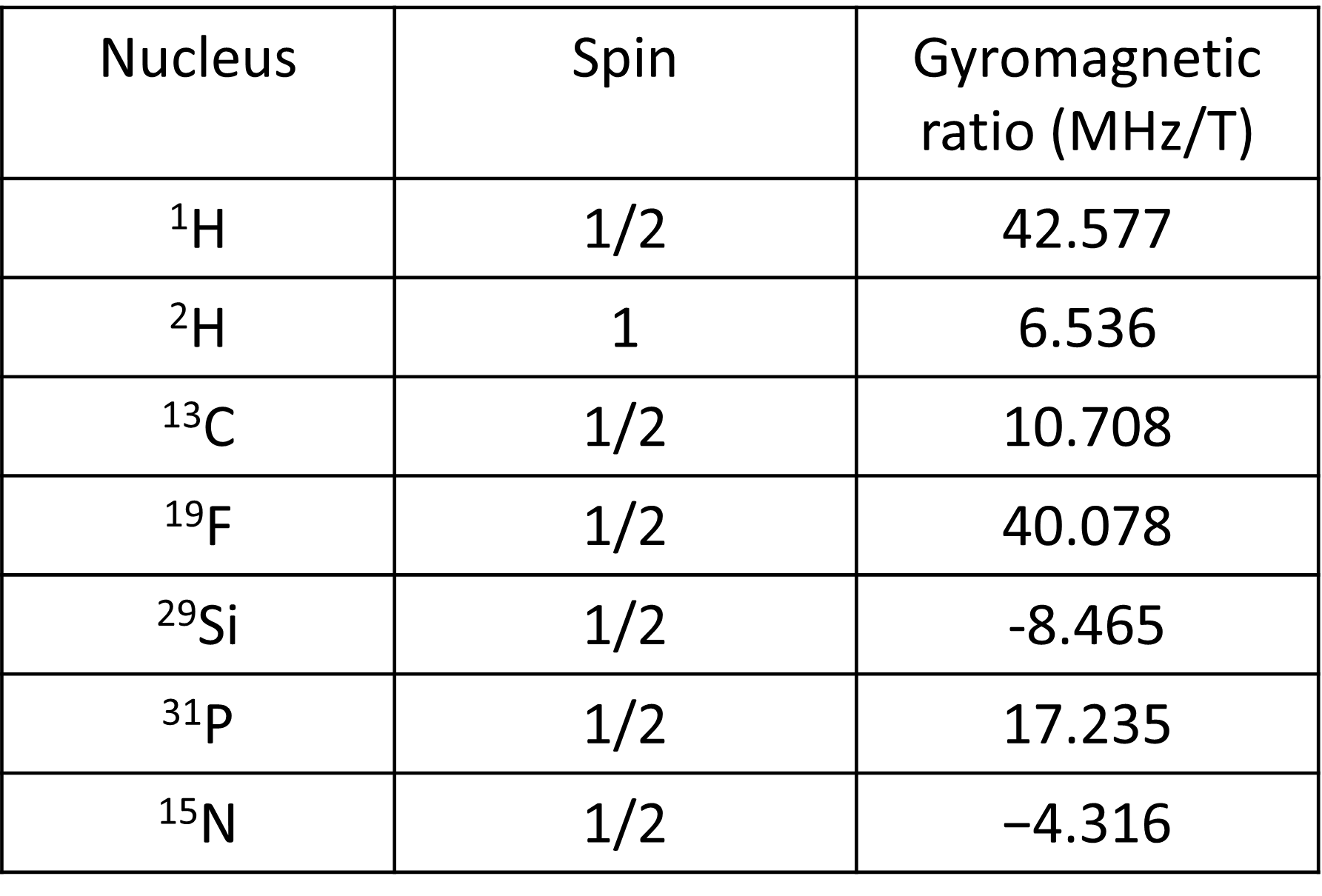}
}
\caption{Gyromagnetic ratios of different nuclear spins. }
\label{s3t1}
\end{figure*}

The magnetism of the nucleus originates from the nuclear spin of the nucleus. This spin can be represented by the symbol $\boldsymbol{I}$ and is present when the number of protons or neutrons in the nucleus is odd. For example, the three isotopes of hydrogen ($^{1}$H, $^{2}$H, and $^{3}$H) and the Carbon-13 isotope (denoted as $^{13}$C) have spin quantum numbers of 1/2, 1, 1, and 1/2, respectively. A nucleus with non-zero spin also has a spin magnetic moment, represented by the symbol $\boldsymbol{\mu}$. The relationship between $\boldsymbol{\mu}$ and $\boldsymbol{I}$ can be expressed as follows: 
\begin{align}
\boldsymbol{\mu} = \gamma \boldsymbol{I}. \label{3.1}
\end{align}

The $\gamma$ here is called the gyromagnetic ratio, and the gyromagnetic ratio is different for each nuclear spin (see Fig. \ref{s3t1}). If the nucleus is placed in an external magnetic field, the spin magnetic moment of the nucleus interacts with the magnetic field. It is difficult to measure the magnetic moment of a single atomic nucleus, but when there are many atoms, the effect will accumulate, and the influence of the spin magnetic moment of all the atomic nuclei to the external magnetic field can then be observed. A collection of such many nuclei is called an ensemble. 

If we compare the spin magnetic moment of an atomic nucleus to a small magnetic needle, then its energy in the magnetic field can be calculated according to the following formula

\begin{align}
E_{\text{mag}} = -\boldsymbol{\mu} \cdot \boldsymbol{B}. \label{3.2}
\end{align}

\begin{figure*}
\centerline{
\includegraphics[width=5.5in]{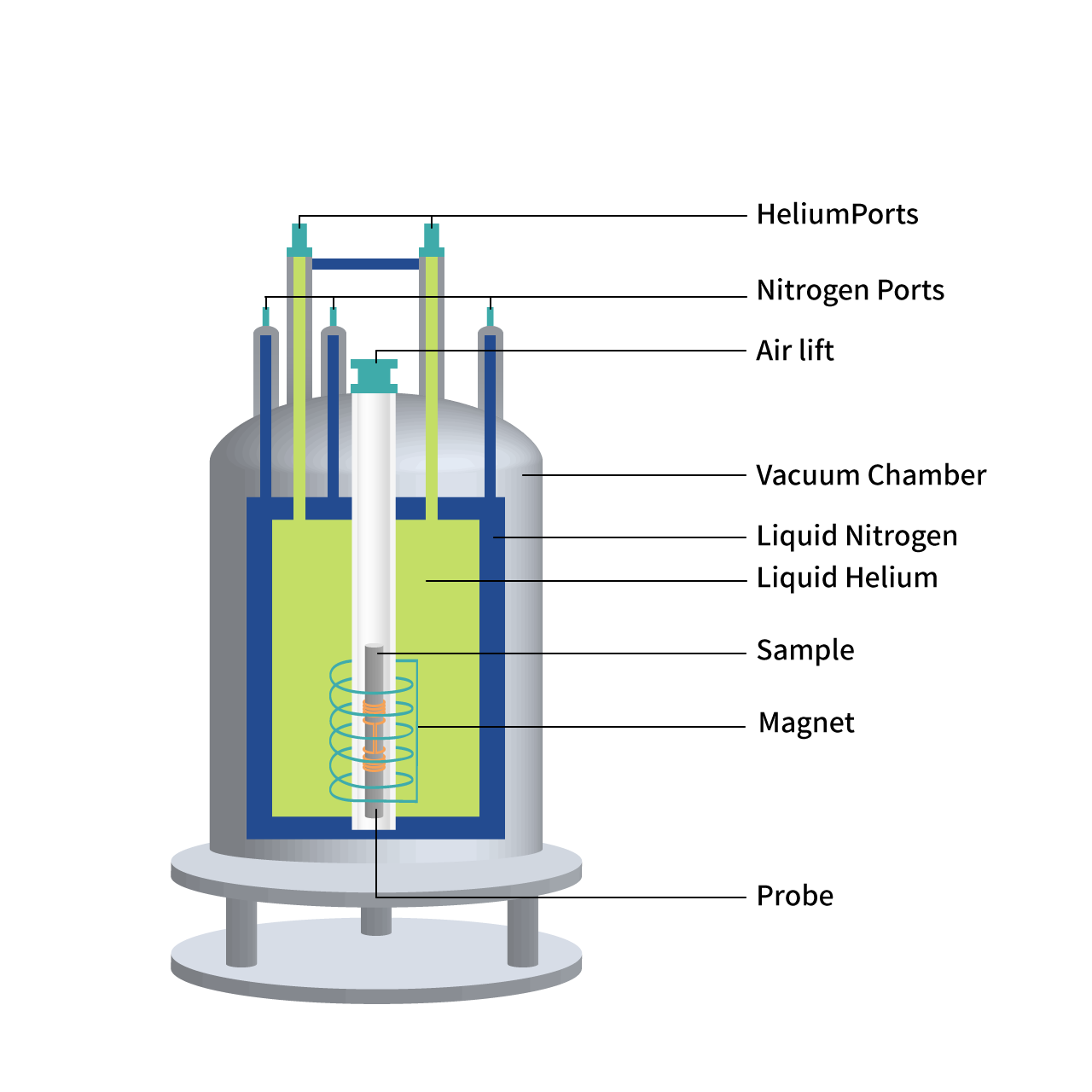}
}
\caption{ Schematic diagram of superconducting magnet NMR spectrometer. The superconducting magnet provides a strong magnetic field for the NMR sample, and the sample is placed in the probe coil, which can apply radio frequency pulses (rf pulses) to the sample and can be used to detect the induced current.}
\label{s3f1}
\end{figure*}

The negative sign in the equation indicates that the energy is lower when the magnetic moment is aligned with the magnetic field. In a state of thermal equilibrium, the magnetic moments of all nuclei tend to align with the direction of the external static magnetic field. If a radio-frequency magnetic field is applied to the sample under the right conditions, the nuclear spin magnetic moment will no longer be aligned with the external magnetic field, but will instead point at an angle and precess around the direction of the external field. The frequency of this precession, called the Larmor frequency, depends on the strength of the external magnetic field ($B_0$) and the gyromagnetic ratio ($\gamma$): $\nu=\gamma B_0$. The precession of the nuclear magnetic moment generates a varying magnetic field, which in turn causes a current oscillating at the Larmor frequency to be induced in a detection coil near the sample. By analyzing this current data, information about the magnetic moment of the entire nuclear ensemble can be obtained. A nuclear magnetic resonance spectrometer (see Fig. \ref{s3f1}) is an instrument that provides a static magnetic field for a sample and also allows for the application of radio-frequency pulses (rf pulses) and the detection of the induced current. 

\subsubsection{NMR system Hamiltonian}

As mentioned previously, the application of a suitable radio frequency pulse to nuclear spins in thermal equilibrium can cause the spins to deviate from the z-axis direction and undergo Larmor precession, which can then be observed. But what makes a pulse suitable, and what is the microscopic principle behind Larmor's precession? In the following sections, we will introduce the nuclear magnetic resonance system from the perspective of the system Hamiltonian.

In Section 2.4, we discussed that the evolution of quantum systems is governed by the Schrödinger equation. An NMR sample includes a large number of electrons and nuclei. In principle, the systematic evolution of the entire sample is described by the following time-dependent Schrödinger equation, namely
\begin{align}
|\dot{\psi}_{\text{full}}\rangle = -i\mathcal{H}_{\text{full}} |\psi_{\text{full}}\rangle. \label{3.3}
\end{align}

Here the Hamiltonian of the entire system $\mathcal{H}_\text{full}$ includes all interactions between electrons, nuclei and magnetic fields ($h$  is omitted here). Although the above equations are complete, it is not practical to study such a complex system of dynamic equations. In order to simplify the problem, in the NMR system, only the nuclear spin part is considered, and the influence of the electrons is included in the nuclear spin Hamiltonian with an average effect. This is the so-called spin Hamiltonian hypothesis \cite{110}. Therefore, the time-dependent Schrödinger equation becomes
\begin{align}
|\dot{\psi}_{\text{spin}}\rangle \approx -i\mathcal{H}_{\text{spin}} |\psi_{\text{spin}}\rangle. \label{3.4}
\end{align}

\begin{figure*}
\centerline{
\includegraphics[width=5in]{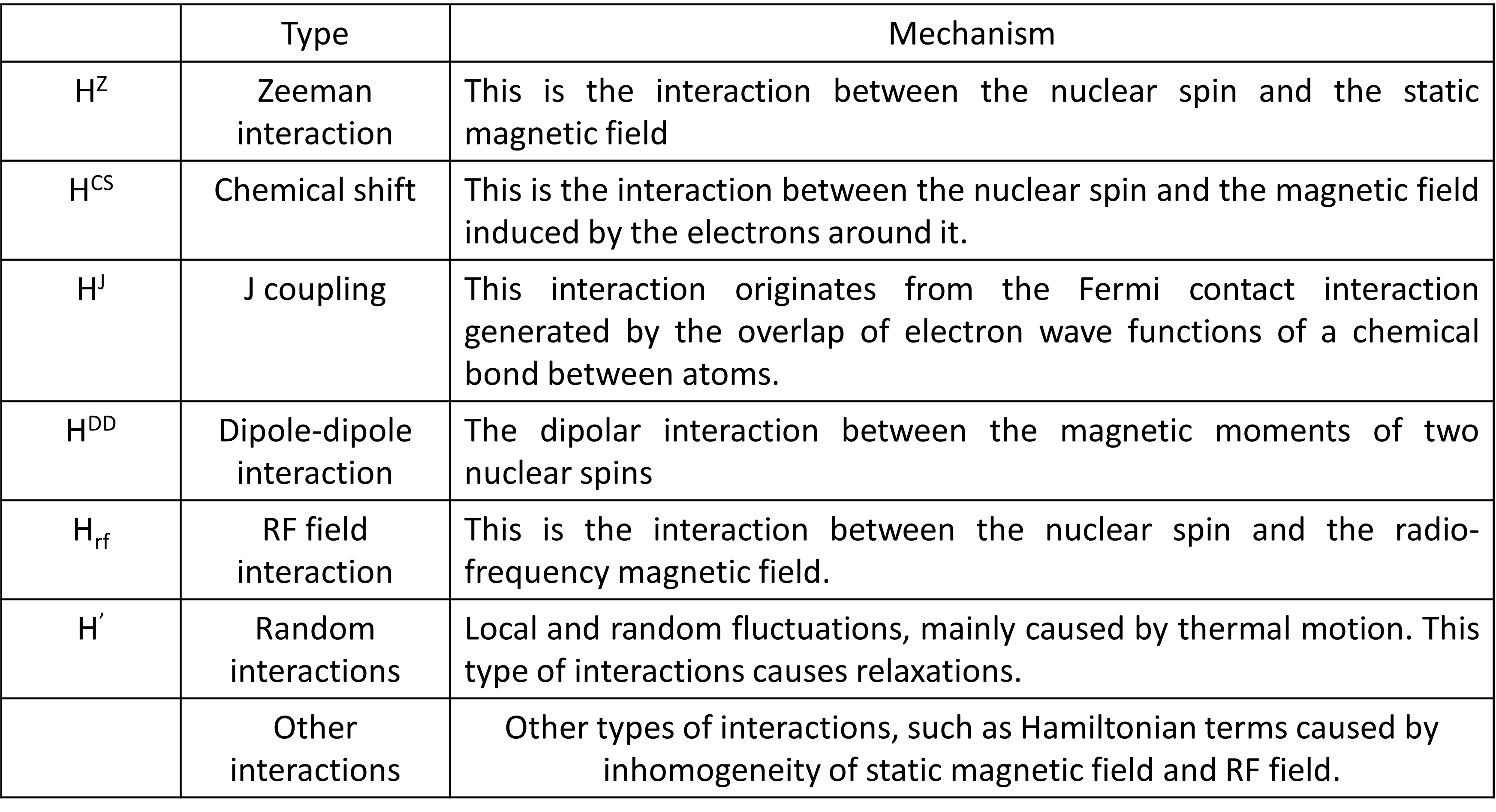}
}
\caption{Various Hamiltonian terms of spin-1/2 NMR systems. Reproduced from Ref. [110]. }
\label{s3t2}
\end{figure*}

\begin{figure*}
\centerline{
\includegraphics[width=4in]{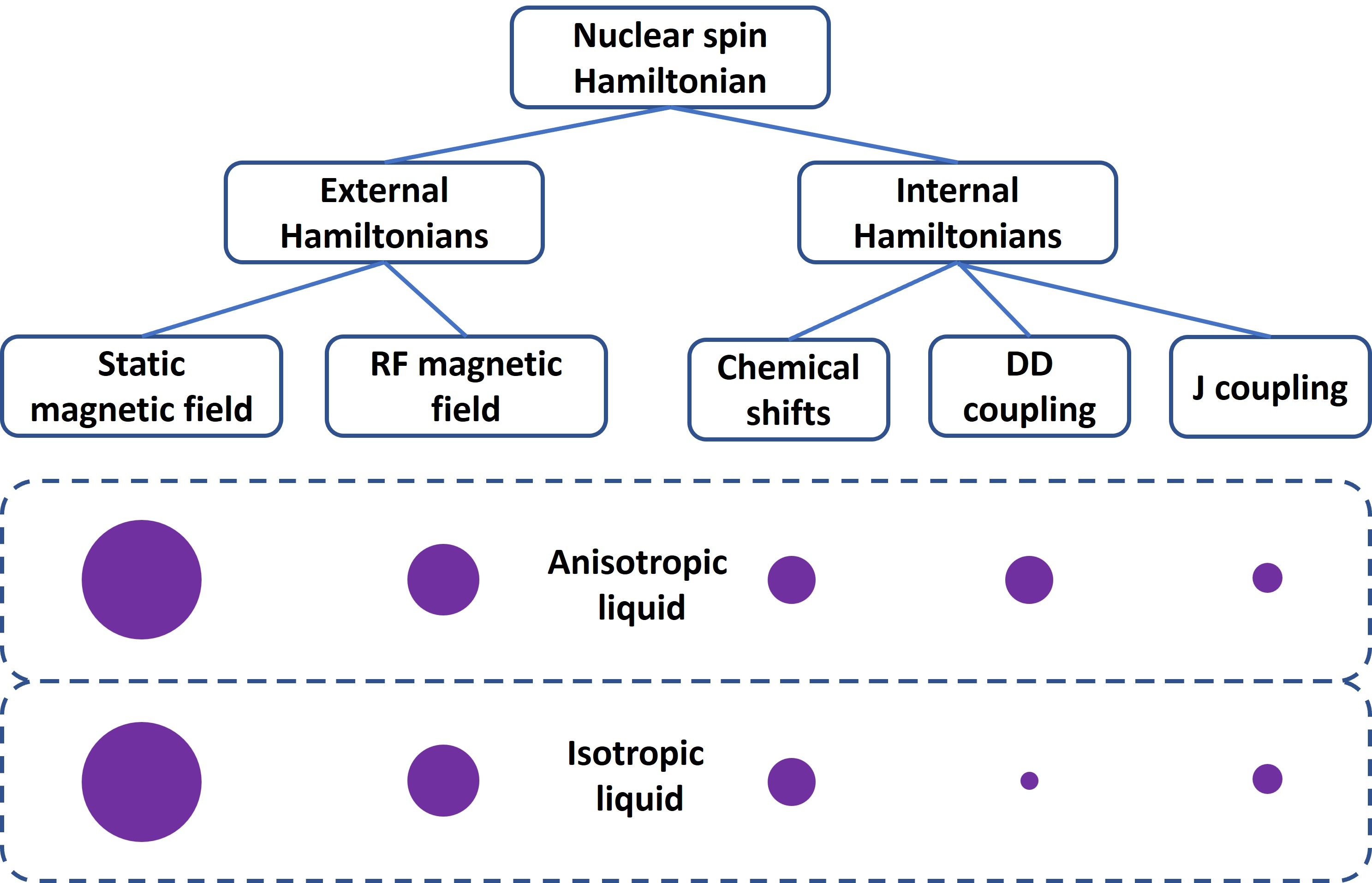}
}
\caption{ The terms of the liquid (spin-1/2) NMR system Hamiltonian and their relative magnitudes (qualitative). The interaction of nuclear spins with the static magnetic field is the strongest term in the Hamiltonian. In isotropic liquids, the dipole-dipole coupling is averaged out and can be approximated as zero. Reproduced from Ref. [110].}
\label{s3f2}
\end{figure*}

From the above equation, it can be seen that to describe the dynamic behavior of nuclear spins well enough and achieve precise quantum control, the specific form of $\mathcal{H}_\text{spin}$ is needed. For convenience, it will be abbreviated as $\mathcal{H}$ in the future. The spin Hamiltonian consists of an external term and an internal term. The external term is mainly dependent on the external electromagnetic field, such as static magnetic fields, radio frequency fields and gradient fields (see Figs. \ref{s3t2}, \ref{s3f2}). The internal term mainly includes the contributions of chemical shift, dipole-dipole coupling and $J$ coupling (see Figs. \ref{s3t2}, \ref{s3f2}) \cite{110}. As we mainly consider the case of spin-1/2 systems, the quadrupole effect can be ignored. We now discuss these Hamiltonian terms in detail item by item.\\
{\bf Static magnetic field}

The interaction of the NMR quantum system with the static magnetic field is the dominant interaction of the system. The static magnetic field $\boldsymbol{B}_0$ in the NMR system is usually generated by superconducting coils or permanent magnets, which is usually a strong magnetic field above 1T. In the laboratory frame, the usual convention is to choose the direction of the static magnetic field $\boldsymbol{B}_0$  as the z -axis direction. Under this convention we can then write the static magnetic field as $\boldsymbol{B}_0= B_0 \hat{z}$ , where $\hat{z}$ is the unit vector along the z direction. The static magnetic field causes Zeeman splitting of the nuclear spin energy level, and the corresponding Hamiltonian is
\begin{align}
\mathcal{H}_j^Z = -\gamma_j B_0 I_z^j = -\omega_j I_z^j, \label{3.5}
\end{align}
which is exactly the form of the Hamiltonian given in Eq. (\ref{3.2}). Here $\gamma_j$  is the gyromagnetic ratio of the $j$th nuclear spin, and $\omega_j=\gamma_j B_0$ is the Larmor frequency. Due to the Zeeman splitting, the spin-½ system then has two eigenstates, denoted by $|0\rangle$ and $|1\rangle$, and the energy splitting is $\omega$.  The $|0\rangle$ state corresponds to the case with the nuclear spin magnetic moment parallel to the static magnetic field and has a lower energy, and the $|1\rangle$ state corresponds to the case that the nuclear spin magnetic moment is antiparallel to the static magnetic field and has a higher energy. In this picture, to move the nuclear spin magnetic moment away from the z-axis is equivalent to drive the system transition between two energy levels, and the energy required is the energy difference between the two energy levels, so the frequency of the applied radio frequency pulse should be the frequency corresponding to the difference between the two energy levels. We'll discuss this in more detail in the RF Pulse Hamiltonian section.

\begin{figure*}
\centerline{
\includegraphics[width=3.5in]{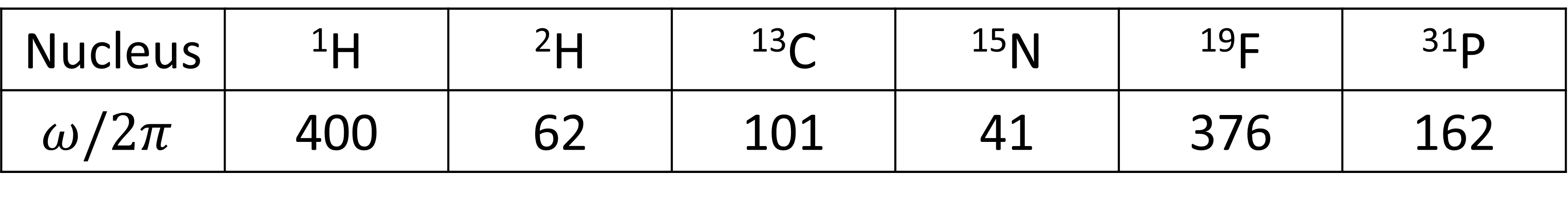}
}
\caption{Larmor frequencies (MHz) of several common nuclei in a static magnetic field of 9.4 T }
\label{s3t3}
\end{figure*}

In the superconducting magnet NMR system, the typical value of $B_0$ is 5-15T, and the nuclear precession frequency is in the order of hundreds of megahertz, which is proportional to the gyromagnetic ratio of the nucleus, so the Larmor frequencies of different nuclei can be very different (see Fig. \ref{s3t3}).\\
{\bf RF field}

We can use the radio frequency field $\boldsymbol{B}_1(t)$ in the x-y plane to achieve the excitation of nuclear spins. A rf field refers to a magnetic field whose frequency falls within the RF region (20kHz-300GHz). A rf field $\boldsymbol{B}_1(t)$  oscillating with a frequency $\omega_{rf}$ which is near the Larmor frequency has a Hamiltonian of the following form
\begin{align}
\mathcal{H}_{rf}(t) = -\sum_j \gamma_j B_1 \left[\cos(\omega_{rf} t + \phi) I_x^j + \sin(\omega_{rf} t + \phi) I_y^j\right].\label{3.6}
\end{align}
Here $B_1$, $\omega_{rf}$ and $\phi$ are the amplitude, frequency, and phase of the RF field, respectively. Generally speaking, $\omega_1=\gamma B_1$ reaches a maximum of 50kHz in liquid-state NMR and several hundred kHz in solid-state NMR.

In the laboratory frame, the motion of nuclear spin under the action of static magnetic field and radio frequency field is very complicated and difficult to describe. Therefore, generally the problem is transformed into a rotating frame that rotates around the z-axis at the speed of $\omega_{rf}$. Considering a single-spin system with Larmor frequency $\omega_0$, in the rotating frame, the state transforms as:
\begin{align}
|\psi\rangle^{rot} = \exp(-i\omega_{rf}t I_z) |\psi\rangle, \label{3.7}
\end{align}
Substituting the above formula into the Schrödinger equation, the Hamiltonian in the rotating frame can be obtained:
\begin{align}
\mathcal{H}_{rf}^{rot} = -(\omega_0 - \omega_{rf}) I_z - \omega_1 [\cos\phi I_x + \sin\phi I_y]. \label{3.8}
\end{align}

\begin{figure*}
\centerline{
\includegraphics[width=2.5in]{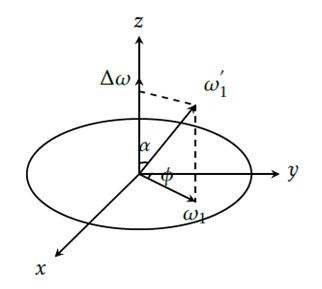}
}
\caption{When the frequency of the RF field is different from the nuclear spin Larmor frequency, the rotation axis of the nuclear spin is determined by the detuning $\Delta\omega $ and the RF field strength $\omega_1$, and the angle deviating from the z direction is $\alpha = \tan^{-1}(\frac{\omega_1}{\Delta\omega})$, rotating frequency is $\omega_1' = \sqrt{(\Delta\omega)^2 + \omega_1^2}$. }
\label{s3f3}
\end{figure*}

It is natural to observe that when the resonance condition is satisfied, i.e. $\omega_0 = \omega_{rf}$, the above formula then describes the interaction between the spin and a 'static magnetic field' $\boldsymbol{B}_1$ in the x-y plane at an angle $\phi$ with the x-axis. Then, similar to the Larmor precession under the action of $\boldsymbol{B}_0$ when the spins are not aligned along $\boldsymbol{B}_0$, the spins will also rotate around $\boldsymbol{B}_1$ in the rotating frame when they are not aligned along $\boldsymbol{B}_1$. If the initial state of the spin magnetic moment is along the z direction, by applying a resonant pulse for a certain time, the spin can be then transformed from the z direction to some other directions. By selecting the phase $\phi$ in Eq. (\ref{3.8}), the rotation axis can also be adjusted. If the resonance condition is not satisfied, then $\omega_0 \neq \omega_{rf}$ as illustrated in Fig. \ref{s3f3}. Let the detuning be $\Delta\omega = \omega_0 - \omega_{rf}$, then the spin will rotate along an axis with an angle $\alpha = \tan^{-1}(\frac{\omega_1}{\Delta\omega})$ from the z direction, with a frequency of $\omega_1' = \sqrt{(\Delta\omega)^2 + \omega_1^2}$.\\
{\bf Chemical shift}

Within the sample, the surrounding electron cloud environment is different for individual nuclei. The distribution and movement of electrons around the nucleus produces a localized magnetic field. This is then the concept of chemical shift, which has important applications in chemistry. The mechanism may be seen as: the applied static magnetic field $\boldsymbol{B}_0$ generates molecular current, and the molecular current in turn generates an induced local magnetic field $\boldsymbol{B}_{induced}$, thus playing a certain shielding effect on the nuclear spin.
Therefore, the total magnetic field experienced by the nuclear spins in the external magnetic field is
\begin{align}
\boldsymbol{B}_{loc} = \boldsymbol{B}_0 + \boldsymbol{B}_{induced}.\label{3.9}
\end{align}
For a reasonable approximation, the induced field depends linearly on the static magnetic field $\boldsymbol{B}_0$, i.e.
\begin{align}
\boldsymbol{B}_{induced} = \boldsymbol{\sigma} \boldsymbol{B}_0,\label{3.10}
\end{align}
where $\boldsymbol{\sigma}$ is called the chemical shift tensor. Typical chemical shift ranges depend on different nuclei, for example for $^1$H it is about 10 ppm (one part per million, a frequency unit commonly used in NMR), and for $^{13}$C and $^{19}$F it is about 200 ppm. When the static magnetic field $B_0$ is 10T, the value of chemical shift is about from several kilohertz to several tens of kilohertz, which is still very small compared to the Larmor frequency (typically the order of hundreds of megahertz). Nevertheless, same nuclear spins with different chemical shifts can be observed on an NMR spectrometer with sufficient frequency accuracy. Since chemical shifts are closely related to molecular structure, information on molecular structure can be obtained by observing the chemical shifts of nuclear spins.\\
{\bf Dipole-dipole coupling}

\begin{figure*}
\centerline{
\includegraphics[width=2.5in]{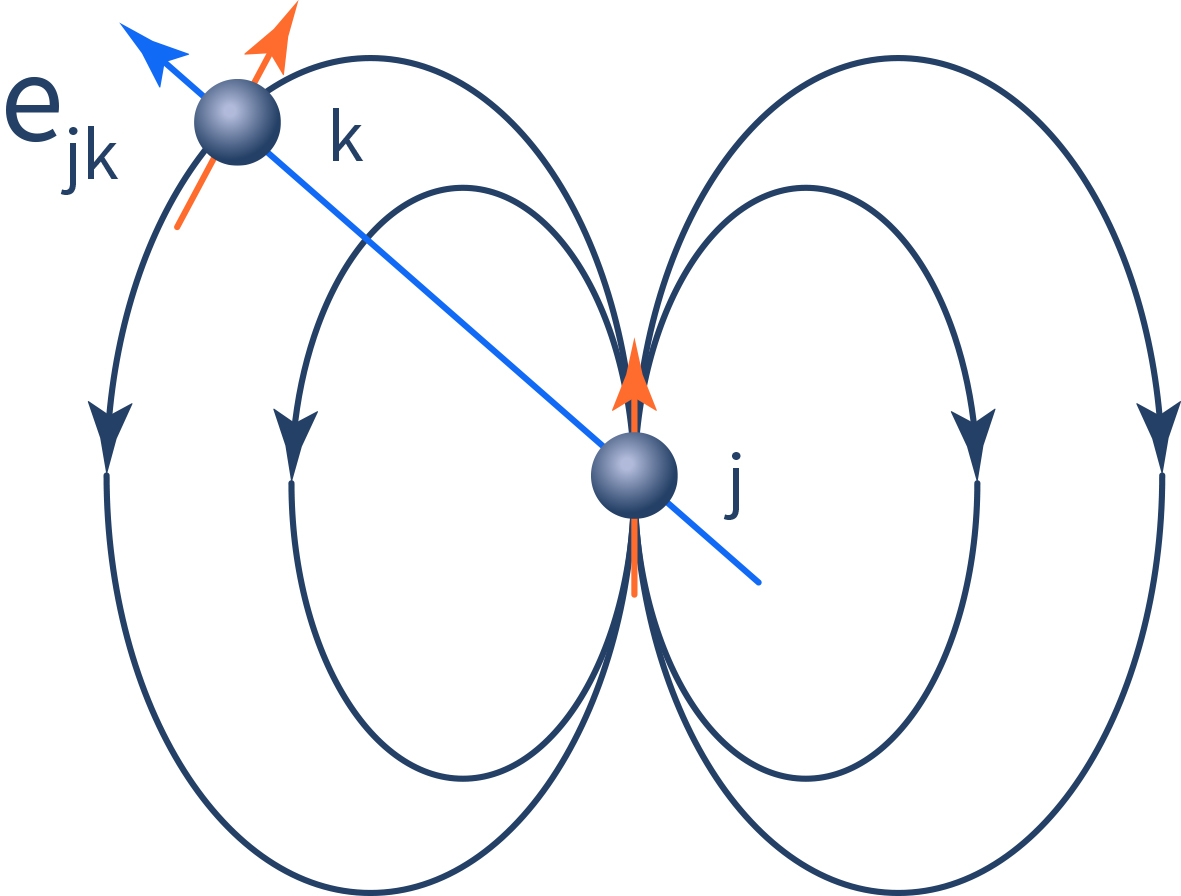}
}
\caption{Each nuclear spin can be viewed as a small magnet, and the magnetic field generated depends on the magnetic moment of that spin. The dipolar interaction between two nuclear magnetic moments depends on the space vector connecting them. Reproduced from Ref. [110]. }
\label{s3f4}
\end{figure*}

Each nuclear spin can be thought of as a small magnet, and the magnetic field generated depends on the spin's magnetic moment. As shown in Fig. \ref{s3f4}, two spins will interact through a magnetic field created by each other, a so-called dipole-dipole coupling. It can be seen that this coupling form is completely determined by the spacing and has nothing to do with any third-party media. Therefore, dipole-dipole coupling is also called as direct coupling. This interaction term can be written as
\begin{align}
\mathcal{H}_{jk}^{DD} = -\frac{\mu_0}{4\pi}\frac{(\gamma_j \gamma_k \hbar)}{r_{jk}^3}[3(\boldsymbol{I}^j \cdot \boldsymbol{e}_{jk})(\boldsymbol{I}^k \cdot \boldsymbol{e}_{jk}) - \boldsymbol{I}^j \cdot \boldsymbol{I}^k].\label{3.11}
\end{align}
Here $\mu_0$ is the vacuum permeability, $\boldsymbol{r}_{jk} = r_{jk} \boldsymbol{e}_{jk}$ is the vector connecting the space locations of spins $j$ and $k$, and $\boldsymbol{I}^j$ ($\boldsymbol{I}^k$) is the spin operator for the spin $j$ ($k$).

In strong static magnetic fields, the non-secular term in the above formula is averaged out, and only the secular term is retained. For homonuclear systems, the above equation can be approximated as
\begin{align}
\mathcal{H}_{jk}^{DD} =& -\frac{\mu_0}{8\pi} \frac{(\gamma_j \gamma_k \hbar)}{(r_{jk}^3)} [3 \cos{\Theta_{jk}} - 1](3I_z^j I_z^k - \boldsymbol{I}^j \cdot \boldsymbol{I}^k), \label{3.12}\\
\cos{\Theta_{jk}} &= \hat{z} \cdot \boldsymbol{e}_{jk}. \label{3.13}
\end{align}

For heteronuclear systems, Eq. (\ref{3.11}) can be approximated as
\begin{align}
\mathcal{H}_{jk}^{DD} = -\frac{\mu_0}{8\pi} \frac{(\gamma_j \gamma_k \hbar)}{r_{jk}^3} [3\cos{\Theta_{jk}} - 1] 2I_z^j I_z^k. \label{3.14}
\end{align}

The magnitude of the dipole-dipole coupling interaction is generally on the order of a few tens of thousands kHz. In isotropic liquid-state NMR, both intermolecular and intramolecular dipole-dipole couplings are averaged out due to the fast rolling of molecules, and hence can be ignored. In solid-state NMR, a simple Hamiltonian form similar to that in the liquid-state case can be achieved by applying multiple pulse sequences or magic-angle spinning techniques.\\
{\bf J Coupling}

J coupling is also known as indirect coupling, because this interaction mechanism originates from the shared electron pair in the chemical bond between atoms, i.e. from the Fermi contact interaction generated by the overlap of electron wave functions. The magnitude depends on the species of interacting nuclei and decreases as the number of chemical bonds increases. The full form of the J-coupling interaction of spins $j$ and $k$ in the same molecule is
\begin{align}
\mathcal{H}_{jk}^{J} = 2\pi \boldsymbol{I}_j \cdot \boldsymbol{J}_{jk} \cdot \boldsymbol{I}_k, \label{3.15}
\end{align}
where $\boldsymbol{J}_{jk}$ is the J-coupling tensor. In isotropic liquids, the J-coupling tensor is averaged by the fast tumbling motion of the molecules. Consequently, the Hamiltonian has a simplified form
\begin{align}
\mathcal{H}_{jk}^{J} = 2\pi J_{jk} \boldsymbol{I}_j \cdot \boldsymbol{I}_k. \label{3.16}
\end{align}
Here $J_{jk}=\frac{J_{jk}^{xx}+J_{jk}^{yy}+J_{jk}^{zz}}{3}$ is the average of the three diagonal terms of the J-coupling tensor, called the isotropic J-coupling or scalar coupling constant. If the system is heteronuclear, the secular approximation can further be used to obtain a simpler Hamiltonian form
\begin{align}
\mathcal{H}_{jk}^{J} = 2\pi J_{jk} I_z^j I_z^k. \label{3.17}
\end{align}

Typical J-coupling strength is usually several Hz to several hundred Hz. For example, the J-coupling strength between two Hydrogen spins with a three-bond distance is about 7 Hz; the J-coupling strength of a single carbon-hydrogen bond is about 135 Hz; the J-coupling strength of a single carbon-carbon bond is about 50Hz.

\subsubsection{Single spin and Larmor precession}

In the previous section, we discussed the main interaction terms in the Hamiltonian of the NMR system. In this and the next sections, we will analyze the specific forms of the NMR system Hamiltonian in the case of a single nuclear spin and two nuclear spins.

For the case of only one nuclear spin in a molecule, there is no inter-spin interaction term in the Hamiltonian. If the rf field is not considered, there is only the Zeeman interaction with the static magnetic field in the single-spin Hamiltonian, which can be written as
\begin{align}
\mathcal{H}_0 = -\gamma B_0 I_z = \omega_0 I_z = \frac{\omega_0 \hbar}{2} \sigma_z, \label{3.18}
\end{align}
which is a 2$\times$2 matrix with two eigenstates $|0\rangle$ and $|1\rangle$, and the corresponding eigen energies are $\frac{\omega_0}{2}$ and $-\frac{\omega_0}{2}$ (ignoring $\hbar$) with an energy difference of $\omega_0$. Notice that we absorb the minus sign of the Hamiltonian in $\omega_0$.

If the initial state of the spin is:
\begin{align}
|\psi(0) \rangle = \frac{|0\rangle + |1\rangle}{\sqrt{2}}, \label{3.19}
\end{align}
Substitute Eq. (\ref{3.18}) and the initial condition Eq. (\ref{3.19}) into Eq. (\ref{3.4}), we obtain the spin state for any time $t$ as:
\begin{align}
|\psi(t) \rangle = \frac{e^{-i\omega_0 t/2} |0\rangle + e^{i\omega_0 t/2} |1\rangle }{\sqrt{2}} = e^{-i\omega_0 t/2} \frac{ |0\rangle + e^{i\omega_0 t} |1\rangle}{\sqrt{2}}, \label{3.20}
\end{align}
where the phase factor $e^{-i\omega_0 t/2}$ is a global phase that can be ignored, as it has no observational effect. This can also be seen from the corresponding density operator of $|\psi(t)\rangle$, i.e., for $|\psi(t)\rangle \langle \psi(t)|$, any global phase factor is canceled out.

\begin{figure*}
\centerline{
\includegraphics[width=3.5in]{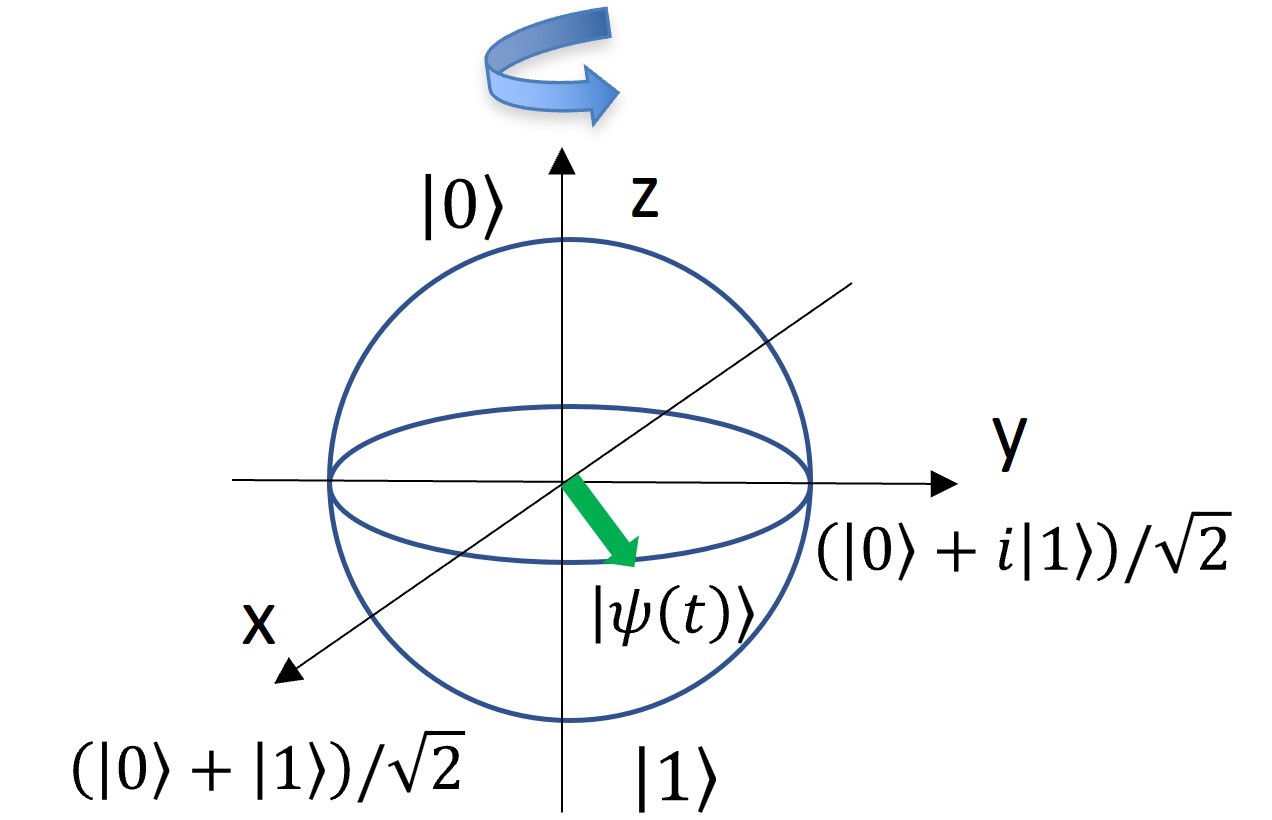}
}
\caption{When only considering Zeeman interaction, the vector corresponding to a single-spin quantum state rotates about the z-axis with frequency $\omega_0$ in the Bloch sphere.}
\label{s3f5}
\end{figure*}

Let us now consider the quantum state in Eq. (\ref{3.20}) from the viewpoint of the Bloch sphere. As already mentioned, in the Bloch sphere, the unit vector pointing to the z direction corresponds to the $|0\rangle$ state, and the unit vector pointing to the negative z direction corresponds to the $|1\rangle$ state. The x direction then corresponds to the $\left( |0\rangle + |1\rangle \right)/\sqrt{2}$ state, and the y direction corresponds to the $\left( |0\rangle + i|1\rangle \right)/\sqrt{2}$ state. And the unit vector on the x-y plane that has an angle $\phi$ with the x direction corresponds to the $\left( |0\rangle + e^{+i\phi} |1\rangle \right)/\sqrt{2}$ state. In this sense, the quantum state given by Eq. (\ref{3.20}) corresponds to a vector in the x-y plane, which is rotating at the frequency $\omega_0$ (Fig. \ref{s3f5}). Since the vector corresponding to the quantum state in the Bloch sphere is also the direction of the spin polarization of the quantum state, the spin polarization also rotates around the z-axis at the frequency $\omega_0$. This then gives the microscopic physical picture of the Larmor precession.

\subsubsection{Energy levels of a two-spin system}

We consider isotropic liquid-state heteronuclear systems. If there are two nuclear spins in the molecule, the system Hamiltonian will have the Zeeman interaction term and the J-coupling interaction terms:
\begin{align}
\mathcal{H}_0 = -\omega_0^1 I_z^1 - \omega_0^2 I_z^2 + 2\pi J I_z^1 I_z^2. \label{3.21}
\end{align}
Here $\omega_0^1$ and $\omega_0^2$ are the Larmor frequencies of the heteronuclear spins.

In Eq. (\ref{3.21}), if the J coupling strength is 0, i.e. $J=0$, then the four eigenstates of the Hamiltonian are $|00\rangle$, $|01\rangle$, $|10\rangle$, and $|11\rangle$, which correspond to the eigenvalues $-(\omega_0^1+\omega_0^2)/2$, $-(\omega_0^1-\omega_0^2)/2$, $(\omega_0^1-\omega_0^2)/2$, $(\omega_0^1+\omega_0^2)/2$ respectively (we ignore $\hbar$). Notice that the signals that can be observed in NMR systems are results of a single spin flip; therefore, in this four-level system, the observable signals correspond to the following four transitions between energy levels: $|00\rangle \leftrightarrow |10\rangle$, $|01\rangle \leftrightarrow |11\rangle$, $|00\rangle \leftrightarrow |01\rangle$, $|10\rangle \leftrightarrow |11\rangle$. The former two transitions correspond to the energy change $\omega_0^1$, and the latter two groups correspond to the energy change $\omega_0^2$. For the case $J=0$, if we initialize the system in a superposition of the four states, the Hamiltonian will cause the two nuclear spins to precess at their respective Larmor frequencies $\omega_0^1$ and $\omega_0^2$.

\begin{figure*}
\centerline{
\includegraphics[width=5.5in]{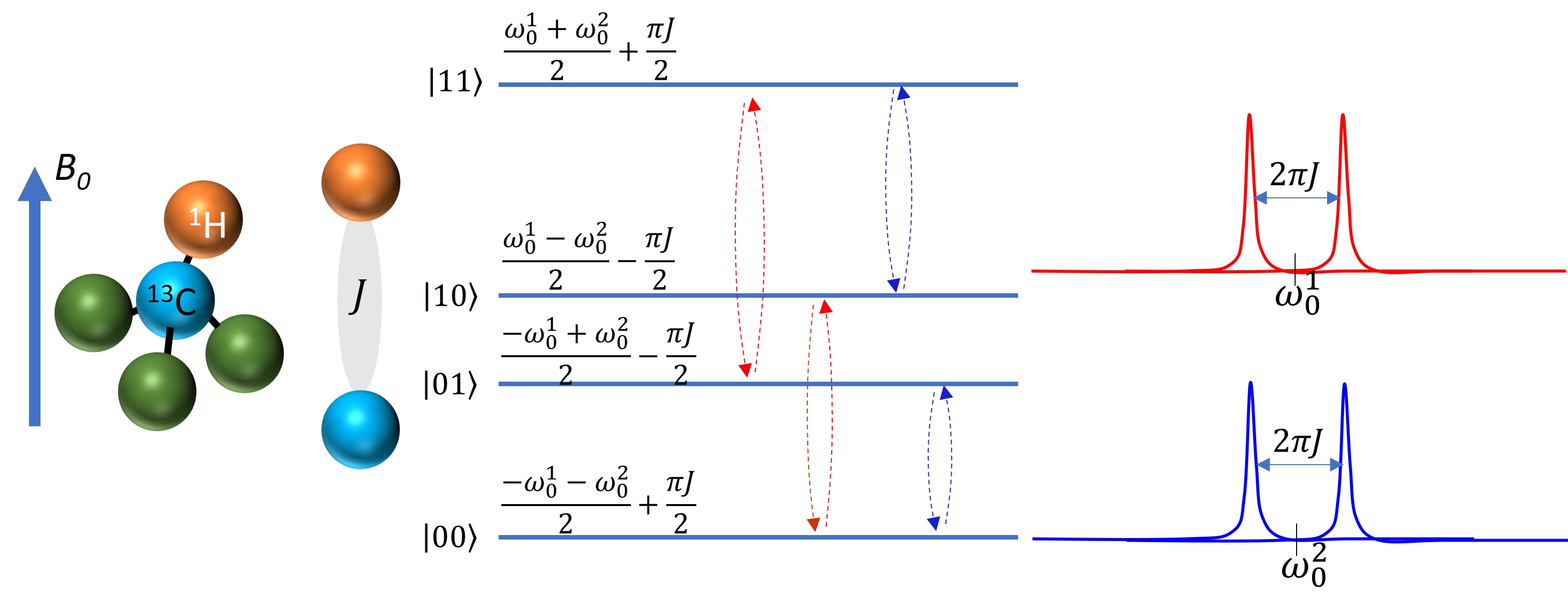}
}
\caption{Energy levels, eigenstates, allowed transitions and the corresponding frequencies of the chloroform molecule.}
\label{s3f6}
\end{figure*}

If the J coupling strength is nonzero, the four eigenstates of the system are still $|00\rangle$, $|01\rangle$, $|10\rangle$, $|11\rangle$, but the corresponding eigenvalues become $-(\omega_0^1+\omega_0^2)/2+\pi J/2$, $-(\omega_0^1-\omega_0^2)/2-\pi J/2$, $(\omega_0^1-\omega_0^2)/2-\pi J/2$, $(\omega_0^1+\omega_0^2)/2+\pi J/2$. In this case, the energy change of the transitions $|00\rangle \leftrightarrow |10\rangle$, $|01\rangle \leftrightarrow |11\rangle$, $|00\rangle \leftrightarrow |01\rangle$, $|10\rangle \leftrightarrow |11\rangle$ are also different: the two degenerate energies in the previous case split to four different energies $\omega_0^1 \pm \pi J$ and $\omega_0^2 \pm \pi J$. If we initialize the system to a superposition state of these four energy eigenstates, the frequency of the signals we can now observe becomes $\omega_0^1 \pm \pi J$ and $\omega_0^2 \pm \pi J$. We can also think about this in the following way: when the second spin is in the $|0\rangle$ state, the first spin precesses with the frequency $\omega_0^1 - \pi J$ ; and when the second spin is in the $|1\rangle$ state, the first spin precesses with the frequency $\omega_0^1 + \pi J$. Similarly, when the first spin is in the $|0\rangle$ state, the second spin precesses with the frequency $\omega_0^2 - \pi J$; and when the first spin is in the $|1\rangle$ state, the second spin precesses with the frequency $\omega_0^2 + \pi J$. Figure \ref{s3f6} is an example of a $^{13}$C-enriched chloroform molecule, which contains a $^1$H nucleus (orange) and a $^{13}$C carbon nucleus (blue), and its Hamiltonian form is as in Eq. (\ref{3.21}). The eigenstates, energy levels, and Fourier transform spectrum of the transitions between them are given in Fig. \ref{s3f6}.

\subsubsection{Longitudinal relaxation process and thermal equilibrium}

In Section 2.5, we mentioned the longitudinal and transverse relaxation processes. We will now briefly introduce the mechanisms of these two relaxations that occur in NMR systems.

\begin{figure*}
\centerline{
\includegraphics[width=2.5in]{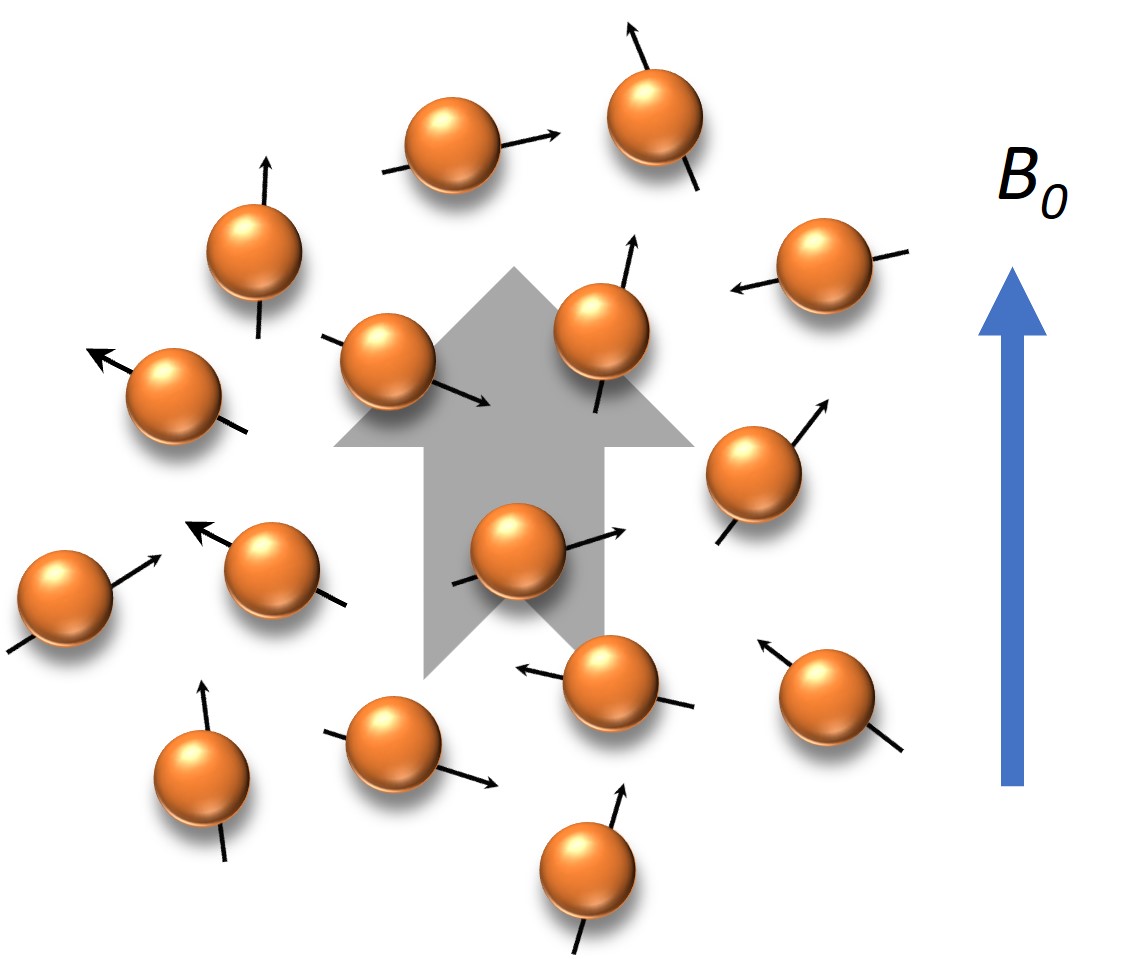}
}
\caption{The thermal equilibrium NMR system has a net magnetic moment along the z direction.}
\label{s3f7}
\end{figure*}

In the previous two sections, when discussing Hamiltonians in case of single or double spins, we only considered systems that are isolated from the environment. In reality, however, no physical systems can be completely isolated. Moreover, NMR samples often contain a large number of molecules, i.e. an ensemble system. As a result, nuclear spins point in all directions in the case of zero magnetic field. Since the direction of the nuclear magnetic moment is parallel to the spin direction ($\gamma>$0) or antiparallel ($\gamma<$0), the direction of the nuclear magnetic moment also points in all directions, thus for the whole ensemble there is no net magnetic moment. When an external magnetic field $\boldsymbol{B}_0$ is applied, all the spins that are not aligned along $\boldsymbol{B}_0$ will precess with their Larmor frequencies around $\boldsymbol{B}_0$. In this situation, the spins are still pointing in all directions, and there is still no net magnetic moment as a whole. From the previous discussion, we know that the frequency of Larmor precession is proportional to the external magnetic field. If the external magnetic field is a constant, then the spin precession will not change. Since there are many nuclear spins in the sample, although the applied external magnetic field $\boldsymbol{B}_0$ is fixed, the local fields generated by different spin magnetic moments are constantly changing. Consequently, the total magnetic field seen by each spin is constantly changing, which then deviates from the direction and size of $\boldsymbol{B}_0$. Due to this change, the precession angle between a spin and the magnetic field $\boldsymbol{B}_0$ also changes gradually \cite{139}. Since the energy of the magnetic moment is lower when the direction of the magnetic moment is the same as that of the magnetic field, the spin magnetic moment is more likely to align with the direction of the magnetic field during the process of spin precession. After a sufficient amount of time, the spin magnetic moments in the entire sample still point in all directions, but the number of spin magnetic moments in the direction of the magnetic field $\boldsymbol{B}_0$ will be greater than those in other directions, so that the entire sample will have a net magnetic moment along the direction of the magnetic field (as shown in Fig. \ref{s3f7}, where $\gamma>$0, the nuclear spin polarization is parallel to the nuclear spin magnetic moment). The magnitude of this net magnetic moment is determined by the temperature and the type of nuclear spin. This equilibrium state reached after a sufficient period of time is called the thermal equilibrium state, which follows the Boltzmann distribution. In this equilibrium state, any particle is in thermal motion, so the local field seen by any one nuclear spin is changing, and the corresponding Larmor precession is also changing. However, as a whole, the sample's net magnetic moment remains the same. This net magnetic moment parallel to the external field is also called longitudinal polarization. This process of returning from a non-equilibrium state to an equilibrium state is called relaxation. In this section, we discuss the process of returning to an equilibrium state of the longitudinal polarization, which is then also called longitudinal relaxation.

The ensemble density matrix of a large number of spins in the thermal equilibrium state has the following form:
\begin{align}
\rho_{eq} = \frac{e^{-\mathcal{H}_S/k_B T}}{\text{Tr}(e^{-\mathcal{H}_S/k_B T})} \approx \frac{1}{2^n} I^{\otimes n} + \sum_{k=1}^n \frac{\epsilon_k \sigma_z^k}{2}, \label{3.22}
\end{align}
Here $n$ is the number of spins in a single molecule, $T$ is the temperature, $\mathcal{H}_S$ is the system Hamiltonian, and $k_B$ is the Boltzmann constant. And since the temperature $T$ is around room temperature, we can use the approximation $||\mathcal{H}_S||/k_B T \approx 10^{-5} \ll 1$. We have discussed in Section 2.2 that the spin polarization along the x, y, z directions are proportional to the measurement expectation values $\langle \sigma_x \rangle$, $\langle \sigma_y \rangle$, $\langle \sigma_z \rangle$. For Eq. (\ref{3.22}), only $\langle \sigma_z \rangle$ is nonzero. That is, there is only net spin polarization along the z direction, which is consistent with the fact that the thermal equilibrium state has a net magnetic moment in the z direction, as discussed in the previous paragraph.

\begin{figure*}
\centerline{
\includegraphics[width=5in]{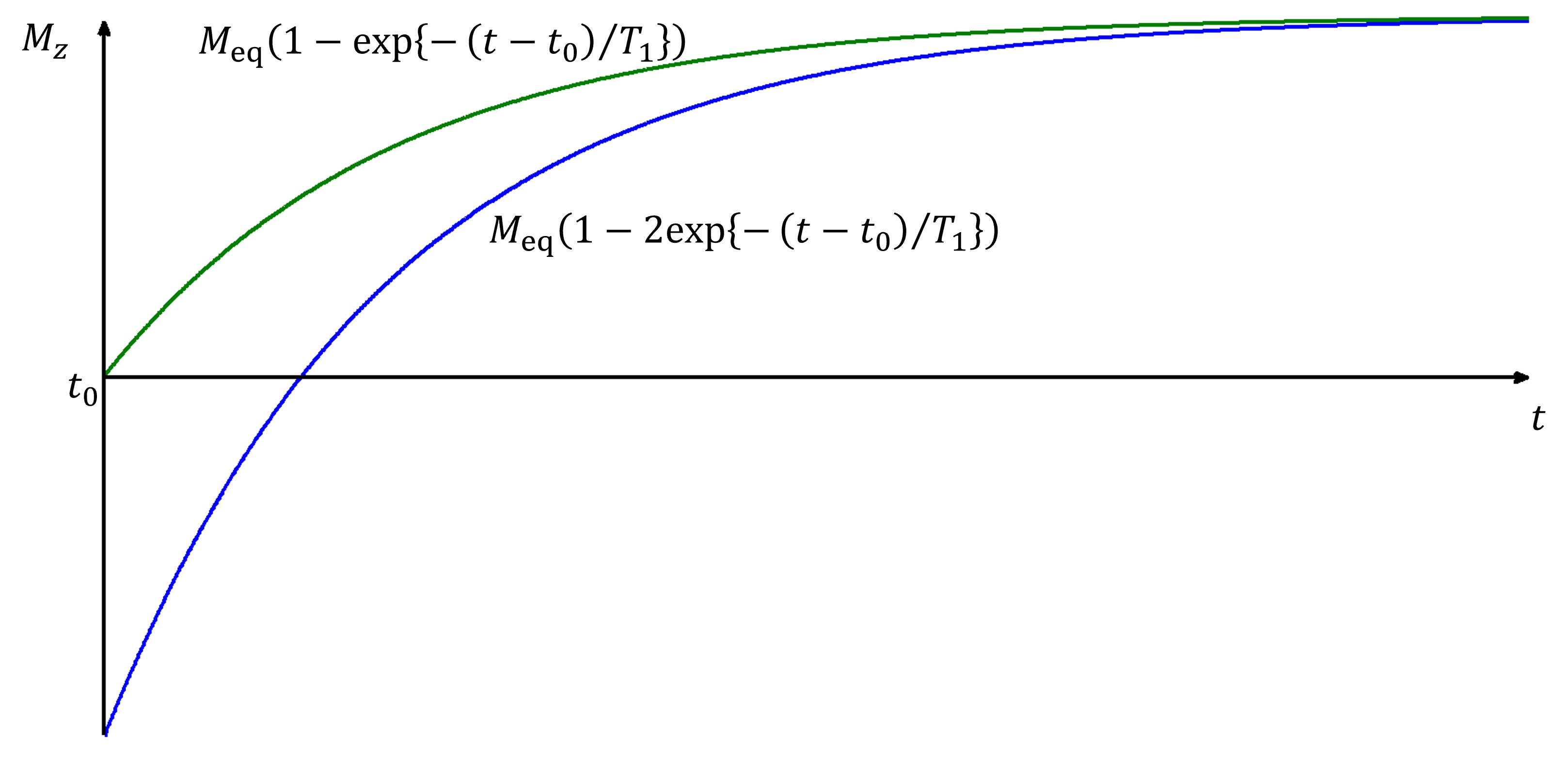}
}
\caption{Longitudinal relaxation process with different initial states.}
\label{s3f8}
\end{figure*}

The relaxation process of the longitudinal polarization is often characterized by an exponential factor. If the initial state of the sample has zero net magnetic moment, then when a static magnetic field is applied along the z direction, the relaxation of the longitudinal polarization has the following form:
\begin{align}
M_z(t) = M_\text{eq} (1 - \exp\{-\frac{(t - t_0)}{T_1}\}), \label{3.23}
\end{align}
Here $M_z (t)$ is the longitudinal magnetic moment of time $t$, which is proportional to $\langle\sigma_z (t)\rangle$. $M_\text{eq}$ is the longitudinal magnetic moment in equilibrium, which is proportional to $\langle\sigma_z\rangle_\text{eq}$. $t_0$ is the initial time, and $T_1$ is the longitudinal relaxation time. The larger $T_1$, the slower the longitudinal relaxation; and the smaller $T_1$, the faster the longitudinal relaxation. In the case where a certain method is used to rotate the polarization in the thermal equilibrium state by 180°, that is, the polarization originally in the positive z direction is rotated to the negative z direction, then the relaxation with this initial state is given by the following formula
\begin{align}
M_z(t) = M_\text{eq} (1 - 2 \exp\{-\frac{(t - t_0)}{T_1}\}). \label{3.24}
\end{align}
The above two equations are illustrated in Fig. \ref{s3f8}.

And we can also combine these two equations as
\begin{align}
M_z(t) = M_\text{eq} + (M_z(t_0) - M_\text{eq}) \exp \{ -\frac{(t-t_0)}{T_1} \}. \label{3.25}
\end{align}

\subsubsection{Transverse relaxation and Fourier transform spectra}

Observing the longitudinal polarization of nuclear spins directly is challenging because in a magnetic field, the electron spins also have a longitudinal magnetic moment in a thermal equilibrium state, similar to nuclear spins. However, the spin magnetic moment of the electron is much larger than that of the nucleus, making it difficult to directly observe the magnetic moment of the nucleus.

NMR uses an ingenious method to observe nuclear spin polarization: use some method to rotate the longitudinal polarization of the nucleus into a plane perpendicular to the external magnetic field, and the nuclear spin polarization is detected in such a plane. The polarization in a direction perpendicular to the external magnetic field is called transverse polarization.

\begin{figure*}
\centerline{
\includegraphics[width=5.5in]{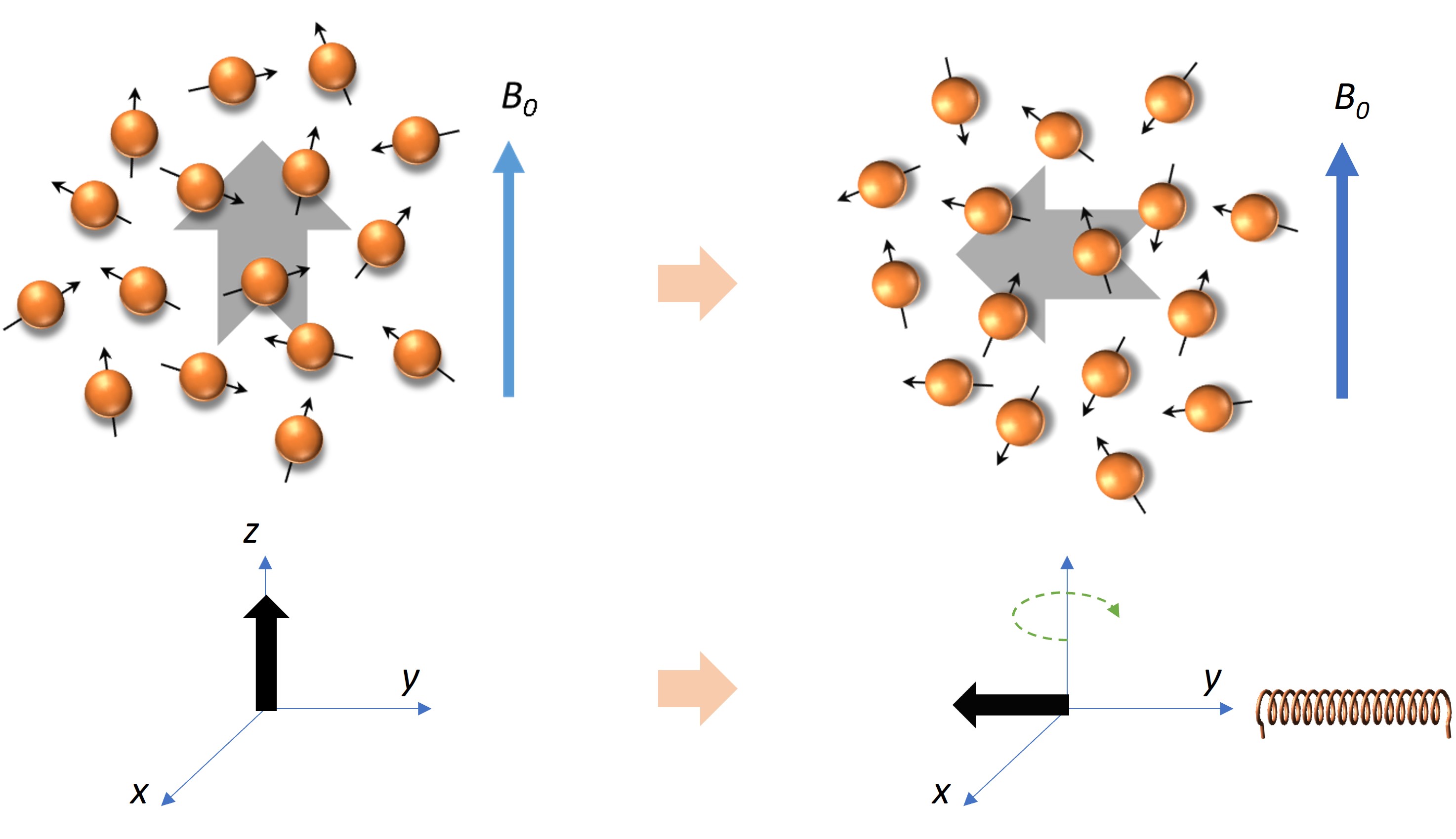}
}
\caption{When the longitudinal polarization of the nucleus is rotated to the x-y plane and becomes the transverse polarization, it will precess around the direction of the external magnetic field at the Larmor frequency. The detection coil perpendicular to the direction of the external magnetic field is used to detect the resulting changing magnetic field due to the rotation of the transverse magnetic moment. The changing magnetic field induces a changing current in the coil, and the frequency of this changing current is the same as the frequency of the changing magnetic field, that is, the Larmor frequency.}
\label{s3f9}
\end{figure*}

\begin{figure*}
\centerline{
\includegraphics[width=5.5in]{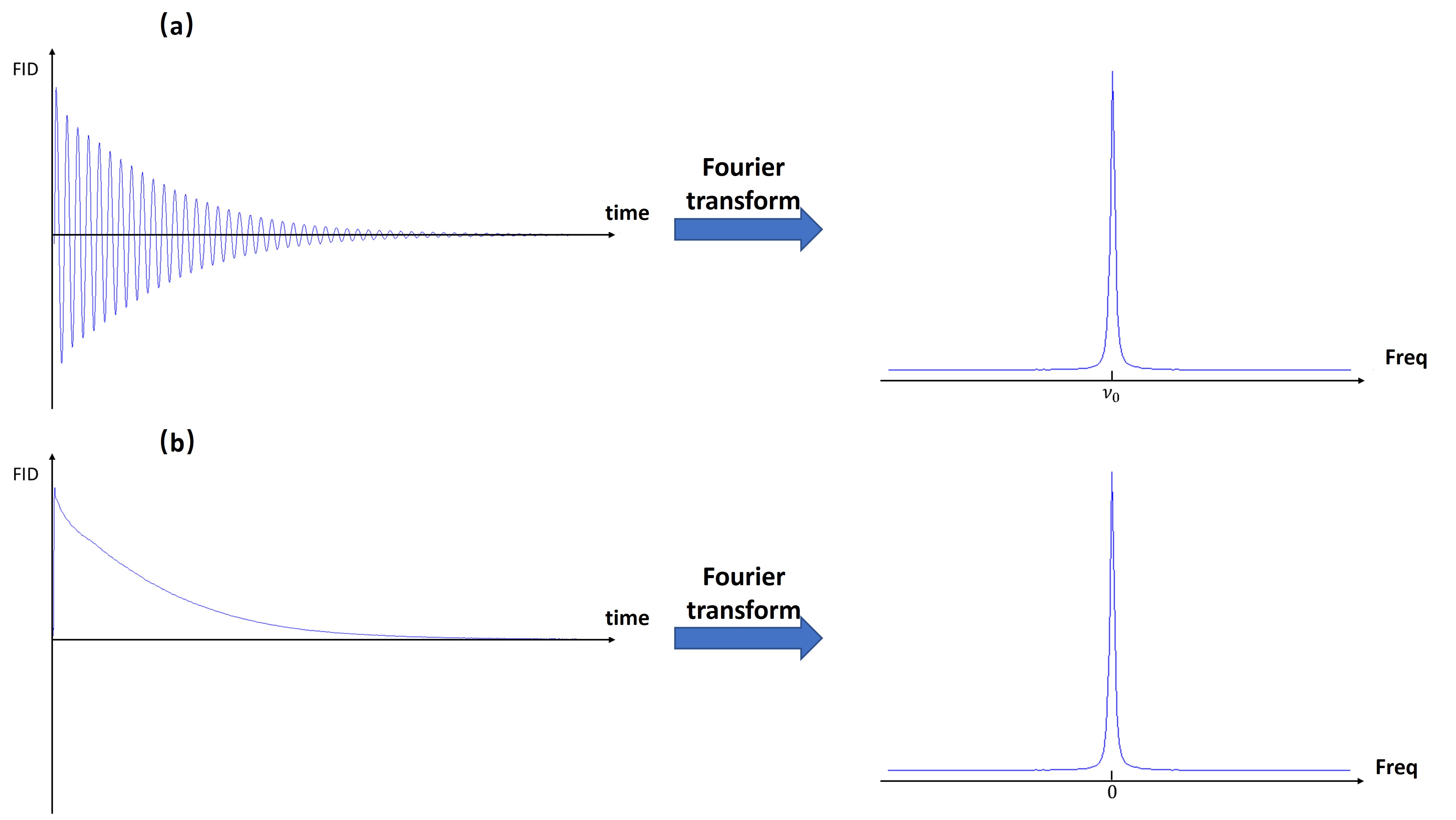}
}
\caption{Schematic diagrams of the free induction decay signal (FID) in the laboratory frame (a) and the rotating frame (b), and their Fourier transform spectra. It is assumed that there is only one type of nuclear spin.}
\label{s3f10}
\end{figure*}

When the longitudinal polarization of the atomic nucleus is rotated to the x-y plane and becomes the transverse polarization, the Larmor precession of each nuclear magnetic moment on the microscopic level will be reflected in the macroscopic transverse polarization, that is, the transverse polarization will also precess around the direction of the external magnetic field, as shown in Fig. \ref{s3f9}. In an NMR spectrometer, a detection coil perpendicular to the direction of the external magnetic field is used to detect the transverse magnetic moment. The rotating transverse magnetic moment provides a changing magnetic field perpendicular to the external field, which induces a changing current in the coil. This changing current has the same frequency as the changing magnetic field, i.e. the Larmor frequency. By design, the detection coil can be very sensitive to signals with Larmor frequency, thus realizing the detection of transverse magnetic moment. This induced current in the coil decays over time and is called the free induction decay (FID) signal (see Fig. \ref{s3f10}).

The free induction decay signal will attenuate. The reason is as follows. At the beginning, each microscopic nuclear magnetic moment performs Larmor precession synchronously, so that the macroscopic transverse net magnetic moment does Larmor precession. But each microscopic nuclear magnetic moment is affected by the local field generated by other nuclei. The local field changes with time. After a period of time, the Larmor precession of each microscopic nuclear magnetic moment is no longer synchronized, so that the macroscopic transverse net magnetic moment gradually decreases to zero. This process is called transverse relaxation. It also has an exponential-factor form:
\begin{align}
M_\perp(t) = M_\perp(t_0) \exp \{ -\frac{(t-t_0)}{T_2} \}. \label{3.26}
\end{align}
Here $M_\perp$ is the transverse magnetic moment, and $M_\perp=\sqrt{M_x^2+M_y^2 }$. Notice that $M_x$ is proportional to $\langle\sigma_x \rangle$, $M_y$ is proportional to $\langle\sigma_y\rangle$ and $T_2$ is the transverse relaxation time.

It is worth noting that if the observer is stationary in the laboratory frame, the transverse polarization will be observed to precess at the Larmor frequency. But if the observer is in a frame that also rotates at the Larmor frequency, the transverse polarization rotation will not be observed, instead only a gradual decrease in the transverse polarization is observed. In NMR signal processing, a rotating frame is often used to process signals (as shown in Fig. \ref{s3f10}(b)).

The frequency information of the nuclear spin precession can be obtained by performing Fourier transform on the FID signal of the nuclear spin. If the FID signal is processed in a rotating frame with the Larmor frequency, the frequency peak after the Fourier transform is at zero.

\subsubsection{Spin echo}

As discussed in the previous section, when the local field where each nuclear spin is located changes slightly over time due to the influence of the surrounding spins, the spins are no longer synchronized to do Larmor precession, so that transverse relaxation occurs, and the FID signal gradually decays. This discussion assumes that the external magnetic field is uniform at different positions of the sample, all of which are $B_0$. If the external magnetic field at each nuclear spin’s location is not the same, it will exacerbate the occurrence of Larmor's precession asynchrony, making the transverse polarization go to zero more quickly. But unlike transverse relaxation, the effects of inhomogeneous static magnetic fields can be counteracted by spin echoes \cite{111}.

To simplify the analysis, we now assume that the local field in which each nucleus is located is not affected by the surrounding spins, that is, there exists only the effect of the external magnetic field. If the external magnetic field is not uniform, some spin precessions are faster, while others are slower. As time progresses, the difference between different spins becomes larger and larger. Spin echo works by applying a pulse that rotates the spins by 180° in the middle of their evolution process, canceling out the effects of inhomogeneous magnetic fields on the spin precession. Consider that both spin {\bf a} and spin {\bf b} are precessing counter-clockwise around the z-axis, and the starting positions are both +x-axis. Assuming that spin {\bf a} rotates faster, then after time $\tau$ spin {\bf a} is ahead of spin {\bf b} by an angle of $\theta$. At this time, if both spins are rotated by 180° around the x-axis, {\bf b} is then positioned ahead of {\bf a} by an angle of $\theta$. Spin {\bf a} continues to rotate at a frequency faster than {\bf b}, and after time $\tau$, spin {\bf a} catches up with {\bf b}, and the two are now synchronized on the +x axis. This is how spin echoes work. That is, by applying a pulse that rotates the spins by 180° in the middle of the evolution process, the effect of the magnetic field inhomogeneity on the spin precession can be canceled.

In practice, the static magnetic field is often inhomogeneous, making it difficult to accurately measure the transverse relaxation time $T_2$ directly using the FID signal. As a result, the spin echo method is often used to measure $T_2$ by canceling out the effects of inhomogeneous magnetic fields on the spin precession. This allows for a more accurate measurement of $T_2$.

\subsection{NMR quantum computer}
\subsubsection{Qubit in NMR quantum computer}

Nuclei with spin 1/2 are used to represent quantum information in liquid-state NMR systems. The two-level system generated by the energy level splitting of the nuclear spin in the presence of a strong external magnetic field is used as a qubit \cite{13}. The two eigenstates of the nuclear spin, denoted by $|0\rangle$ and $|1\rangle$, correspond to the spin being parallel or antiparallel to the external magnetic field, respectively. These eigenstates form the basis of the qubit state space.

\begin{figure*}
\centerline{
\includegraphics[width=3in]{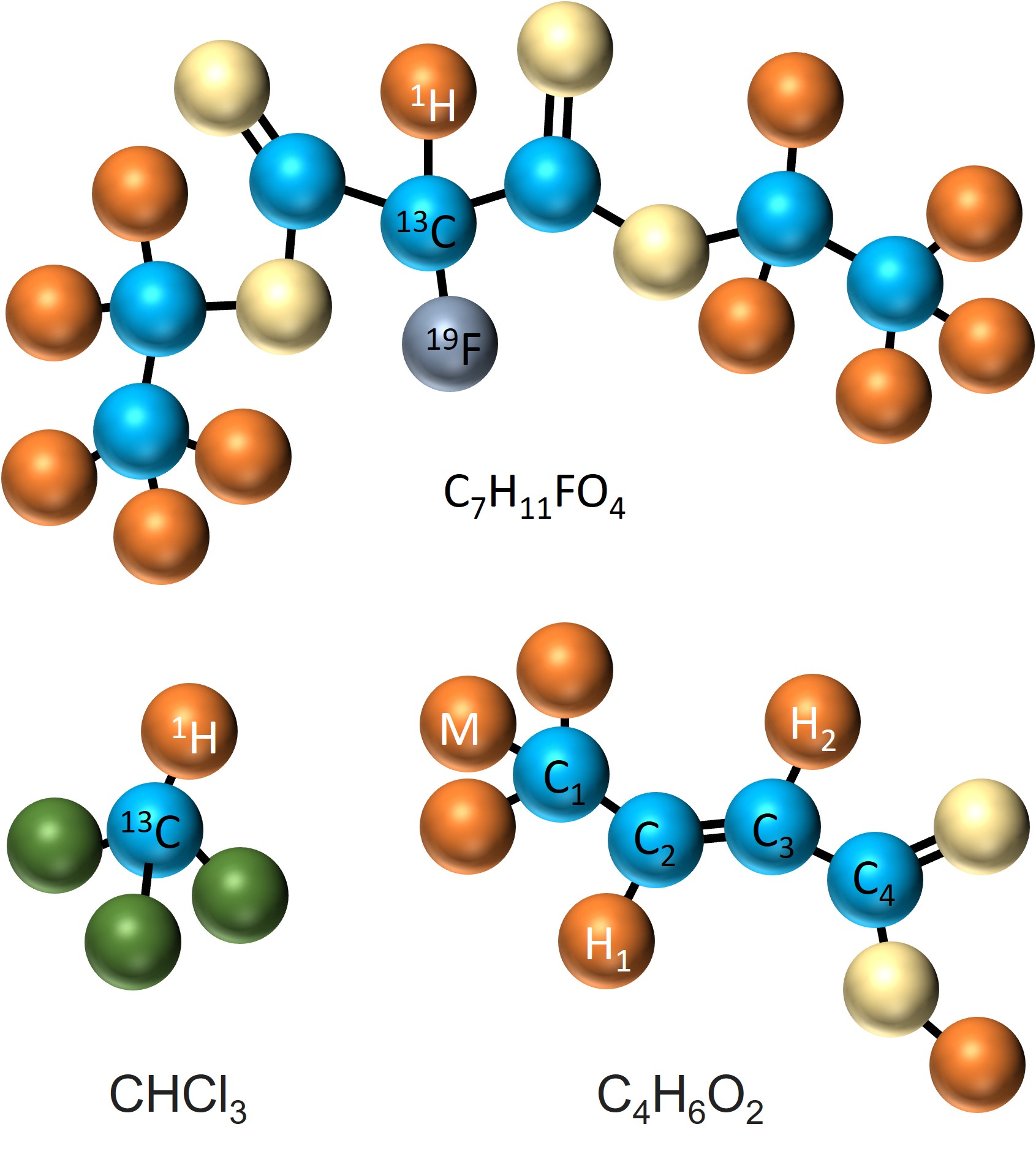}
}
\caption{Molecular Structure of the Quantum computing sample: Chloroform (CHCl$_3$), Crotonic acid (C$_4$H$_6$O$_2$), Diethyl fluoromalonate (C$_7$H$_{11}$FO$_4$).  For Chloroform [112],  $^1$H and $^{13}$C can be used as two qubits. For Diethyl fluoromalonate [90]  only the C that is connected to $^{19}$F is $^{13}$C, all the other Cs are $^{12}$C with spin 0. Despite there are 11 $^1$H, only the $^1$H connected to $^{13}$C can be used as a qubit. Therefore, Diethyl fluoromalonate is a three-qubit sample. For Crotonic acid [80], all the four Cs are $^{13}$C. There are three $^1$H connected to C1, denoted by M, which form two subspaces of spin 3/2 and spin 1/2. One can selectively operate only the spin-1/2 subspace by certain pulse sequences. So, M can be treated as a single qubit. Crotonic acid is a seven-qubit sample.}
\label{s3f11}
\end{figure*}

In NMR systems, it is common to treat the nuclear spin state in its rotating frame, which is defined by the Larmor frequency of the nuclear spin. The Hamiltonian of the single-qubit system in this rotating frame is described by Eq. (\ref{3.8}). In systems with multiple types of nuclei, different rotating frames can be chosen for each nuclear spin, with the frequency of each rotating frame corresponding to the Larmor frequency of the corresponding nucleus. This coordinate system is referred to as a multiple rotating frame. 

In a system with multiple spin-½ nuclei, the differences in the Larmor frequencies of each nucleus can be used to distinguish between different qubits during measurement, control, and other operations. Different types of nuclei have very different gyromagnetic ratios and thus very different Larmor frequencies, making it easy to distinguish them as different qubits. Even within the same type of nucleus, the Larmor frequencies can differ due to the presence of chemical shift. Under certain conditions, such as the resolution of the NMR spectrum and the frequency selectivity of the radio frequency pulse are good enough, it may be possible to distinguish between these nuclei and use them as different qubits. A molecular system with n distinguishable spin-1/2 nuclei can function as an n-qubit quantum information processor. For example, the commonly used $^{13}$C-labeled chloroform molecule (see Fig. \ref{s3f11}) can be used as a two-qubit quantum information processor.

\subsubsection{NMR quantum gates}
{\bf Single-qubit gates are implemented by radio frequency pulses}

Let the Larmor frequency of a single-qubit spin be $\omega$, and the system Hamiltonian is written as $\mathcal{H}_S=\omega\sigma_z/2$. Consider a square wave pulse of duration $t_p$ that resonates with the spin, i.e. the pulse amplitude $\omega_1 (t)=\omega_1$, and the pulse phase $\phi(t) = \omega t + \varphi \quad (0 \leq t \leq t_p)$
. Then in the rotating frame of the spin itself, the time evolution operator of the spin can be written as 
\begin{align}
U(t_p) = \exp \left[ -i \omega_1 t_p \left( \frac{\cos \varphi \sigma_x}{2} + \frac{\sin \varphi  \sigma_y}{2} \right) \right].\label{NMRsingle}
\end{align}
 If we choose $\varphi= 0^\circ$ or $\varphi= 90^\circ$, then the evolution operator corresponds to the rotation operation about the x or y axis, respectively. Adjusting time $t_p$ and intensity $\omega_1$ can control the angle of rotation, i.e. $\theta=\omega_1 t_p$. In this sense, radio frequency pulses that resonate with the spins can be used to rotate the spins by any angle around the x or y axis.
For any single-qubit transformation $U$, according to Bloch theorem, there exist real numbers $\alpha$, $\beta$, $\gamma$, $\delta$, such that \cite{4,12}
\begin{align}
U = e^{i \alpha} R_x(\beta) R_y(\gamma) R_x(\delta). \label{3.27}
\end{align}
Here $R_x$ and $R_y$ are rotations about the x and y axis, respectively. Notice that only rotations about x and y axis are required to implement an arbitrary single-qubit gate. Therefore, any single-bit gate can be realized using RF pulses.\\
{\bf Two-qubit gates are implemented by RF pulses and J coupling}

In order to implement a two-qubit gate, these two qubits need to interact with each other. In the liquid NMR system, qubits interact with each other via J couplings. 

Take the CNOT gate as an example. Considering two nuclear spins with J-coupling interactions, the system Hamiltonian in the double rotating frame is $\mathcal{H}_S=\pi J\sigma_z^1 \sigma_z^2  /2$. The evolution operator of the J coupling term is $U_J(t) = \exp \left[ -i \frac{\pi Jt \sigma_z^1 \sigma_z^2}{2} \right]
$, i.e.
\begin{align}
U_J(t) = \left(\begin{array}{cccc}
e^{-i \frac{\pi Jt}{2}} & 0 & 0 & 0 \\
0 & e^{i \frac{\pi Jt}{2}} & 0 & 0 \\
0 & 0 & e^{i \frac{\pi Jt}{2}} & 0 \\
0 & 0 & 0 & e^{-i \frac{\pi Jt}{2}}
\end{array}\right) \label{3.28}
\end{align}

\begin{figure*}
\centerline{
\includegraphics[width=5.5in]{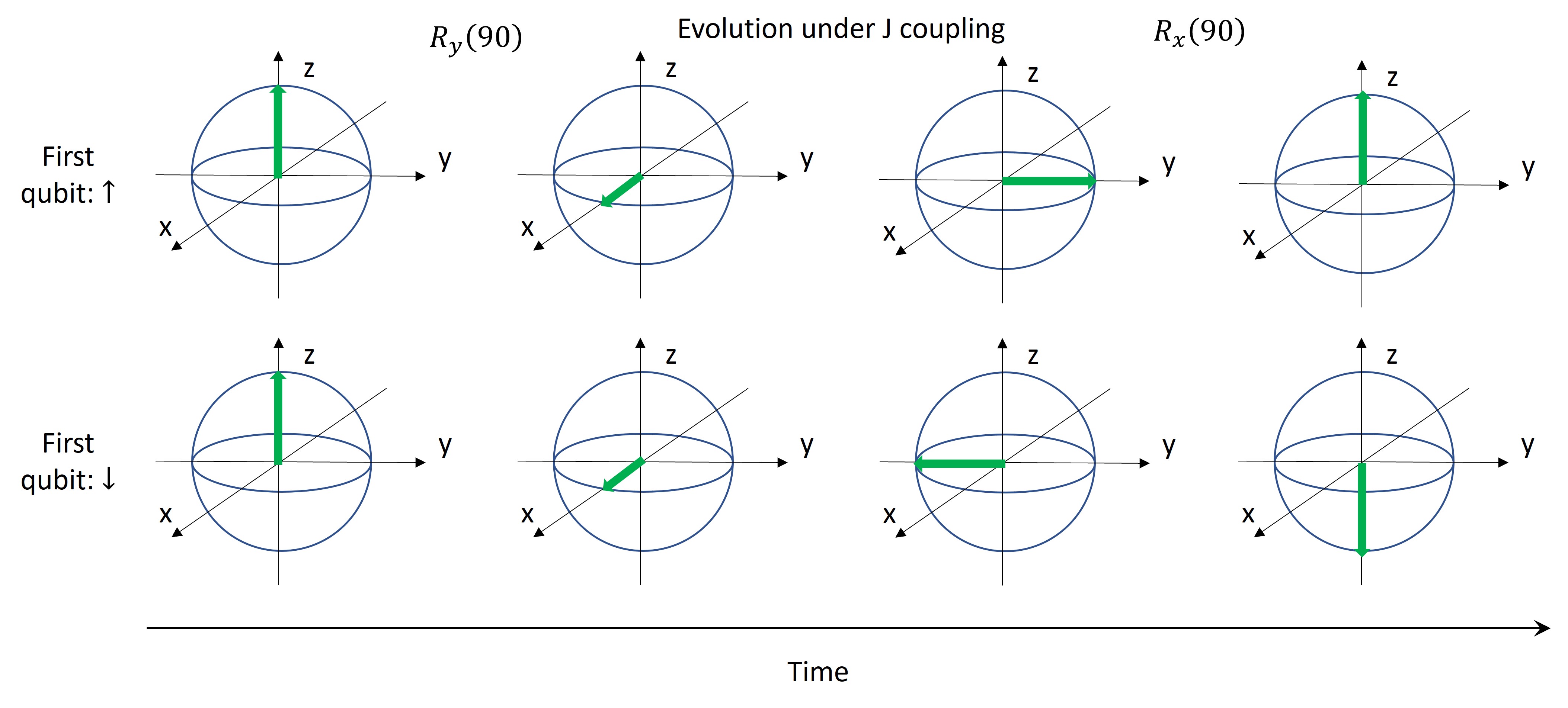}
}
\caption{A simple schematic diagram of the spin evolution of the second qubit when the CNOT gate is implemented according to the Eq. (\ref{3.29}).}
\label{s3f12}
\end{figure*}

It is straightforward to verify that, through the J-coupling evolution and single-qubit rotations, a two-qubit CNOT gate can be realized \cite{110}:
\begin{align}
U_\text{CNOT} = \sqrt{i} R_z^1 \left(\frac{\pi}{2}\right) R_z^2 \left(\frac{\pi}{2}\right) R_x^2 \left(\frac{\pi}{2}\right) \nonumber\\\cdot U_J \left(\frac{1}{2J}\right) R_y^2 \left(\frac{\pi}{2}\right). \label{3.29}
\end{align}
This CNOT gate takes the first qubit as the control bit and the second qubit as the target bit. Figure \ref{s3f12} shows a simple schematic diagram of the second-qubit evolution of the above-mentioned CNOT gate method, and the rotation operation about the z direction is omitted here.

It should be mentioned, as introduced earlier, J couplings are carried by the shared electrons in the chemical bond between atoms, so the closer these two nuclear spins, the greater the J coupling. The coupling between two nuclear spins with a connection of more than 3-4 chemical bonds can be neglected. $U_J \left(\frac{1}{2J}\right)$ plays the key role in the CNOT gate and the magnitude of $J $ decides the speed of CNOT. Therefore, the CNOT gate between nuclear spins with a very  small J coupling might take very long time and is not pratical anymore. In such situations, one may need to make use  of another nuclear spin which has larger couplings with both of the two spins to help in composing the CNOT gate.

Given the method of the CNOT gate implementation, other two-qubit gates can then be implemented with a combination of CNOT gates and single-qubit gates.

\subsubsection{Pseudo-pure state in NMR}

In general, quantum computing requires the initialization of all qubits to some suitable pure state, which is typically chosen as $|00...0\rangle$. NMR systems are ensemble systems, and NMR quantum computing uses the nuclear spins in all molecules of the system that have the same chemical environment and are indistinguishable in frequency as a quantum bit.  Any measurement result of the system is given as an ensemble mean. As mentioned in section 4.1.4, when no radio frequency (RF) pulse is applied to the NMR sample, the system tends to be in a thermal equilibrium state. Therefore, the initial state of an NMR quantum computer is generally the thermal equilibrium state of the system $\rho_\text{eq}$ (Eq. (\ref{3.22})). The second term in $\rho_\text{eq}$  is the polarization term, which can lead to measurable NMR signal. For $\rho_\text{eq}$, the polarization strength is very week ($\epsilon\sim10^{-5}$). In other words, $\rho_\text{eq}$ is a highly mixed state, which is not suitable as an initial state for quantum computing.

\begin{figure*}
\centerline{
\includegraphics[width=3.5in]{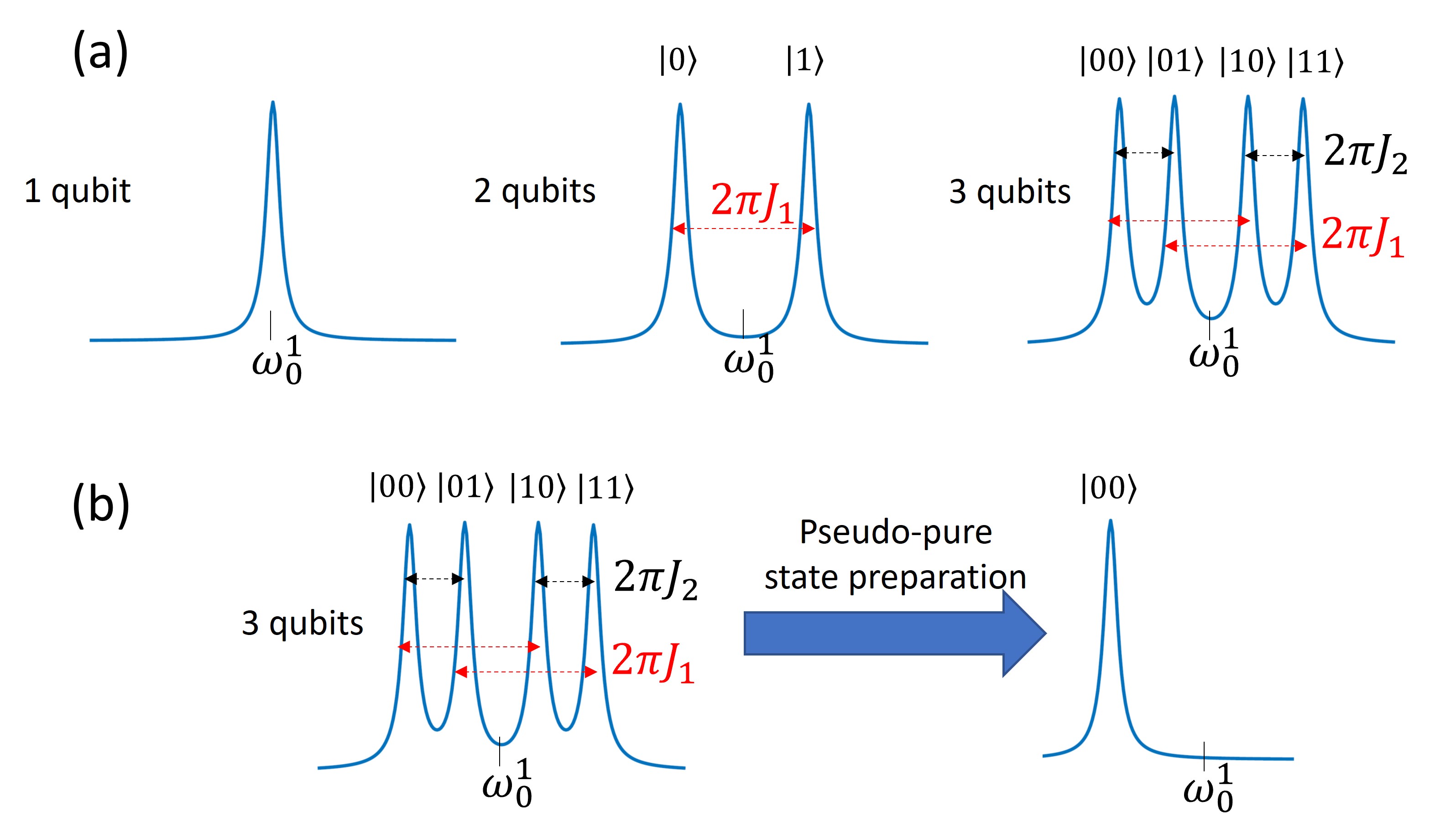}
}
\caption{(a) Simulated examples of peak splitting in NMR spectra. For an $n$-qubit sample, observing the thermal equilibrium spectrum of one qubit, with sufficient resolution and suitable J couplings, 2$^{n-1}$ peaks will be observed, each peak representing an eigenstate of the other $(n-1)$-qubits. (b) A simulated example of the pseudo-pure state spectrum of the three-qubit case in (a). The pseudo-pure state spectrum has only one peak, which corresponds to the  $|00\rangle$ state of the other two qubits.}
\label{s3f13}
\end{figure*}

To solve this issue, the concept of Pseudo-pure state is proposed (see Fig. \ref{s3f13}) \cite{8,9,10}, i.e.
\begin{align}
\rho_\text{pps} = \frac{(1-\eta)}{2^n} I^{\otimes n} + \eta | 00 \cdots 0 \rangle\langle 00\cdots0 |. \label{3.31}
\end{align}
Here the polarization factor $\eta$ is called the effective purity of the pseudo-pure state. The concept of pseudo-pure state is proposed based on the following two important points: First, the NMR signal is only related to the difference between the population of different energy levels, and has nothing to do with the absolute value of the population. That is, only the trace zero part of the density matrix will contribute, and the remaining part proportional to the identity matrix has no signal. The trace-zero part of the density matrix is defined as the deviation density matrix \cite{6}. Second, the unitary operation has no effect on the identity matrix, so after a series of unitary operations is performed on the system, the measurement result of the final state still only depends on the dynamic evolution result of the initial deviation density matrix. Two states, as long as they have the same deviation density matrix, will have the same dynamic behavior and measurement results. Notice that, except for the polarization factor, the evolution of $\rho_\text{pps}$ (under any unitary transformation) and the observed effect are completely equivalent to the pure state $|00...0\rangle$. Therefore, the pseudo-pure state can be used as the benchmark initial state for the NMR ensemble quantum computation.\\
{\bf Pseudo-pure state preparation by spatial averaging method}

The goal of pseudo-pure state preparation is to prepare from thermal equilibrium $\rho_\text{eq}$ to pseudo-pure state $\rho_\text{pps}$. $\rho_\text{pps}$ only has non-zero diagonal elements, and only one diagonal element is not equal to the other diagonal elements. The diagonal elements of $\rho_\text{eq}$ are generally all different. The key from $\rho_\text{eq}$ to $\rho_\text{pps}$ is how to adjust the size of the diagonal elements and remove the off-diagonal elements generated in this process. Pseudo-pure state preparation methods include time average method \cite{10}, spatial average method \cite{9,77}, logical labeling method \cite{113}, cat state preparation method \cite{80}, etc.

\begin{figure*}
\centerline{
\includegraphics[width=2.5in]{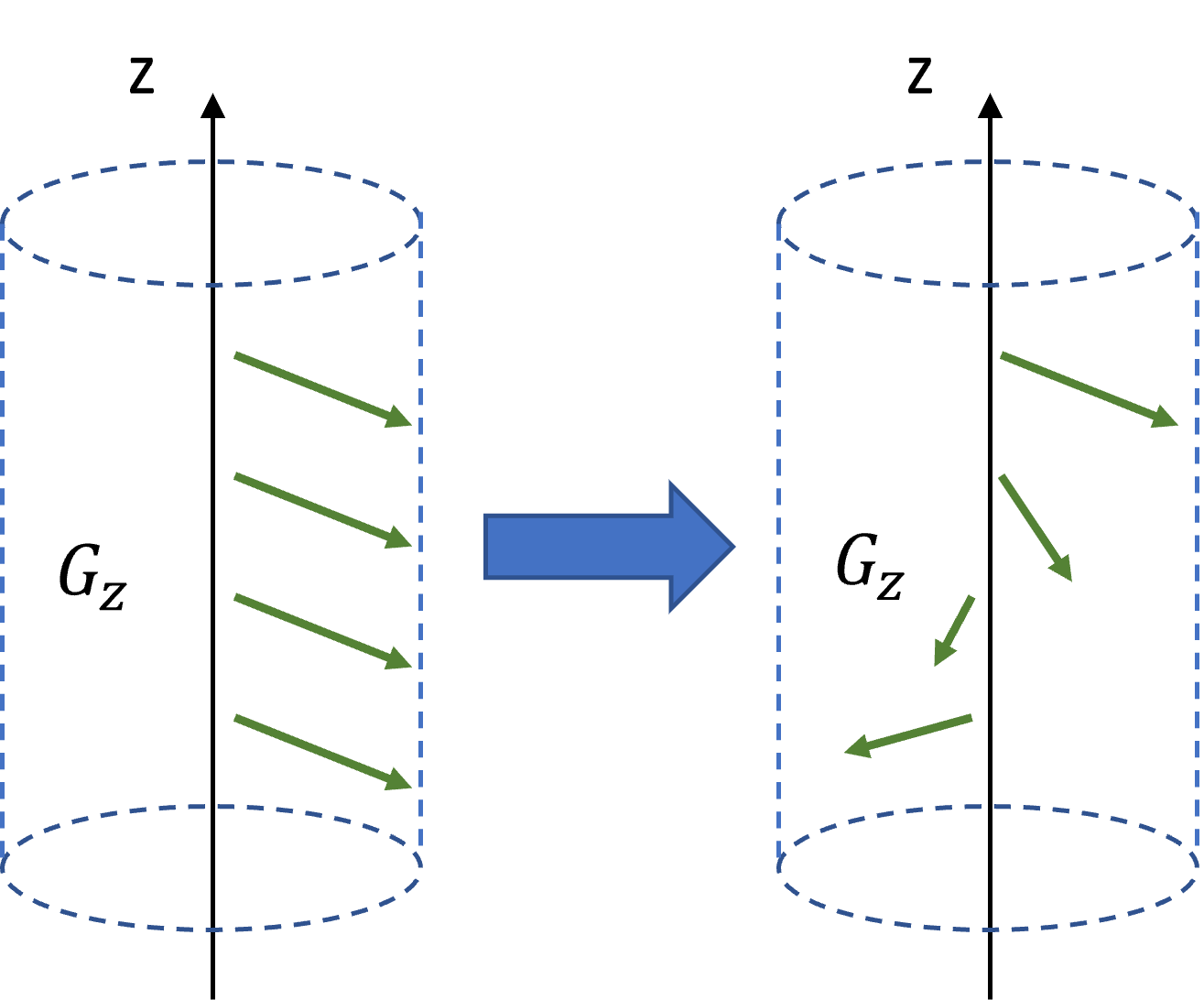}
}
\caption{Dephasing under gradient field.}
\label{s3f14}
\end{figure*}

As an example, we use the spatial averaging method to realize the steps for the preparation of homonuclear two-qubit pseudo-pure state. To introduce the spatial averaging method, let us first briefly introduce a control technique in NMR: the gradient field method. The gradient field has the form $G_z=Az$, where $A$ is the strength of the applied gradient field. That is, the gradient field is not a uniform magnetic field: along the z-direction, the strength of this magnetic field is position-dependent. The gradient field results in different precession frequencies of the nuclear spins at different positions along the z direction. Assuming that the system has transverse magnetization before the gradient field is applied, after a period of time, the transverse magnetization vectors at different positions along the z direction rotate by different angles. The average effect of the entire ensemble is that the transverse magnetization decreases. If the gradient field is strong enough, and the time that the gradient field applied is long enough, the transverse magnetization of the system will become zero. This process is called dephasing, as shown in Fig. \ref{s3f14}.

In the spatial averaging method, the gradient field is used to eliminate the off-diagonal elements (that is, transverse polarization) generated in the process of adjusting the diagonal elements of the density matrix.

The density matrix of the thermal equilibrium state of the two-qubit homonuclear system is $\sigma_z^1+\sigma_z^2$ (ignore the identity matrix and the polarization coefficients). Here are the steps of the spatial averaging method, combining RF pulses, J-coupling evolution, and gradient fields:
\begin{align}
\rho_\text{eq}=&\sigma_z^1+\sigma_z^2\nonumber\\
\xrightarrow{R_x^2(\frac{\pi}{3})}\sigma_z^1 + \frac{1}{2} \sigma_z^2 &- \frac{\sqrt{3}}{2} \sigma_y^2 \xrightarrow{G_z}  \sigma_z^1 + \frac{1}{2} \sigma_z^2 \nonumber\\
\xrightarrow{R_x^1 \left( \frac{\pi}{4} \right) }\frac{\sqrt{2}}{2} \sigma_z^1 + \frac{1}{2}& \sigma_z^2 - \frac{\sqrt{2}}{2} \sigma_y^1  \xrightarrow {U_J \left( \frac{1}{2J} \right)} \frac{\sqrt{2}}{2} \sigma_z^1 \nonumber\\+& \frac{1}{2} \sigma_z^2 + \frac{\sqrt{2}}{2} \sigma_x^1 \otimes \sigma_z^2 \nonumber\\
\xrightarrow {R_y^1 \left( -\frac{\pi}{4} \right) }\frac{1}{2} \sigma_z^1 + \frac{1}{2} &\sigma_z^2 - \frac{1}{2} \sigma_x^1 + \frac{1}{2} \sigma_x^1 \otimes \sigma_z^2 \nonumber\\&+ \frac{1}{2} \sigma_z^1 \otimes \sigma_z^2 \nonumber\\
\xrightarrow {G_z}\frac{1}{2} \sigma_z^1 + \frac{1}{2}& \sigma_z^2 + \frac{1}{2} \sigma_z^1 \otimes \sigma_z^2\label{3.32}
\end{align}
$R_{x,y}^{1,2}$ is the RF pulse that rotates the 1 and 2 nuclei around the x and y axes.  $U_J$ is the evolution under the action of J coupling. The final state $\frac{1}{2} \sigma_z^1 + \frac{1}{2} \sigma_z^2 + \frac{1}{2} \sigma_z^1 \otimes \sigma_z^2$ is then a pseudo-pure state, which can be written as $2(\lvert 00 \rangle \langle 00 \rvert
- I^{\otimes 2}/4)$.

\subsubsection{Measurements in NMR}

In a liquid-state NMR system, a quantum register is not defined by a single molecule, but by an ensemble of many identical molecules. That is to say, in NMR quantum computing, any quantum state has many copies. Any measurement result obtained is an ensemble average.

The measurable physical quantity in the NMR system is the transverse spin magnetization of the sample, that is, the magnetization in the x-y plane. Or in other words, the measurable quantity is the spin angular momentum in the x-y plane. As the transverse magnetization precesses about the z-axis, the precession induces a current in the detection coil in the x-y plane. Since the sample system will eventually return to a thermal equilibrium state, the transverse magnetization will eventually decrease to zero, and the induced current in the coil thus will eventually become zero. This signal is the free induction decay signal (FID) we mentioned earlier. In the laboratory frame, the relationship between the FID signal and the system density matrix can be expressed as \cite{13}
\begin{align}
S(t) \propto \mathrm{Tr} \left( e^{-i\mathcal{H}t/\hbar} \rho e^{i\mathcal{H}t/\hbar} \sum_{k=1}^n \left( \sigma_x^k - i\sigma_y^k \right) \right). \label{3.33}
\end{align}
For a single-qubit system,
\begin{align}
S(t) \propto \mathrm{Tr} \left( \rho e^{i\omega t} \left( \sigma_x - i\sigma_y \right) \right).\label{3.34}
\end{align}
Performing Fourier transform on the above signal will obtain a peak at the frequency of Larmor frequency $\omega$, whose amplitude is determined by $\mathrm{Tr} \left( \rho \left( \sigma_x - i\sigma_y \right) \right)$. Therefore, by measuring the amplitude of the peaks of the Fourier transform spectrum, one can obtain the values of $\langle \sigma_x \rangle$ and $\langle \sigma_y \rangle$.
For multi-qubit systems, $\langle \sigma_x \rangle$ and $\langle \sigma_y \rangle$ for each qubit can be obtained from the peaks at its corresponding Larmor frequency.\\
{\bf Observables in NMR}

As mentioned in Section 2.2, any single-qubit density matrix can be written as a linear superposition of the Pauli matrices and the identity matrix
\begin{align}
\rho = & \frac{1}{2} I +\frac{1}{2} \langle \sigma_x \rangle \sigma_x + \frac{1}{2} \langle \sigma_y \rangle \sigma_y + \frac{1}{2} \langle \sigma_z \rangle \sigma_z, \label{3.35}\\
& I = \begin{pmatrix} 1 & 0 \\ 0 & 1 \end{pmatrix} \label{3.36}\\
\langle \sigma_x \rangle =&  \mathrm{Tr}(\rho \sigma_x), \langle \sigma_y \rangle = \mathrm{Tr}(\rho \sigma_y), \langle \sigma_z \rangle = \mathrm{Tr}(\rho \sigma_z). \label{3.37}
\end{align}
For a quantum state, by measuring all three quantities in the Eq. (\ref{3.37}) and substituting them into the Eq. (\ref{3.35}), the density matrix of the quantum state can be obtained. This process is called quantum state tomography \cite{6,7}. This single-qubit quantum state tomography method can be extended to multi-qubit,
\begin{align}
\rho = \frac{1}{2^n} I^{\otimes n} + \frac{1}{2^n} \sum_{i_1, i_2, \cdots, i_n} c_{i_1, i_2, \cdots, i_n} \sigma_{i_1}^1 \sigma_{i_2}^2 \cdots \sigma_{i_n}^n,\nonumber\\ i_k = 0, x, y, z, \text{ and } (i_1, i_2, \cdots, i_n) \neq (0, 0, \cdots, 0) \label{3.38}
\end{align}
Here $\sigma_0$ is the identity matrix, $n$ is the number of qubits, and $c_{i_1, i_2, \cdots, i_n}=\text{Tr}(\rho \sigma_{i_1}^1 \sigma_{i_2}^2\cdots\sigma_{i_n}^n)$. Obtaining the Pauli coefficients $c_{i_1, i_2, \cdots, i_n}$ by measuring the corresponding multi-qubit Pauli matrices, one can then obtain the density matrix.

However, not all $c_{i_1, i_2, \cdots, i_n}$ can be read directly from the NMR spectrum. As mentioned earlier,  $\langle \sigma_x \rangle$ and $\langle \sigma_y \rangle$ of each qubit can be measured from the FID signal. That is to say, the first-order coherence terms of the quantum state can be directly readout. However, the zeroth-order and higher-order coherent terms cannot be read directly. Intuitively, the first-order coherence terms correspond to the transitions of one nuclear spin, and the higher-order coherence terms correspond to the transitions of multiple nuclear spins. In terms of Pauli matrices, the density matrix terms that can induce free induction decay signals have the following form
\begin{align}
&\sigma_{i_1}^1 \sigma_{i_2}^2\cdots\sigma_{i_{n-1}}^{n-1}\sigma_{i_n}^n,\nonumber\\
 i_k \in \{i_1,\cdots,&i_n\}, i_k= x \,\text{or} \, y \,\text{is true for only one} \, i_k.\label{3.39}
\end{align}
If more than one of $\{i_1,\cdots,i_n\}$ are x or y, then this matrix element cannot be measured by the FID signal. So how should the coefficients of these matrix elements be measured? One will need to use RF pulses to rotate some of the $\sigma_x$ and $\sigma_y$ to $\sigma_z$, such that the term will only involve one $\sigma_x$ and $\sigma_y$. As a result, the term after rotation can induce FID signal, so that the coefficient of the term can then be measured.\\
{\bf Examples of NMR observables}

\begin{figure*}
\centerline{
\includegraphics[width=3.3in]{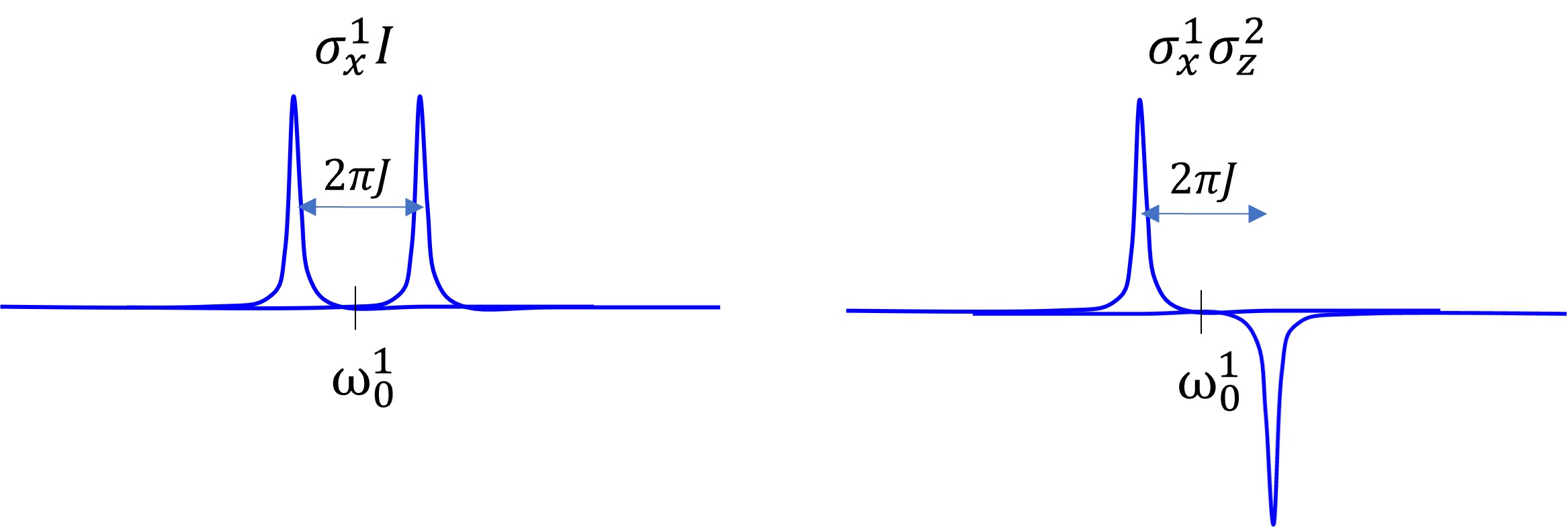}
}
\caption{Spectra of $\sigma_x^1 I$ and $\sigma_x^1 \sigma_z^2$.}
\label{s3f15}
\end{figure*}

In real experiments, once a spectrum is obtained, how can we obtain the coefficients of the Pauli matrix required in the Eq. (\ref{3.38})? Take two qubits as an example here. Suppose the Larmor frequencies of the two qubits are $\omega_0^1$ and $\omega_0^2$ respectively, then the spectrum near $\omega_0^1$ will be determined by the coefficients $c_{x,0}$, $c_{y,0}$, $c_{x,z}$, $c_{y,z}$ of $\sigma_x^1I$, $\sigma_y^1I$, $\sigma_x^1\sigma_z^2$, $\sigma_y^1 \sigma_z^2$. The spectrum near $\omega_0^2$ will be determined by the coefficients $c_{0,x}$, $c_{0,y}$, $c_{z,x}$, $c_{z,y}$ of $I\sigma_x^2$, $I\sigma_y^2$, $\sigma_z^1 \sigma_x^2$, $\sigma_z^1 \sigma_y^2$. The spectra of $\sigma_x^1 I$ and $\sigma_x^1 \sigma_z^2$ are given in Fig. \ref{s3f15}.

\begin{figure*}
\centerline{
\includegraphics[width=5in]{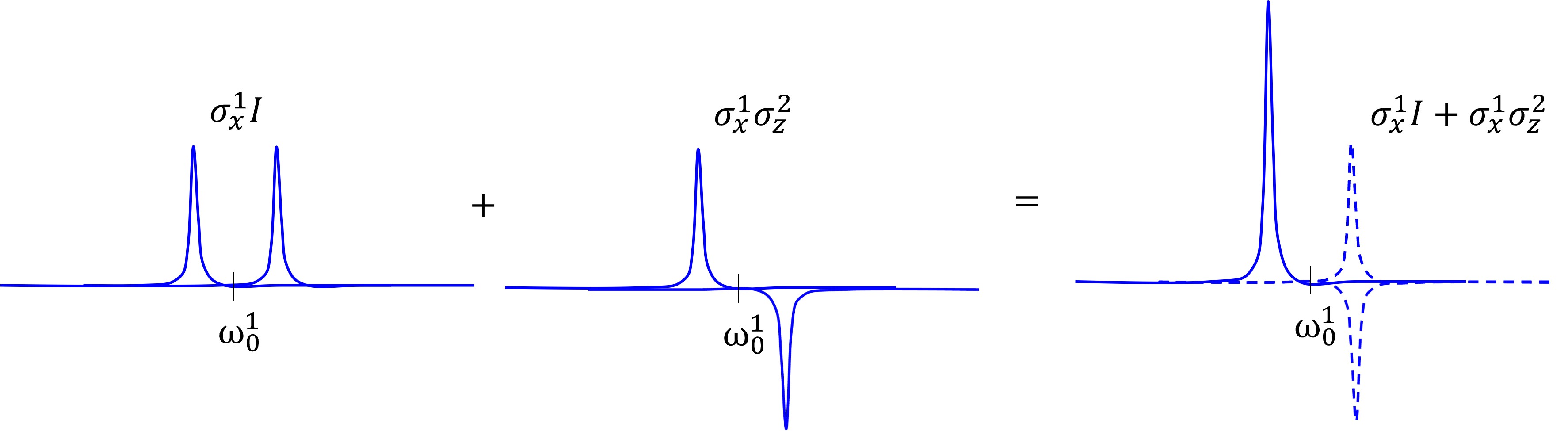}
}
\caption{Spectrum of $\sigma_x^1 I+\sigma_x^1 \sigma_z^2$.}
\label{s3f16}
\end{figure*}

If the density matrix of the quantum state is  $\sigma_x^1 I+\sigma_x^1 \sigma_z^2$, the corresponding spectrum is shown in Fig. \ref{s3f16}.

\begin{figure*}
\centerline{
\includegraphics[width=5in]{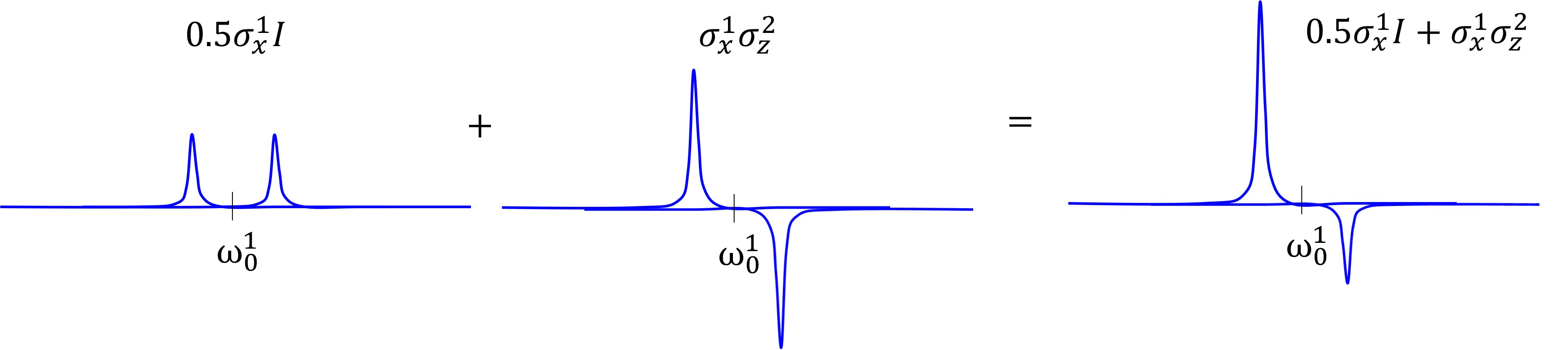}
}
\caption{Spectrum of $0.5\sigma_x^1 I+\sigma_x^1 \sigma_z^2$.}
\label{s3f17}
\end{figure*}

Notice that the right peak is cancelled, resulting in a single peak for $\sigma_x^1 I+\sigma_x^1 \sigma_z^2$. Similarly, spectrum of other superposition of  $\sigma_x^1 I$ and $\sigma_x^1 \sigma_z^2$ can also be obtained, for example $0.5\sigma_x^1 I+\sigma_x^1 \sigma_z^2$ (see Fig. \ref{s3f17}).

Notice that the spectrum of $\sigma_y^1 I$ and $\sigma_y^1 \sigma_z^2$ and that of $\sigma_x^1 I$ and $\sigma_x^1 \sigma_z^2$ will only differ in phase. Therefore, by analyzing the amplitude and phase of the peaks near $\omega_0^1$, the coefficients $c_{x,0}$, $c_{y,0}$, $c_{x,z}$, $c_{y,z}$ can be obtained. Similarly, by analyzing the amplitude and phase of the peaks near $\omega_0^2$, the coefficients $c_{0,x}$, $c_{0,y}$, $c_{z,x}$, $c_{z,y}$ can be obtained.

\subsubsection{Mixed state quantum computing}

\begin{figure*}
\centerline{
\includegraphics[width=4.5in]{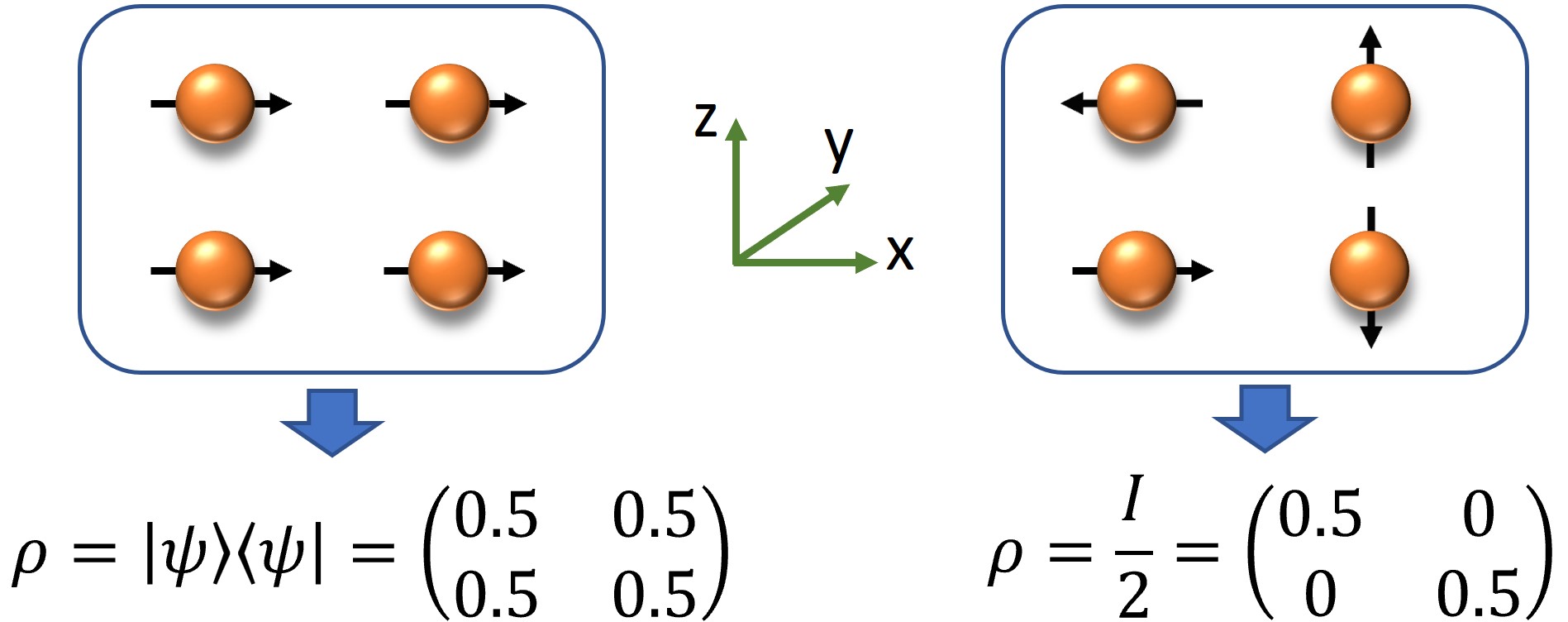}
}
\caption{Illustration of pure and mixed state systems. Taking a single-spin system as an example here, each system has many single-spin microscopic particles. Left: a pure state with spin polarization in the positive x direction, and the state of all microscopic particles is a quantum superposition state $|\psi\rangle=\sqrt{2}/2(|0\rangle+|1\rangle)$. Measuring this system, with probability 1/2 one will obtain $|0\rangle$, and probability 1/2 one will obtain $|1\rangle$. Right: the microscopic particles in this system have spins in all directions. It is impossible to use a state vector  $|\psi\rangle$ to describe this system. However, it can be described by the ensemble density matrix $\rho$. The diagonal elements represent the probability for a spin being in the $|0\rangle$  state and $|1\rangle$ state, respectively. This is a maximum mixed state, and the two diagonal elements of the density matrix are equal. Notice that the diagonal elements of the two density matrices are the same, which means that when measuring in the $|0\rangle$, $|1\rangle$ basis, the two states have the same probability of obtaining the $|0\rangle$ state and $|1\rangle$ state. However, the off-diagonal elements of the pure state in the left graph are non-zero. These off-diagonal elements of the density matrix contain important information to distinguish between pure and mixed states.}
\label{s3f18}
\end{figure*}

As mentioned above, in the thermal equilibrium state of the NMR system, the nuclear spins point to all directions. The number of spins in the direction along the magnetic field is slightly more than in other directions, thus forming the polarization along the direction of the magnetic field. Starting from this part of the polarization, a pseudo-pure state can be prepared as the initial state for quantum computing. Since it is a pseudo-pure state, it is not a real pure state, and the corresponding quantum state is in fact a mixed state. That is, an NMR system is a mixed-state system. Figure \ref{s3f18} demonstrates the difference between a pure state and a mixed state. In a pure state system, the quantum states of all microscopic particles are the same, while in a mixed state system, the states of microscopic particles are not uniform.

The state of the NMR system at room temperature is a state that is very close to the maximally mixed state, and the thermal equilibrium state is $\rho_{eq}$, as in Eq. (\ref{3.22}). The second term in $\rho_{eq}$, $\sum_{k=1}^n \frac{\epsilon_k \sigma_z^k}{2}$ is the sum of polarizations of different spins. As mentioned above, $\epsilon_k$ is determined by temperature, magnetic field strength, and the gyromagnetic ratio of nuclear spins, which is very small at room temperature, i.e. on the order of $10^{-5}$. After preparing the pseudo-pure state, the density matrix is then 
\begin{align}
\rho = \frac{(1 - \epsilon')}{2^n} I^{\otimes n} + \epsilon' \lvert \phi \rangle \langle \phi \rvert,\label{NMR}
\end{align}
where $\lvert \phi \rangle \langle \phi \rvert$ is some pure state. This is a more general expression than Eq. (\ref{3.31}).  Since some polarization will be lost during the preparation of the pseudo-pure state, usually $\epsilon' < \epsilon_k$. In Ref. \cite{114}, it is pointed out that, for a quantum state of the form $\rho$ in the above quation, even if  $\lvert \phi \rangle \langle \phi \rvert$ is entangled, $\rho$ has no quantum entanglement for $\epsilon'$ less than some threshold. For the current NMR experiment for quantum computing, $\epsilon_k$ is less than such a threshold. In this sense, there is in fact no entanglement. On the other hand, why should one care about entanglement? In some sense, entanglement has always been considered to be an important resource for quantum systems to be superior to classical systems in communication and computation. It is generally believed that in pure-state quantum computing, entanglement must exist if there could be an exponential speedup over classical computing \cite{115,116}. However, NMR system uses mixed states for quantum computing, so is it possible for mixed states without entanglement to be capable of performing computational tasks exponentially faster than classical computing?

\begin{figure*}
\centerline{
\includegraphics[width=5.5in]{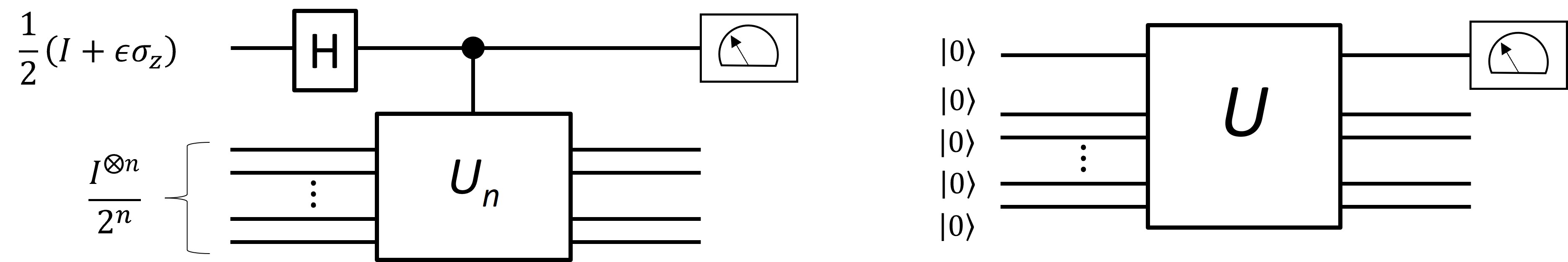}
}
\caption{DQC1 circuit diagram (left) compared with pure state quantum computing circuit diagram (right). For pure state quantum computing, the initial state is a pure state, generally all bits are in the $|0\rangle$ state. After the quantum gate operation $U$, the result can be obtained by measuring one or more qubits. In the initial state of DQC1, only one qubit has a non-zero polarization, and the other qubits are in the maximally mixed state. At the end of the calculation, the first qubit is measured.}
\label{s3f19}
\end{figure*}

The answer is affirmative. In 1998, Knill and Laflamme proposed a model of quantum computation based on mixed states, known as the "magic power of one bit" model, or DQC1 for short (see Fig. \ref{s3f19}) \cite{117}. This is an $(n+1)$-qubit model. For the initial state, only one qubit has non-zero polarization and is in state$1/2 (I+\epsilon \sigma_z )$, and the other $n$ qubits are in the maximally mixed state $I^{\otimes n}/2^n$. This initial state is common in NMR systems. For example, it can be prepared starting from the thermal equilibrium state of an $(n+1)$-qubit, rotating $n$-qubit polarizations to the x-y plane and eliminating them with a gradient field. The DQC1 scheme is to perform a Hadamard gate on the first qubit, followed by performing a controlled $U_n$ gate, with the first qubit as the control qubit. The controlled $U_n$ gate means that when the first bit is $|0\rangle$, $U_n$ is not executed on the remaining $n$ qubits, and when the first bit is $|1\rangle$, $U_n$ is executed.  Knill and Laflamme proved that, for this system, one only needs to perform two experiments to calculate the trace of $U_n$ (summation of the diagonal elements of $U_n$). This is regardless of the dimension of $U_n$, i.e., the same protocol works for $2\times2$ or $2^{64}\times2^{64}$. The task of calculating the trace of $U_n$ cannot be computed efficiently with a classical computer. Imagine that one can easily calculate the sum of the diagonal elements of a $2\times2$ $U_n$ with a desktop computer, as after all a $2\times2$-dimensional matrix has only 4 elements and only two diagonal elements. However, a $2^{64}\times2^{64}$-dimensional matrix $U_n$ has $3.4\times10^{38}$ elements and $1.8\times10^{19}$ diagonal elements. Such a large amount of data would need a supercomputer to handle.

DQC1 is a very important model, and its proposal has inspired researchers to reconsider what makes quantum computing different from classical computing. Clearly, quantum entanglement cannot explain the quantum advantage of DQC1 here. Later, a quantum correlation called quantum discord was proposed \cite{118}. Quantum entanglement is a type of quantum discord. Some researchers have found that quantum discord can often be observed in the final state of the DQC1 model, so quantum discord is likely to be the key to explaining the quantum advantage of DQC1. There are also researchers who have different opinions. They found that not all $U_n$ will get a final state with quantum discord, and quantum discord cannot be used to explain the acceleration of the trace calculation of $U_n$. It has also been argued that the advantage of DQC1 stems from the entangling power of its quantum gate operation to entangle certain states of the first qubit and the maximal mixed state of other qubits. To date, what exactly gives the quantum computing model represented by DQC1 an advantage over classical computing, and what exactly is the most important quantum computing resource, remain a mystery that has not been fully solved, and researchers are continuing exploring this mystery.

\subsection{Quantum optimal control algorithms and GRAPE}

Quantum optimal control is a branch of quantum control \cite{119,120,121,122}, and quantum control theory is an extension of the control theory from the classical world to the quantum world. In the world dominated by classical laws of physics, the concepts, theories, and methods of system control are widely used in machinery, chemistry, electric power, aerospace, and other industries, which is almost ubiquitous. In recent decades, with the increasing level of control over microscopic particle systems, it is natural that many quantum system control problems have emerged. In particular, the ultimate goal of quantum computing, i.e. building practical quantum computers, has greatly contributed to the development of quantum control. Quantum computing puts forward higher requirements for the ability to control quantum systems. One of the most typical control problems is how to realize high-precision quantum logic gates. As mentioned above, in the NMR system, the RF pulse can realize the rotation of the nuclear spin qubits around the x and y axes by any angle, and then can realize any single-qubit gate. Combined with the coupling evolution between the nuclear spins, arbitrary multi-qubit gates can be implemented. However, the specific implementation methods of complex quantum gates, such as how to split into single-qubit rotation operations and J-coupling evolution, or how to implement a quantum gate in the shortest time possible, are all worthy of study. These tasks can be summarized as the following control problem: design the shape of the electromagnetic wave pulse to achieve the desired quantum gate operation for the system.

    Similar to the classical world that the motion of objects follows Newton's second law, in the quantum world, the dynamic evolution of quantum systems follows the Schrödinger equation \cite{11}. In Newton's second law, force plays a decisive role in the evolution of the system. In the Schrödinger equation, the physical quantity that acts similarly to force is the Hamiltonian, which is usually the total energy of the system. The Hamiltonian of the system itself is represented by $\mathcal{H}_0$, and the Hamiltonian generated by the control pulse is represented by $u(t)\mathcal{H}_1$, where $u(t)(0\leq t\leq t_p)$ represents the amplitude of the time-containing pulse. The evolution of the system is completely determined by $\mathcal{H}_0+u(t)\mathcal{H}_1$ and the initial state of the system. The evolution of a quantum system is usually represented by a unitary operator, called the evolution operator. Obviously, at time $t$=0, the evolution operator of the system is $U(0)=I^{\otimes n}$, where $I$ represents the identity transformation, i.e. nothing happens. Under the RF pulse drive, the evolution operator of the system is finally $U(t_p)$. Assuming that the target quantum logic gate is $\bar{U}$, we use a function $F$ called quantum gate fidelity to measure the closeness between the actual transformation $U(t_p)$ and the target transformation $\bar{U}$, namely $F(U(t_p), \bar{U}) = \left| \mathrm{Tr} \left( U(t_p) \cdot \bar{U}^{\dagger} \right) \right|^2 / 2^{2n}$, where $n$ is the number of qubits. $F = 1$ indicates that the control pulse can well realize the target logic gate, and the smaller $F$ the greater the difference between $U(t_p)$ and the target transformation $\bar{U}$. To summarize, we need to solve the time-containing pulse $u(t)$, such that $F(U(t_p), \bar{U})$ reaches the maximum value, and $U(t)$ satisfies the Schrödinger equation
\begin{align}
\frac{dU(t)}{dt} = -i \left( \mathcal{H}_0 + u(t) \mathcal{H}_1 \right) U(t). \label{3.40}
\end{align}

When the number of qubits is small, it is relatively easy to obtain an accurate solution to the above problem; when the number of qubits exceeds 2, it is difficult to obtain a mathematical analytical solution. Therefore, in practical situations, it is often necessary to search for the desired pulse by methods of numerical optimization. In 2005, N. Khaneja and S. J. Glaser et. al proposed the method of gradient ascent pulse engineering (GRAPE) \cite{123}. This method applies to real experiments effectively and has become the most commonly used quantum control optimization algorithm. In the following we discuss how to obtain optimal pulses to achieve the target operation based on the GRAPE algorithm.

\begin{figure*}
\centerline{
\includegraphics[width=4.5in]{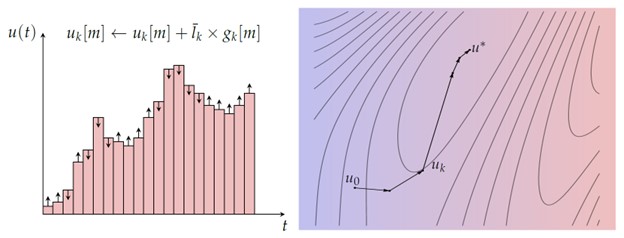}
}
\caption{Left: a schematic diagram of the pulse parameter (pulse amplitude) iterating along the direction of gradient, the subscript $k$ indicates that this is the $k$-th iteration; Right: after continuous iterations, the algorithm finally tends to a local optimal control solution $u^*$.}
\label{s3f20}
\end{figure*}

    In the GRAPE numerical optimization, we divide the total evolution time of the pulse equally into $N$ discrete segments. The duration of each segment is $\Delta t = t_p/N$, and the amplitude in the $j$th segment is denoted by $u(j)$. Since the amplitude of the RF field in each segment is a fixed value, the evolution operator $U_j$ of each segment of the pulse can be easily obtained according to the Schrödinger equation, so that the corresponding evolution of the entire pulse is $U(t_p)=U_N\cdots U_1$. Therefore, $U(t_p)$ is the total effect of the segment-by-segment evolution of the system driven by the RF pulses. The essence of the GRAPE algorithm is to regard the fidelity $F$ as a multivariate function of the parameter set ${u(j)}$, thereby transforming the RF pulse search into an extreme value optimization problem of a multivariate function (see Fig. \ref{s3f20}). In calculus, we know that a multivariate function value changes fastest by changing its parameters along the direction of its gradient. Therefore, the most important thing in the algorithm is to find the gradient of $F$ with respect to each parameter ${u(j)}$, such that ${u(j)}$ can be changed in a certain step according to the direction of the gradient, and then a new value of gradients of the $F$ function is calculated, and then iterate. Estimated to the first order, the gradients of $F$ are
\begin{align}
\frac{\partial F}{\partial u(j)} = -2 \mathrm{Re} \left[ \mathrm{Tr} \left( \bar{U}^{\dagger} \cdot U_N \cdots (-i \Delta t \mathcal{H}_1) U_j \cdots U_1 \right) \right]. \label{3.41}
\end{align}

The steps of the gradient ascent algorithm (GRAPE) are as follows:\begin{enumerate}
\item{} Initialization: set the initial value of the parameter ${u(j)}$, which can be randomly generated, or a set of parameters from a pulse sequence that is quite different from the target logic gate;

\smallskip

\item{} Find the gradient direction: for $j =1,...,N$, calculate the gradient $g(j)=  \partial F/(\partial u(j))$;
\smallskip

\item{} Find the step size: search along the gradient direction to find the step size $l$ that maximizes the rise of the function;

\smallskip

\item{} Change the parameters along the gradient direction, that is, calculate $F (u + lg)$. If the target requirement is not met, go back to step 2.

\end{enumerate}

The process will stop if that $F$ reaches the target requirement (for example, 0.9999), or the change of $F$ value before and after the iteration is less than a given small value (that is, the local optimal solution has been reached).

As an algorithm based on gradient optimization, the GRAPE algorithm can only give local extrema in principle, that is to say, it cannot guarantee to find the best pulse, but only the optimal pulse within a certain range of parameters. However, practice shows that the GRAPE algorithm performs well and can often give satisfactory solutions. It has become one of the core control techniques of NMR quantum computing and quantum simulation experiments.

\begin{figure*}
\centerline{
\includegraphics[width=4in]{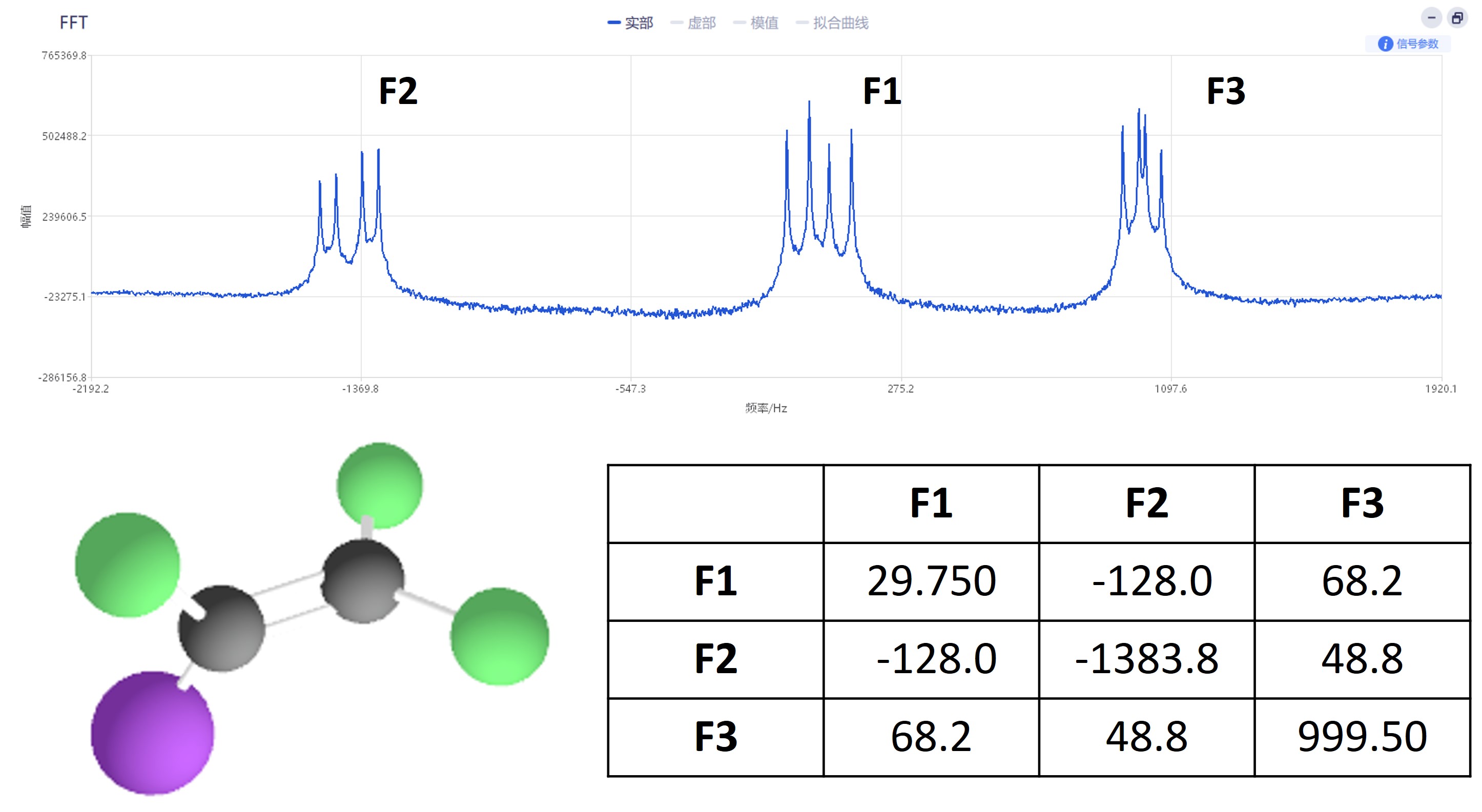}
}
\caption{The molecule structure, its parameter table, and the Fourier transform spectrum of $^{19}$F of a three-qubit NMR system. In the molecule, there are three $^{19}$F nuclear spins (green) used as three qubits. The diagonal elements of the table are the chemical shifts of the three $^{19}$F. The off-diagonal elements are their J couplings. All the values in the table are in unit of Hz [124].}
\label{s3f21}
\end{figure*}

Here we consider the Grape algorithm in a three-qubit NMR system as an example. This three-qubit system is the system used in the SpinQ Triangulum desktop platform \cite{124}. The molecule structure and NMR spectrum is shown in Fig. \ref{s3f21}. The internal Hamiltonian of this system is
\begin{align}
\mathcal{H}_0 = &2\pi \left( \nu_1 I_z^1 + \nu_2 I_z^2 + \nu_3 I_z^3 \right)\nonumber\\ &+ \sum_{k=x, y, z} \left[ 2\pi (J_{1,2} I_k^1 I_k^2 + J_{1,3} I_k^1 I_k^3 + J_{2,3} I_k^2 I_k^3) \right]. \label{3.42)}
\end{align}
Here, $\nu_1$, $\nu_2$ and $\nu_3$ are chemical shifts of the three $^{19}$F qubits. $J_{1,2}$, $J_{1,3}$ and $J_{2,3}$ are the J coupling constants between them. Those six parameters are given in Fig. \ref{s3f21}. For this system, because of the small frequency difference between the three $^{19}$F qubits, it is difficult to realize single-qubit operations using simple pulse shapes. Thus, Grape algorithm is implemented for pulse shape engineering. The RF Hamiltonian of this system is
\begin{align}
\mathcal{H}_1 = 2\pi u_x \left( I_x^1 + I_x^2 + I_x^3 \right) + 2\pi u_y \left( I_y^1 + I_y^2 + I_y^3 \right). \label{3.43}
\end{align}
The control parameter set ${u(j)}$ contains two elements at each segment $u(j)=\{u_x (j),u_y (j)\}$. Their gradients can be expressed according to Eq. (\ref{3.41}) as follows:
\begin{align}
\frac{\partial F}{\partial u_x(j)} &= -2 \mathrm{Re} [ \mathrm{Tr} ( \bar{U}^{\dagger} \cdot U_N  \nonumber\\&\cdots (-i \Delta t 2\pi \left( I_x^1 + I_x^2 + I_x^3 \right)) U_j \cdots U_1) ], \label{3.44}\\
\frac{\partial F}{\partial u_y(j)} &= -2 \mathrm{Re} [ \mathrm{Tr} ( \bar{U}^{\dagger} \cdot U_N \nonumber\\& \cdots (-i \Delta t 2\pi \left( I_y^1 + I_y^2 + I_y^3 \right)) U_j \cdots U_1 )]. \label{3.45}
\end{align}
Implementing the Grape algorithm, single-qubit rotation pulses of 1~2 ms length can be obtained with high fidelity $>$99.5\%.

     There are also protocols combining classical and quantum resources for pulse engineering. For example, gradients of a pulse can be measured directly from the quantum system under control \cite{83,125,126}. Such closed-loop optimization can take into consideration uncertainties in the system Hamiltonian and/or the distortion of the control pulse by the transmission line. In addition, such a hybrid optimization method can probably solve the problem that when a system is large the gradients cannot be calculated efficiently using classical computers.

It is worth noting that the formulation of the quantum optimal control problem and the numerical optimization algorithm introduced here are not limited to a specific system. These concepts and methods were initially developed in the context of quantum computing using NMR, but they can be applied more broadly. In this sense, the NMR system is a well-established and concise system for building quantum computers and can be used to develop a variety of experimental methods for other experimental quantum computing platforms.

\section{NMR quantum computing examples}

\begin{figure*}
\centerline{
\includegraphics[width=5.5in]{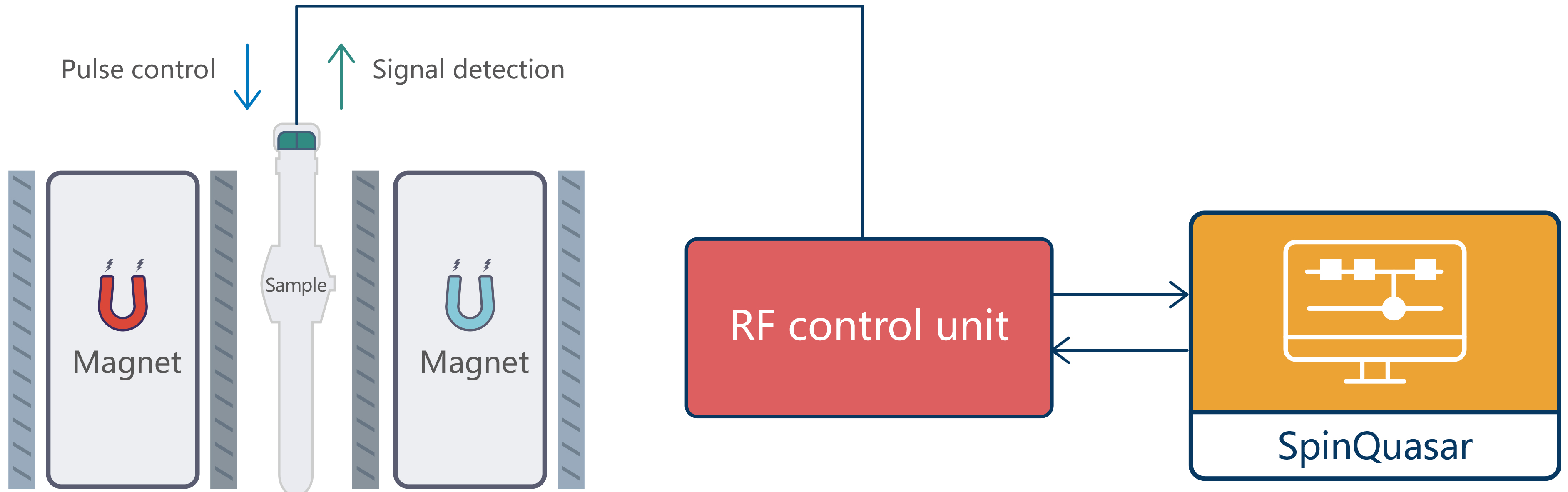}
}
\caption{Schematic structure of Gemini.}
\label{s4f1}
\end{figure*}

In this section, some basic quantum computing examples are introduced. They are demonstrated using the desktop NMR platforms \cite{124,127} developed and manufactured by SpinQ technology. The desktop NMR platforms include Gemini, Gemini Mini, and Triangulum. Gemini and Gemini Mini both accormodate a two-qubit system. Triangulum accormodates a three-qubit system. Other than the qubit number, Gemini and Triangulum have similar hardware structures.  Both Gemini and triangulum utilize a pair of permanent magnets to provide the static magnetic field required by NMR quantum computing (Fig. \ref{s4f1}). Quite different from traditional high field NMR spectrometers, Gemini and triangulum does not contain cryogenic systems to provide magnetic fields. Thus, no liquid nitrogen or liquid helium refills are needed. In addition, Gemini and triangulum realize a compact structure in which the magnets, probe, pulse generators and signal amplifiers are assembled in the main unit of the size 0.8*0.6*0.35m3 and 0.6*0.5*0.3m3.  Their weights are 50 Kg and 44 Kg. A PC acts as the interface between the user and Gemini or Triangulum, sending computing commands and displaying computing results. Gemini Mini has a more compact structure than Gemini and Triangulum. Smaller permanent magnets are used in Gemini Mini. In addition, a PC is not needed to control Gemini Mini. A touchpad is integrated in Gemini Mini as its user interface. Gemini Mini has a small size of 0.2*0.26*0.35 m3 and its weight is only ~14Kg.

\begin{figure*}
\centerline{
\includegraphics[width=5.5in]{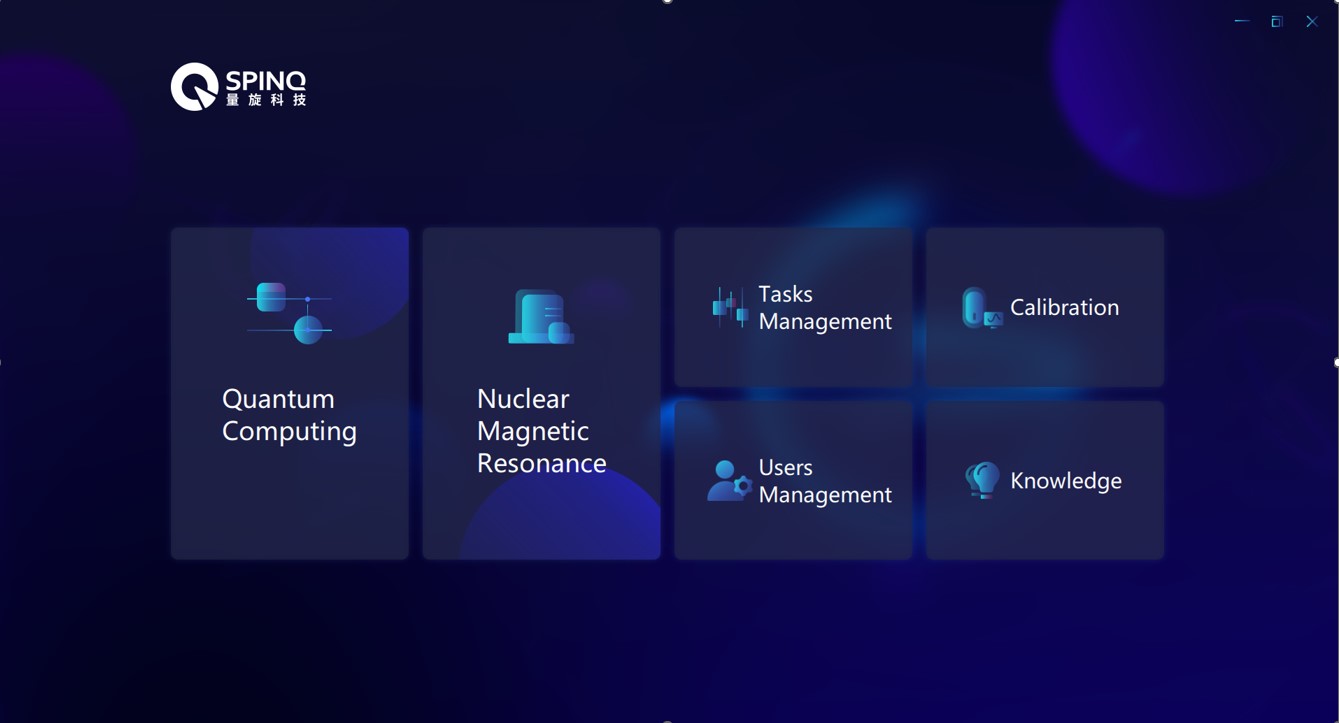}
}
\caption{SpinQuasar interface.}
\label{s4f2}
\end{figure*}

The softwere interface installed on Gemini and Triangulum is SpinQuasar (Fig. \ref{s4f2}). Spinquarsar includes several modules which are “Quantum computing”, “Nuclear magnetic resonance”, “Task management”, “Calibration”, etc. In the “Quantum computing” module, there are some built-in examples of famous quantum algorithms such as Grover algorithm. Users can also compose and implement their own circuits. In the nuclear magnetic resonance module, users can observe NMR FID signals and Rabi oscillations. In the calibration module, users can calibrate pseudo-pure states, etc. SpinQuasar also include a simulator which can simulate the quantum circuit for comparison with the experiments. 

\begin{figure*}
\centerline{
\includegraphics[width=3.5in]{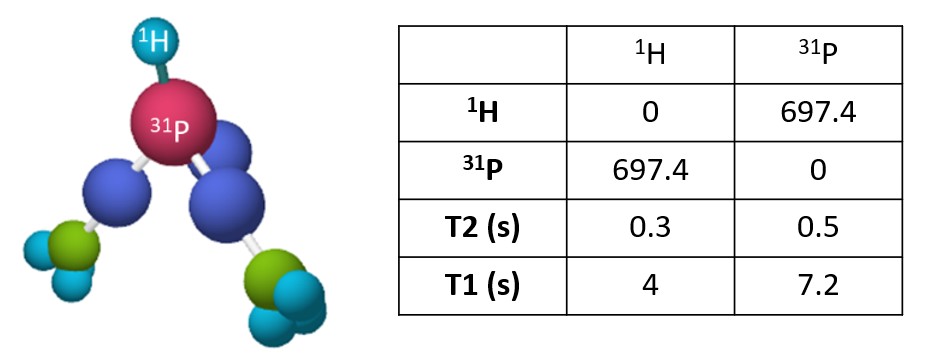}
}
\caption{The molecule structure (left) and its parameter table (right) in Gemini and Gemini Mini [127]. The J coupling between the $^1$H and $^{31}$P nuclear spins is 697.4 Hz. The control pulses are on resonance with $^1$H and $^{31}$P spins and thus their frequency offsets are both 0 Hz. }
\label{s4f3}
\end{figure*}

In this section, we introduce how to realize quantum computation using Gemini. Figure \ref{s4f3} shows the molecule structure and the Hamiltonian parameters of the quantum system in Gemini. The $^{31}$P and $^1$H are used as the two qubits.

\subsection{Observing quantum states}
It is very convenient to use SpinQuasar to observe single-qubit and two-qubit states.
\subsubsection{Experimental implementation}
\begin{figure*}
\centerline{
\includegraphics[width=5.5in]{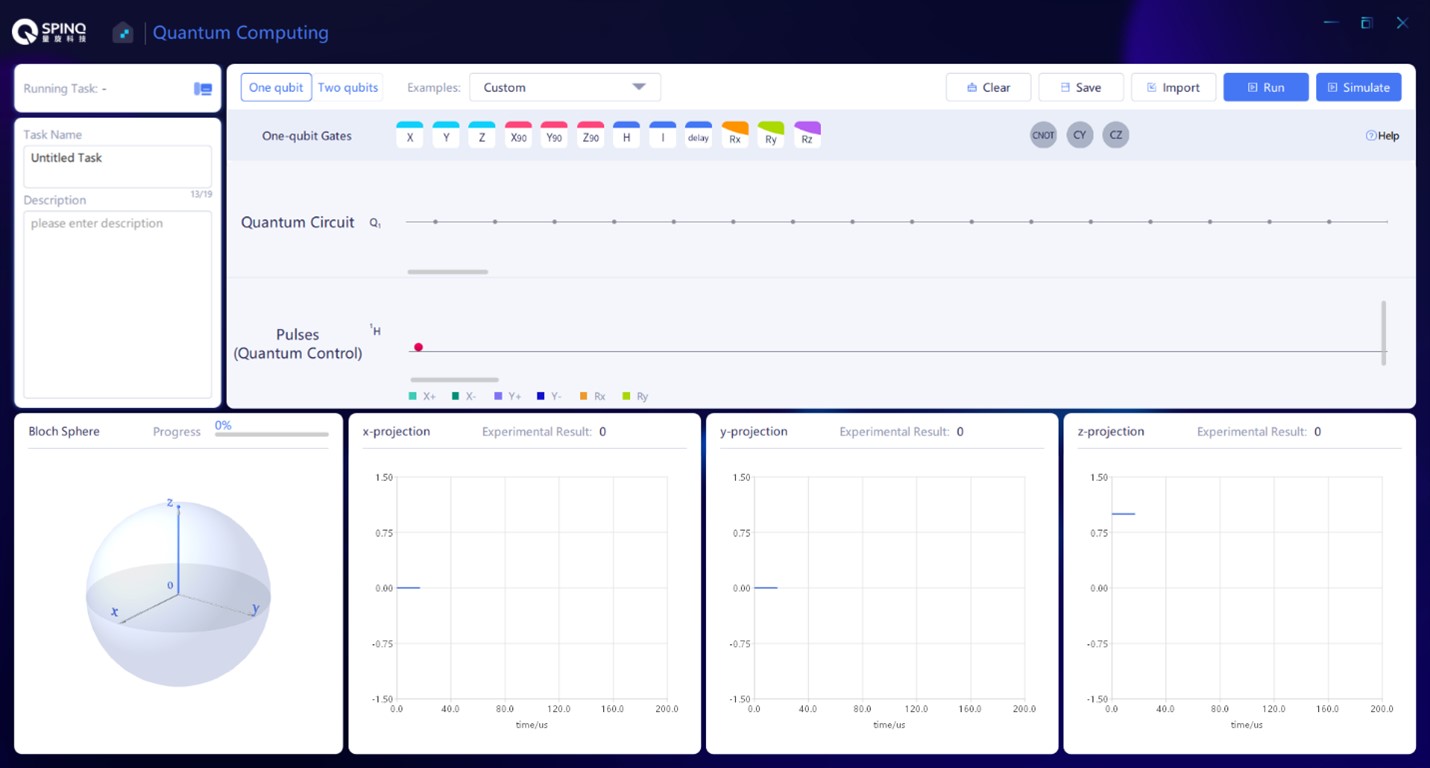}
}
\caption{SpinQuasar interface for single-qubit experiment.}
\label{s4f4}
\end{figure*}

\begin{figure*}
\centerline{
\includegraphics[width=5.5in]{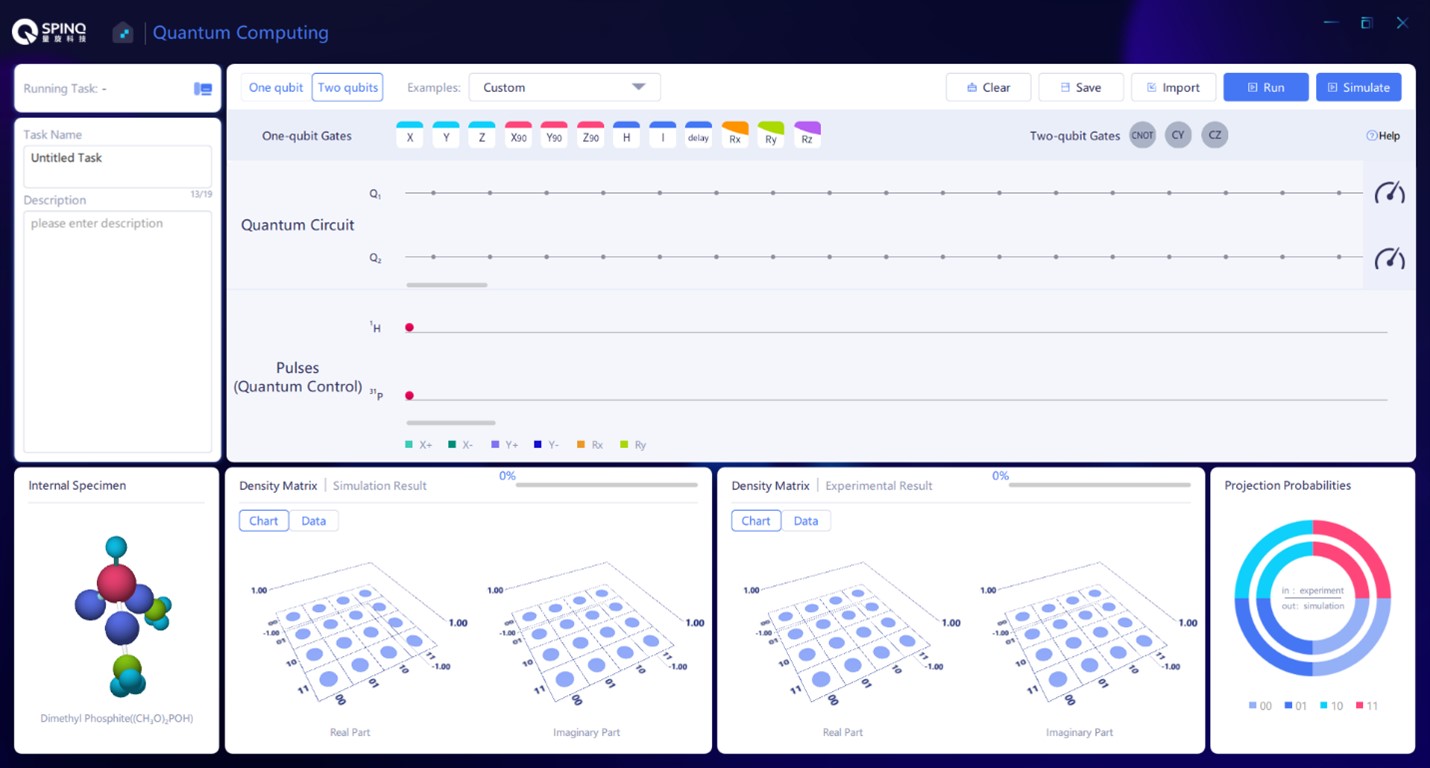}
}
\caption{SpinQuasar interface for two-qubit experiment.}
\label{s4f6}
\end{figure*}

Figures \ref{s4f4} and \ref{s4f6} shows the SpinQuasar interfaces for single-qubit experiment and two-qubit experiment on Gemini. 

In both interfaces, there are two buttons with “Run” and “Simulate” on them, which can be pressed to respectively activate the built-in experimental module and built-in numerical simulation module in SpinQuasar.

\begin{figure*}
\centerline{
\includegraphics[width=5.5in]{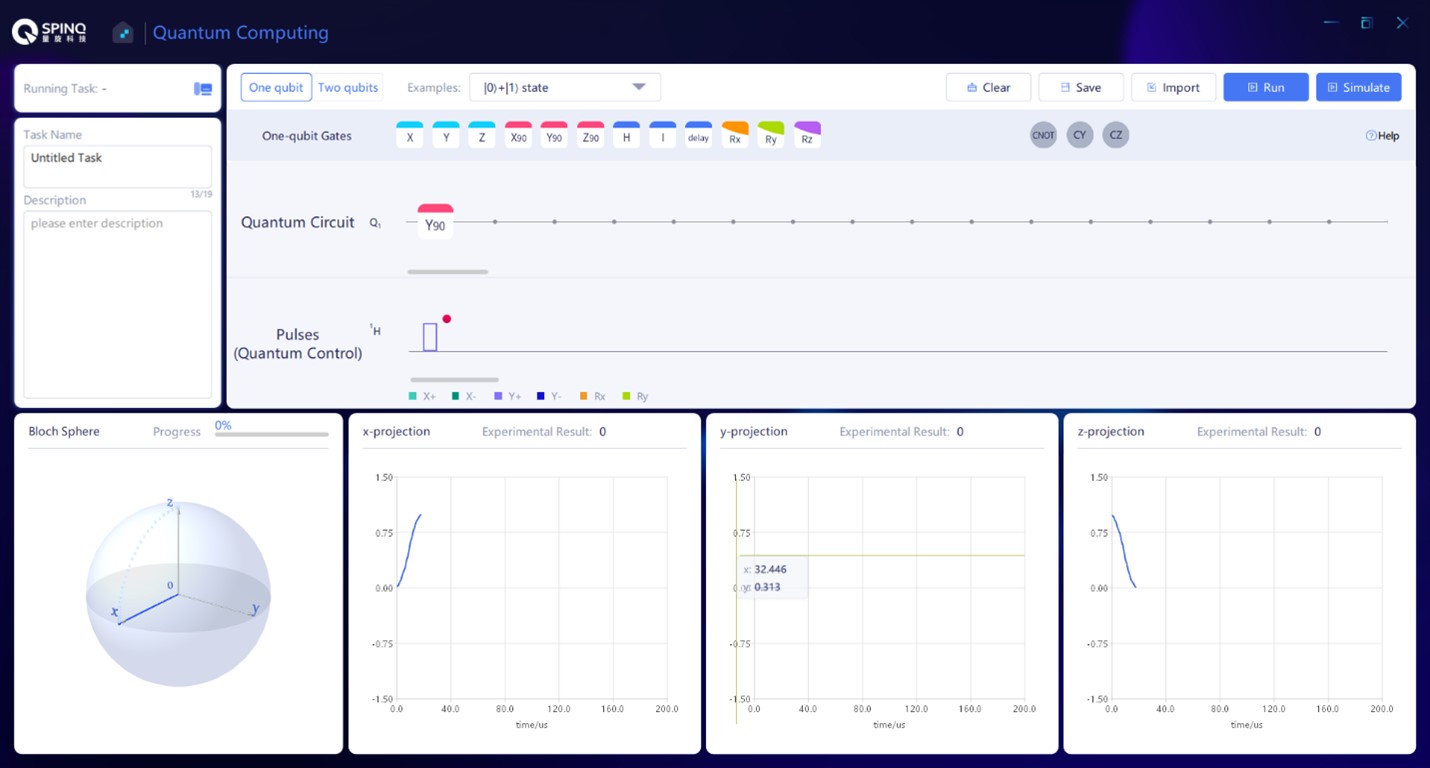}
}
\caption{The circuit of single-qubit state $\frac{1}{\sqrt{2}} |0\rangle + \frac{1}{\sqrt{2}} |1\rangle
$.}
\label{s4f5}
\end{figure*}

In the single-qubit experiment window (Fig. \ref{s4f4}), in the lower left half, there is a Bloch sphere, which can display the position of the quantum state in it. On the right side of the Bloch sphere, the projections of the quantum state in x, y and z directions are separately given (i.e., angular momentums in x, y and z directions). Users can choose a designated state from the drop-down lists. In Fig. \ref{s4f5} we use the state $\frac{1}{\sqrt{2}} |0\rangle + \frac{1}{\sqrt{2}} |1\rangle
$ as an example.  

\begin{figure*}
\centerline{
\includegraphics[width=5.5in]{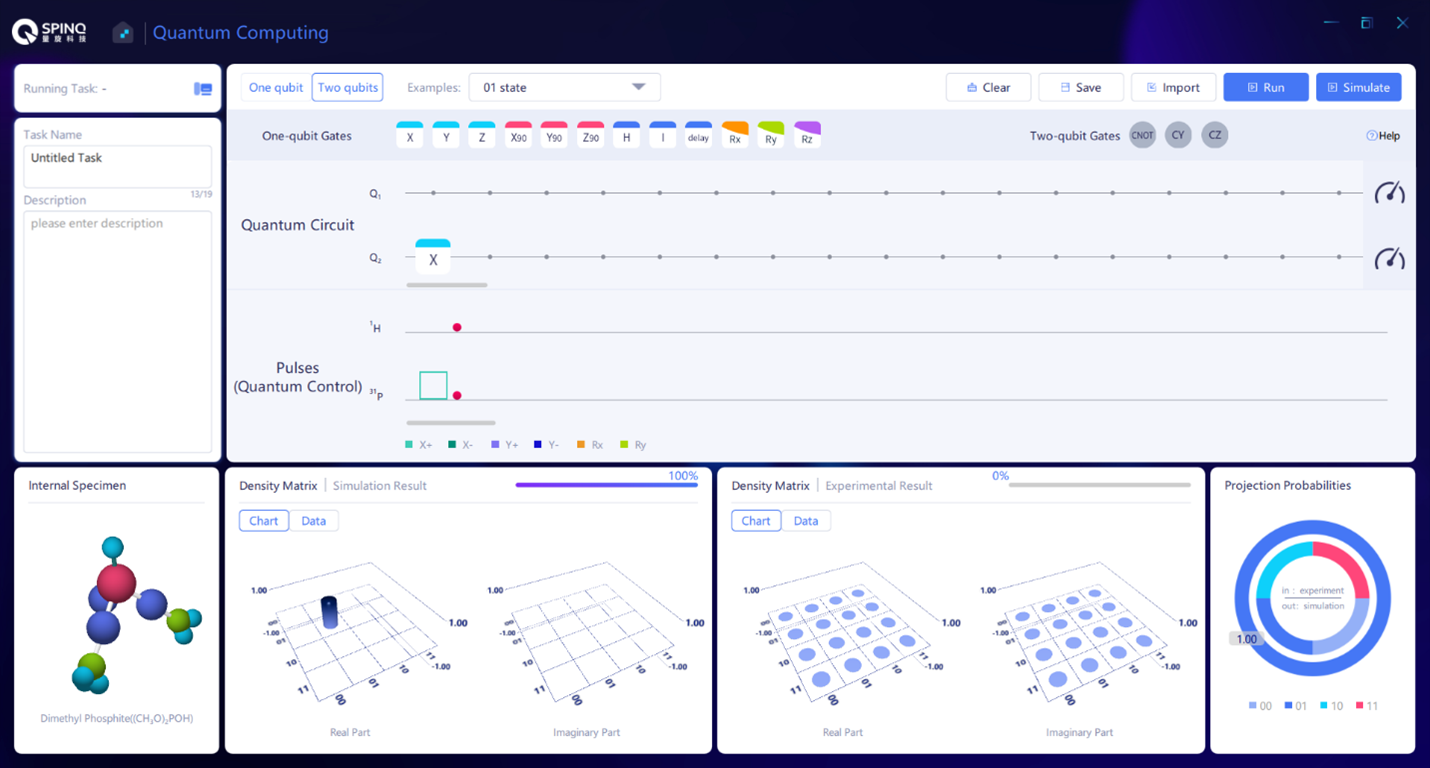}
}
\caption{The circuit of two-qubit basis state $ |01\rangle$.}
\label{s4f7}
\end{figure*}

In the two-qubit experiment window (Fig. \ref{s4f6}), in the lower left half, we can see the molecular structure of the two-qubit sample, and the diagrams on the right side of the molecular structure are the simulation density matrix, experimental density matrix and probability graph of the quantum state in the four basis states (see the probability by placing the cursor on the probability ring), respectively. We have introduced that the single-qubit basis states are $ |0\rangle$ and $ |1\rangle$. For multiple-qubit systems, its basis states are the direct product of single-qubit basis states. For the two-qubit system, the bases of its state space are $ |00\rangle$, $ |01\rangle$, $ |10\rangle$, $ |11\rangle$. Any two-qubit pure state is a unit vector in the four-dimensional complex vector space span by these four basis vectors. And the density matrices of the two-qubit system are thus $4\times4$ matrices, as shown in Fig. \ref{s4f6}. Users can also choose a designated two-qubit state from the drop-down lists. In Fig. \ref{s4f7} we use the state $ |01\rangle$ as an example.

\subsubsection{Discussions}

Readers can choose from the drop-down lists to realize different single-qubit and two-qubit states, for example, $ |0\rangle$ and $ |1\rangle$, and their uniform suppperpostion states,$ |00\rangle$, $ |01\rangle$, $ |10\rangle$, $ |11\rangle$ and their uniform superposition states. It would be interesting for the readers to explore the following question: In the single-qubit experiment and two-qubit experiment, which quantum state has the experimental result closest to the simulation result, and which quantum state has the largest difference between the experimental result and simulation result. This question is usually equivalent to the question: For those prepared states, whose gate sequence has the best and whose has the worst performance. And the performance of a gate sequence is determined by the number and quality of gates used.

\subsection{Single-qubit gates}

In classical computation, logic gates can be used to achieve binary addition and multiplication, etc., and then achieve various complex computation functions. Quantum computer uses quantum logic gates, which transform one quantum state to another quantum state. Quantum computation tasks can be carried out through a series of quantum gate operations on the quantum state. In quantum computation, single-qubit logic gates and the double-qubit CNOT logic gate constitutes a universal quantum gate set, i.e. the combination of single-qubit gates and CNOT gate can be used to achieve any quantum gate operation. The focus of this section is to give more details on basic single-qubit gates in the NMR system, and realize single-qubit gates using Gemini. With examples of single-qubit gate implementation, we also emphasize that any single-qubit gate can be achieved by combinations of rotation gates in x and y directions.

\subsubsection{Quantum state evolution and quantum gates}

We have introduced quantum gates in Section 2.4. Quantum gates are unitary operations, i.e. $UU^\dagger=I^{\otimes n}$, and transform a quantum state $|\psi(t_1) \rangle$ to another quantum state $|\psi(t_2) \rangle$, $|\psi(t_2) \rangle=U|\psi(t_1) \rangle$. The unitarity guarantees that the quantum state vector after $U$ is still normalized. The $U$ acting on a single qubit are known as single-qubit quantum gates, and that acting on multi qubits are known as multi-qubit quantum gates. The matrix form of the gate $U$ which acts on $n$ qubits has a dimension of $2^n\times2^n$. 

Assuming the Hamiltonian $\mathcal{H}$ does not change over time, By solving the Schrodinger equation in Eq. (\ref{1.16}), the following can be obtained
\begin{align}
|\psi(t_2)\rangle = e^{-i(t_2-t_1)\mathcal{H}/\hbar}|\psi(t_1)\rangle, \label{4.13}
\end{align}
Therefore, the evolution operator is $U = e^{-i(t_2-t_1)\mathcal{H}/\hbar}$. Specific quantum logic gates can be achieved by controlling $\mathcal{H}$ and $t_2-t_1$. In the next section, we will give more details about the Hamiltonian $\mathcal{H}$ in NMR that can realize single-qubit gates.  And the constant $\mathcal{H}$ assumption applies in the Gemini case.

The matrix form of several important single-qubit gates, such as Hadamard gate and P($\pi/4$) gate, is given in Eqs. (\ref{1.21}). Another important type of single-qubit gates is Pauli gates (Eq. (\ref{1.6})). $\sigma_x$ is the NOT gate, which changes $|0\rangle$ to $|1\rangle$ and changes $|1\rangle$ to $|0\rangle$. $\sigma_y$ not only flips the bit, but also changes the phase. $\sigma_z$ only changes the relative phase between $|0\rangle$ and $|1\rangle$ , and increase the relative phase by $\pi$. Based on Pauli gates, we can further obtain a very important type of gates, i.e. rotation gates as in Eq. (\ref{1.22}). Here we explicitly give the rotation gates around x, y and z axes:
\begin{align}
R_x(\theta) = e^{-i \frac{\theta \sigma_x}{2}},R_y(\theta) = e^{-i \frac{\theta \sigma_y}{2}},R_z(\theta) = e^{-i \frac{\theta \sigma_z}{2}}. \label{4.14}
\end{align}
The operations given in the above equation can be visualized in Bloch sphere, i.e. they can rotate the state vectors (which are also the vectors of spin angular momentum) by an angle of $\theta$ around x, y and z axes. It is easy to verify that a rotation of $\pi$ around x axis is equivalent to Pauli gate  $\sigma_x$, upon a global phase. Similarly, a rotation of $\pi$  around y (z) axis is equivalent to Pauli gate $\sigma_y$ ($\sigma_z$). 

For any single-qubit gate $U$, the Bloch theorem (Eq. (\ref{3.27})) exits, guaranteeing that only the rotations around x and y axes are required for achieving any single-qubit gate.

\subsubsection{Single-qubit gates in NMR}
As introduced in Section 4, the two-level nuclear spin system in the magnetic field is used as a qubit. The direction of the external static magnetic field $\boldsymbol{B}_0$ is usually along z axis by default. The matrix form of the Hamiltonian $\mathcal{H}_0$ for the interaction between the spin and $\boldsymbol{B}_0$ is given in Eq. (\ref{3.18}). The Hamiltonian has two eigenstates, i.e. $|0\rangle$ and $|1\rangle$, with the energy levels of $\frac{\omega_0 \hbar}{2}$
and  $-\frac{\omega_0 \hbar}{2}$, respectively. Since $|0\rangle$ and $|1\rangle$ are the eigenstates of $\mathcal{H}_0$, the system evolution under the effect of $\mathcal{H}_0$ does not change the probability distribution of spins on  $|0\rangle$ and $|1\rangle$.  Substitute the Hamiltonian in Eq. (\ref{4.13}) with $\mathcal{H}_0$, one gets
\begin{align}
U = e^{-i(t_2-t_1)\omega_0 I_z} = \begin{pmatrix} e^{-i(t_2-t_1)\omega_0/2} & 0 \\ 0 & e^{i(t_2-t_1)\omega_0/2} \end{pmatrix}. \label{4.20}
\end{align}
Assuming that the initial state is $|\psi(t_1)\rangle = a|0\rangle + b|1\rangle$, $|a|^2+|b|^2=1$, then $|\psi(t_2)\rangle$ can be obtained
\begin{align}
|\psi(t_2)\rangle = U|\psi(t_1)\rangle &= ae^{-i(t_2-t_1)\omega_0/2}|0\rangle+ be^{i(t_2-t_1)\omega_0/2}|1\rangle  \nonumber\\&= e^{-i(t_2-t_1)\omega_0/2}(a|0\rangle + be^{i(t_2-t_1)\omega_0}|1\rangle) \label{4.21}
\end{align}
 This is a more general form of Eq. (\ref{3.20}). $e^{-i(t_2-t_1)\omega_0/2}$ is the global phase and has no observable physical effects. It clearly shows, as $t_2$ gradually increases, the vector corresponding to $|\psi(t_2)\rangle $ does Larmor precession and the probability distribution of the spin on $|0\rangle$ and $|1\rangle$ does not change over time. In other words, if the qubit is in $|0\rangle$, it will always be in $|0\rangle$ under the effect of $\mathcal{H}_0$; if it is in $|1\rangle$ from the very beginning, it will always be in $|1\rangle$ under the effect of $\mathcal{H}_0$. Therefore, if only $\mathcal{H}_0$ exists, the switch between $|0\rangle$ and $|1\rangle$ cannot be achieved, let alone quantum computation.

\begin{figure*}
\centerline{
\includegraphics[width=3.5in]{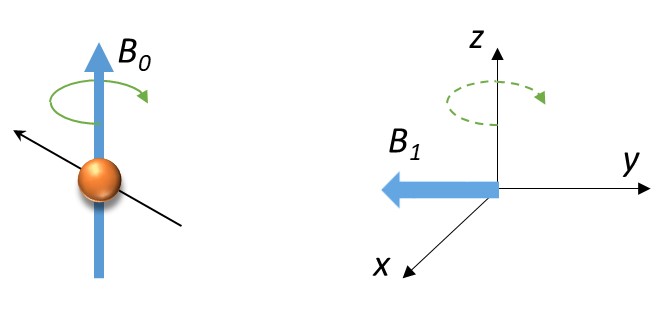}
}
\caption{The RF field $\boldsymbol{B}_1$ rotates around z axis at the same frequency as nuclear spin.}
\label{s4f8}
\end{figure*}

\begin{figure*}
\centerline{
\includegraphics[width=5.5in]{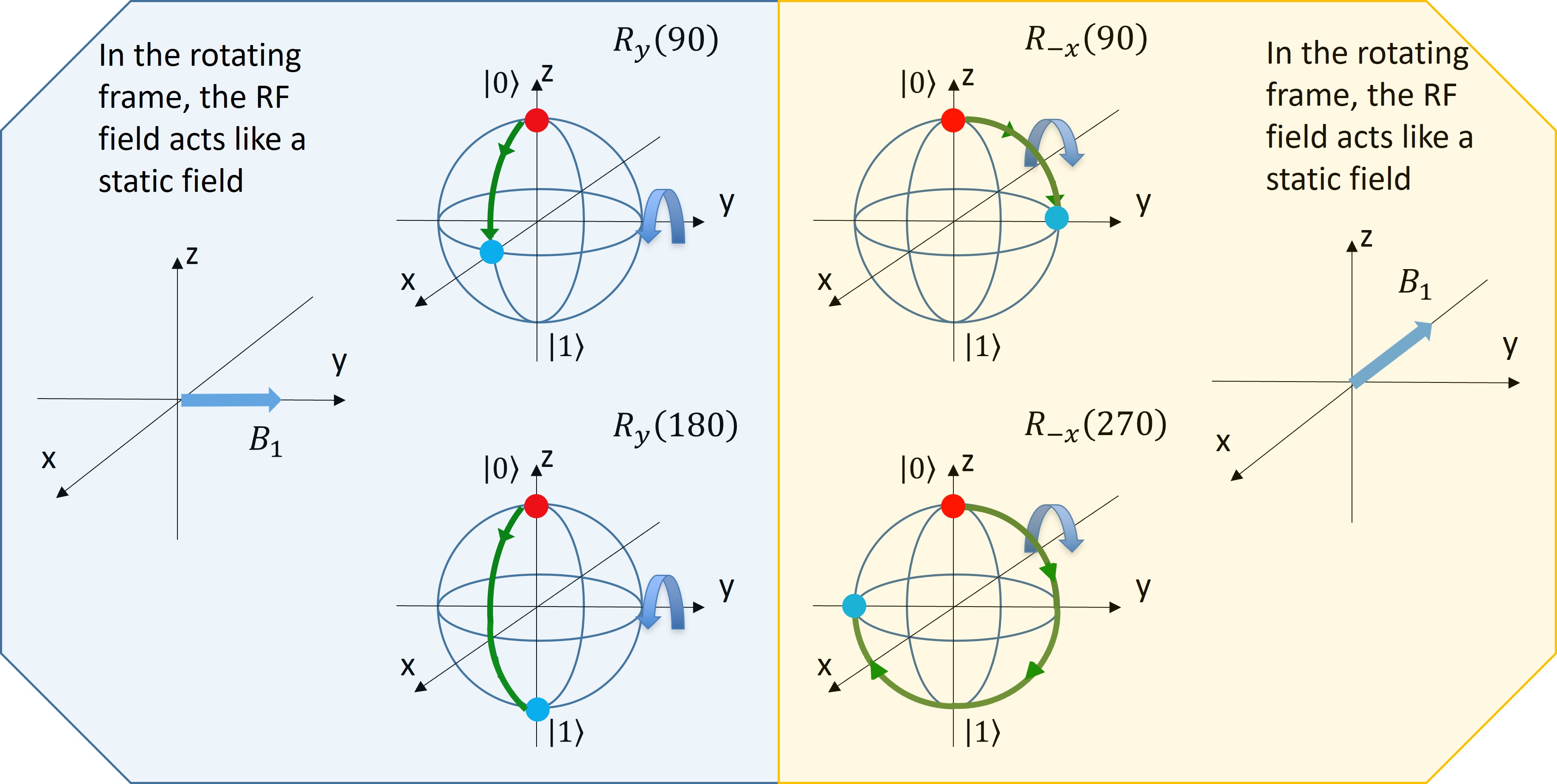}
}
\caption{$\boldsymbol{B}_1$ rotates spins with different phases and angles.}
\label{s4f9}
\end{figure*}

To achieve active control over nuclear spins, it is necessary to introduce the RF field $\boldsymbol{B}_1 (t)$ within x-y plane. $\boldsymbol{B}_1 (t)$ usually rotates around z axis at the same frequency as the Larmor frequency of the nuclear spin, namely to rotate at the same speed as nuclear spin, as shown in Fig. \ref{s4f8}. In the rotating frame that also rotates around z axis at the Larmor frequency, the direction of $\boldsymbol{B}_1 (t)$ is still. The Hamiltonian of this RF pulse in the rotating frame,  which is the on resonance case of Eq. (\ref{3.8}),  is as follows:
\begin{align}
\mathcal{H}_{rf}^{rot} = \omega_1 [\cos(\phi) I_x + \sin(\phi) I_y]. \label{4.16}
\end{align}
It is a Hamiltonian that makes the spin rotate around the direction of $\boldsymbol{B}_1 (t)$, and the phase $\phi$ can be chosen properly to set the rotation axis. As mentioned earlier, in Gemini, RF control feilds have constant amplitudes, namely, square pulse shapes. Substitute the Hamiltonian in Eq. (\ref{4.13}) with $\mathcal{H}_{rf}^{rot}$ in the above equation, one can obtain the expression in Eq. (\ref{NMRsingle}). When $\phi$=0, the spin rotates around x axis, and when $\phi$=90°, the spin rotates around y axis. $\omega_1 = \gamma B_1$ determines the rotation frequency. Figure \ref{s4f9} shows some examples: The first rotation is $\phi$=90°,  $\omega_1 t_p$=90°; the second rotation is $\phi$=180°, $\omega_1 t_p$=90°; the third rotation is $\phi$=90°, $\omega_1 t_p$=180°; the fourth rotation is $\phi$=180°, $\omega_1 t_p$=270°.

If the spin rotates around x or y axes, the probability distribution of the spin on $|0\rangle$ and $|1\rangle$ can be changed. Therefore, when the RF pulse is on resonance with the spin, the pulse can realize the switch of the nuclear spin between $|0\rangle$ and $|1\rangle$. In NMR, RF field is the most important means to control spins.

As analyzed above, it can be seen that as RF pulses can realize the single-qubit rotation gates around x and y axes, they can be used to achieve any single-qubit gate according to Equation (\ref{3.27}).

\subsubsection{Experimental implementation}

\begin{figure*}
\centerline{
\includegraphics[width=5.5in]{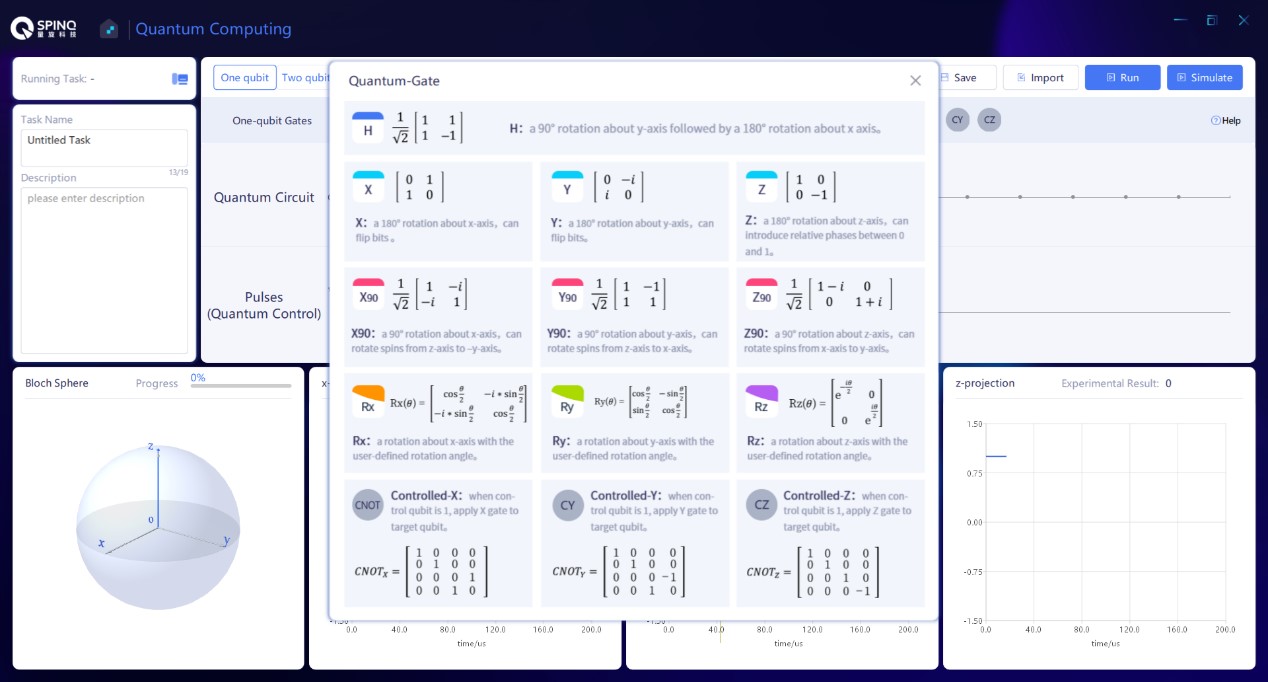}
}
\caption{Single-qubit gates in SpinQuasar.}
\label{s4f10}
\end{figure*}

\begin{figure*}
\centerline{
\includegraphics[width=5.5in]{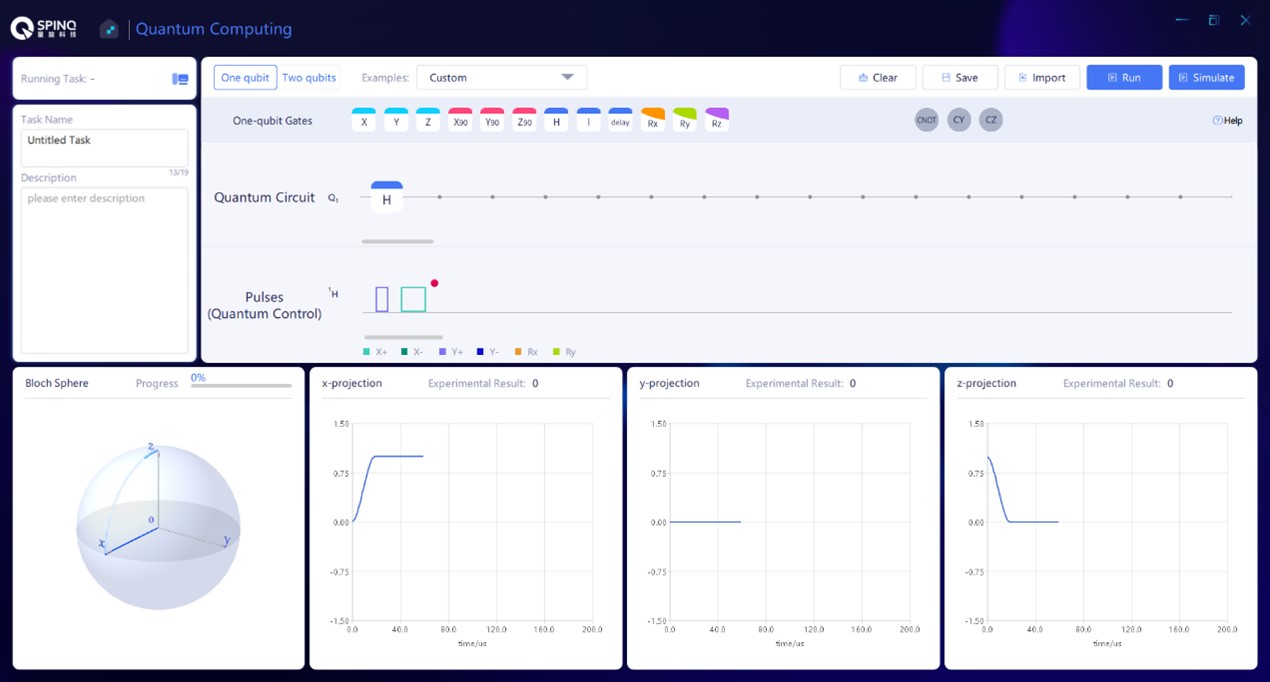}
}
\caption{H gate and its simulation result.}
\label{s4f11}
\end{figure*}

\begin{figure*}
\centerline{
\includegraphics[width=5.5in]{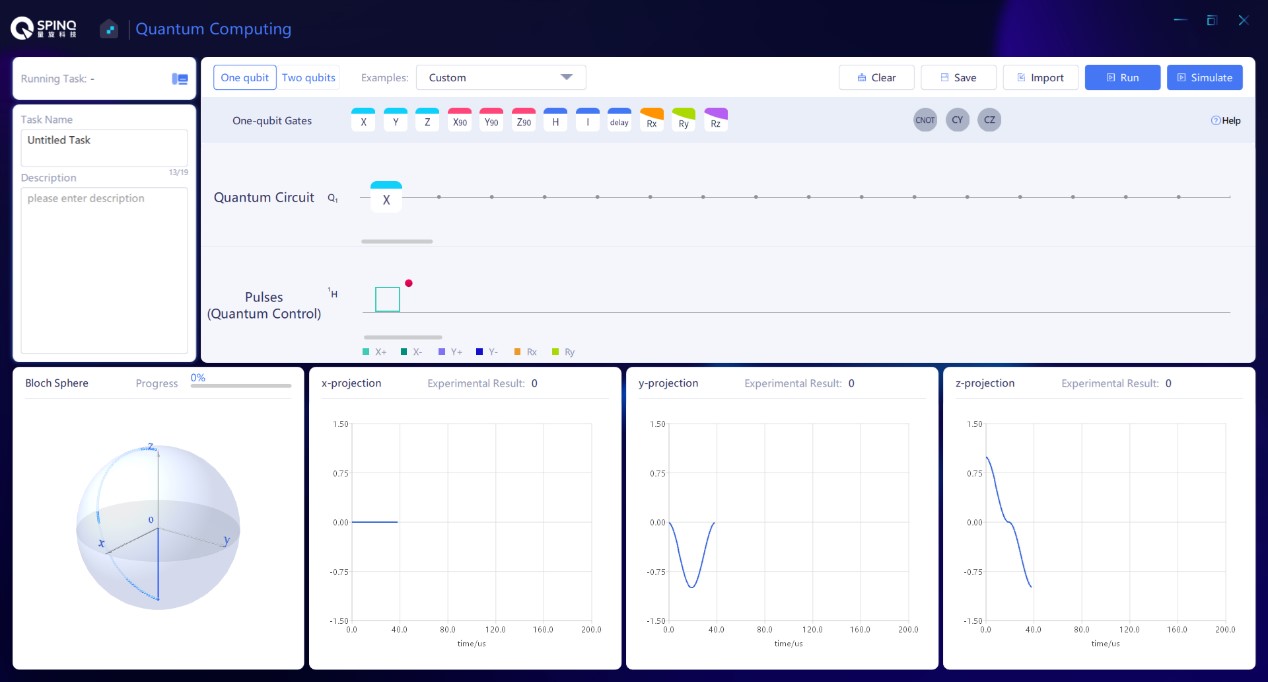}
}
\caption{X gate and its simulation result.}
\label{s4f12}
\end{figure*}

\begin{figure*}
\centerline{
\includegraphics[width=5.5in]{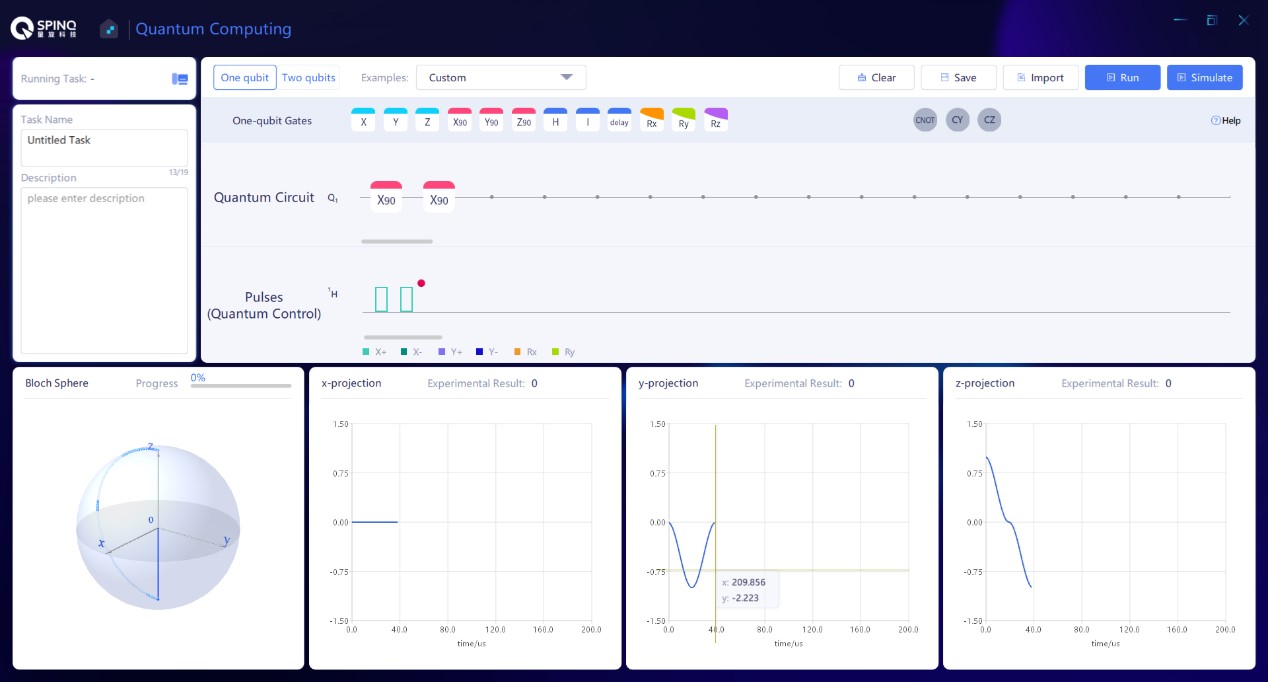}
}
\caption{Two X90 gates and their simulation result.}
\label{s4f13}
\end{figure*}

It is easy to implement different gates using Spinquasar. See Fig. \ref{s4f10} for the built-in single-qubit gates in Spinquasar. Here we use H, X90 and X gates as examples. As discssed above, H gate can create uniform superposition states of $|0\rangle$ and $|1\rangle$:
\begin{align}
\text{H}|0\rangle = \frac{|0\rangle + |1\rangle}{\sqrt{2}}, \quad
\text{H}|1\rangle = \frac{|0\rangle - |1\rangle}{\sqrt{2}}. \label{4.17}
\end{align}
They are called uniform superposition states because the probabilities are both 1/2 on $|0\rangle$ and $|1\rangle$. The difference between them lies in the relative phase of $|0\rangle$ and $|1\rangle$. 

X90 represents the gate that makes the spin rotate by 90° around x axis, $R_x(\pi/2)$, and its matrix form is as follows:
\begin{align}
\text{X90} = \frac{1}{\sqrt{2}} \begin{pmatrix} 1 & -i \\ -i & 1 \end{pmatrix}. \label{4.18}
\end{align}
It is easy to verify that X90 can also create a uniform superposition state starting from $|0\rangle$ or $|1\rangle$. The difference between the uniform superposition states created by X90 and that by H also lies in the relative phase between $|0\rangle$ and $|1\rangle$. In fact, all 90° rotation gates starting from $|0\rangle$ or $|1\rangle$ can create uniform superposition states, since all the states in the x-y plane of the Bloch sphere have equal probabilities of $|0\rangle$ and $|1\rangle$.

X gate, or $\sigma_x$ gate, is equivalent to the gate that makes the spin rotate by 180° around x axis, $R_x(\pi)$.  It is also equivalent to the operation of X90 twice, and the matrix form is as follows:
\begin{align}
\text{X}= \begin{pmatrix} 0 & 1 \\ 1 & 0 \end{pmatrix}. \label{4.19}
\end{align}

In SpinQuasar, not only the gate sequence is shown, but also the pulse sequence. According to Eq. (\ref{3.27}), any single-qubit gate can be achieved only by rotating the spins around x and y axes. The H, X90 and X gates are simple examples, and their pulse sequences only contain x and y rotation pulses, which is demonstrated in the “Pulses” panel below the quantum circuit (Figs. \ref{s4f11}-\ref{s4f13}). Also, we can use the following way to do a simple demonstration: to use X90 gate to achieve X gate. Theoretically, X can be achieved by two X90 gates. Readers are also advised to verify it in experiment.

All gates discussed above are single-qubit gates, while the spin system we use is a two-qubit system, whose initial state is in $|00\rangle$ state. We will use the first qubit only for the demonstration of single-qubit gates.

Figures \ref{s4f11}, \ref{s4f12}, \ref{s4f13} illustrate the realization of H, X, and two X90 gates. Their pulse sequences are shown in the “Pulses” panel. It clearly shows that all the required pulses are x and y rotation pulses. Specificly, for a H gate, a short y rotation pulse ($R_y(\pi/2)$) and a long x rotation pulse ($R_x(\pi)$) are implemented.  Comparing Fig. \ref{s4f12}, and Fig. \ref{s4f13}, the simulation results of one X gate and two X90 gates are the same, as expected.

\subsubsection{Discussions}

Different quantum computing platforms have different basic gates. For example, x and y rotation gates (pulses) are basic NMR operations. When composing a quantum circuit, it is usually necessary to optimize the basic gate (pulse) sequence for the best performance of the hardware, e.g. to reduce the number of basic gates. Here is an example. To implement a combination of gates, H, X90, X90, H, if an X180 has better performance than two X90 gates on a particular platform, then the squence should be optimized to H, X180, H. Readers can consider whether H, X180, H can be further optimized after translating H to basic gates of NMR quantum computing.

\subsection{Rabi oscillation observation and pulse calibration}

Rabi oscillation is a very important physical phenomenon, which refers to the behavior of a two-level quantum system under the action of a periodic driving field. It widely exists in condensed matter physics, atomic physics, high-energy physics, and other fields. In the NMR system, Rabi oscillations of nuclear spins can be observed under the action of RF pulses. In quantum computation, Rabi oscillation is particularly important for calibrating quantum gates, and only by calibrating the quantum gates can quantum computation be successfully realized.

\subsubsection{Rabi oscillation of nuclear spins}

\begin{figure*}
\centerline{
\includegraphics[width=3.5in]{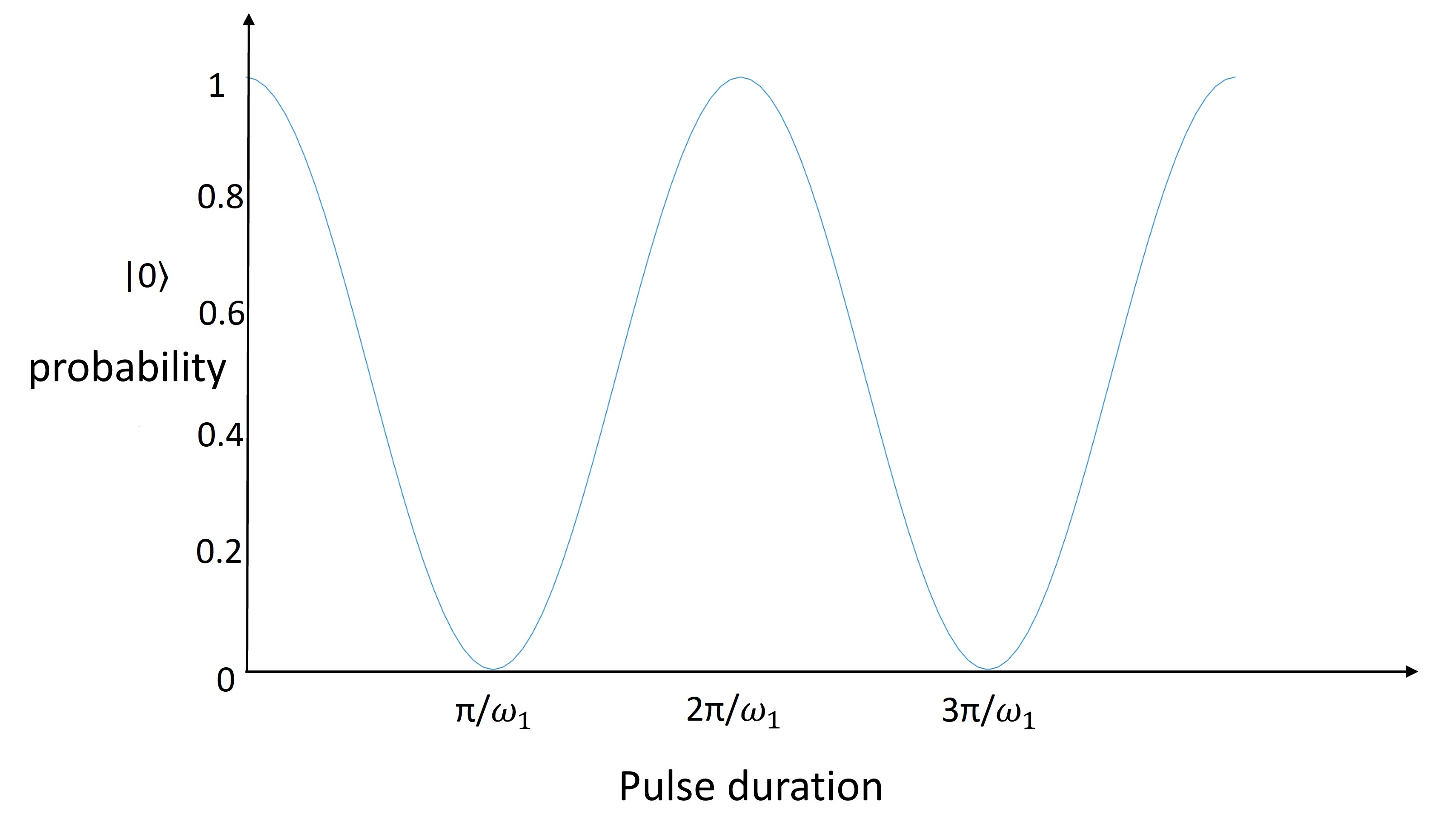}
}
\caption{The propabilities of $|0\rangle$ in Rabi oscillations.}
\label{s4f14}
\end{figure*}

As explained in last section, RF fields are used to realize single-qubit operations. The RF field Hamiltonian $\mathcal{H}_{rf}^{rot}$ is a Hamiltonian that makes spin rotate around the direction of $\boldsymbol{B}_1$. The phase $\phi$ determines the rotation axis, and $\omega_1=\gamma B_1$ determines the rotation frequency. The angle of rotation is denoted as $(t_2-t_1 ) \omega_1$. Assuming that $\phi=0$, $|\psi(t_1)\rangle=|0\rangle$, the following can be obtained by solving the Schrodinger equation:
\begin{align}
|\psi(t_2)\rangle = \cos\left(\frac{(t_2 - t_1)\omega_1}{2}\right)|0\rangle - i \sin\left(\frac{(t_2 - t_1)\omega_1}{2}\right)|1\rangle. \label{4.22}
\end{align}
Therefore, at $t_2$ moment, the probability that the quantum state is in $|0\rangle$ is $\cos^2\left(\frac{(t_2 - t_1)\omega_1}{2}\right)$, and the probability that the quantum state is in $|1\rangle$ is $\sin^2\left(\frac{(t_2 - t_1)\omega_1}{2}\right)$. It can be seen that the probabilities change over time, and the quantum state is oscillating between $|0\rangle$ and $|1\rangle$ states, which is Rabi oscillation (as shown in Fig. \ref{s4f14}). When $(t_2-t_1 ) \omega_1=\pi/2$, $|\psi(t_2)\rangle$ is the uniform superposition state of $|0\rangle$ and $|1\rangle$, $|\psi(t_2)\rangle = \frac{1}{\sqrt{2}}(|0\rangle - i |1\rangle)$. In Bloch sphere, its spin vector is along -y axis, that is to say, the RF pulse rotates the spin around x axis by 90° from z axis. When $(t_2-t_1 ) \omega_1=\pi$, $|\psi(t_2)\rangle$ is $|1\rangle$, which means the spin is rotated by 180° around x axis from z axis to reach -z axis.

\subsubsection{Pulse calibration}

\begin{figure*}
\centerline{
\includegraphics[width=3.5in]{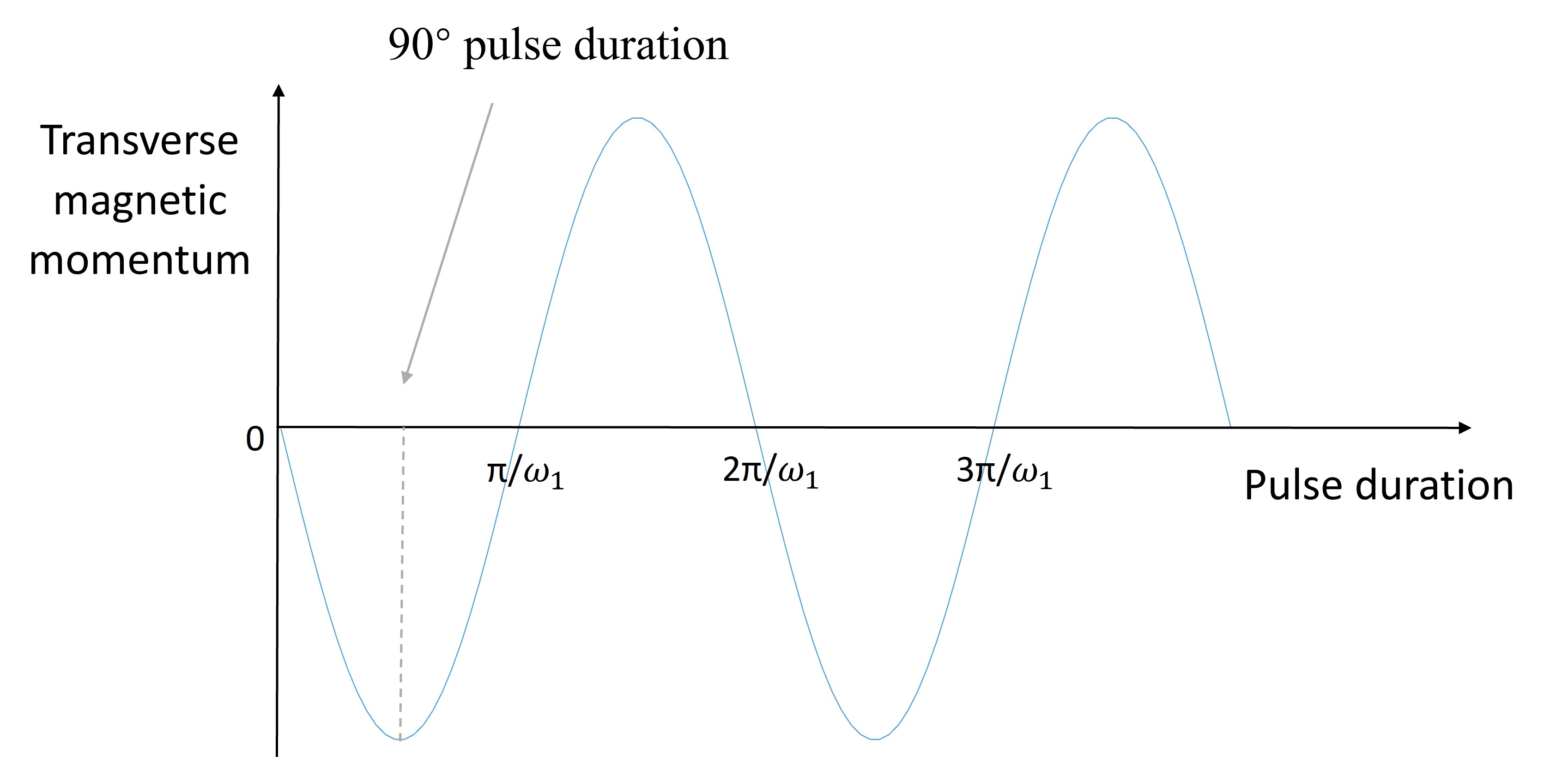}
}
\caption{The transverse magnetic momentum in Rabi oscillations.}
\label{s4f15}
\end{figure*}

In this section, we will consider how we should set the amplitude and duration of a RF pulse to achieve a specific rotation gate. The method is to calibrate the pulse by measuring Rabi oscillation. In the NMR system, the intensity of the nuclear spin signal is proportional to the transverse component of the magnetic moment of the nuclear spin, i.e. the magnetic moment within x-y plane. If the nuclear spin is in $|0\rangle$ state (z direction) or $|1\rangle$ state (-z direction), there is no signal. We have given the oscillation of the probability of $|\psi(t_2)\rangle$ in  $|0\rangle$ state with the application of a pulse in Fig. \ref{s4f14}. Figure \ref{s4f15} shows the oscillation of the transverse magnetic moment of  $|\psi(t_2)\rangle$. It is not difficult to understand that when the $|0\rangle$ state probability is 1 or 0, the absolute value of the transverse magnetic moment is the minimum, and when the $|0\rangle$ state probability of $|\psi(t_2)\rangle$ is 1/2, the absolute value of the transverse magnetic moment is the maximum. When calibrating a pulse and starting from $|0\rangle$ state, the RF pulse with a definite power and resonant with the spin is applied. The pulse duration is incremented to measure the curve as shown  in Fig. \ref{s4f15}. Then we will know the value of $(t_2-t_1)$ that can achieve a 90° or 180° rotation, or the rotation of any angle. This is the method for calibrating the pulse duration while fixing the pulse power. Also, we can fix the pulse duration and change the pulse power to measure the Rabi oscillation curve, which is the method for calibrating the pulse power while fixing the pulse duration. The choice of a specific method for pulse calibration is subject to the specific experimental requirements and purposes.

It is also important to note that the pulse calibration of the NMR system is not required to start from $|0\rangle$ state. The thermal equilibrium state can be used as the starting point, because the spin momentum of the thermal equilibrium state is also in the z direction, and the oscillation curve measured after a resonant pulse is applied is the same as the curve shown in Fig. \ref{s4f15}.

\subsubsection{Experimental implementation}

\begin{figure*}
\centerline{
\includegraphics[width=5.5in]{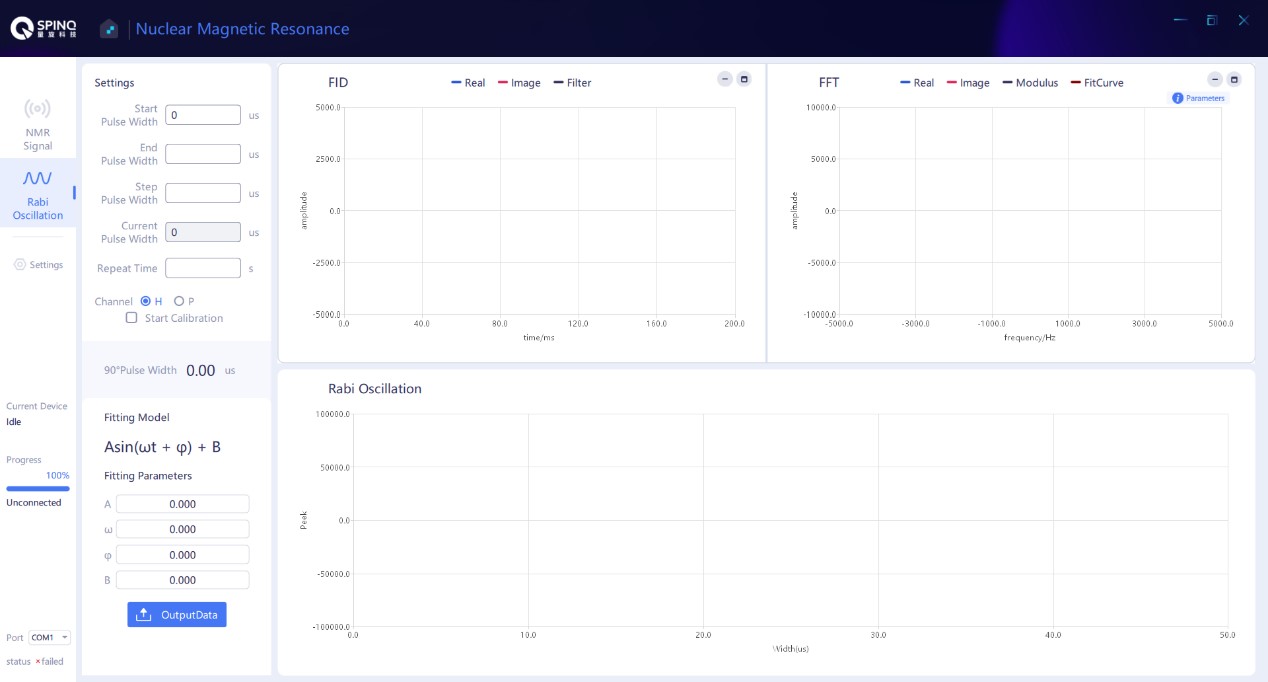}
}
\caption{Rabi oscillation interface in SpinQuasar.  Either $^1$H or $^{31}$P can be chosen by clicking the button in the pannel on the left. }
\label{s4f16}
\end{figure*}

In Spinquasar, Rabi oscillations can be observed (see Fig. \ref{s4f16}). The pulse power is fixed, and the pulse duration is changed to measure the curve of Rabi oscillation. A fitting of the observed data to a sine function is implemented to get the pulse lengths of 90° and 180° pulses of $^1$H and $^{31}$P nuclear spins.

\subsubsection{Discussion}
Theoretically, the width of 180° pulse is twice that of 90° pulse for a fixed amplitude, assuming the pulse shape is square. However, in experiments the pulses are usually not in ideal square shapes. For example, pulses have a finite rise time before reaching the desired amplitude and a finite tail before reaching zero amplitude. Therefore, the width of 180° pulse may not be exactly twice that of 90° pulse.

\subsection{Measurement of relaxation time}

Relaxation is known as an important phenomenon in quantum systems, which refers to the process of the system returning to the state of thermal equilibrium. Relaxation time refers to the characteristic time of the process of returning to the state of thermal equilibrium, which can be used to characterize the speed of the process. In NMR spectroscopy, the relaxation time measurement of nuclear spins has many uses, such as determining material composition, studying molecular structure and studying chemical reactions, etc. In quantum computation, relaxation time usually corresponds to the life of the qubit. 

In Section 2.5, we have introduced transverse and longitudinal relaxations and their charatistic times $T_2$ and $T_1$. In Sections 4.1.4 and 4.1.5, we also briefly discussed the mechanism of the two processes in NMR systems. Transverse relaxation makes a pure state degrade to a mixed state. It can reduce the fidelity of quantum gates or damage the stored quantum information. Thus, transverse relaxation is usually harmful to quantum computation. Longitudinal relaxation can change the population distribution on states $|0\rangle$  and $|1\rangle$ . Therefore, Longitudinal relaxation can also have a negative effect on quantum states and quantum gates. However, longitudinal relaxation can also be used to initialize a qubit by improving the polarization of the qubit to the value allowed by the environment, namely to obtain $|\langle \sigma_z\rangle|$ as large as possible, so as to make preparations for the following quantum computation.

\subsubsection{Measurement method of relaxation times}

As shown in Fig. \ref{s1f6}, in the longitudinal relaxation process and transverse relaxation process of nuclear spins, the change of the spin polarization generally happens in accord with an exponential function. If we can measure the longitudinal polarization ($\langle \sigma_z\rangle$) and transverse polarization ($\langle \sigma_x\rangle$ and $\langle \sigma_y\rangle$) of the spin in these two processes and fit with the exponential function, we can estimate $T_1$ and $T_2$.

\begin{figure*}
\centerline{
\includegraphics[width=3.5in]{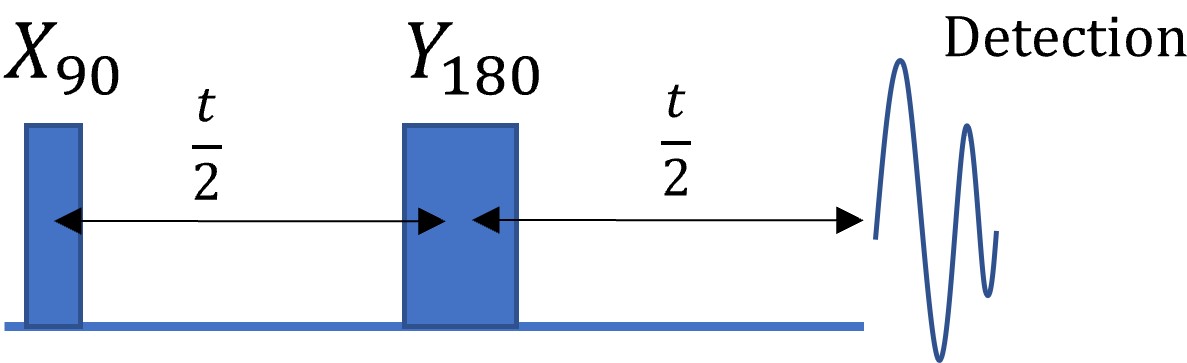}
}
\caption{The sequence for measuring $T_2$. }
\label{s4f17}
\end{figure*}

The pulse sequence shown in Fig. \ref{s4f17} is generally used to measure transverse relaxation time $T_2$, which is usually known as the pulse sequence of spin echo. The first 90° pulse refers to the process of rotating the spin from z direction in the state of thermal equilibrium to x-y plane. Then the spin polarization that gradually decreases with time within x-y plane is measured. It should be noted that the second pulse is a 180° pulse, and the spin polarization within x-y plane is still in x-y plane after the pulse. The function of this pulse is to eliminate the effect of static magnetic field inhomogeneity on the spin polarization. After the curve of transverse polarization over time is measured, the equation $Ae^{-t/T_2}$ shall be used for fitting, so as to estimate $T_2$.

\begin{figure*}
\centerline{
\includegraphics[width=3in]{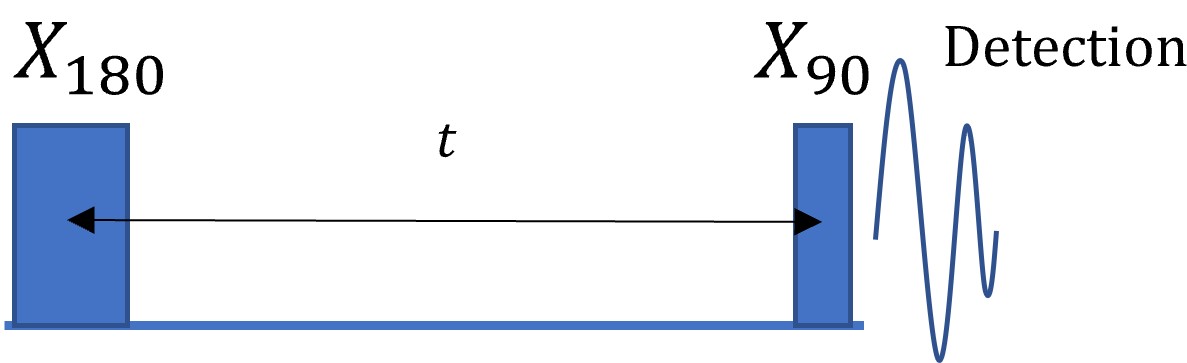}
}
\caption{The sequence for measuring $T_1$. }
\label{s4f18}
\end{figure*}

The pulse sequence shown in Fig. \ref{s4f18} is usually used to measure longitudinal relaxation time $T_1$. The first pulse is a 180° pulse, which rotates the spin from z direction in the state of thermal equilibrium to -z direction, so that it gradually changes over time under the effect of longitudinal relaxation. After a period, to measure the spin polarization, the spin should be rotated from z axis to x-y plane, so a 90° pulse is required. After the curve of longitudinal polarization changing over time is measured, the equation $B(1-2e^{-t/T_1})$ shall be used for fitting to estimate $T_1$. This equation for fitting is from Eq. (\ref{3.24}) and different from that given in Fig. \ref{s1f6}, because the initial value of longitudinal polarization in Fig. \ref{s1f6} is zero, while the initial value of longitudinal polarization measured in Fig. \ref{s4f18} is the opposite value of the polarization in the thermal equilibrium state.

\subsubsection{Experimental implementation}

\begin{figure*}
\centerline{
\includegraphics[width=5.5in]{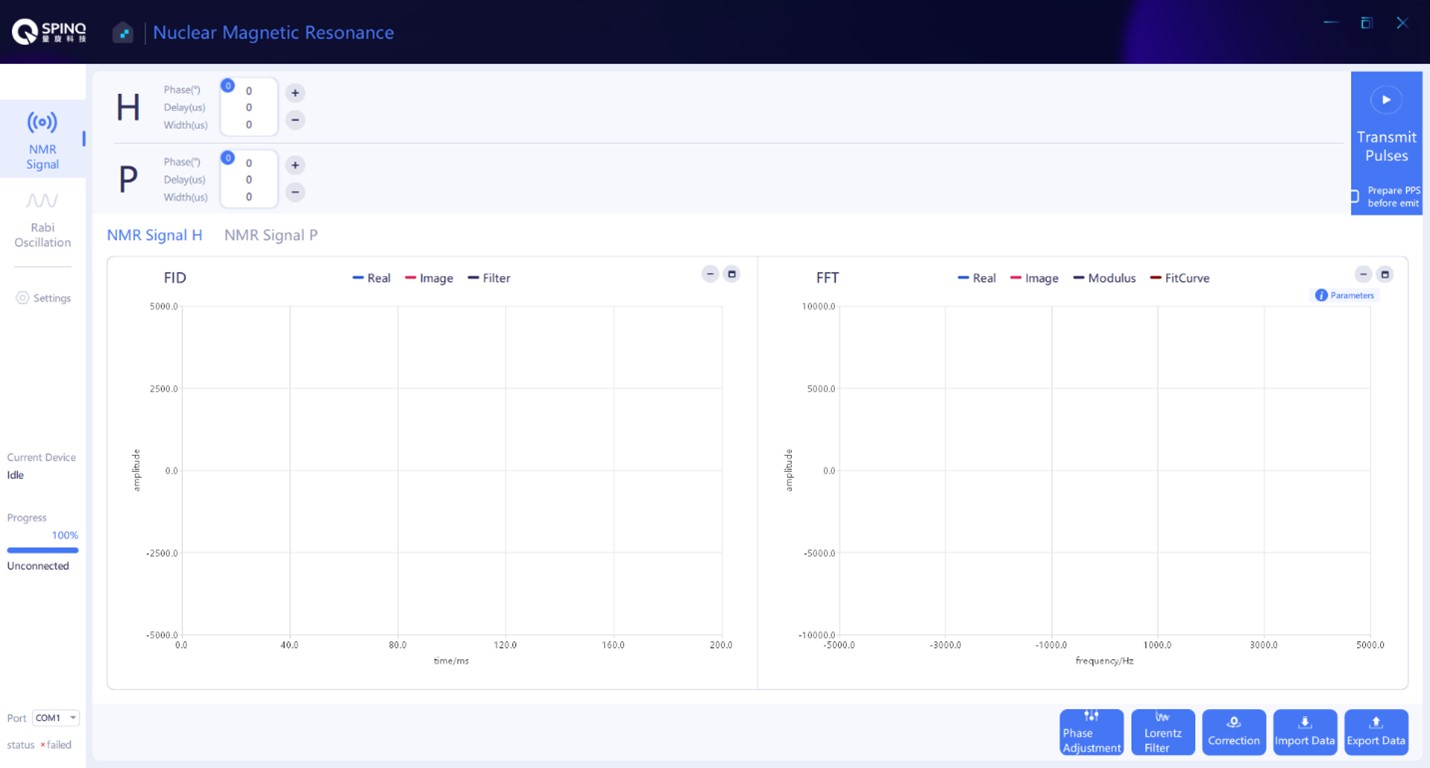}
}
\caption{The interface for composing the sequences of relaxation time measurements.}
\label{s4f19}
\end{figure*}

Figure \ref{s4f19} is the interface for composing the sequences to measure $T_1$ and $T_2$.

\begin{figure*}[!htbp]
\centerline{
\includegraphics[width=3in]{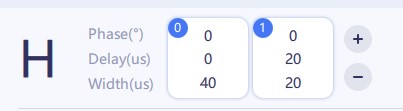}
}
\caption{Pulse sequence for $T_1$ measurement of $^1$H. }
\label{s4f20}
\end{figure*}

In the upper half of the interface, there is the panel for setting pulse sequences. Here, we suppose the calibration of 90° and 180° rotation pulses for $^1$H and $^{31}$P has been commpleted, and 90° and 180° rotation pulses have a duration of 20 us and 40 us for $^1$H  and 40 us and 80 us for $^{31}$P. 
Set the pulse sequence for measuring the longitudinal relaxation time of $^1$H, as in Fig. \ref{s4f20}.

\begin{figure*}[!htbp]
\centerline{
\includegraphics[width=3in]{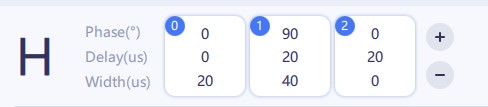}
}
\caption{Pulse sequence for $T_2$ measurement of $^1$H. }
\label{s4f21}
\end{figure*}

\begin{figure*}[!htbp]
\centerline{
\includegraphics[width=3in]{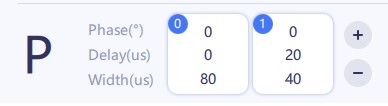}
}
\caption{Pulse sequence for $T_1$ measurement of $^{31}$P. }
\label{s4f22}
\end{figure*}

\begin{figure*}[!htbp]
\centerline{
\includegraphics[width=3in]{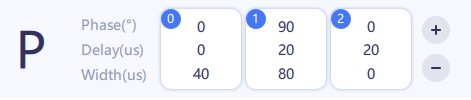}
}
\caption{Pulse sequence for $T_2$ measurement of $^{31}$P. }
\label{s4f23}
\end{figure*}

As for Pulse No. 0, its pulse width is 40 us and hence it is a 180° rotation pulse. Its phase is 0 (rotating around x axis), and the delay before this pulse is set to be 0. As for Pulse No. 1, its pulse width is 20 us hence it is a 90° rotation pulse. Its phase is 0 (rotating around x axis), and the delay before this pulse is $t$. In order to measure the longitudinal relaxation, several experiments need to be carried out with the delay before Pulse No. 1, $t$, increasing. As an example, we set $t$ in experiments to be 20 us, 50 us, 100 us, 200 us, 400 us, 1.2 ms, 4 ms, 12 ms, 50 ms, 200 ms, 1 s, 4 s and 15 s. For each value of $t$, click “Transmit pulse” on the right side to start the experiment. The experimental results will be displayed in the window below the pulse sequence pannel. Record the $^1$H signal strength obtained in each experiment.

Set the pulse sequence for measuring the transverse relaxation time of $^1$H, as in Fig. \ref{s4f21}.

Pulse No. 0 is a 90° rotation pulse with a zero phase  (rotating around x axis), and the delay before this pulse is set to be 0. Pulse No. 1 is a 180° pulse with a 90° phase (rotating around y axis), and the delay before this pulse is $t/2$. As for Pulse No. 2, the pulse width is set to 0 and the prepulse delay is $t/2$, which means Pulse No. 2 is not a real pulse but a delay of $t/2$. In order to measure the transverse relaxation, several experiments need to be carried out with $t/2$ increasing. As an example,  $t/2$ can be set as 10 us, 20 us, 40 us, 80 us, 160 us, 500 us, 1.5 ms, 5 ms, 20 ms, 80 ms, 320 ms and 1.5 s. For each value of $t/2$, click “Transmit pulse” on the right side to start the experiment. Record the $^1$H signal strength obtained in each experiment.

The pulse sequences for measuring longitudinal and transverse relaxation times of $^{31}$P can be set similarly as above. Figure \ref{s4f22} is the pulse sequence for measuring the longitudinal relaxation time of $^{31}$P.

Pulse No. 0 is a 180° rotation pulse, its phase is 0, and the delay before this pulse is set to be 0. Pulse No. 1 is a 90° rotation pulse, its phase is 0, and the delay before this pulse is $t$. Run experiments with $t$ changing. Here is an example sequence for $t$: 20 us, 50 us, 100 us, 200 us, 400 us, 1.2 ms, 4 ms, 12 ms, 50 ms, 250 ms, 1.2 s, 6 s and 20 s. For each value of $t$, record the $^{31}$P signal strength obtained in the experiment. Here $t$ values different from the $^1$H $T_1$ measurement are chosen, because $^{31}$P in Gemini has a longer $T_1$ than $^1$H does.

Set the pulse sequence for measuring the transverse relaxation time of  $^{31}$P, as in Fig. \ref{s4f23}.

Pulse No. 0 is a 90° rotation pulse, its phase is 0, and the delay before this pulse is set to be 0. Pulse No. 1 is a 180° pulse, its phase is 90°, and the delay before this pulse is $t/2$. Pulse No. 2 has a width of 0 and the prepulse delay is $t/2$. Run experiments with $t/2$ changing. For an example, $t/2$ can be chosen as 10 us, 20 us, 40 us, 80 us, 160 us, 500 us, 1.5 ms, 5 ms, 20 ms, 80 ms, 320 ms and 1.5 s. For each value of $t/2$, record the $^{31}$P signal strength obtained in the experiment.

$Ae^{-t/T_2}$ and $B(1-2e^{-t/T_1})$ shall be used to fit the transverse relaxation curve and longitudinal relaxation curve of $^1$H and $^{31}$P to estimate the values of $T_1$ and $T_2$.

\subsection{Measurment of truth table of CNOT gate}

In quantum computation, multi-qubit gates are also required because complex tasks cannot be accomplished by single-qubit gates only. There are many types of multi-qubit gates, and the most important one is the type of controlled-$U$ gates, namely to perform certain unitary operations on the target qubit/qubits conditioned upon certain states of the control qubit/qubits. CNOT gate is a controlled-$U$ gate in the case of two bits. One of the reasons that CNOT gate plays an important role is that single-qubit gates and CNOT constitute a universal gate set, i.e. the combination of single-qubit gates and CNOT can be used to achieve any quantum gate operation. Truth table is a concept frequently used in logic operation and classical computation, which is used to describe all possible states of input and output. In this section, we use the concept of truth table to describe the correspondence between different input basis states and output basis states of CNOT gate.

\subsubsection{Truth table of CNOT$_{12}$ and CNOT$_{21}$}

CNOT gate is a two-qubit gate, among these two qubits, one is known as the control qubit and the other is known as the target qubit. The effect of CNOT gate depends on the state of the control qubit, i.e. when the control qubit is in $|0\rangle$ state, no operation will be carried out on the target qubit; when the control qubit is in $|1\rangle$  state, Not gate will be carried out on the target qubit. When bit 1 is the control qubit and bit 2 is the target qubit, we denote the CNOT gate as CNOT$_{12}$.  The matrix form of CNOT$_{12}$ is given in Eq. (\ref{1.23}) and its effect is shown in Eqs. (\ref{1.24},\ref{1.25}).
Figure \ref{s1t1} can be considered as the truth table of  CNOT$_{12}$ with all the four basis states listed in the left columns as input states and their correspoding output states  in the right columns.

When bit 2 is the control qubit and bit 1 is the target qubit, we denote the CNOT gate as CNOT$_{21}$. Its effect is as follows:
\begin{align}
\text{CNOT}_{21} |00\rangle &= |00\rangle, \text{CNOT}_{21} |01\rangle = |11\rangle \label{4.25}\\
\text{CNOT}_{21} |10\rangle &= |10\rangle, \text{CNOT}_{21} |11\rangle = |01\rangle. \label{4.26}
\end{align}
Its matrix form is  
\begin{align}
\text{CNOT}_{21} = \begin{pmatrix} 1 & 0 & 0 & 0 \\ 0 & 0 & 0 & 1 \\ 0 & 0 & 1 & 0 \\ 0 & 1 & 0 & 0 \end{pmatrix}.\label{CNOT21}
\end{align}

\begin{figure*}[!htbp]
\centerline{
\includegraphics[width=3.5in]{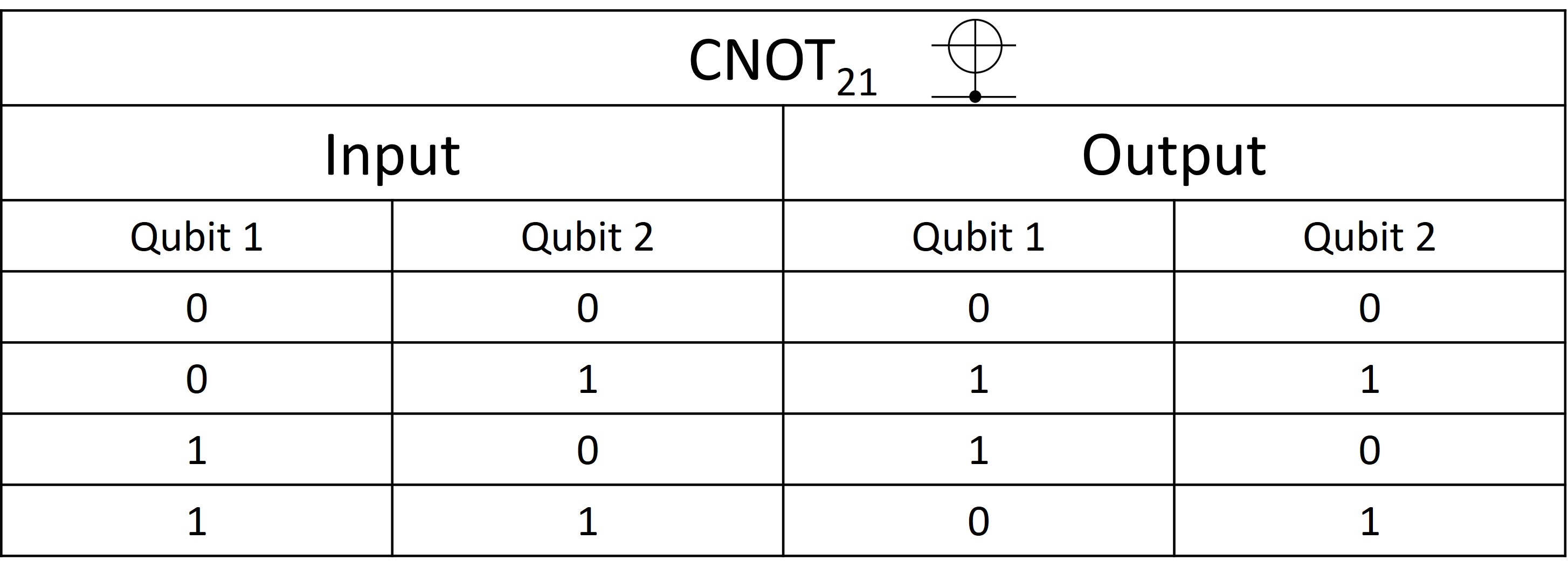}
}
\caption{Truth table of CNOT$_{21}$.}
\label{s4t2}
\end{figure*}

The truth table of CNOT$_{21}$ are shown in Fig. \ref{s4t2}.

\subsubsection{NMR realization of CNOT gate}

The molecule we use in Gemini is dimethyl phosphite, where $^{31}$P and $^1$H are used as two qubits (Fig. \ref{s4f3}). The coupling between these two nuclei is about 700Hz. As we mentioned before, the evolution under the J-coupling interation, $U_J(t)$ in Eq. (\ref{3.28}), plays a crucial role in realization of CNOT gates.  CNOT gate can be achieved by combining $U_J$ and some RF pulses \cite{110}:
\begin{align}
\text{CNOT}_{12} = &e^{i \frac{\pi}{4}} R_x^1 (\frac{\pi}{2}) R_y^1 (\frac{\pi}{2}) R_{-x}^1 (\frac{\pi}{2}) \nonumber\\\cdot &R_x^2 (\frac{\pi}{2}) R_{-y}^2 (\frac{\pi}{2}) U_J \left(\frac{1}{2J}\right) R_y^2 (\frac{\pi}{2}). \label{4.28}
\end{align}

This equation is the further decomposition of Eq. (\ref{3.29}). Here all the  single-qubit gates are x and y rotation pulses, e. g., $R_y^2 (\pi/2)$ refers to the pulse applied on the second qubit ($^{31}$P) to rotate it around y axis by 90°. $U_J (1/2J)$ refers to the free evolution under the effect of J coupling for a duration of $1/2J$, which is 720us for our system. According to the realization of CNOT$_{12}$ mentioned above, it is easy to obtain the realization of CNOT$_{21}$ (the second qubit is the control qubit):
\begin{align}
\text{CNOT}_{21} =& e^{i \frac{\pi}{4}} R_x^2 (\frac{\pi}{2}) R_y^2 (\frac{\pi}{2}) R_{-x}^2 (\frac{\pi}{2}) \nonumber\\\cdot &R_x^1 (\frac{\pi}{2}) R_{-y}^1 (\frac{\pi}{2}) U_J \left(\frac{1}{2J}\right) R_y^1 (\frac{\pi}{2}) .\label{4.29}
\end{align}

\subsubsection{Experimental implementation}

To verify the truth tables of CNOT$_{12}$ and CNOT$_{21}$, four experiments shall be carried out for each of them. During these four experiments, $|00\rangle$, $|01\rangle$, $|10\rangle$ and $|11\rangle$ are used as initial states, and then CNOT$_{12}$ and CNOT$_{21}$ act on these four initial states. The density matrices of final states are measured to compare with the theoretical results. 

\begin{figure*}[!htbp]
\centerline{
\includegraphics[width=5.5in]{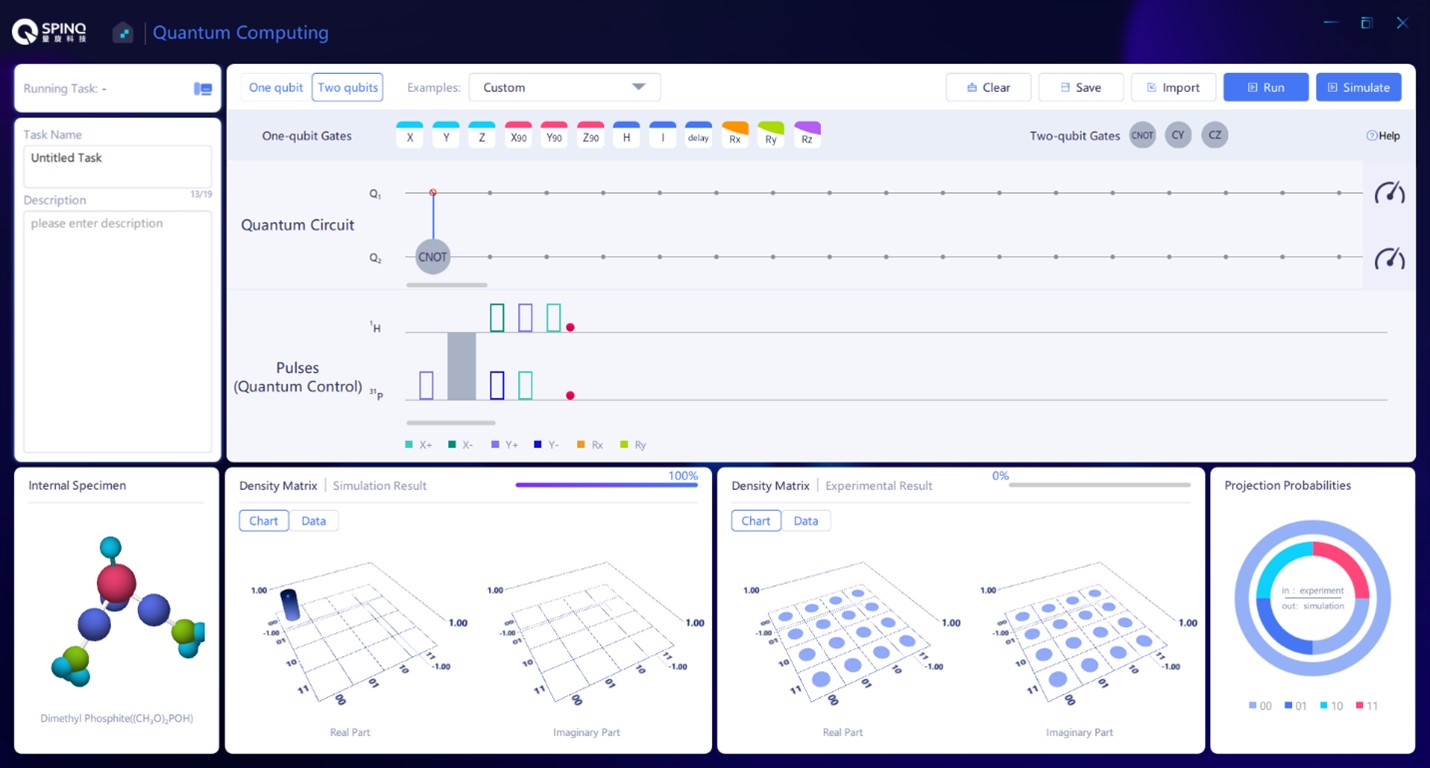}
}
\caption{The circuit to verify CNOT$_{12}$ $|00\rangle=|00\rangle$.}
\label{s4f25}
\end{figure*}

\begin{figure*}[!htbp]
\centerline{
\includegraphics[width=5.5in]{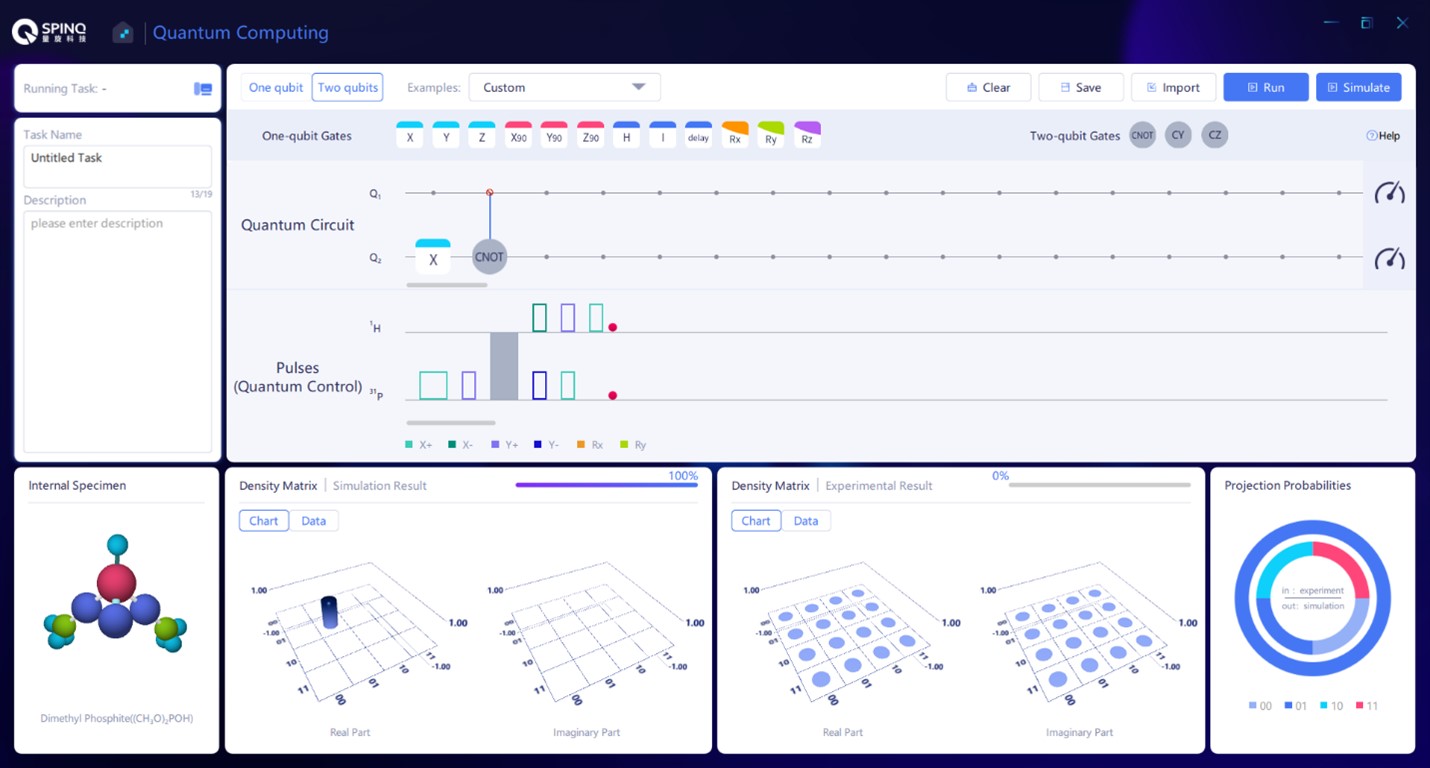}
}
\caption{The circuit to verify CNOT$_{12}$ $|01\rangle=|01\rangle$.}
\label{s4f26}
\end{figure*}

\begin{figure*}[!htbp]
\centerline{
\includegraphics[width=5.5in]{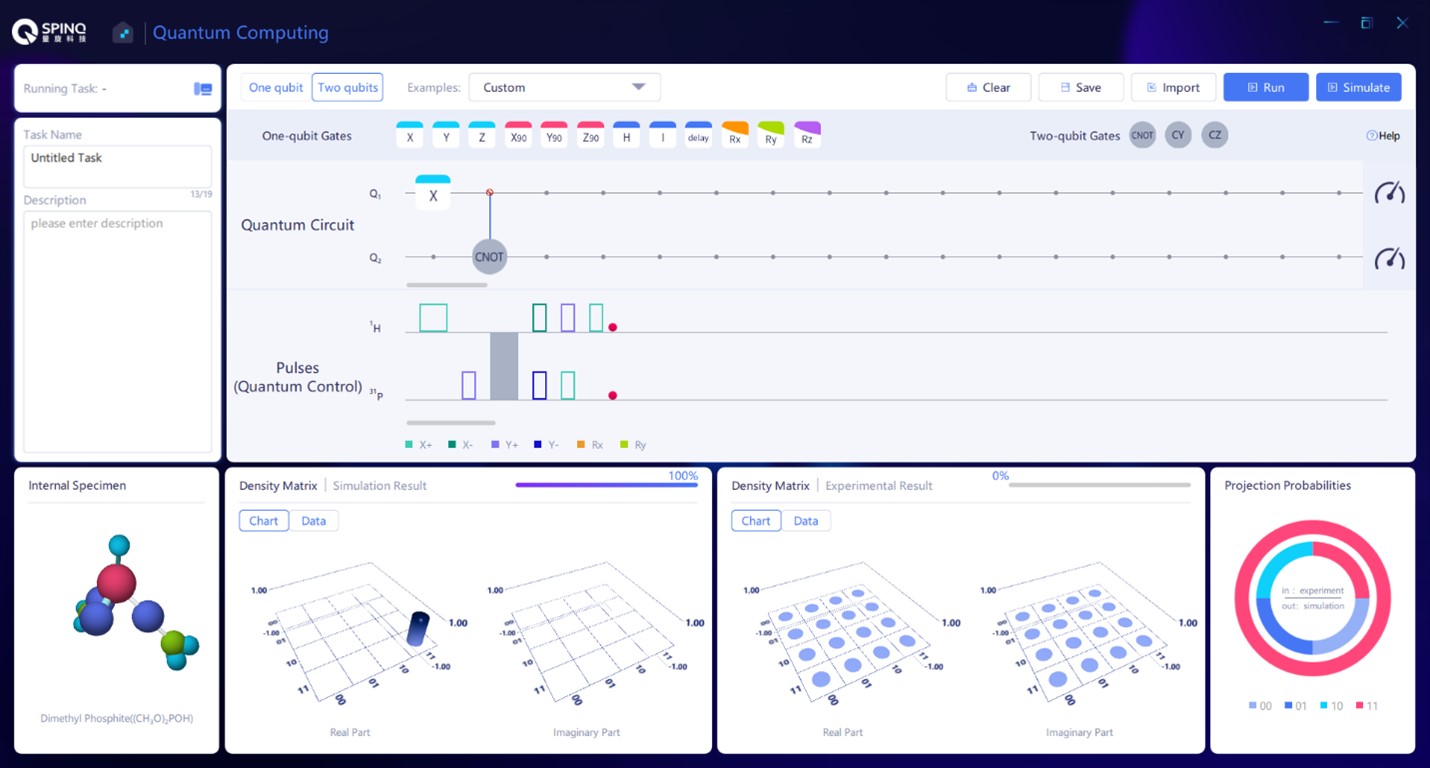}
}
\caption{The circuit to verify CNOT$_{12}$ $|10\rangle=|11\rangle$.}
\label{s4f27}
\end{figure*}

\begin{figure*}[!htbp]
\centerline{
\includegraphics[width=5.5in]{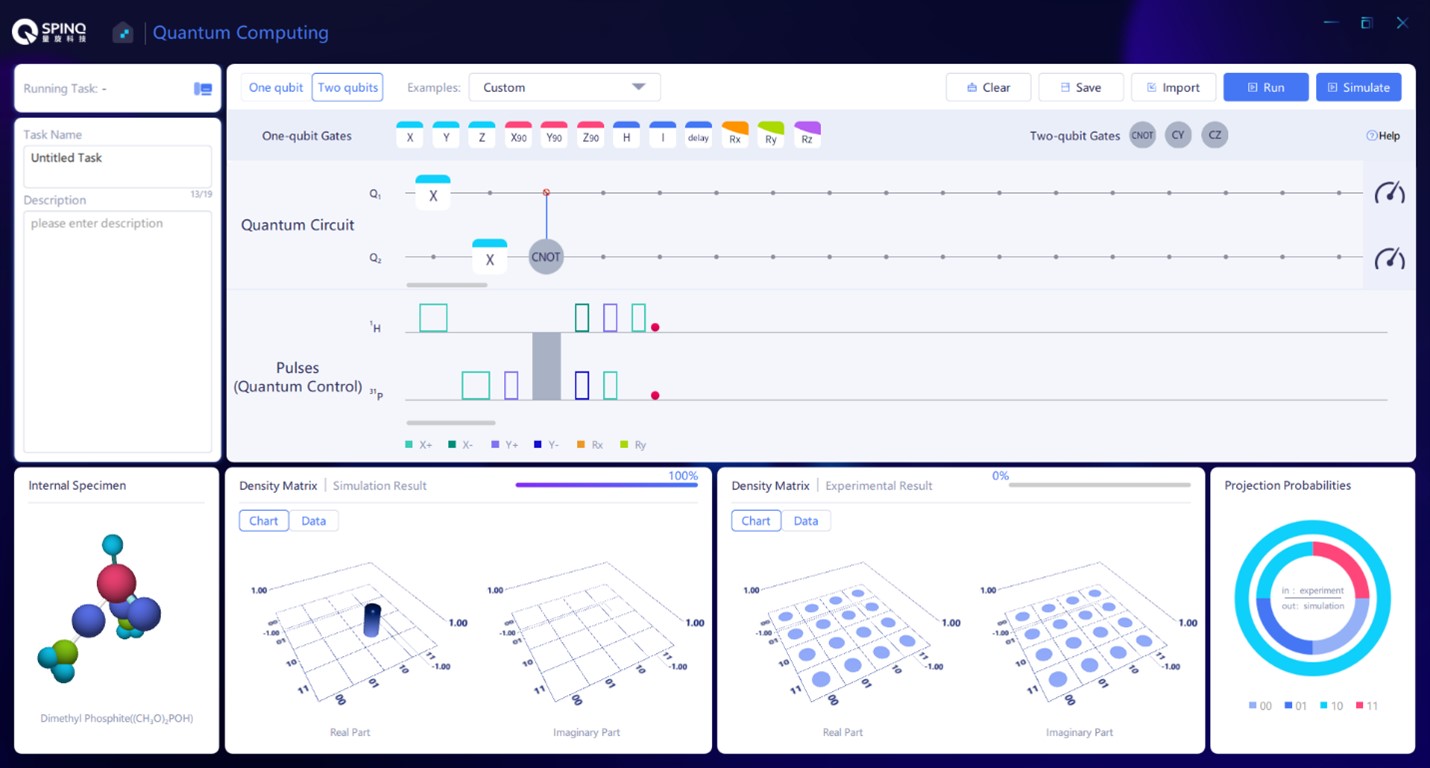}
}
\caption{The circuit to verify CNOT$_{12}$ $|11\rangle=|10\rangle$.}
\label{s4f28}
\end{figure*}

Figure \ref{s4f25} shows the circuit to verify CNOT$_{12}$ $|00\rangle=|00\rangle$. The initial state of the system is $|00\rangle$ by default. Drag CNOT gate and put it on the two lines in “Quantum Circuit”, with the dot on the first line and the big circle on the second line, which means that CNOT$_{12}$ acts on $|00\rangle$ and the first qubit is the control qubit.

Figure \ref{s4f26} is the circuit to verify CNOT$_{12}$ $|01\rangle=|01\rangle$. Drag X gate to the second line in “Quantum Circuit”, which changes the initial state from $|00\rangle$ to  $|01\rangle$. Then drag CNOT gate and put it on the two lines, with the dot on the first line and the big circle on the second line, which means that CNOT$_{12}$ gate acts on   $|01\rangle$ state.

Figure \ref{s4f27} is the circuit to verify CNOT$_{12}$ $|10\rangle=|11\rangle$. Drag X gate to the first line in “Quantum Circuit” to prepare the initial state $|10\rangle$ . Then drag CNOT gate and put it on the two lines.

Figure \ref{s4f28} is the circuit to verify CNOT$_{12}$ $|11\rangle=|10\rangle$. Drag X gate to both lines in “Quantum Circuit”, which changes the initial state from $|00\rangle$ to $|11\rangle$. Then drag CNOT gate and put it on the two lines.

The above steps can be repeated to verify CNOT$_{21}$ $|00\rangle=|00\rangle$, CNOT$_{21}$ $|01\rangle=|11\rangle$, CNOT$_{21}$ $|10\rangle=|10\rangle$ and CNOT$_{21}$ $|11\rangle=|01\rangle$. It should be noted that here the dot is on the second line and the big circle is on the first line when dragging CNOT gate and putting it on the two lines in “Quantum Circuit”, which means that the second qubit is the control qubit.

\subsubsection{Discussion}

It would be interesting for readers to test, in each CNOT gate, which initial state has the experimental final state that is closest to the simulated final state, which is the most different, and what is the reason. Usually, the final results reflect the quality of both the gate and the initial states. 
Another interesting question would be, for these two CNOT gates, which one has better experimental results. The difference in the result quality of the two gates can reflect the difference of the control quality of the two qubits.

\subsection{Preparation of Bell state}

In quantum information, both quantum superposition and quantum entanglement are important resources. The parallelism of quantum computation comes from quantum superposition. Quantum entanglement is also the reason why many quantum algorithms are better than the classical algorithm, and has important applications in quantum communication, quantum sensing and other fields. In this section, Bell states \cite{129} will be prepared. Bell states are a set of two-qubit superposition states, which are also the maximally entangled quantum states in the case of two qubits \cite{130}.  In addition, they were the first entangled states studied in history. The preparation of Bell states requires a two-qubit gate, i.e. CNOT gate. 

\subsubsection{Bell states}

Bell state is a set of four quantum states, separately are
\begin{align}
\Psi^+ &= \frac{1}{\sqrt{2}} (|00\rangle + |11\rangle), \label{4.30} \\
\Psi^- &= \frac{1}{\sqrt{2}} (|00\rangle - |11\rangle), \label{4.31}\\
\Phi^+ &= \frac{1}{\sqrt{2}} (|01\rangle + |10\rangle), \label{4.32} \\
\Phi^- &= \frac{1}{\sqrt{2}} (|01\rangle - |10\rangle). \label{4.33}
\end{align}

\begin{figure*}[!htbp]
\centerline{
\includegraphics[width=5.5in]{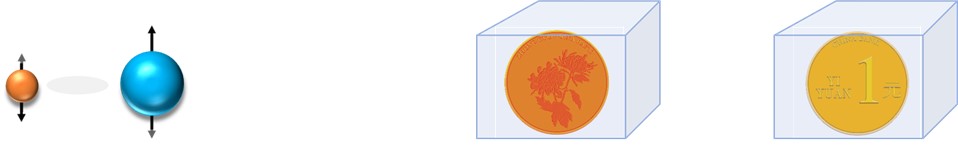}
}
\caption{Quantum entanglement and classical correlation.}
\label{s4f29}
\end{figure*}

These four states are all the superposition states of two-qubit basis vector states. Taking $\Phi^{-}$ as an example, these two qubits are in $|01\rangle$ with a probability of  $\left(\frac{1}{\sqrt{2}}\right)^2 = \frac{1}{2}$ and in  $|10\rangle$ with a probability of 1/2. Supposing that $|0\rangle$ and $|1\rangle$ correspond to spin-up and spin-down states of a qubit in z direction, then if we measure the first qubit in z direction and the result is spin-up $|0\rangle$, the second qubit is certainly in the state of spin-down $|1\rangle$; if the measurement result of the first qubit is spin-down $|1\rangle$, the second qubit is certainly in the state of spin-up $|0\rangle$ (as shown in the left part of Fig. \ref{s4f29}). The state of the first qubit is correlated with that of the second qubit, just as the states of Schrodinger’s Cat and the atomic nucleus mentioned before. In addition, $\Phi^{-}$ has an important feature, i.e. as long as the direction selected for measuring the two spins is the same (not limited to z direction), the states of these two qubits are always opposite (anticorrelation).

When you first see the correlation between these two qubits, it seems to also exist in the classical world, as shown in the right of Fig. \ref{s4f29}. If it is known that the coins in two boxes face opposite directions and are in different colors (yellow and red), open the first box: If the coin in the first one is found to be face up then the coin in the second box is certainly back up; if the first coin is in red, the second one must be in yellow. However, $\Phi^{-}$ has the property that the coin model does not have. Supposing that we measure different properties of these two spins in $\Phi^{-}$, such as measuring the spin in different directions, spin of the first qubit in x direction and that of the second qubit in z direction, the anticorrelation of spins mentioned above disappears. No matter the measurement result of the first qubit is spin-up (+x direction) or spin-down (-x direction), the second qubit is always spin-up (+z direction) or spin-down (-z direction) with equal probabilities of 1/2. It is quite different from the coin model, and just imagine that whether we choose to observe the direction (compare with the spin measurement in z direction) or the color (compare with the spin measurement in x direction) of the coin in the first box will not affect the direction of the second coin (compare with the spin measurement of the second qubit in z direction). Therefore, the fact when measuring $\Phi^{-}$, the choice of measurement method (measurement in x or z direction) of the first qubit has an impact on the measurement result of the second one is an important manifestation that quantum entanglement is quite different from classical correlation. Moreover, such impact caused by entanglement will always exist regardless of the distance between these two qubits.

The two qubits in Bell states have important applications in quantum communication, and they are usually distributed to both parties who need to transfer messages. With the help of quantum gate and quantum measurement, as well as classical information that needs to be transferred sometimes, both parties achieve information transmission, which is safer than classical communication.

\subsubsection{Bell state preparation}

We will focus on the preparation of $\Phi^{-}$. Here, we will use a controlled-unitary which derives from the standard CNOT gate:
\begin{align}
\text{CY} = \begin{pmatrix} 
1 & 0 & 0 & 0 \\ 
0 & 1 & 0 & 0 \\ 
0 & 0 & 0 & -1 \\ 
0 & 0 & 1 & 0 
\end{pmatrix}. \label{4.34}
\end{align}
This is a quantum gate whose first qubit is the control qubit. Considering the initial states are $|00\rangle$, $|01\rangle$, $|10\rangle$, $|11\rangle$, respectively:
\begin{align}
\text{CY} |00\rangle &= |00\rangle, \text{CY} |01\rangle = |01\rangle \label{4.35}\\
\text{CY} |10\rangle &= |11\rangle, \text{CY} |11\rangle = -|10\rangle. \label{4.36}
\end{align}
If the initial state of the two qubits is $\frac{1}{\sqrt{2}} \left( |01\rangle + |11\rangle \right)
$, after the operation CY, the state of $\Phi^{-}$ as we want can be obtained. It should be noted here that $\frac{1}{\sqrt{2}} \left( |01\rangle + |11\rangle \right)$ is not an entangled state, which can be written as $\frac{1}{\sqrt{2}} \left( |0\rangle + |1\rangle \right) |1\rangle$, that is to say, the first qubit is in $\frac{1}{\sqrt{2}} \left( |0\rangle + |1\rangle \right) $ state and the second is in $|1\rangle$ state, and there is no correlation between these two qubits. It thus can be seen that here quantum entanglement is generated by CY, which reflects the importance of the CNOT-like gate.

The specific process of starting from $|00\rangle$ state to prepare $\Phi^{-}$  is as follows: 
\begin{align}
|00\rangle&\xrightarrow{\text{H on the first qubit}}\frac{1}{\sqrt{2}} \left( |0\rangle + |1\rangle \right) |0\rangle \nonumber\\&\xrightarrow{\text{X on the second qubit}}\frac{1}{\sqrt{2}} \left( |0\rangle + |1\rangle \right) |1\rangle\xrightarrow{\text{CY}}\Phi^{-}.\label{4.37}
\end{align}

\begin{figure*}[!htbp]
\centerline{
\includegraphics[width=5.5in]{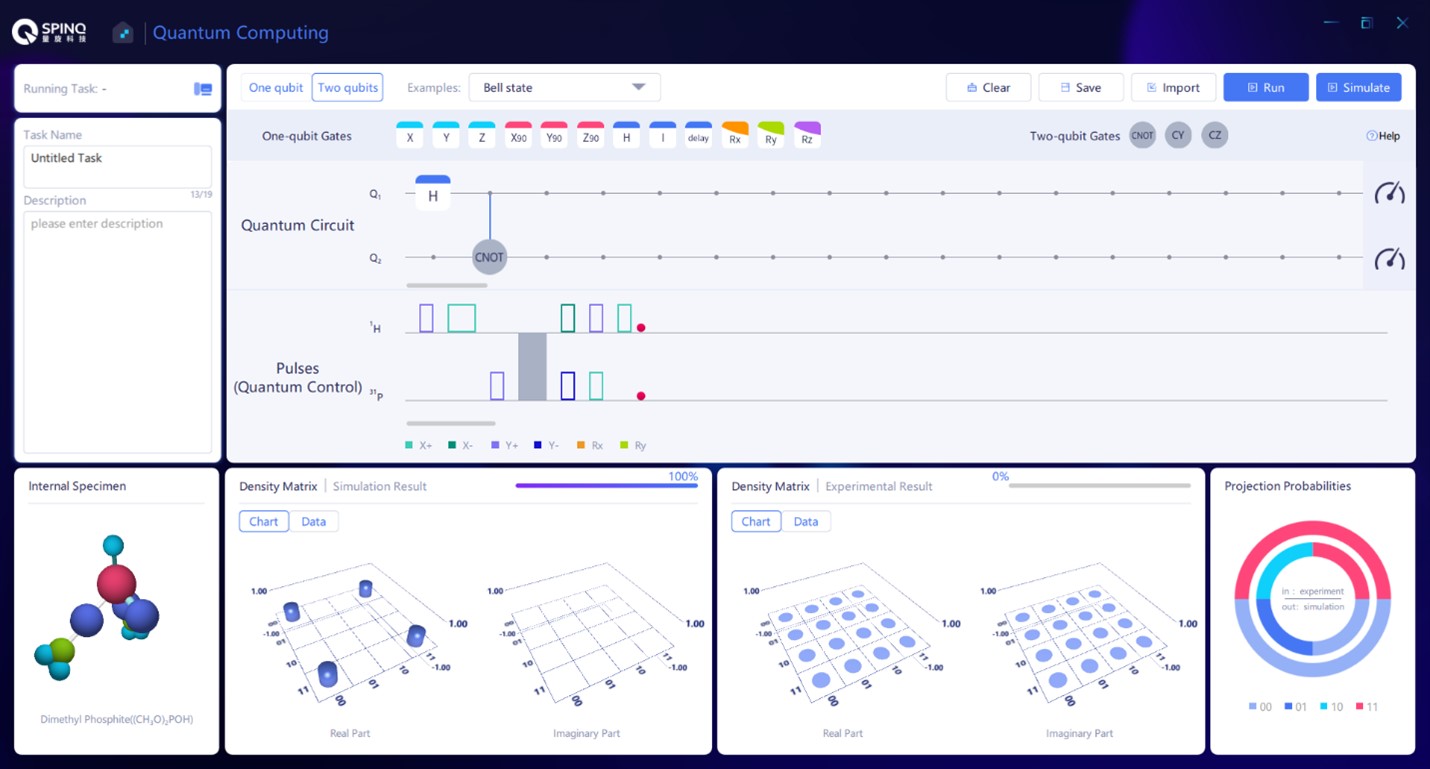}
}
\caption{The built-in example of a Bell state.}
\label{s4f30}
\end{figure*}

\subsubsection{Experimental implementation}

In Spinquasar, there is a built-in sequence for Bell state preparation as shown in Fig. \ref{s4f30}. Readers can try to identify which one of the four Bell states is this state.

\begin{figure*}[!htbp]
\centerline{
\includegraphics[width=5.5in]{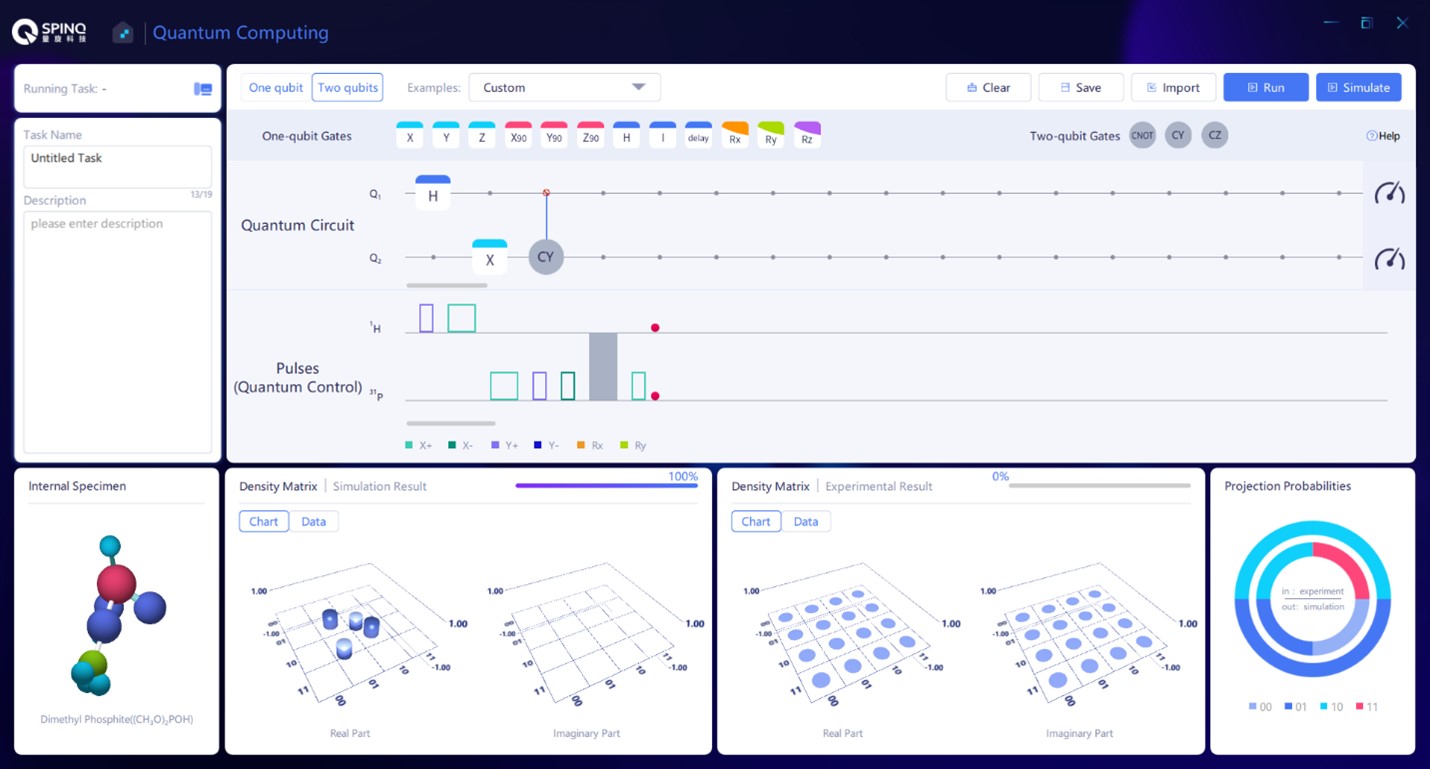}
}
\caption{The circuit to prepare $\Phi^{-}$.}
\label{s4f31}
\end{figure*}

For the $\Phi^{-}$ state, the quantum circuit can be composed as in Fig. \ref{s4f31}. Here we give the theoretical density matrix of $\Phi^{-}$ state:
\begin{align}
|\Phi^-\rangle\langle\Phi^-| = \frac{1}{2} \left(\begin{array}{c} 0 \\ 1\\-1\\0 \end{array} \right)\left(0\; 1\; -1 \; 0 \right) = \frac{1}{2} \left(\begin{array}{cccc} 0 & 0 & 0& 0\\ 0 & 1  & -1 & 0\\ 0 & -1  & 1 & 0\\ 0 & 0  & 0 & 0 \end{array} \right). \label{4.38}
\end{align}
The basis vectors used in this density matrix are the eigen states of z angular momentum of two qubits, i.e. $ |00\rangle$, $ |01\rangle$,  $|10\rangle$ and $ |11\rangle$. The first diagonal element of the matrix refers to the probability in $ |00\rangle$ state, the second refers to the probability in $ |01\rangle$ state, the third refers to the probability in $ |10\rangle$ state and the fourth refers to the probability in $ |11\rangle$ state. The simulation results should be the same as the above matrix.

Here, we can use the fidelity defined in Eq. (\ref{1.27}) to evaluate how close the theoretical state and the experimentally prepared state are. If they are exactly the same, the fidelity is 1; the smaller the fidelity, the greater the difference between these two quantum states. We use this equation instead of Eqs. (\ref{1.26})  and (\ref{1.28}) to evaluate the quality of the prepared $\Phi^{-}$ state, because the $\Phi^{-}$ state we want to prepare is a pure state, while the experimental result is usually a mixed state.

\subsubsection{Data analysis}

We want to calculate the fidelity of  $\Phi^{-}$  state prepared in the experiment using Eq. (\ref{1.27}).
The real part and imaginary part of the density matrix of experimental results returned by Spinquasar should be written into one matrix to get $\rho$. The method is to multiply the matrix of imaginary part by the imaginary unit $i$ and then add it to the real part. Then $F(|\Phi^{-}\rangle, \rho) = \langle \Phi^{-} | \rho | \Phi^{-} \rangle$ can be calculated.

\subsubsection{Discussion}

Is there any other method that can be used to prepare  $\Phi^{-}$? The answer is YES, and the method given below can also be used to prepare  $\Phi^{-}$ state.
\begin{align}
|00\rangle&\xrightarrow{\text{H on the first qubit}}\frac{1}{\sqrt{2}} \left( |0\rangle + |1\rangle \right) |0\rangle\nonumber\\&\xrightarrow{\text{X on the second qubit}}\frac{1}{\sqrt{2}} \left( |0\rangle + |1\rangle \right) |1\rangle\nonumber\\&\xrightarrow{\text{Z on the first qubit}}\frac{1}{\sqrt{2}} \left( |0\rangle - |1\rangle \right) |1\rangle\xrightarrow{\text{CNOT}}\Phi^{-}\label{4.39}
\end{align}
Here $ \text{CNOT}=\left(\begin{array}{cccc} 1 & 0 & 0& 0\\ 0 & 1  & 0 & 0\\ 0 & 0  & 0 & 1\\ 0 & 0  & 1 & 0 \end{array} \right)$ and the first qubit is the control qubit.

Readers can try to implement the above method of preparing  $\Phi^{-}$ state and compare the final state fidelity with that of the methond in Eq. (\ref{4.37}).  The methods to prepare  $\Phi^{-}$ are not limited to the two we introduced here. Which method provides the best result usually depends on the number and quality of gates used.

\subsection{Deutsch algorithm}

To solve a problem using a computer, the first step is to decide what algorithm is to be used. When people say quantum computers are superior to classical computers, they mean that in solving some problems, quantum algorithms have the advantages of using less steps or using less resources. 

Several most widely known quantum algorithms are Shor’s algorithm for performing prime factorization of integers, Grover’s algorithm for unstructured search, Deutsch’s algorithm for determining whether a function is constant or balanced, etc. Currently, the RSA cryptosystem is widely used in daily secure data transmission. Shor’s algorithm can be used to attack this cryptosystem effectively. This means, when a universal quantum computer which can run Shor algorithm is built, data secured by RSA will not be safe anymore. Unstructured search is an important problem in data processing. Many problems in computer science can be reduced to this problem. If a specific entry in an unstructured database with N elements is wanted, and there is no hint on which is more likely to be it, then classical algorithms have to check the entries one by one. In the worst case, this trial-and-error method takes N steps. But Grover algorithm only need the order of $\mathcal{O}$($\sqrt{N}$)  trials.  When N is large, the is a big advantage over classical algorithms. Deutsch algorithm \cite{23,72,131} is another algorithm which can demonstrate the quantum advantage. For a function with one-bit input and one-bit output, if one is interested in whether its two outputs are the same (constant function) or opposite (balanced function), one needs to evaluate the function twice to compare the results if he/she uses a classical computer. If a quantum computer is used, by implementing Deutsch algorithm, only one evaluation is needed to tell if it is a balanced or constant function.

In this section, Deutsch algorithm will be implemented in the NMR system as an example of quantum algorithm implementations for beginners.

\subsubsection{Theory of Deutsch algorithm}

Given a function $f(x)$, $x\in\{0,1\}$, and $f(x)\in\{0,1\}$, what we are interested in is whether $f(0)$ and $f(1)$ are the same or different. This means we want to differentiate between the following situations: (1) $f(0)$ and $f(1)$ are both 0 or both 1, and we call it a constant function; (2) $f(0)$ and $f(1)$ are 0 and 1 or 1 and 0, and we call it a balanced function. In classical computers, two evaluations of this function to calculate $f(0)$ and $f(1)$ separately are needed to get an answer.

\begin{figure*}[!htbp]
\centerline{
\includegraphics[width=2.5in]{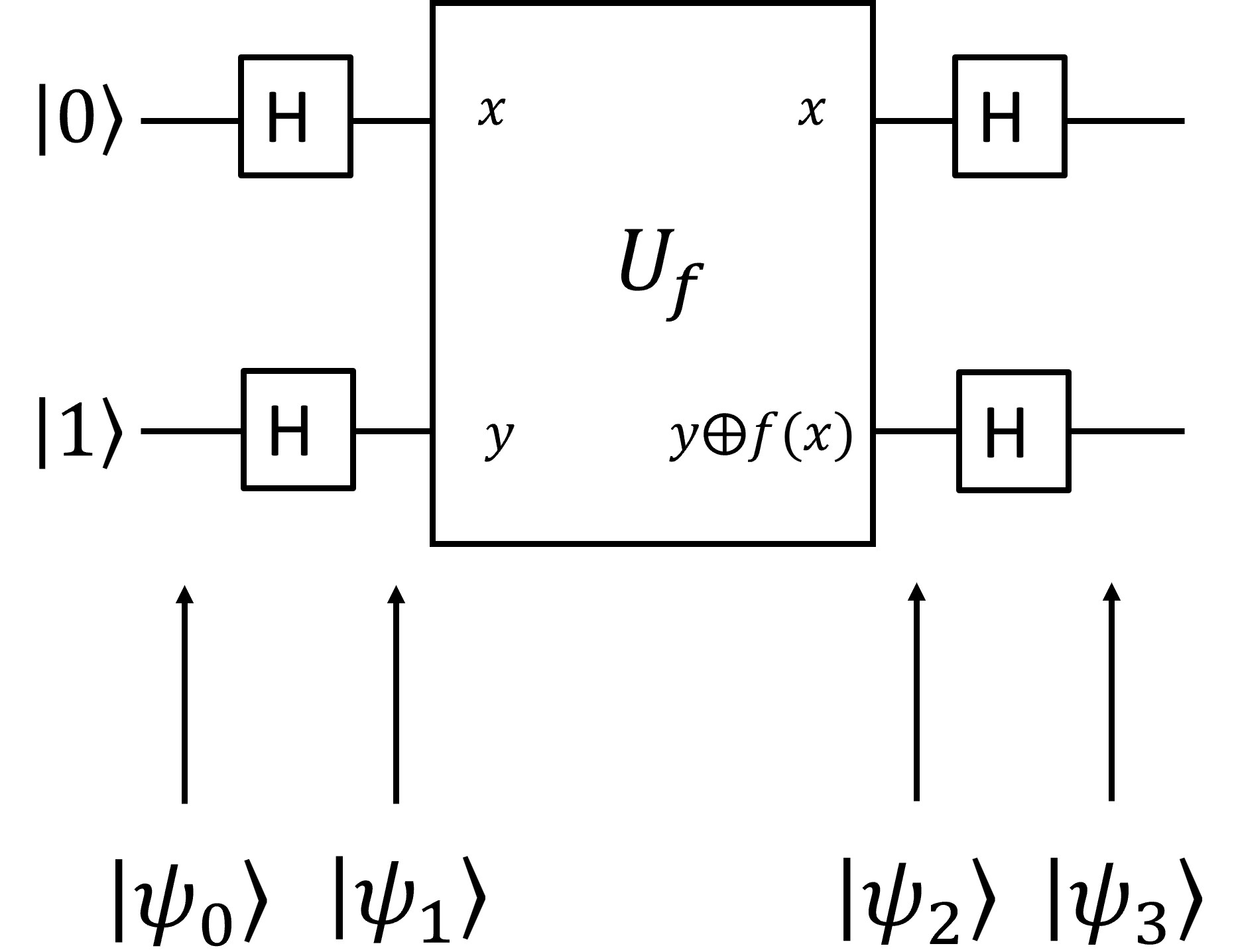}
}
\caption{Deutsch Algorithm circuit.}
\label{s4f32}
\end{figure*}

In Deutsch algorithm, two qubits are used to solve this problem. One qubit is used to store the input information while the other qubit encodes the output information. The quantum circuit is shown in Fig. \ref{s4f32}.

Here, $U_f$ is the key operation. It keeps the first qubit which is in the state $|x\rangle$  unchanged while encodes $f(x)$ to the second qubit. The encoding uses modulo 2 addition: $0 \oplus 0 = 1 \oplus 1 = 0$, $0 \oplus 1 = 1 \oplus 0 = 1$. The other important operation is the first Hadamard gate (H) on the first qubit, which creates the superposition state of  $|0\rangle$ and  $|1\rangle$. This means the state of the first qubit contains two inputs 0 and 1. By inputting this superposition state to $U_f$, the outputs corresponding to 0 and 1 can be evaluated at the same time. This parallelism provides the speedup of Deutsch algorithm. The H gates on the second qubits assist in differentiating between constant and balanced functions. Next, we analyze the state evolution in this algorithm step by step:
\begin{align}
|\psi_0 \rangle = |0\rangle |1\rangle; \quad |\psi_1 \rangle = \frac{1}{2} (|0\rangle + |1\rangle) (|0\rangle - |1\rangle). \label{4.40}
\end{align}
$|\psi_1 \rangle $ can be written as $|\psi_1 \rangle = \frac{1}{2} (|00\rangle + |10\rangle-|01\rangle - |11\rangle)$. \\
(1) If $f(0)=f(1)$ which means $f$ is a constant function, when $f(0)=f(1)=0$, the effect of $U_f$ is
\begin{align}
|0\rangle |0\rangle \xrightarrow{U_f} |0\rangle |0\rangle \label{4.41} \\
|0\rangle |1\rangle \xrightarrow{U_f} |0\rangle |1\rangle \label{4.42}  \\
|1\rangle |0\rangle \xrightarrow{U_f} |1\rangle |0\rangle \label{4.43}  \\
|1\rangle |1\rangle \xrightarrow{U_f} |1\rangle |1\rangle. \label{4.44} 
\end{align}
Therefore, $|\psi_2\rangle = \frac{1}{2} \left( |0\rangle |0\rangle + |1\rangle |0\rangle - |0\rangle |1\rangle - |1\rangle |1\rangle \right) = \frac{1}{2} \left( |0\rangle + |1\rangle \right) \left( |0\rangle - |1\rangle \right)$, $|\psi_3\rangle = |0\rangle |1\rangle$.\\
When $f(0)=f(1)=1$, the effect of $U_f$ is
\begin{align}
|0\rangle |0\rangle \xrightarrow{U_f} |0\rangle |1\rangle \label{4.45} \\
|0\rangle |1\rangle \xrightarrow{U_f} |0\rangle |0\rangle \label{4.46}  \\
|1\rangle |0\rangle \xrightarrow{U_f} |1\rangle |1\rangle \label{4.47}  \\
|1\rangle |1\rangle \xrightarrow{U_f} |1\rangle |0\rangle. \label{4.48} 
\end{align}
Therefore, $|\psi_2\rangle = \frac{1}{2} \left( |0\rangle |1\rangle + |1\rangle |1\rangle - |0\rangle |0\rangle - |1\rangle |0\rangle \right) = -\frac{1}{2} \left( |0\rangle + |1\rangle \right) \left( |0\rangle - |1\rangle \right)$, $|\psi_3\rangle = -|0\rangle |1\rangle$.\\
(2) If $f(0)\neq f(1)$ which means $f$ is a balanced function, when $f(0)=0$, $f(1)=1$, the effect of $U_f$ is
\begin{align}
|0\rangle |0\rangle \xrightarrow{U_f} |0\rangle |0\rangle \label{4.49} \\
|0\rangle |1\rangle \xrightarrow{U_f} |0\rangle |1\rangle \label{4.50}  \\
|1\rangle |0\rangle \xrightarrow{U_f} |1\rangle |1\rangle \label{4.51}  \\
|1\rangle |1\rangle \xrightarrow{U_f} |1\rangle |0\rangle. \label{4.52} 
\end{align}
Therefore, $|\psi_2\rangle = \frac{1}{2} \left( |0\rangle |0\rangle + |1\rangle |1\rangle - |0\rangle |1\rangle - |1\rangle |0\rangle \right) = \frac{1}{2} \left( |0\rangle - |1\rangle \right) \left( |0\rangle - |1\rangle \right)$, $|\psi_3\rangle = |1\rangle |1\rangle$.\\
When $f(0)=1$, $f(1)=0$, the effect of $U_f$ is
\begin{align}
|0\rangle |0\rangle \xrightarrow{U_f} |0\rangle |1\rangle \label{4.53} \\
|0\rangle |1\rangle \xrightarrow{U_f} |0\rangle |0\rangle \label{4.54}  \\
|1\rangle |0\rangle \xrightarrow{U_f} |1\rangle |0\rangle \label{4.55}  \\
|1\rangle |1\rangle \xrightarrow{U_f} |1\rangle |1\rangle. \label{4.56} 
\end{align}
Therefore, $|\psi_2\rangle = \frac{1}{2} \left( |0\rangle |1\rangle + |1\rangle |0\rangle - |0\rangle |0\rangle - |1\rangle |1\rangle \right) = -\frac{1}{2} \left( |0\rangle - |1\rangle \right) \left( |0\rangle - |1\rangle \right)$, $|\psi_3\rangle = -|1\rangle |1\rangle$.

There is only a sign difference between $|0\rangle |1\rangle$ and $-|0\rangle |1\rangle$, as well as between $|1\rangle |1\rangle$ and $-|1\rangle |1\rangle$. This difference is a global phase difference and is not observable. However, we do not need to differentiate within the two pairs, we only need to differentiate between constant and balanced functions. By observing the final states of the two qubits, if they are both $|1\rangle$, then the function is balanced; otherwise the function is constant. This means, one evaluation of the function is enough for differentiating between constant and balanced using Deutsch algorithm.

\subsubsection{Quantum gates in Deutsch algorithm}

\begin{figure*}[!htbp]
\centerline{
\includegraphics[width=3.5in]{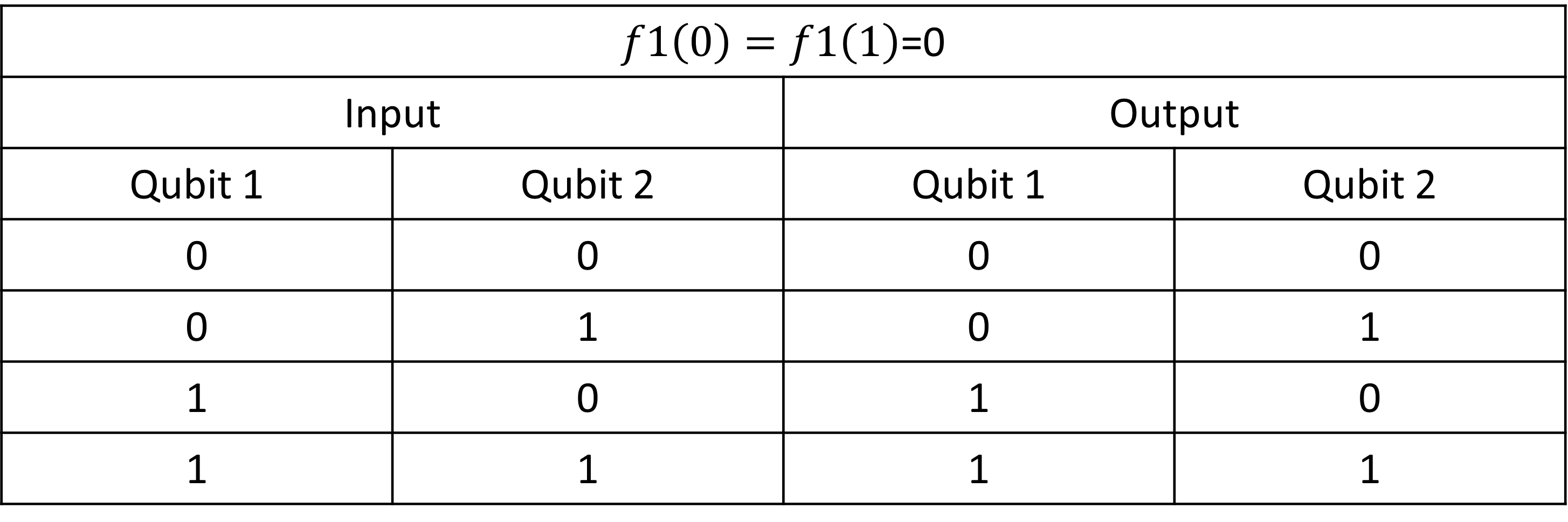}
}
\caption{Truth table of $U_f$ in the case of $f(0)=f(1)=0$.}
\label{s4t3}
\end{figure*}

\begin{figure*}[!htbp]
\centerline{
\includegraphics[width=3.5in]{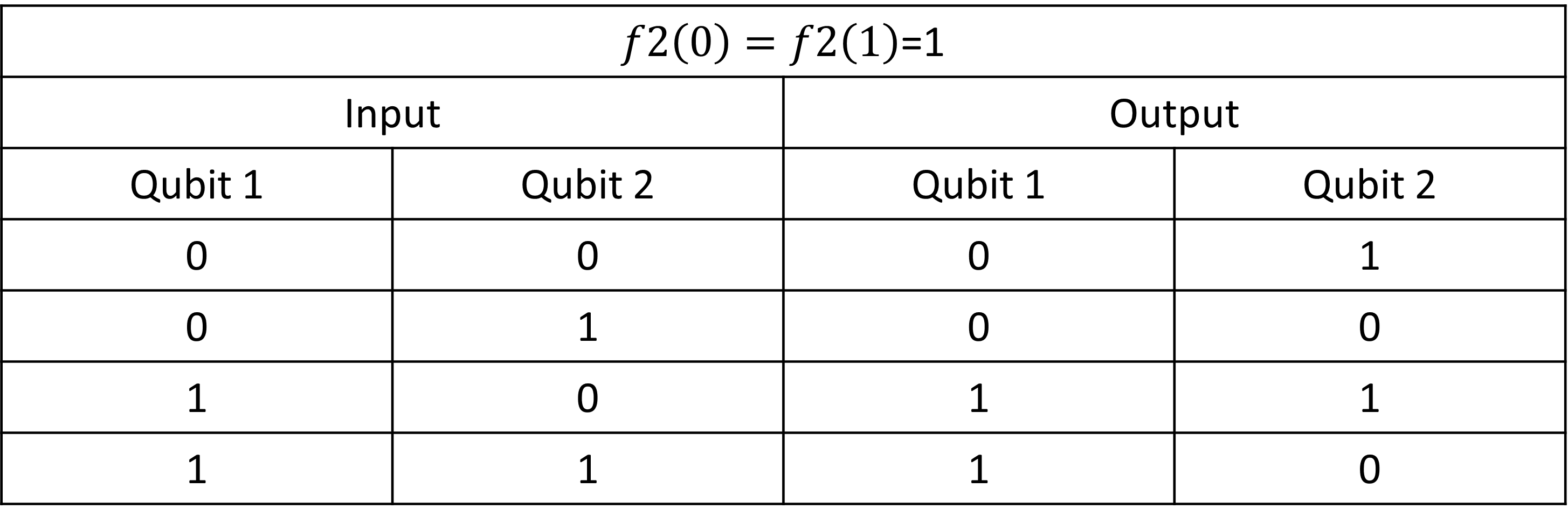}
}
\caption{Truth table of $U_f$ in the case of $f(0)=f(1)=1$.}
\label{s4t4}
\end{figure*}

\begin{figure*}[!htbp]
\centerline{
\includegraphics[width=3.5in]{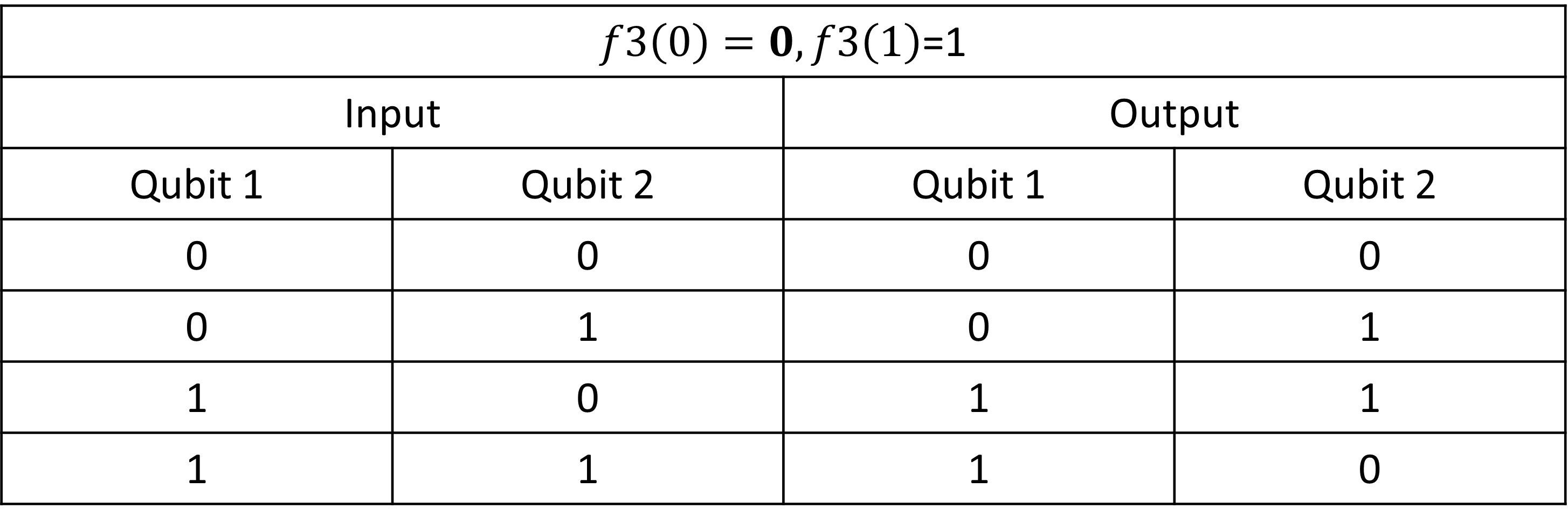}
}
\caption{Truth table of $U_f$ in the case of $f(0)=0$, $f(1)=1$.}
\label{s4t5}
\end{figure*}

\begin{figure*}[!htbp]
\centerline{
\includegraphics[width=3.5in]{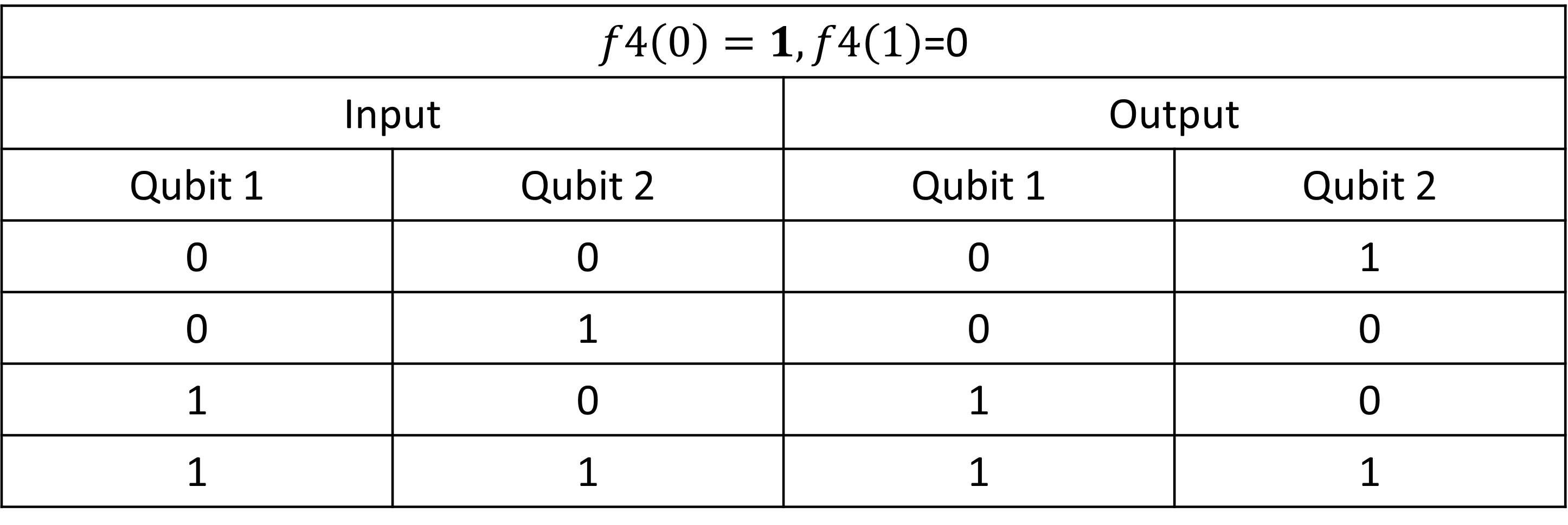}
}
\caption{Truth table of $U_f$ in the case of $f(0)=1$, $f(1)=0$.}
\label{s4t6}
\end{figure*}

The H gate is already introduced in previous sections. Here we focus on how to realize $U_f$. $U_f$ is a two-qubit gates with the first qubit of the input state to be the input of $f$. The second qubit is an ancilla qubit. The output of $f$ is encoded on the second qubit of the output state of $U_f$. This encoding does not mean the second qubit is in the state $|f(x)\rangle$. Rather, the second qubit is in the state of the modulo 2 addition result of $f(x)$ and the input state of the second qubit. The realization of $U_f$ is related with the form of $f$. Figures \ref{s4t3}, \ref{s4t4}, \ref{s4t5} and \ref{s4t6} show the truth tables for $U_f$ gates in different cases of $f$.

\begin{figure*}[!htbp]
\centerline{
\includegraphics[width=5.5in]{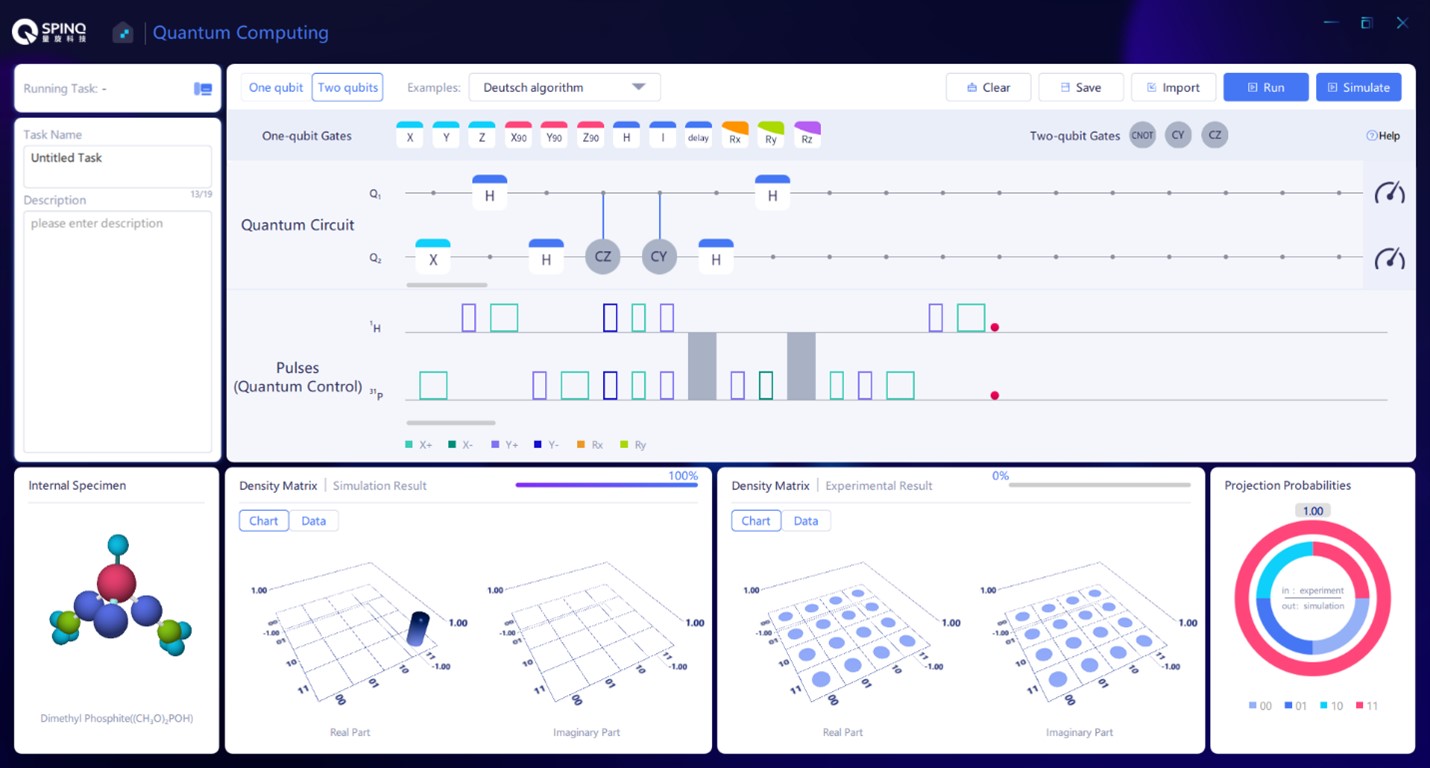}
}
\caption{The built-in example of Deutsch algorithm.}
\label{s4f33}
\end{figure*}

\begin{figure*}[!htbp]
\centerline{
\includegraphics[width=5.5in]{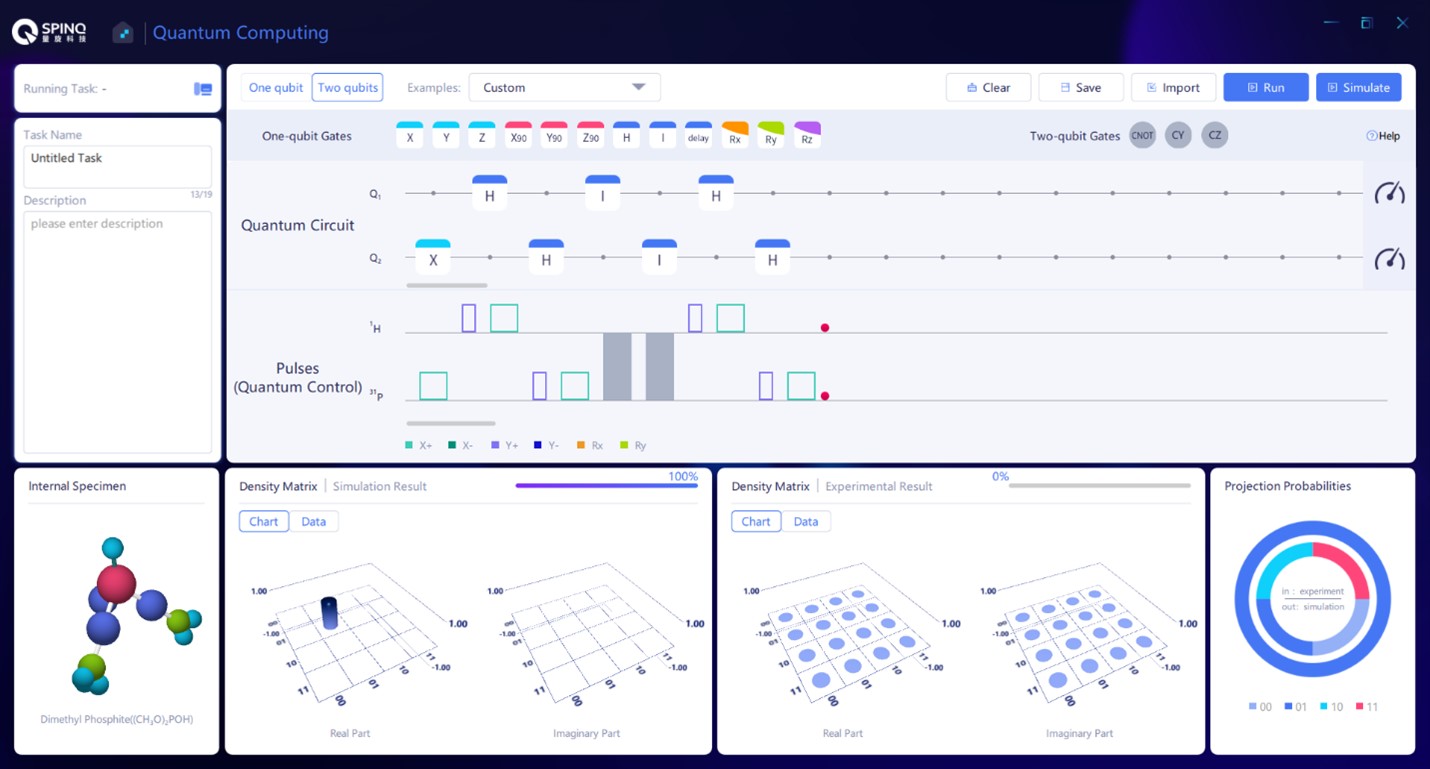}
}
\caption{The circuit to realize $f1$.}
\label{s4f34}
\end{figure*}

\begin{figure*}[!htbp]
\centerline{
\includegraphics[width=5.5in]{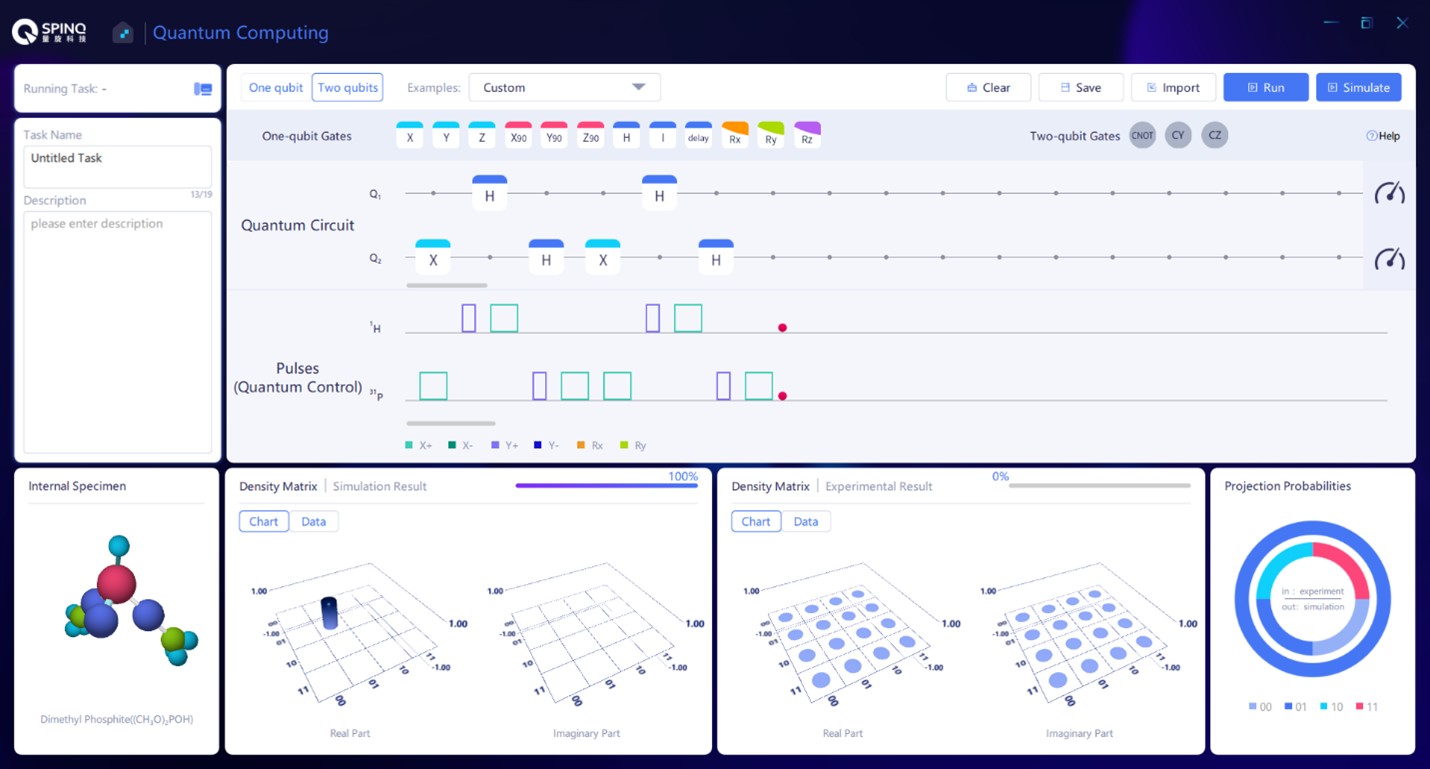}
}
\caption{The circuit to realize $f2$.}
\label{s4f35}
\end{figure*}

\begin{figure*}[!htbp]
\centerline{
\includegraphics[width=5.5in]{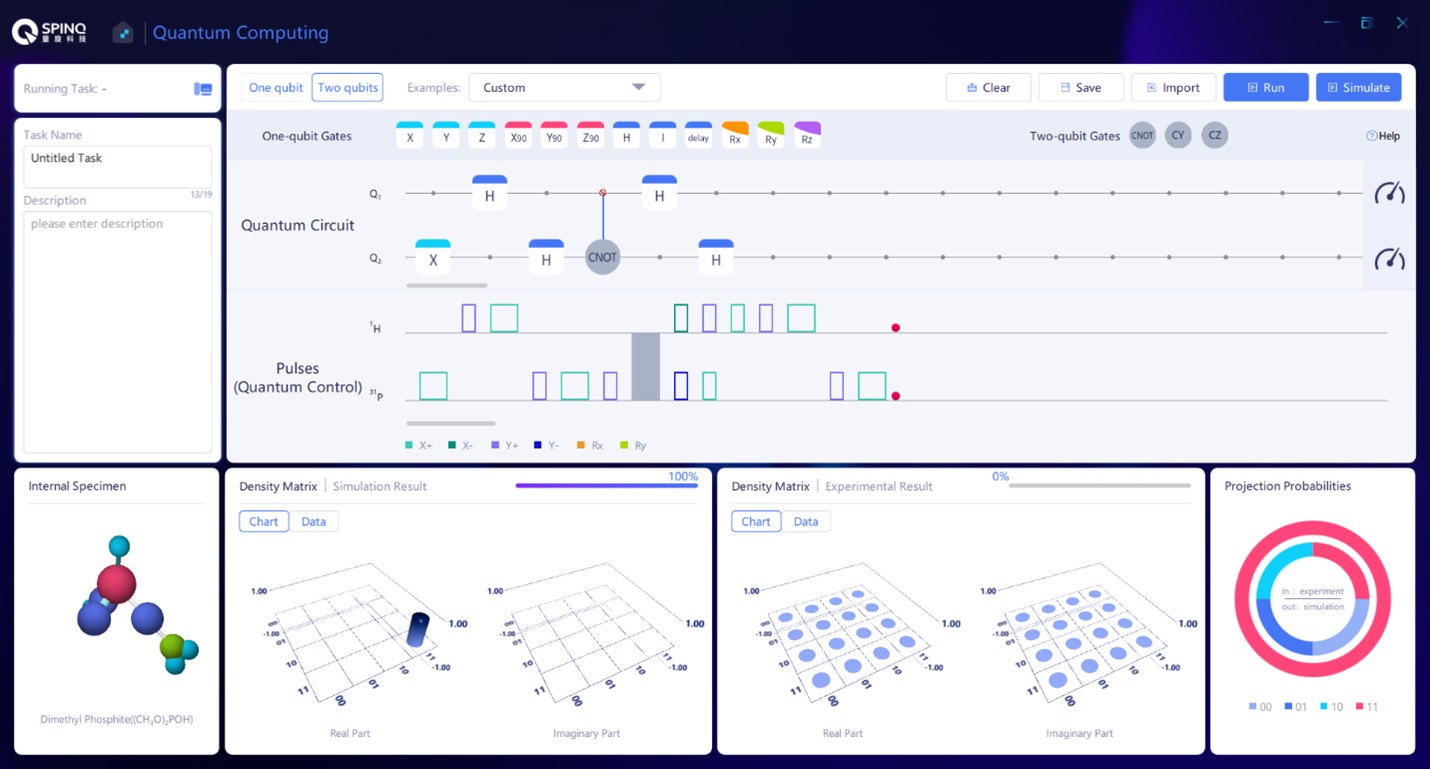}
}
\caption{The circuit to realize $f3$.}
\label{s4f36}
\end{figure*}

\begin{figure*}[!htbp]
\centerline{
\includegraphics[width=5.5in]{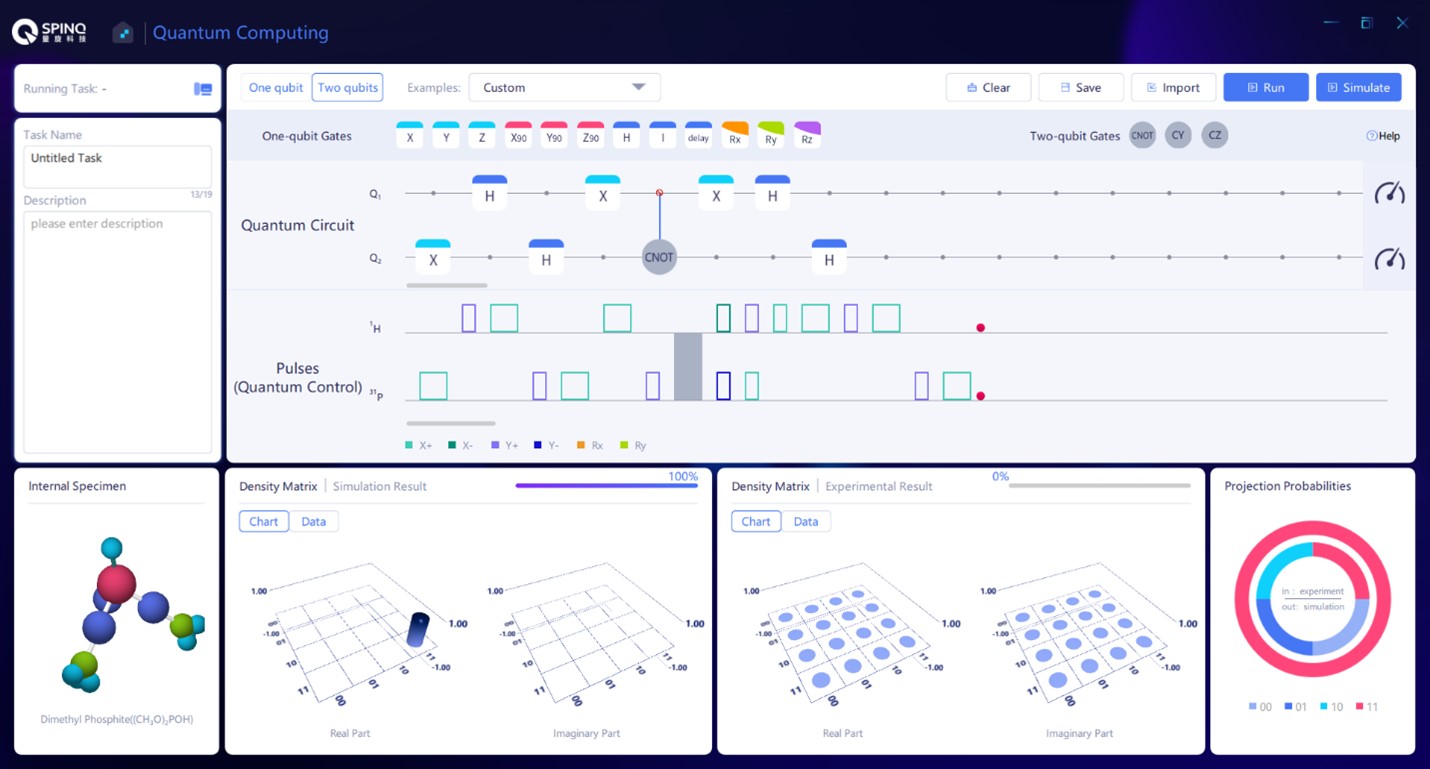}
}
\caption{The circuit to realize $f4$.}
\label{s4f37}
\end{figure*}

$f1$ and $f2$ are constant functions, $f3$ and $f4$ are balanced functions. From the tables above, we can tell the $U_f$ gate corresponding to $f1$ is the Identity gate, i.e. $\begin{pmatrix}
1 & 0 & 0 & 0 \\
0 & 1 & 0 & 0 \\
0 & 0 & 1 & 0 \\
0 & 0 & 0 & 1 \\
\end{pmatrix}$
. The $U_f$ gate of $f2$ is the Not gate (X gate) on the second qubit. The $U_f$ gate of $f3$ is the CNOT gate, and the first qubit is the control qubit. The $U_f$ gate of $f4$ is $|0\rangle$-controlled CNOT gate with the first qubit as the control qubit: When the first qubit is in $|0\rangle$, implement X gate on the second qubit, otherwise implement Identity on the second qubit. The $|0\rangle$-controlled CNOT gate can be decomposed as follows: First, implement X gate on the first qubit; second implement the CNOT gate with the first qubit as the control qubit; third, implement X gate on the first qubit again.

\subsubsection{Experimental implementation}

In the built-in examples of SpinQuasar, there is an implementation of Deutsch algorithm (Fig. \ref{s4f33}). Readers can try to simulate the sequence used there and find out whether this $U_f$ corresponds to a constant or balanced function.

Figures \ref{s4f34}-\ref{s4f37} shows the construction of the corresponding circuits needed to realize the two-qubit Deutsch algorithm introduced in last section. It should be noted that the Deutsch algorithm needs  $|0\rangle |1\rangle$ as the initial state, but the default state in SpinQuasar is $|0\rangle |0\rangle$. Hence in the circuit, before the implementation of $U_f$ an X gate on the second qubit is needed to change it from $|0\rangle$ to $|1\rangle$.

\subsection{Grover algorithm}

In this section, Grover algorithm \cite{26,73,132} will be implemented in the NMR system as an example of quantum algorithm implementations.

\subsubsection{Theory of Grover algorithm}

The problem considered here is to search for a specific entry in an unstructured database. For example, in a database the saved entries are the colors of $N$ balls, of which one is red and the others are green. Consider the $N$ entries are labelled as 1 to $N$. And suppose by evaluating a function $\text{F}(x)$, the $x$th entry can be checked. When F$(x)$ is 0, the color is green; when $\text{F}(x)$ is 1, the color is red. If the index $x_0$ of the red ball is wanted, and there is no information on where it is more likely to be, by using classical algorithms the entries need to be checked one by one (see the left part in Fig. \ref{s1f11}). In the worst case, $N$ trials are needed to find the red color. The complexity of this process is $\mathcal{O}$($\sqrt{N}$).

Grover algorithm can speed up this process greatly. Because of quantum superposition, the $N$ indices of the data can be stored in one quantum state of log$_2N$ qubits. Using this quantum state as the input, one evaluation of F$(x)$ can produce the outputs corresponding to all the $N$ indices (see the right part in Fig. \ref{s1f11}). This means one evaluation obtains all the information on colors.  However, all the information is stored in the output quantum state with equal probabilities. If the log$_2N$ qubits are measured now, the probability to get the correct answer $x_0$  is $1/N$. Therefore, further operations are needed to increase the probability to get $x_0$ (see Fig. \ref{s1f12}). Upon $\mathcal{O}$($\sqrt{N}$) repetitions of this operation, the probability to get $x_0$ can be very close to 1. When $N$ is large, the probability is larger than $1- 1/N$.

\begin{figure*}[!htbp]
\centerline{
\includegraphics[width=5.5in]{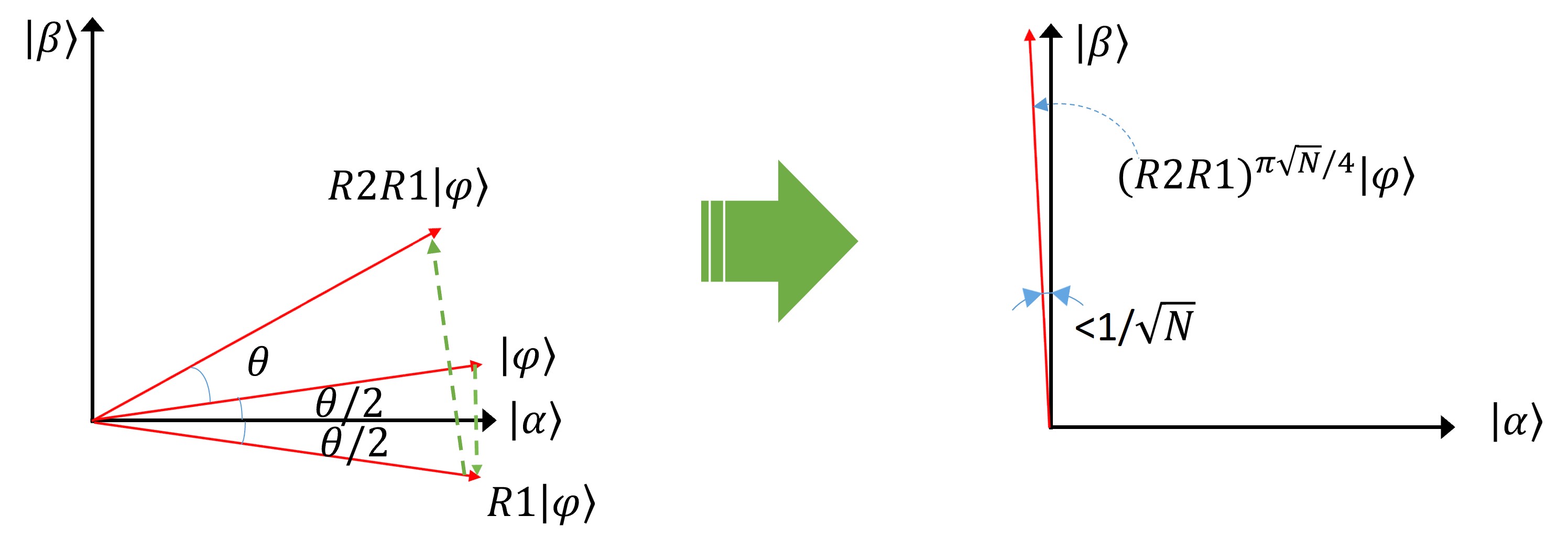}
}
\caption{Grover algorithm rotates the system towards the target state. Around $(\pi\sqrt{N})/4$ repititios of this rotation can transform the system to the desired state with a large probability close to 1.}
\label{s4f38}
\end{figure*}

The quantum circuit is shown in Fig. \ref{s1f13}. Here $n=\text{log}_2 N$ qubits are used. The first operation H$^{\otimes n}$ is the Hadamard operation on all qubits. By this operation, the indices from 1 to $N$ are stored in this superposition state. Let $|\varphi\rangle$ denote the state after H$^{\otimes n}$, then 
\begin{align}
|\varphi\rangle = \frac{1}{\sqrt{N}} (|1\rangle + |2\rangle + |3\rangle + \dots + |N\rangle) = \frac{1}{\sqrt{N}} \sum_{x=1}^{N} |x\rangle.\label{uninform_super}
\end{align}
$|x\rangle$ is the basis state of this $n$ qubits system, and encodes the index $x$ of the database. The probabilities for all the  $N$ basis states are the same and are $1/N$. The $G$ operation is used to improve the probability of $|x_0\rangle$ . $G$ can be decomposed into two steps. $R1$ is the operation that evaluates the F function for all indices and store the results in the superposition state:  $|x\rangle \rightarrow (-1)^{\text{F}(x)} |x\rangle$. If the color is green, $ \text{F}(x)=0$ and $|x\rangle$ stays the same; if the color is red, $\text{F}(x_0 )=1$, $|x_0\rangle$ gains a minus sign. Such an operation can be expressed as an operator $R1=I^{\otimes n}-2|x_0\rangle\langle x_0|$. This operator can be understood intuitively: $ I^{\otimes n}$ is the indentity operator and means nothing happens, following it is the action of subtracting two $|x_0\rangle$, it naturally comes to the result that only the $|x_0\rangle$  state changes its sign. The $R2$ operation can be expressed as $R2=2|\varphi\rangle\langle \varphi|-I^{\otimes n}$. This operation also has an intuitive explanation: $-I^{\otimes n}$ means all the basis states change their signs, and then adding two $|\varphi\rangle$ states means only the $|\varphi\rangle$ basis keeps its original sign. To get an idea of why such a $R2$ operation is needed, we will resort to the following picture (Fig. \ref{s4f38}).

\begin{figure*}[!htbp]
\centerline{
\includegraphics[width=5.5in]{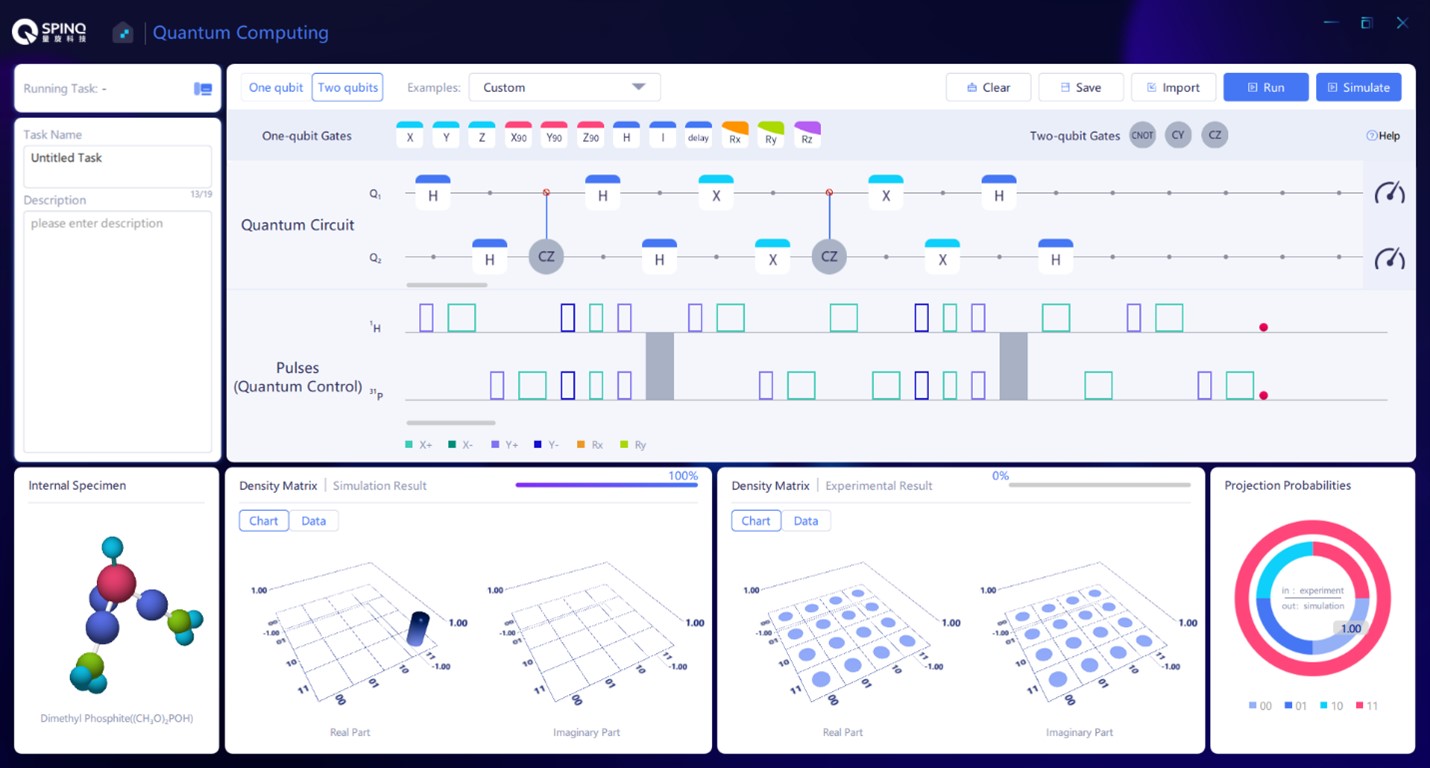}
}
\caption{Realization of Grover algorithm on SpinQuasar.}
\label{s4f39}
\end{figure*}

We can consider the Hilbert space of the $n$ qubits spanned by two vectors $|\alpha\rangle$ and $|\beta\rangle$. $|\beta\rangle$ is the desired result, $|\beta\rangle=|x_0\rangle$, $|\alpha\rangle$ is the superposition of all the basis states except $|x_0\rangle$,  $|\alpha\rangle = \frac{1}{\sqrt{N-1}} \sum_{x \neq x_0} |x\rangle$ . $|\alpha\rangle$ and $|\beta\rangle$ are orthonormal. The initial superposed state used in Grover algorithm can be expressed as $|\varphi\rangle = \sqrt{\frac{N-1}{N}} |\alpha\rangle + \frac{1}{\sqrt{N}} |\beta\rangle$. $|\varphi\rangle$ has a projection of $1/\sqrt{N}$ on the $|\beta\rangle$ axis. The corresponding vectors of $|\alpha\rangle$, $|\beta\rangle$ and $|\varphi\rangle$ are given in Fig. \ref{s4f38}. After $R1$, the sign of $|x_0\rangle$ in the state changes and the quantum state changes from $|\varphi\rangle$ to $ \sqrt{\frac{N-1}{N}} |\alpha\rangle - \frac{1}{\sqrt{N}} |\beta\rangle$ with a projection of -$1/\sqrt{N}$ on the $|\beta\rangle$ axis. Hence $R1$ is an operation to flip the vector of $|\varphi\rangle$ with $|\alpha\rangle$ to be the symmetry axis. From the analysis above, we know that $R2$ keeps the $|\varphi\rangle$ basis unchanged while changing the signs of all the other base. This is an operation to flip the quantum state with $|\varphi\rangle$ to be the axis of symmetry. The net outcome of $G$ is the rotation of the quantum state from $|\varphi\rangle$ towards $|\beta\rangle$, and the rotation angle satisfies $\sin \theta = \frac{2\sqrt{N-1}}{N}$. When $N$ is large, $\theta\approx\sin \theta = \frac{2\sqrt{N-1}}{N}\approx\frac{2}{\sqrt{N}}$. Therefore, rotating the quantum state roughly $(\pi/2)/(2/\sqrt{N})=(\pi\sqrt{N})/4$ times can make the state reach $|\beta\rangle$ with a large propability, and the error is smaller than $1/N$.

\subsubsection{Quantum gates when $N$=4}

We use the case $N$=4 as an example. In this situation, two qubits are needed. If our goal is $|x_0\rangle=|4\rangle$ which has a binary form of $|11\rangle$, then $R1$ and $R2$ can be expressed in the matrix form:
\begin{align}
R1 &= \begin{pmatrix}
1 & 0 & 0 & 0 \\
0 & 1 & 0 & 0 \\
0 & 0 & 1 & 0 \\
0 & 0 & 0 & -1
\end{pmatrix},\label{4.57}\\
R2 &= \text{H}^{\otimes 2}\begin{pmatrix}
1 & 0 & 0 & 0 \\
0 & -1 & 0 & 0 \\
0 & 0 & -1 & 0 \\
0 & 0 & 0 & -1
\end{pmatrix}\text{H}^{\otimes 2}\nonumber\\&=\text{H}^{\otimes 2}\text{X}^{\otimes 2}\begin{pmatrix}
1 & 0 & 0 & 0 \\
0 & 1 & 0 & 0 \\
0 & 0 & 1 & 0 \\
0 & 0 & 0 & -1
\end{pmatrix}\text{X}^{\otimes 2}\text{H}^{\otimes 2}.\label{4.58}
\end{align}

$R1$ is the controlled-Z operation (CZ), which implements a Z gate on the target qubit when the control qubit is in  $|1\rangle$. Here the control qubit is the first qubit. X is the Not gate. Hence $R2$ can be realized in this way: First, apply H gate on both qubits; second, apply Not gate on both qubits; third, apply the controlled-Z gate; fourth, Not gate on both qubits; last, H gate on both qubits. It can be easily proved that when $N$=4 one implementation of  $G$ can realize a successful search with probability 1.

\subsubsection{Experimental implementation}

Figure \ref{s4f39} shows the implementation of two-qubit Grover algorithm using SpinQuasar.

\subsubsection{Discussion}

In the above sections, we discuss Grover algorithm in the two-qubit case with  $|x_0\rangle=|4\rangle$ to be the target state. Now, if the target state is  $|x_0\rangle=|2\rangle$, what does the quantum circuit look like?  In this case, $R1$ needs to be changed to$R1 = \begin{pmatrix}
1 & 0 & 0 & 0 \\
0 & -1 & 0 & 0 \\
0 & 0 & 1 & 0 \\
0 & 0 & 0 & 1
\end{pmatrix}$, which is a $|0\rangle$-controlled-Z gate. The control qubit is the first qubit, and when the first qubit is $|0\rangle$, implement Z gate on the second qubit. Readers can consider the cases when $|x_0\rangle=|1\rangle$ or $|x_0\rangle=|3\rangle$.

\subsection{Quantum approximate counting}

Grover algorithm can realize a search in an unstructured database. In last section we discussed the case when there is only one target entry. When there are more than one entries that satisfy the given conditions, the searched result is the uniform superposition of the quantum states corresponding to these entries. However, to realize such a successful search, the quantity of the entries that satisfy the conditions is required beforehand. Quantum counting algorithm can find out the quantity of the entries that satisfy the conditions. Quantum approximate counting algorithm \cite{133,134} can find out an approximate of this quantity with a wanted accuracy. Like Grover algorithm, this algorithm has a $\sqrt{N}$ speed-up compared to the classical algorithms.

\subsubsection{Theory of Quantum approximate counting}

In last section, we introduced if there is only one target in an unstructured database then how to search for it. When there are more than one target entries, say $M$ entries, the procedure of Grover algorithm is the same as in the one-target case, while the final result is the uniform superposition of the basis states that correspond to those entries. Here, we introduce the multi-target case of Grover algorithm briefly. The quantum circuit is the same as the one-target case as in Fig. \ref{s1f11}.

The first operation H$^{\otimes n}$ stores all the 1 to $N$ indices in the $n$-qubit state $|\varphi\rangle $ as in Eq. (\ref{uninform_super}). The $G$ operation following it can also be decomposed to two steps, $R1$ and $R2$. $R1$ realizes the transformation $|x\rangle \rightarrow (-1)^{\text{F}(x)} |x\rangle$. R2 can be expressed as $R2=2|\varphi\rangle\langle \varphi|-I^{\otimes n}$. As stated above, all the steps are the same as in the case of one target, except that the realization of $R1$ gate is different because here the transformation $|x\rangle \rightarrow (-1)^{\text{F}(x)} |x\rangle$ changes the signs of more than one basis states.

We can also use the help of Fig. \ref{s4f38} to understand the process. The state space can be span by two orthonormal bases $|\alpha\rangle$ and $|\beta\rangle$. Now $|\beta\rangle$ is the uniform superposition of those basis states, $|\beta\rangle=\frac{1}{\sqrt{M}} \sum_{x=x_t} | x \rangle$, where $x_t$ are the indices of the target entries. $|\alpha\rangle$ is the uniform superposition of all the other basis states $|\alpha\rangle = \frac{1}{\sqrt{N-M}} \sum_{x \neq x_t} |x\rangle$ . The uniform superposition of all the basis states can be expressed as $|\varphi\rangle= \sqrt{\frac{N-M}{N}} |\alpha\rangle +\sqrt{\frac{M}{N}} |\beta\rangle$. The effect of $G$ is to rotate  $|\varphi\rangle$ towards $|\beta\rangle$ and the rotation angle $\theta$ satisfies  $\sin \frac{\theta}{2 }= \sqrt{\frac{M}{N}}$. If we let [ ] denote an integer that is closest to a given real number, then repeating the rotation for $\left[ \arccos(\sqrt{M/N})/\theta \right]$ times can make the quantum state very close to $|\beta\rangle$.  In other words, by repeating $G$ for $\left[ \arccos(\sqrt{M/N})/\theta \right]$ times one can find the target state $|\beta\rangle$ with a very high probability. It can be easily proved that if $M \leq N/2$, the needed repetition is fewer than $\pi/4 \sqrt{N/M}$. It is clear here that the quantity of the target entries, $M$, is required to implement the search successfully. If $M$ is unknown, quantum counting algorithm should be implemented first.

\begin{figure*}
\centerline{
\includegraphics[width=4in]{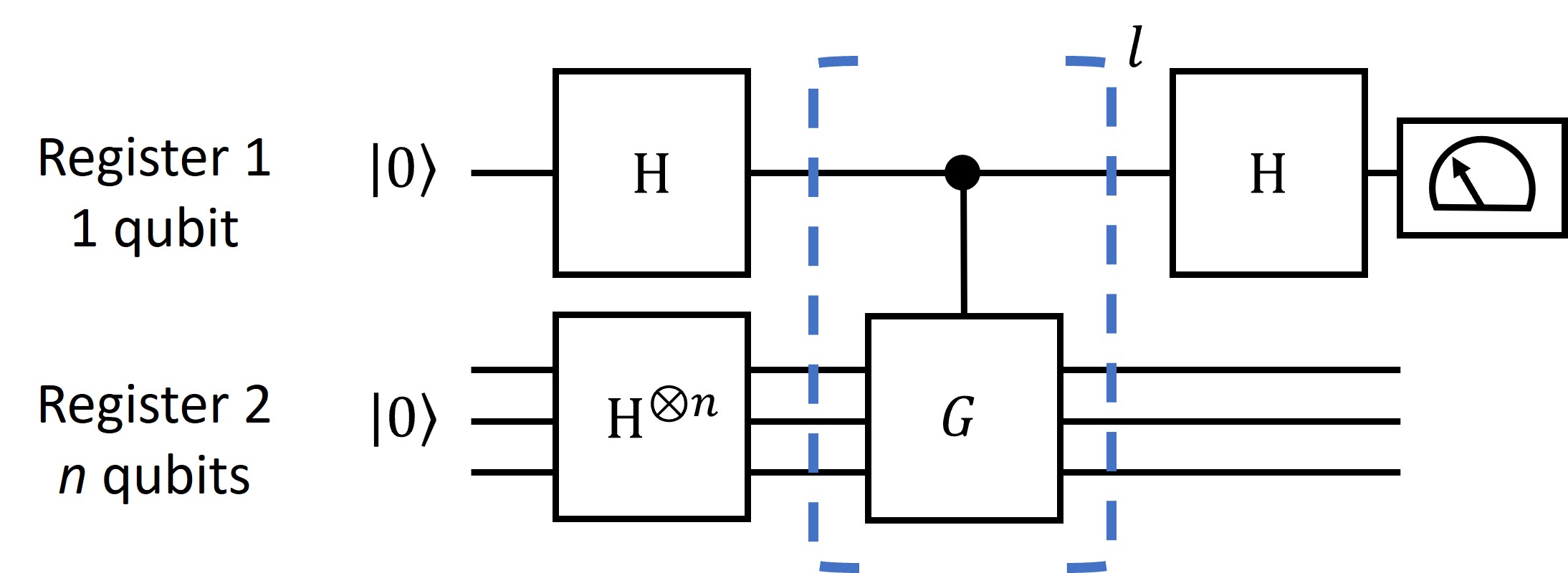}
}
\caption{The circuit for quantum approximate counting.}
\label{s4f40}
\end{figure*}

In Grover algorithm, one register with $n$ qubits is used. In quantum approximate counting, two registers are needed, as shown  in Fig. \ref{s4f40}. One register has $n$ qubits and is used to store the quantum states that correspond to the indices of all the entries in the database. The qubits in the other register are used as the control qubits—if the control qubits are in the state  $|1\rangle$ then the $G$ operation is implemented on the $n$ qubits in the register mentioned earlier. In short, some of the change of the $n$ qubits which is caused by $G$ can be passed to the control qubits and by measuring the control qubits the estimation of how many target entries there are can be done. Quantum phase estimation is a choice to realize such a controlled-operation and estimation process. In Quantum phase estimation, the more the control qubits are, the higher the estimation accuracy is. Here, we introduce a simpler method which is proposed in Ref. \cite{135}, where one control qubit is used.

As mentioned above, $G$ can be considered as a rotation operation in the space span by $|\alpha\rangle$ and $|\beta\rangle$, and can be expressed as a 2$\times$2 matrix,
\begin{align}
G = \begin{pmatrix} \cos\theta & -\sin\theta \\ \sin\theta & \cos\theta \end{pmatrix}. \label{4.59}
\end{align}
Here $\theta$ is the rotated angle and $\sin \frac{\theta}{2 }= \sqrt{\frac{M}{N}}$.  $G$ has two eigen states and two eigen values. The detailed form of the eigen states is not important. Here we use $|\Psi_1\rangle$ and $|\Psi_2\rangle$ to denote the two eigen states, and their eigen values are $e^{i\theta}$ and $e^{i(2\pi-\theta)}$,
\begin{align}
G | \Psi_1 \rangle &= e^{i \theta} | \Psi_1 \rangle, \label{4.60}\\
G | \Psi_2 \rangle &= e^{i (2 \pi - \theta)} | \Psi_2 \rangle \label{4.61}.
\end{align}
$|\varphi\rangle$ can be expressed as the superposition of $|\Psi_1\rangle$ and $|\Psi_2\rangle$, $|\varphi\rangle=a|\Psi_1\rangle+b|\Psi_2\rangle$. If $G$ is implemented to $|\varphi\rangle$ for $l$ times, then
\begin{align}
G^l | \varphi \rangle = ae^{il \theta} | \Psi_1 \rangle + be^{il(2\pi - \theta)} | \Psi_2 \rangle. \label{4.62}
\end{align}
If the control qubit is in   $\frac{1}{\sqrt{2}} \left( |0 \rangle + |1 \rangle \right)$, after $l$ controlled-$G$ gates, the state of the control qubit and the $n$ qubits is
\begin{align}
\frac{1}{\sqrt{2}} \left( \left( |0 \rangle + e^{il\theta} |1 \rangle \right) a | \Psi_1 \rangle + \left( |0 \rangle + e^{il(2\pi - \theta)} |1 \rangle \right) b | \Psi_2 \rangle \right). \label{4.63}
\end{align}
It is clear that because of the implementation of the controlled-$G$ gates, there is a phase factor in front of the state $ |1 \rangle $ of the control qubit. For convenience’s sake, we introduce  $|\phi_1\rangle$ and $|\phi_2\rangle$:
\begin{align}
| \phi_1 \rangle = \frac{1}{\sqrt{2}} \left( |0 \rangle + e^{il\theta} |1 \rangle \right), \quad | \phi_2 \rangle = \frac{1}{\sqrt{2}} \left( |0 \rangle + e^{il(2\pi - \theta)} |1 \rangle \right). \label{4.64}
\end{align}
Then the state after $l$ controlled-$G$ gates and after the H gate on the control qubit can be written as
\begin{align}
\text{H}(a | \phi_1 \rangle | \Psi_1 \rangle + b | \phi_2 \rangle | \Psi_2 \rangle). \label{4.65}
\end{align}
The density matrix of the above state is
\begin{align}
\rho =& \text{H}(aa^* | \phi_1 \rangle | \Psi_1 \rangle \langle \Psi_1 | \langle \phi_1 | + bb^* | \phi_2 \rangle | \Psi_2 \rangle \langle \Psi_2 | \langle \phi_2 | \nonumber\\&+ ab^* | \phi_1 \rangle | \Psi_1 \rangle \langle \Psi_2 | \langle \phi_2 | + a^* b | \phi_2 \rangle | \Psi_2 \rangle \langle \Psi_1 | \langle \phi_1 |) \text{H}.\label{4.66}
\end{align}
Next, the control qubit is observed. In a complex system, if only a sub system is observed, the observed density matrix is a partial trace of the whole density matrix. This partial trace density matrix contains all the information that can be obtained by measuring this sub system and is usually called as the reduced density matrix. To get a partial trace is to calculate $\sum_k \langle k | \rho | k \rangle$  where $|k\rangle$ is the basis of the subsystems that are not observed. Now we calculate the reduced density matrix of the control qubit, which means $|k\rangle$ is the basis of the $n$-qubit register:
\begin{align}
\rho_c = \text{Tr}_n \rho =& \langle \Psi_1 | \rho | \Psi_1 \rangle + \langle \Psi_2 | \rho | \Psi_2 \rangle = aa^* \text{H} | \phi_1 \rangle \langle \phi_1 | \text{H} \nonumber\\&+ bb^*\text{H} | \phi_2 \rangle \langle \phi_2 | \text{H}. \label{4.67}
\end{align}
$\rho_c$ is the density matrix that is to be observed. $\text{H} | \phi_1 \rangle \langle \phi_1 | \text{H}$ and $\text{H} | \phi_2 \rangle \langle \phi_2 | \text{H}$ have expressions as follows
\begin{align}
\text{H} | \phi_1 \rangle \langle \phi_1 | \text{H}&=\frac{1}{2} \begin{pmatrix} 1 + \cos l\theta  & i \sin l\theta  \\ -i \sin l\theta  & 1 - \cos l\theta  \end{pmatrix},\label{4.68}\\
\text{H} | \phi_2 \rangle \langle \phi_2 | \text{H}&=\frac{1}{2} \begin{pmatrix} 1 + \cos l(2\pi-\theta) & i \sin l(2\pi-\theta) \\ -i \sin l(2\pi-\theta) & 1 - \cos l(2\pi-\theta) \end{pmatrix}\nonumber\\
&=\frac{1}{2} \begin{pmatrix} 1 + \cos l\theta  & -i \sin l\theta  \\ i \sin l\theta  & 1 - \cos l\theta  \end{pmatrix}.\label{4.69}
\end{align}
The diagonal elements of the above density matrices are the same. Hence the diagonal elements of $\rho_c$ are $( 1 + \cos l\theta)/2$ and $( 1 - \cos l\theta)/2$ as well. If we measure the control qubit’s angular momentum along the z direction which is $\langle \sigma_z\rangle=\text{Tr}(\sigma_z \rho_c )=\cos l\theta$, we can estimate the value of $\theta$, and further we can get an estimate of $M$. To increase the accuracy of the estimation of $\theta$, one way is to measure the angular momentum multiple times and increase $l$ each time. This way measures the curve of the diagonal elements of  $\rho_c$ changing along with $l$. $\theta$ can be estimated by Fourier transformation or data fitting of this curve. Another way is to increase the number of control qubits to implement quantum phase estimation. Here we will not discuss this method further.

\subsubsection{Quantum approximate counting when N=2}

\begin{figure*}
\centerline{
\includegraphics[width=4in]{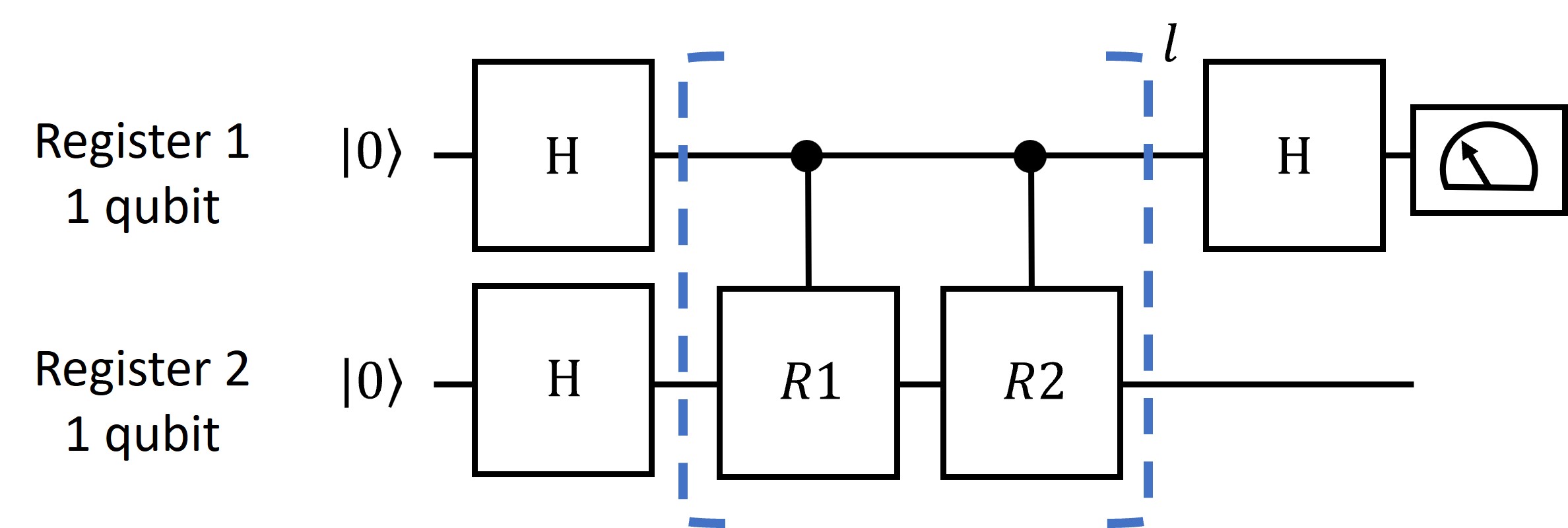}
}
\caption{The circuit for quantum approximate countingin the case of $N$=2.}
\label{s4f41}
\end{figure*}

Here we analyze the situation when the database contains two entries, $N=2$. Then $n$=1 and the circuit is shown in the Fig. \ref{s4f41}.

We want to know how many target entries there are. Because the database has only two entries in total, there are only four possible cases: there are 0 target entry; one target entry and it is the first entry; one target entry and it is the second entry; two target entries. Next, we give the quantum gates used in the four cases. In the four cases, the $R2$ operations are all same, which is to change the sign of all the states that are orthogonal to $|\varphi\rangle$,
\begin{align}
R2 = \text{H} \begin{pmatrix}
1 & 0 \\
0 & -1
\end{pmatrix} \text{H}= \begin{pmatrix}
0 & 1 \\
1 & 0
\end{pmatrix} = \sigma_x \label{4.70}
\end{align}
$R1$ are $I_2$, $-\sigma_z$, $\sigma_z$ and -$I_2$ in the four cases. Here $I_2$ is the $2\times2$ identity matrix.
\begin{align}
I_2 &= \begin{pmatrix}
1 & 0 \\
0 & 1
\end{pmatrix},\label{4.71}\\
-\sigma_z &= \begin{pmatrix}
-1 & 0 \\
0 & 1
\end{pmatrix},\label{4.72}\\
\sigma_z &= \begin{pmatrix}
1 & 0 \\
0 & -1
\end{pmatrix},\label{4.73}\\
-I_2 &= \begin{pmatrix}
-1 & 0 \\
0 & -1
\end{pmatrix}.\label{4.74}
\end{align}

\begin{figure*}
\centerline{
\includegraphics[width=5.5in]{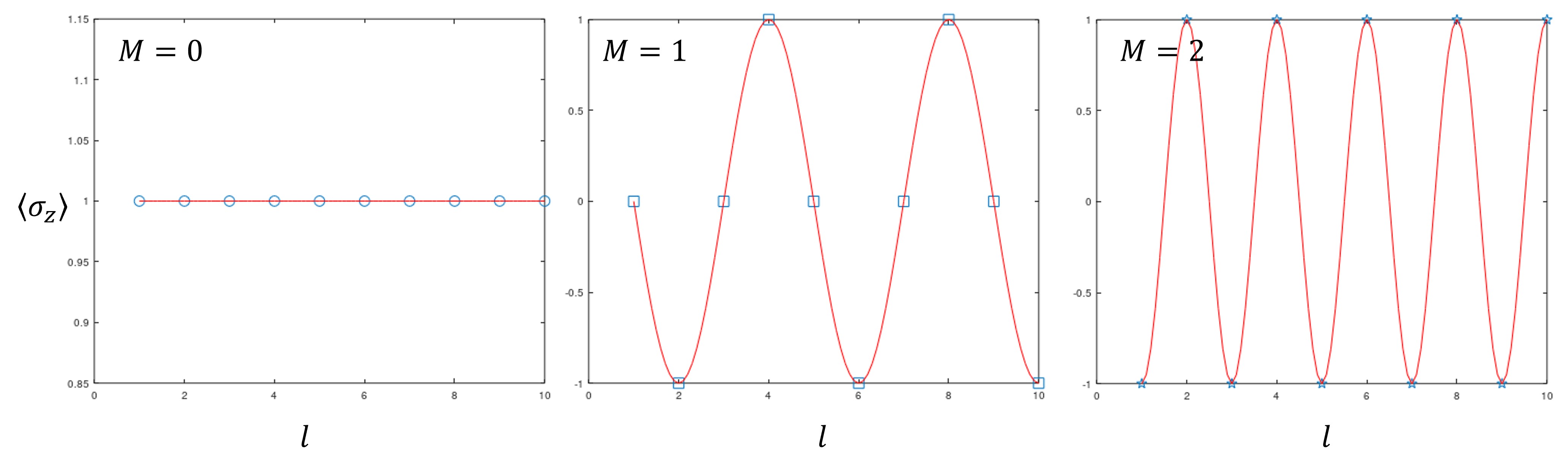}
}
\caption{The theoretical curves of $\sigma_z$ when $M=0$, $M=1$ and $M=2$.}
\label{s4f42}
\end{figure*}

It should be noted that the gates are implemented in the controlled-unitary form. The matrix form of the controlled-$R2$ gate is
\begin{align}
\text{control}-R2=\begin{pmatrix}
1 & 0 & 0 & 0\\
0 & 1 & 0 & 0\\
0 & 0 & 0 & 1\\
0 & 0 & 1 & 0
\end{pmatrix}=\text{CNOT}.\label{4.75}
\end{align}
In the four cases, the matrix forms of the controlled-$R1$ gate are
\begin{align}
\text{control}-I_2&=\begin{pmatrix}
1 & 0 & 0 & 0\\
0 & 1 & 0 & 0\\
0 & 0 & 1 & 0\\
0 & 0 & 0 & 1
\end{pmatrix}=I_4,\label{4.76}\\
\text{control}-(-\sigma_z)&=\begin{pmatrix}
1 & 0 & 0 & 0\\
0 & 1 & 0 & 0\\
0 & 0 & -1 & 0\\
0 & 0 & 0 & 1
\end{pmatrix}\nonumber\\&=\sigma_x^2\begin{pmatrix}
1 & 0 & 0 & 0\\
0 & 1 & 0 & 0\\
0 & 0 & 1 & 0\\
0 & 0 & 0 & -1
\end{pmatrix}\sigma_x^2=\sigma_x^2\text{CZ}\sigma_x^2,\label{4.77}\\
\text{control}-(\sigma_z)&=\begin{pmatrix}
1 & 0 & 0 & 0\\
0 & 1 & 0 & 0\\
0 & 0 & 1 & 0\\
0 & 0 & 0 & -1
\end{pmatrix}=\text{CZ},\label{4.78}\\
\text{control}-(-I_2)&=\begin{pmatrix}
1 & 0 & 0 & 0\\
0 & 1 & 0 & 0\\
0 & 0 & -1 & 0\\
0 & 0 & 0 & -1
\end{pmatrix}=\sigma_z^1.\label{4.79}
\end{align}
Here, $I_4$ is the $4\times4$ identity matrix; CNOT gate flips the target qubit (namely, implement $\sigma_x$ operation) when the control qubit is in $|1\rangle$; CZ is the controlled-phase gate, which implement $\sigma_z$ operation to the target qubit when the control qubit is in $|1\rangle$; $\sigma_x^2$ is the $\sigma_x$ operation on the second qubit; $\sigma_z^1$ is the $\sigma_z$ operation on the first qubit.

\begin{figure*}
\centerline{
\includegraphics[width=5.5in]{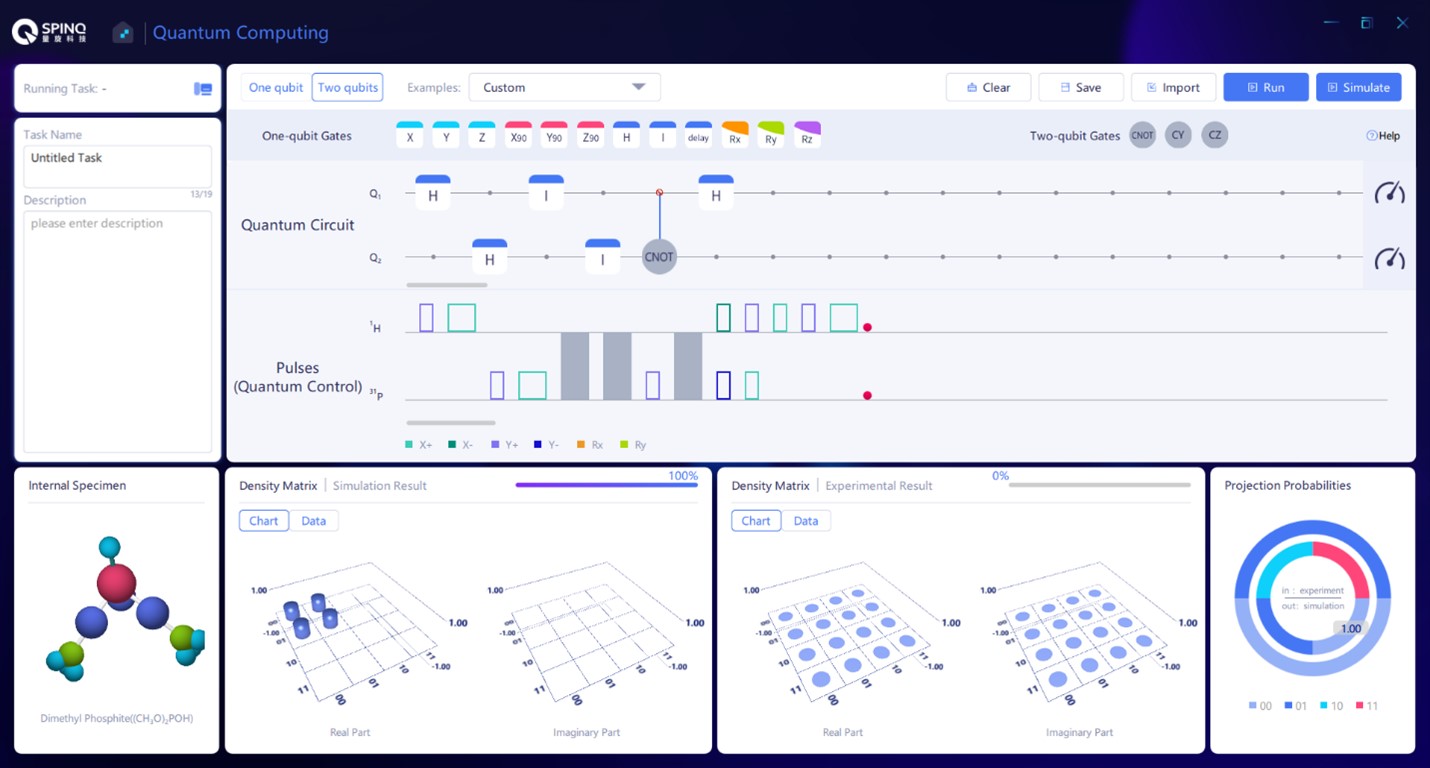}
}
\caption{The circuit for the case when there is no target entry in the database.}
\label{s4f43}
\end{figure*}

\begin{figure*}
\centerline{
\includegraphics[width=5.5in]{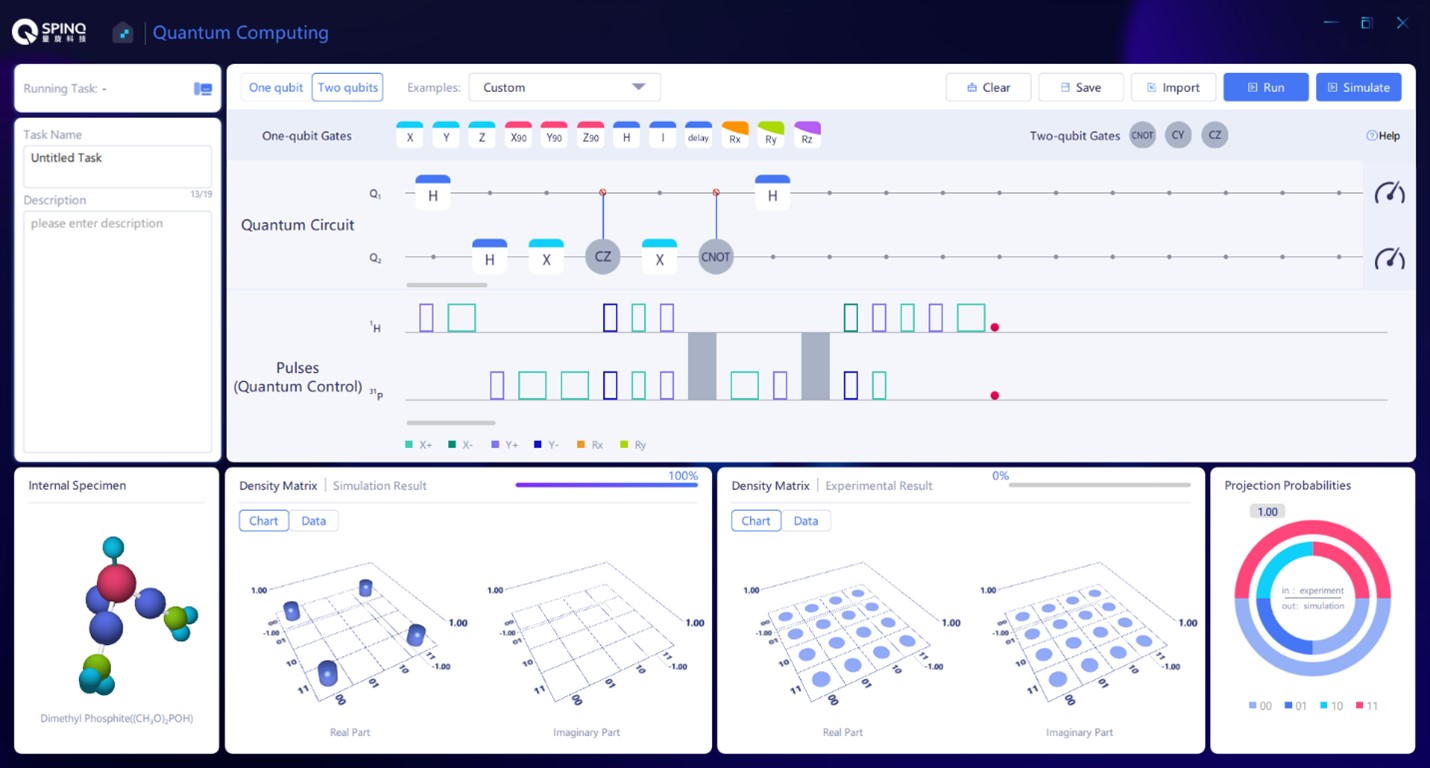}
}
\caption{The circuit for the case when the first entry is the target entry.}
\label{s4f44}
\end{figure*}

\begin{figure*}
\centerline{
\includegraphics[width=5.5in]{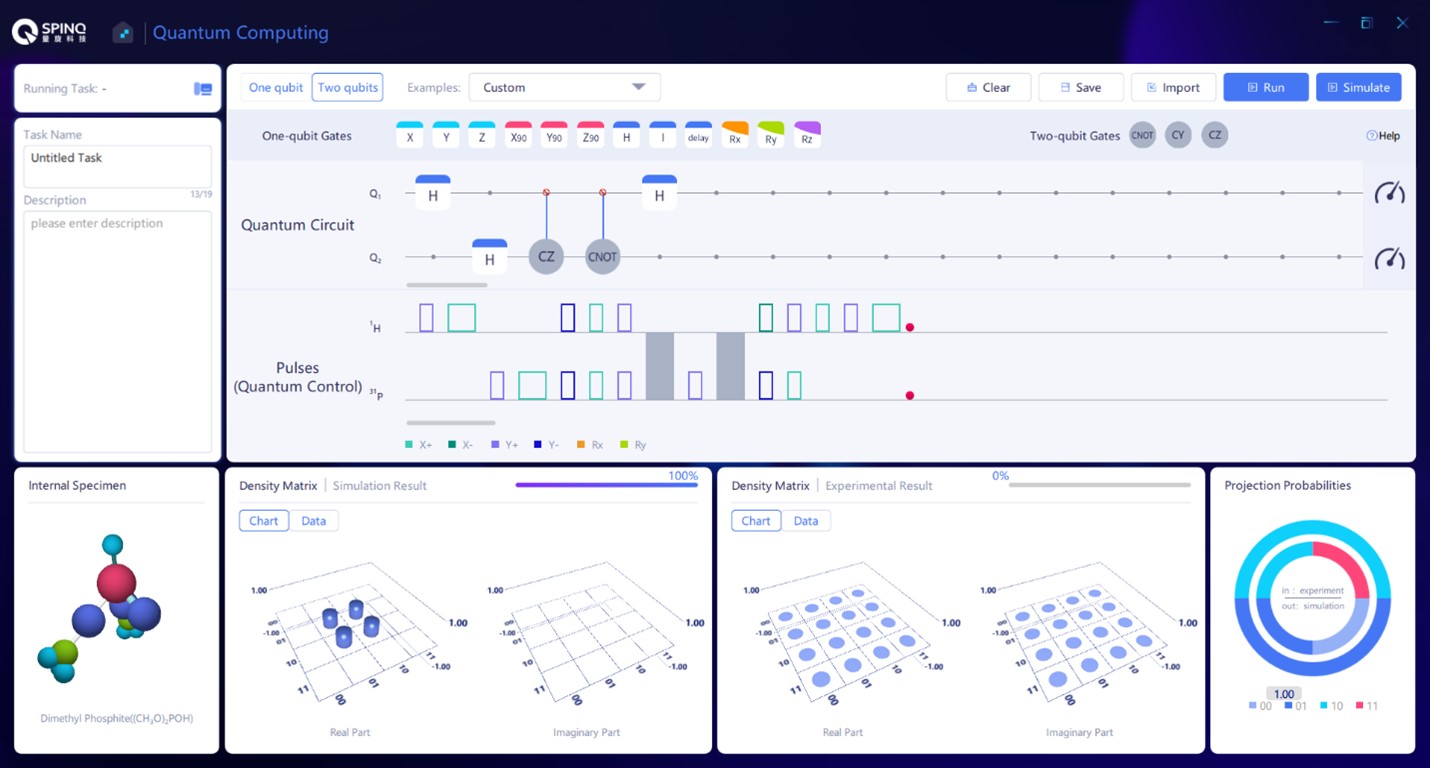}
}
\caption{The circuit for the case when the second entry is the target entry.}
\label{s4f45}
\end{figure*}

\begin{figure*}
\centerline{
\includegraphics[width=5.5in]{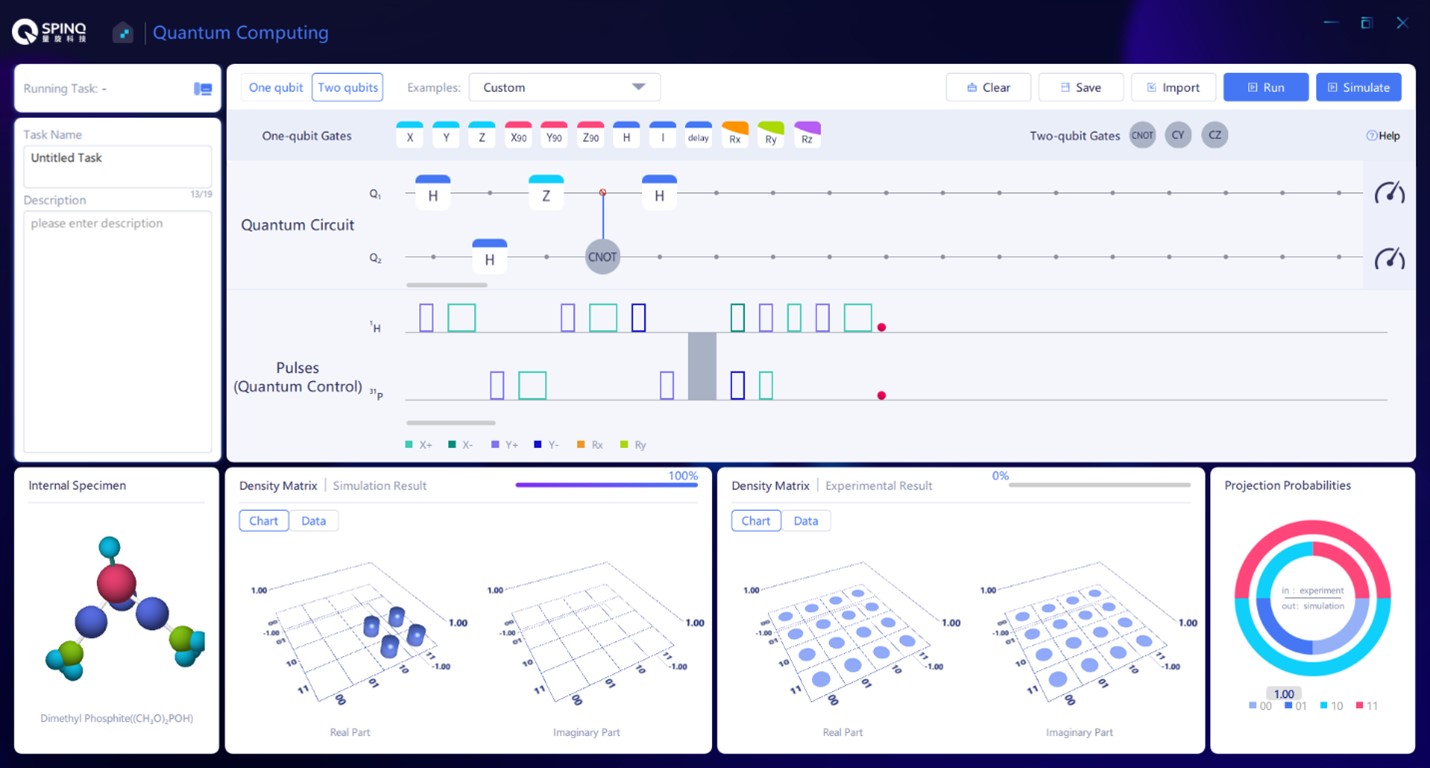}
}
\caption{The circuit for the case when both the two entries are target entries.}
\label{s4f46}
\end{figure*}

From the analysis above, we need to observe the z-direction angular momentum of the control qubit, $\langle\sigma_z\rangle=\cos l\theta$. Usually, we need to change $l$ and measure the curve of $\langle\sigma_z\rangle$ changing with $l$. This curve is a cosine curve with its frequency determined by $\theta$. When  $n$=1, the values of $\theta = 2\arcsin\sqrt{M/N}$  are 0, $\pi$/2, $\pi$/2, $\pi$ in the four cases. Figure \ref{s4f42} shows the theoretical curves when $\theta$ is 0, $\pi$/2 and $\pi$ (solid red lines). The theoretical values of  $\langle\sigma_z\rangle$  when $l$=1,2,…,9,10 are marked using circles, squares and stars. If in experiments, a certain pattern of the curve is measured, for example, no oscillation is observed, then $M$=0, or the oscillation frequency is close to $\pi$/2, then $M$=1, or the oscillation frequency is close to $\pi$, then $M$=2. Actually, because it is a simple case when $n$=1, one measurement of $\langle\sigma_z\rangle$ with $l$=1 can give the correct value of $M$ with large probabilities. If the measured $\langle\sigma_z\rangle$ is close to 1, then $M$=0; if the measured $\langle\sigma_z\rangle$ is close to 0, then $M$=1; if the measured $\langle\sigma_z\rangle$ is close to -1, then $M$=2.

\subsubsection{Experimental implementation}

Here we implement quantum approximate counting algorithm for the $N$=2 case. Figures \ref{s4f43}-\ref{s4f46} are the circuits for the four cases: (1) There is no target entry in the database; (2) the first entry is the target entry; (3) the second entry is the target entry; (4) both the two entries are target entries. The first qubit ($^1$H) is used as the control qubit; the second qubit ($^{31}$P) is used to store the quantum states corresponding to the entry indices. 

\subsubsection{Data analysis}

Here we extract the reduced density matrix of the control qubit (the first qubit). The way to calculate a reduced density matrix from a whole system density matrix is to calculate  $\sum_k \langle k | \rho | k \rangle$ where $| k \rangle$ are the base of the subsystem that is not observed. For a 4$\times$4 density matrix $\rho$, the reduced density matrix of the first qubit can be expressed as
\begin{align}
\begin{pmatrix}
\rho_{11}+\rho_{22} & \rho_{13}+\rho_{24} \\
\rho_{31}+\rho_{42} & \rho_{33}+\rho_{44}
\end{pmatrix} \label{4.80}
\end{align}
$\theta$ and $M$ can then be derived from the reduced density matrices.

\subsection{Bernstein-Vazirani algorithm}

Bernstein-Vazirani algorithm \cite{23} solves such a problem, i.e. how an unknown data sequence can be determined by parity check. In classical algorithms, if an $n$-bit binary number needs to be determined by parity check, $n$ queries are required, while if Bernstein-Vazirani quantum algorithm is used, only one query is required for obtaining this unknown data sequence. Here we will introduce an variant \cite{137} based on Bernstein-Vazirani algorithm, which is characterized by no need for the participation of quantum entanglement. This algorithm is an example of quantum speedup provided by quantum superposition.

\subsubsection{Theory of Bernstein-Vazirani algorithm}

\begin{figure*}[!htbp]
\centerline{
\includegraphics[width=3.5in]{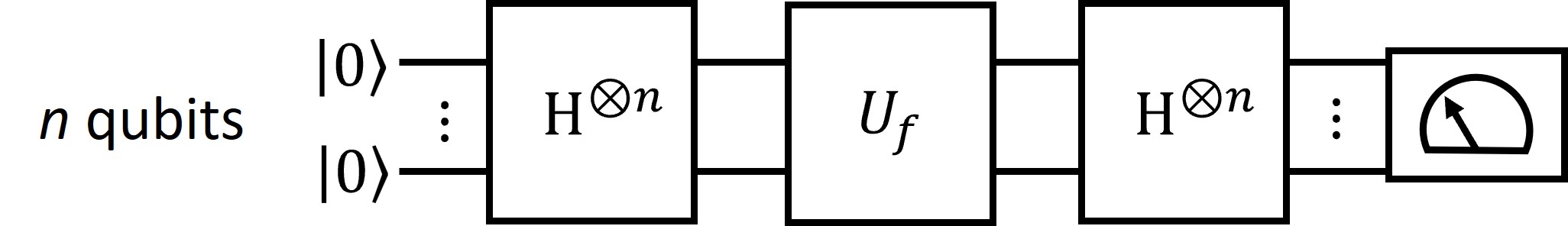}
}
\caption{The circuit of the variant Bernstein-Vazirani algorithm introduced here.}
\label{s4f47}
\end{figure*}

A database has an $n$-bit binary number $\boldsymbol{a}\in\{0,1\}^n$. We use $a_i$ to denote the $i$th bit of $\boldsymbol{a}$ ($i$=1,…,$n$). Assume that we can query this database. By selecting any $n$-bit binary number $\boldsymbol{x}\in\{0,1\}^n$ during each query, the query returns $\boldsymbol{a}\cdot \boldsymbol{x}(\text{mod}\; 2)$, i.e. the modulo two addition of those bits which are non-zero for both $\boldsymbol{a}$ and $\boldsymbol{x}$. Therefore, the query function can be denoted as $f(\boldsymbol{x})$:
\begin{align}
f(\boldsymbol{x}) = \boldsymbol{a} \cdot \boldsymbol{x} \pmod{2} = \sum_{i=1}^{n} a_i x_i \pmod{2}. \label{4.81}
\end{align}
This function can return 0 or 1 only. In case of 0, it means that the sum of bits where both $\boldsymbol{a}$ and $\boldsymbol{x}$ are 1 is an even number; in case of 1, it means that the sum of bits where both $\boldsymbol{a}$ and $\boldsymbol{x}$ are 1 is an odd number.

With such a query function, if we use classical algorithms, we can confirm what $\boldsymbol{a}$ is after $n$ queries, and the $\boldsymbol{x}$ values used in $n$ queries are expressed as follows:
\begin{align}
\begin{matrix}
1 & 0 & 0 & \ldots & 0 & 0 \\
0 & 1 & 0 & \ldots & 0 & 0 \\
0 & 0 & 1 & \ldots & 0 & 0 \\
  &   &   & \vdots & &  \\
0 & 0 & 0 & \ldots & 1 & 0 \\
0 & 0 & 0 & \ldots & 0 & 1\label{4.82}
\end{matrix}
\end{align}
We can confirm one bit of  $\boldsymbol{a}$ after each query.

Bernstein-Vazirani quantum algorithm can be used to solve the problem, so that $\boldsymbol{a}$ can be obtained by only one query. The core of the quantum algorithm is to prepare a uniform superposition state of $n$ qubits, use the superposition state as the input of the query function $f(\boldsymbol{x})$, which means that a parallel query is carried out making use of the nature of quantum superposition, and then code all query results to this uniform superposition state. The only difference between the original Bernstein-Vazirani quantum algorithm and the algorithm variant to be introduced here is that the original Bernstein-Vazirani algorithm uses additional qubits to assist in realizing the coding process, which we will not go into details. The circuit of the algorithm is shown in Fig. \ref{s4f47}.

The initial state of the n qubits is $|000\cdots 00\rangle$, and the $n$ qubits are in a uniform superposition state after Hadamard gate H$^{\otimes n}$,
\begin{align}
\left| \varphi \right\rangle = \frac{1}{\sqrt{2^n}} \sum_{\boldsymbol{x} \in \{0,1\}^n} \left| \boldsymbol{x} \right\rangle.\label{4.83}
\end{align}
It should be pointed out that the effect of Hadamard gate H$^{\otimes n}$ on any initial state can be expressed as follows:
\begin{align}
\text{H}^{\otimes n} \left| \boldsymbol{y} \right\rangle = \frac{1}{\sqrt{2^n}} \sum_{\boldsymbol{x} \in \{0,1\}^n} (-1)^{\boldsymbol{y} \cdot \boldsymbol{x}} \left| \boldsymbol{x} \right\rangle,\label{4.84}
\end{align}
The form of $\left| \varphi \right\rangle$ in Eq. (\ref{4.83})  can be derived from the above equation.

After H$^{\otimes n}$, it is a $U_f$ gate, which is used to achieve query by $f(\boldsymbol{x})$ and store the query results into the quantum state. The method is to realize $\left| \boldsymbol{x} \right\rangle \rightarrow (-1)^{f(\boldsymbol{x})} \left| \boldsymbol{x} \right\rangle
$. $\left| \varphi \right\rangle$ will be transformed as follows:
\begin{align}
\left| \varphi \right\rangle &\xrightarrow{U_f} \frac{1}{\sqrt{2^n}} \sum_{\boldsymbol{x} \in \{0,1\}^n} (-1)^{f(\boldsymbol{x})} \left|\boldsymbol{ x} \right\rangle \nonumber\\&= \frac{1}{\sqrt{2^n}} \sum_{\boldsymbol{x} \in \{0,1\}^n} (-1)^{\boldsymbol{a} \cdot \boldsymbol{x}} \left| \boldsymbol{x} \right\rangle.\label{4.85}
\end{align}
By comparing the expression on the right side of the above equation with that on the right side of Eq. (\ref{4.84}), it can be found that they share the same form. As H$^{\otimes n}$ is a unitary operation which is reversible, and is its own inverse operation, $\left| \boldsymbol{a} \right\rangle$ will be obtained by applying another H$^{\otimes n}$ gate on the right side of the above equation:
\begin{align}
\text{H}^{\otimes n}\frac{1}{\sqrt{2^n}} \sum_{\boldsymbol{x} \in \{0,1\}^n} (-1)^{\boldsymbol{a} \cdot \boldsymbol{x}} \left| \boldsymbol{x} \right\rangle=\left| \boldsymbol{a} \right\rangle.\label{4.86}
\end{align}
Therefore,  $\left| \boldsymbol{a} \right\rangle$ state will be obtained by measuring all qubits at this moment, which makes it possible to obtain  $ \boldsymbol{a} $ by using the function $f(\boldsymbol{x})$ once only.

\begin{figure*}[!htbp]
\centerline{
\includegraphics[width=5in]{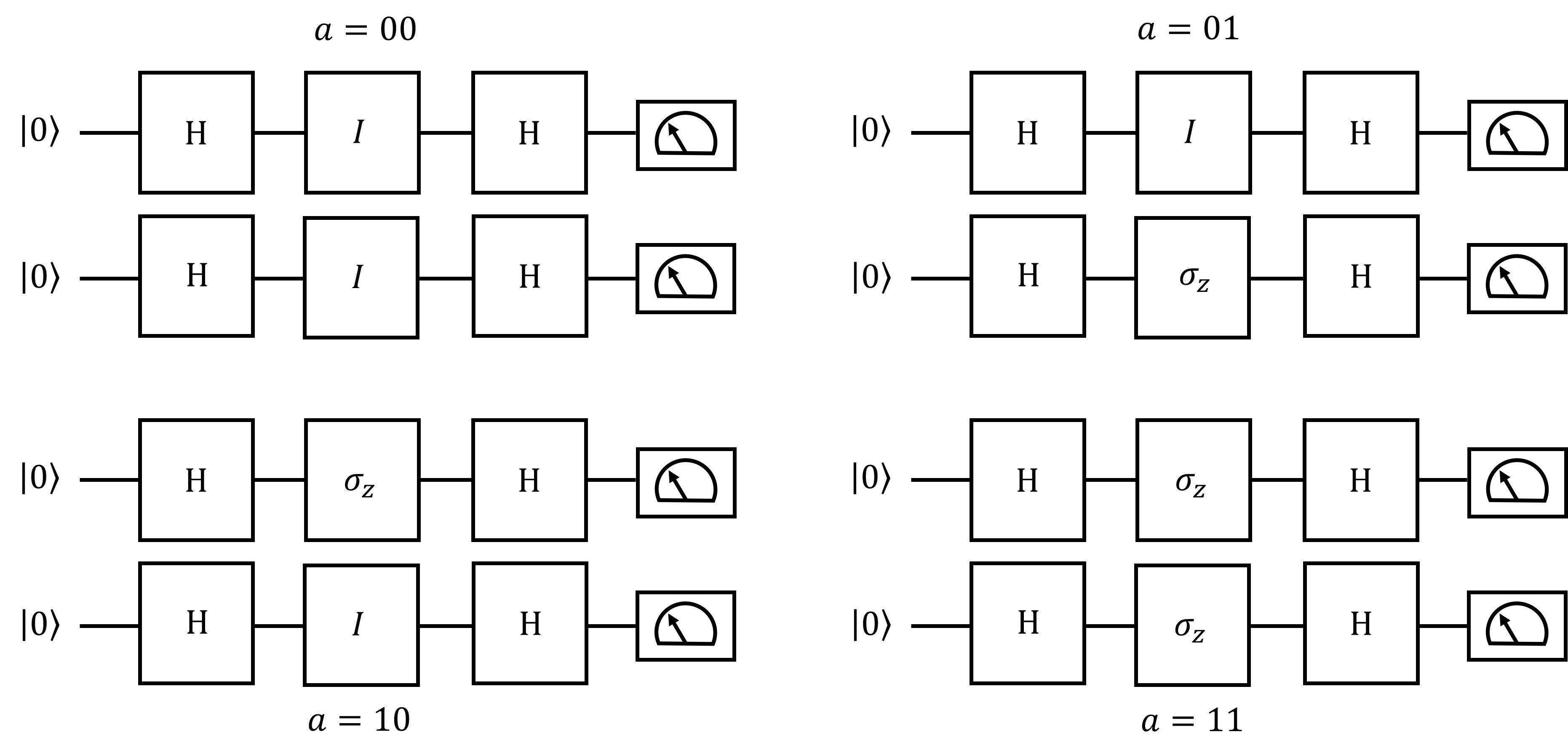}
}
\caption{The quantum circuits of the algorithm in the two-qubit case for different $\boldsymbol{a}$ values.}
\label{s4f48}
\end{figure*}

\subsubsection{Quantum gates used in the algorithm}

Now, we will take a look at how $U_f$ gate is realized. In fact, it can be decomposed into the direct product of $n$ single-qubit gates.
\begin{align}
U_f = U_1 \otimes U_2 &\otimes \cdots \otimes U_{n-1} \otimes U_n,\nonumber\\
U_i = \begin{cases}
I, & \text{if } a_i = 0 \\
\sigma_z, & \text{if } a_i = 1
\end{cases}, \quad &I = \begin{pmatrix}
1 & 0 \\
0 & 1
\end{pmatrix}, \quad \sigma_z = \begin{pmatrix}
1 & 0 \\
0 & -1
\end{pmatrix}.\label{4.87}
\end{align}
Implementing $U_f$ is equivalent to implementing single-qubit gate $U_i$ on each qubit. H$^{\otimes n}$ gate is also the direct product of single-qubit gates H. Therefore, the quantum gates required for the algorithm are all direct products of single-qubit gates, and these gates cannot generate entanglement. Moreover, the initial state of the algorithm is  $|000\cdots 00\rangle$, and there is no entanglement at the very beginning. All this means that the algorithm does not apply any quantum entanglement. The quantum advantage of this algorithm over the classical counterpart comes from the quantum superposition.

\begin{figure*}[!htbp]
\centerline{
\includegraphics[width=5.5in]{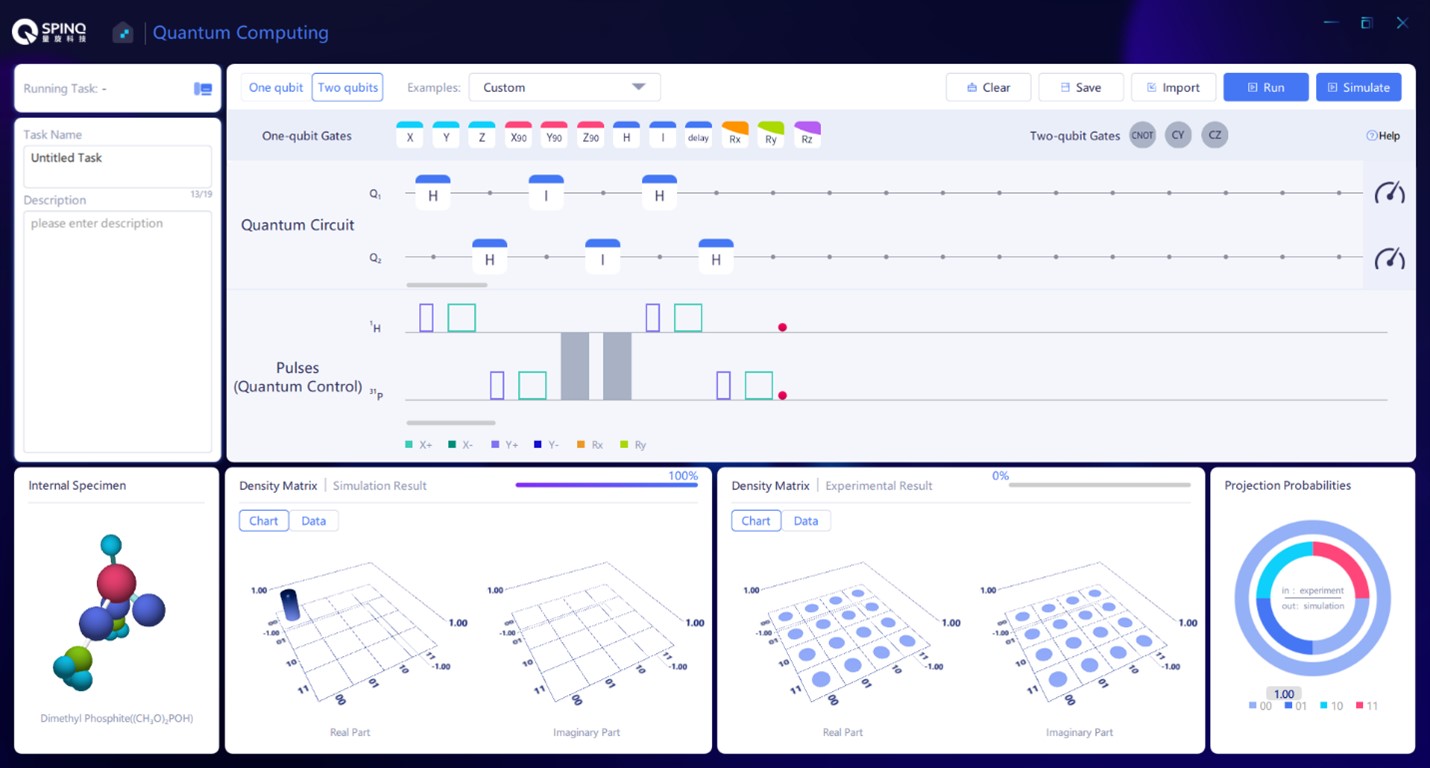}
}
\caption{The circuit when  $\boldsymbol{a}$ is 00.}
\label{s4f49}
\end{figure*}

\begin{figure*}[!htbp]
\centerline{
\includegraphics[width=5.5in]{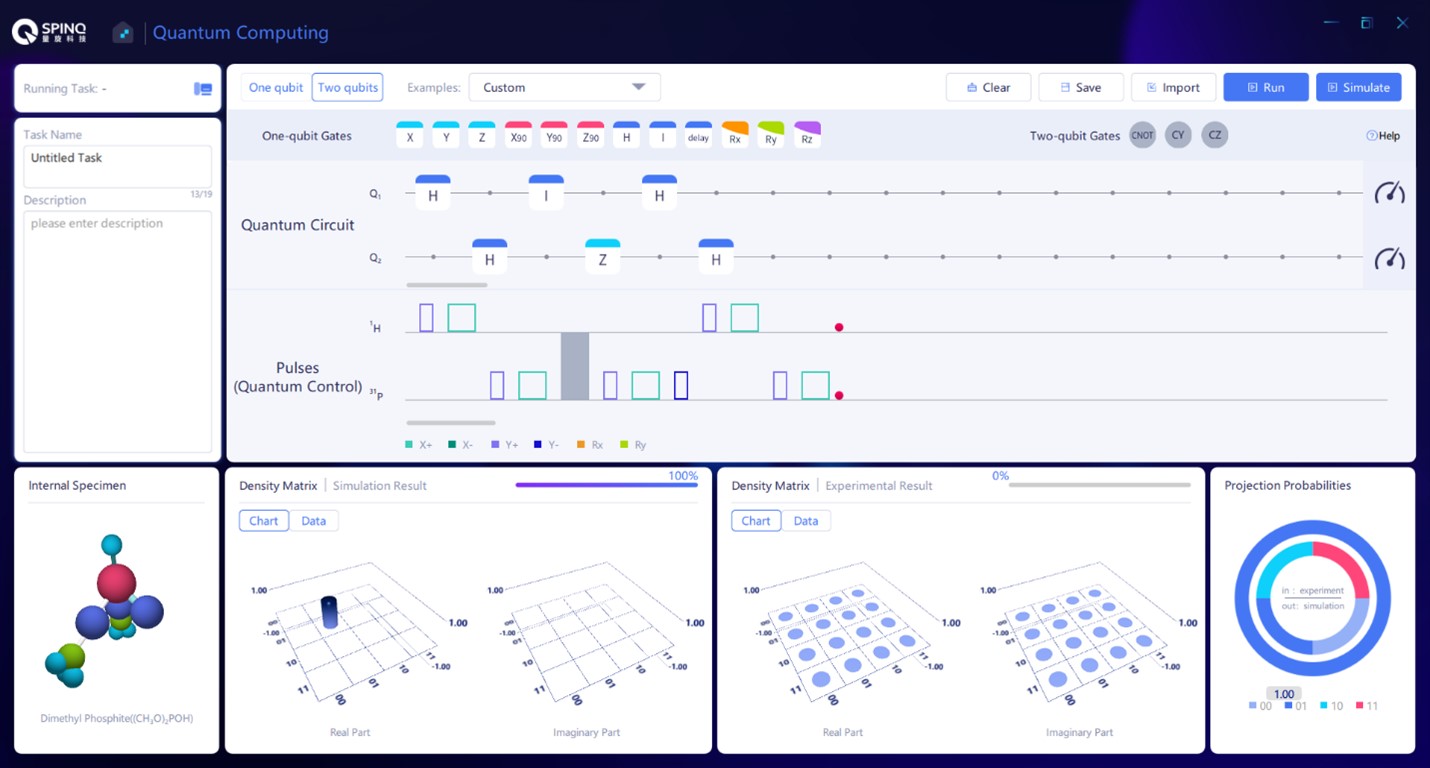}
}
\caption{The circuit when  $\boldsymbol{a}$ is 01.}
\label{s4f50}
\end{figure*}

\begin{figure*}[!htbp]
\centerline{
\includegraphics[width=5.5in]{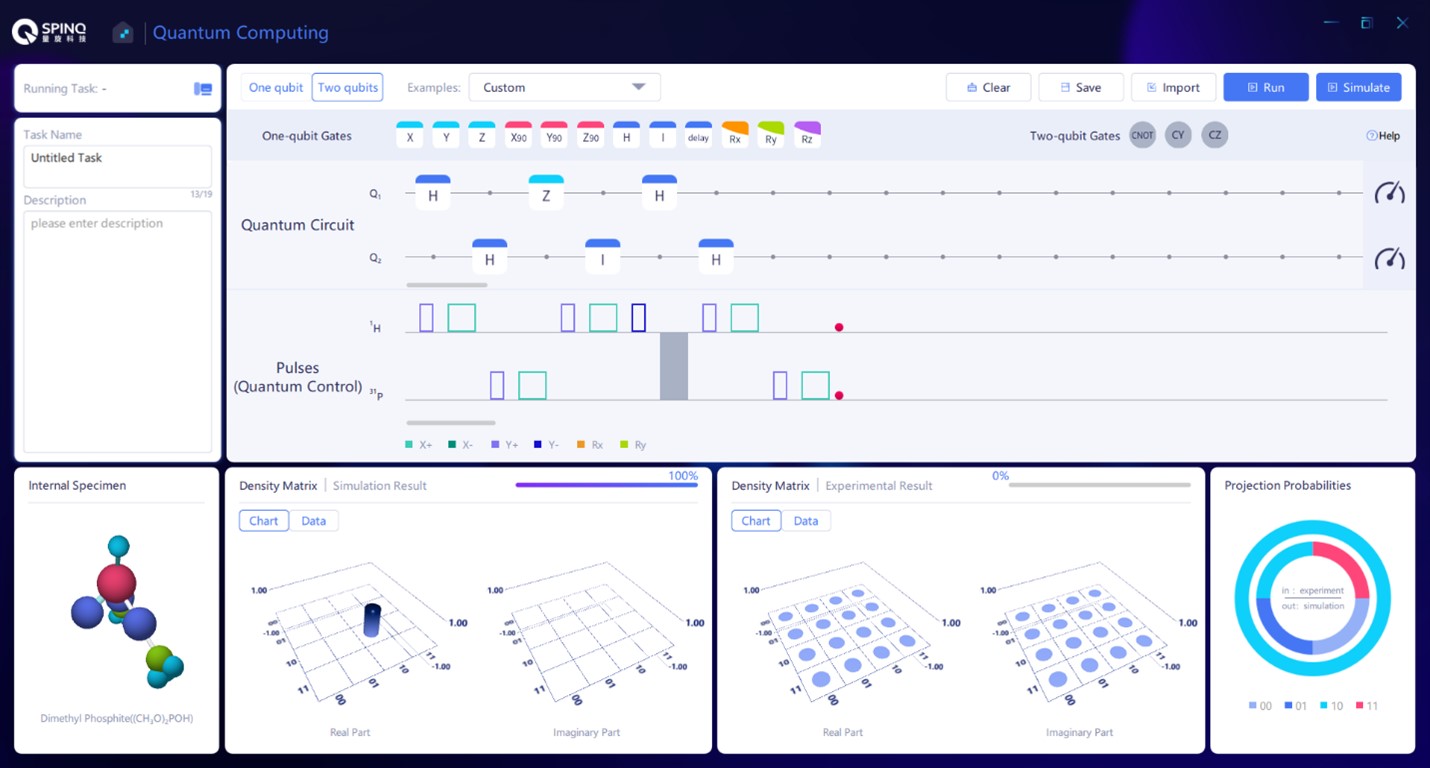}
}
\caption{The circuit when  $\boldsymbol{a}$ is 10.}
\label{s4f51}
\end{figure*}

\begin{figure*}[!htbp]
\centerline{
\includegraphics[width=5.5in]{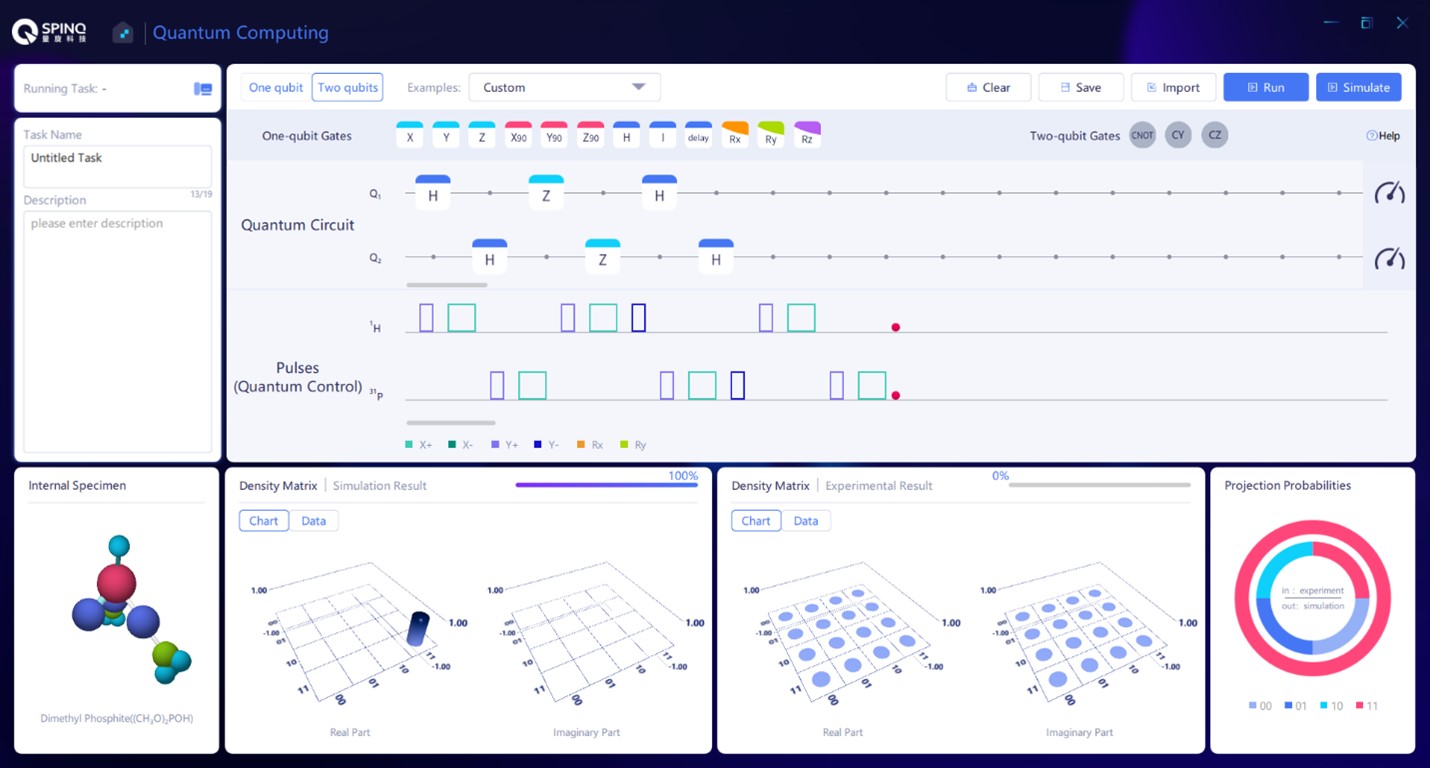}
}
\caption{The circuit when  $\boldsymbol{a}$ is 11.}
\label{s4f52}
\end{figure*}

We will implement the algorithm in a two-qubit case. In the two-qubit case, $\boldsymbol{a}$ has four possible values, i.e. 00, 01, 10 and 11. The quantum circuits of the algorithm in these four cases are shown in Figure \ref{s4f48}.

\subsubsection{Expriemntal implementation}

Here we implement the variant of Bernstein-Vazirani algorithm introduced above in the two-qubit system. Figures \ref{s4f49}-\ref{s4f52} show the quantum circuits in SpinQuasar to realize the algorithm in which $\boldsymbol{a}$ is 00, 01, 10 and 11, respectively.

\subsection{Quantum simulation of quantum harmonic oscillator}

Quantum simulation \cite{1,14,15,16} uses a controllable quantum system to simulate another quantum system, so as to solve some problems of the simulated system. Due to the exotic properties of quantum systems, it is difficult to simulate them with classical computers. Quantum simulation is a very important topic in the study of quantum computation. In this section, a two-qubit system is used to simulate the four-level quantum harmonic oscillator \cite{138}. This experiment is one of the earliest quantum simulations realized in history. Through the experiment, readers can have a preliminary understanding of the methods and steps of quantum simulation.

\subsubsection{Quantum simulation}

Quantum systems are quite different from classical systems. A quantum system can be described by a state in Hilbert space, whose number of dimensions exponentially increases with the number of particles in the quantum system. The state of the quantum system can be the superposition state of any basis vectors in Hilbert space, which is quantum superposition. Also, there exist correlations in the quantum system which are different from those in classical systems, i.e. quantum correlation (quantum discord and quantum entanglement). In view of the above, it is difficult to simulate a quantum system with classical computers. Even with the help of today’s supercomputers, we can simulate only a few tens of qubits. However, many problems can be solved only by effectively simulating the quantum system, such as understanding a lot of physical phenomena in multi-body problems and the design of new quantum materials, etc. Feynman proposed a possible solution in 1982, namely to use a controllable quantum system to simulate another quantum system with problems to be solved, which is known as quantum simulation, and the controllable quantum system is known as a quantum simulator. The schematic diagram of the quantum simulation process is given in Fig. \ref{s1f9}. The quantum system with problems to be solved is usually an uncontrollable or unrealizable quantum system under the current laboratory conditions. The initial state of the quantum system is coded to that of the quantum simulator, and the evolution process of the quantum system is simulated by the evolution process of the quantum simulator. After the initial state of the quantum simulator is prepared and evolves under controllable conditions, some characteristics of the final state of the simulated quantum system can be obtained by measuring the final state of the quantum simulator. However, it should be noted that the correspondence between these two systems does not require exactly the same evolution time.

\subsubsection{Quantum harmonic oscillator}

Quantum harmonic oscillator is an important model in quantum mechanics, which is one of a few models with exact analytical solutions. Moreover, any smooth potential well can be approximated by a harmonic oscillator potential well near the stable equilibrium point. The quantum simulation of the quantum harmonic oscillator model is a proof of principle test to verify the feasibility of quantum simulation.

The Hamiltonian of a quantum harmonic oscillator is expressed as follows:
\begin{align}
\mathcal{H}_{\text{QHO}} = \hbar \Omega \left(\hat{N}+ \frac{1}{2}\right) = \sum_n \hbar \Omega \left(n+\frac{1}{2}\right) \left| n \rangle \langle n \right|.\label{4.88}
\end{align}
Wherein, $\Omega$ refers to the frequency of the quantum harmonic oscillator, and $\hat{N}$ refers to the number operator, whose physical meaning in the quantum harmonic oscillator model is the number of energy quanta. $| n \rangle$ refers to the $n$th eigenstate of  $\hat{N}$,  $\hat{N}| n \rangle=n| n \rangle$, and $| n \rangle$ is also the $n$th eigenstate of $\mathcal{H}_\text{QHO}$,  $\mathcal{H}_\text{QHO}| n \rangle=\hbar \Omega \left(n+\frac{1}{2}\right)| n \rangle$. Therefore, $| n \rangle$ state has $n$ energy quanta, and the energy of each energy quanta is $\hbar \Omega$, so that the total energy of $| n \rangle$ is $\hbar \Omega \left(n+\frac{1}{2}\right)$. It should be noted that when $n$=0, the energy of the system is $1/2 \hbar \Omega$, not 0, and this energy is known as zero-point energy. Next, we will see what changes will take place in the quantum harmonic oscillator after evolution for $t$. First of all, assuming that the initial state $|\Phi(0)\rangle$ of the quantum harmonic oscillator is the eigenstate $| n \rangle$ of Hamiltonian, then the quantum state gets a global phase change $\exp(-it\Omega(n+1/2))$ after evolution, which cannot be observed in the experiment, so the observation result is that the system is still in $| n \rangle$. If the initial state of the quantum harmonic oscillator is in the superposition state of different eigenstates, such as $1/\sqrt{2}(| n \rangle+| m \rangle)$, the coherence terms $|n\rangle\langle m|$ and $|m\rangle\langle n|$ can be found non-zero in its density matrix. After evolution for $t$, the coherent terms will get a phase change determined by the difference in the energy levels of the two eigenstates and change to  $\exp (-it\Omega(n-m))|n\rangle\langle m|$ and $\exp(-it\Omega(m-n))|m\rangle\langle n|$, and such change in phase can be observed.

The quantum harmonic oscillator has an infinite number of energy levels, but the quantum simulator we use is a two-qubit system, which has only four energy levels. So we will simulate a truncated quantum harmonic oscillator model. The correspondence between the eigenstates of the quantum simulator to be used in the experiment and those of the quantum harmonic oscillator is given in the equation below:
\begin{align}
\left| n=0 \right\rangle& \leftrightarrow \left| \uparrow\uparrow \right\rangle \nonumber\\
\left| n=1 \right\rangle& \leftrightarrow \left| \downarrow\uparrow \right\rangle\nonumber \\
\left| n=2 \right\rangle& \leftrightarrow \left| \downarrow\downarrow \right\rangle\nonumber \\
\left| n=3 \right\rangle& \leftrightarrow \left| \uparrow\downarrow \right\rangle.\label{4.89}
\end{align}
We simply denote $\left| n=0 \right\rangle$, $\left| n=1 \right\rangle$, $\left| n=2 \right\rangle$ and $\left| n=3 \right\rangle$ as $\left|0 \right\rangle$, $\left|1 \right\rangle$, $\left|2 \right\rangle$ and $\left|3 \right\rangle$, and please note that $\left|0 \right\rangle$ and $\left|1 \right\rangle$ are not the two states of a qubit expressed by us as usual, and we denote the two states of a qubit as $|\uparrow\rangle$ and $| \downarrow\rangle$. There is another important point to be noted that the energy of $\left| \uparrow\uparrow \right\rangle $, $\left| \downarrow\uparrow \right\rangle$, $\left| \downarrow\downarrow \right\rangle$ and $\left| \uparrow\downarrow \right\rangle$ is not incremental like $\left|0 \right\rangle$, $\left|1 \right\rangle$, $\left|2 \right\rangle$ and $\left|3 \right\rangle$, and even so, the mapping given in the above equation can complete the quantum simulation. According to the mapping in the above equation, the evolution operator $U_\text{q}$ of the quantum harmonic oscillator can be mapped to that of the quantum simulator:
\begin{align}
U_q =& \exp\left[-\frac{i\mathcal{H}_{\text{QHO}}t}{\hbar}\right] \nonumber\\=& \exp[-i\Omega t(\frac{1}{2}\left|0\rangle\langle0\right|+\frac{3}{2}\left|1\rangle\langle1\right|\nonumber\\&+\frac{5}{2}\left|2\rangle\langle2\right|+\frac{7}{2}\left|3\rangle\langle3\right|)]\nonumber\\
\rightarrow U_\text{qs}=& \exp[-i\Omega t(\frac{1}{2}\left|\uparrow\uparrow\rangle\langle\uparrow\uparrow\right|+\frac{3}{2}\left|\downarrow\uparrow\rangle\langle\downarrow\uparrow\right|\nonumber\\&+\frac{5}{2}\left|\downarrow\downarrow\rangle\langle\downarrow\downarrow\right|+\frac{7}{2}\left|\uparrow\downarrow\rangle\langle\uparrow\downarrow\right|)].\label{4.90}
\end{align}
$U_\text{qs}$ can be expressed in the form of spin operators of the quantum simulator as follows:
\begin{align}
U_\text{qs} = \exp\left[i\Omega t\left(\sigma_z^2\left(1 + \frac{1}{2}\sigma_z^1\right)\right)\right]\exp\left[-i2\Omega t\right].\label{4.91}
\end{align}
There is a global phase in $U_\text{qs}$, which cannot be observed. Therefore, we focus only on how we can realize $\exp\left[i\Omega t\left(\sigma_z^2\left(1 + \frac{1}{2}\sigma_z^1\right)\right)\right]$, and we will omit the global phase and denote $U_\text{qs}$ as $\exp\left[i\Omega t\left(\sigma_z^2\left(1 + \frac{1}{2}\sigma_z^1\right)\right)\right]$ in later analysis. $\exp\left[i\Omega t\left(\sigma_z^2\left(1 + \frac{1}{2}\sigma_z^1\right)\right)\right]$ can be decomposed into basic quantum gates of NMR system:
\begin{align}
&\exp\left[i\Omega t\left(\sigma_z^2\left(1 + \frac{1}{2}\sigma_z^1\right)\right)\right]\nonumber\\=&\exp\left[i\Omega t\sigma_z^2\right]\exp\left[i\Omega t\frac{1}{2}\sigma_z^1\sigma_z^2\right]\nonumber\\
=&\exp\left[i\Omega t\sigma_z^2\right]\exp\left[i \frac{\pi}{2}\sigma_x^1\right]\exp\left[-i\Omega t\frac{1}{2}\sigma_z^1\sigma_z^2\right]\exp\left[-i \frac{\pi}{2}\sigma_x^1\right]\nonumber\\
=&\exp\left[i \frac{\pi}{4}\sigma_x^2\right] \exp\left[-i\Omega t\sigma_y^2\right] \exp\left[-i \frac{\pi}{4}\sigma_x^2\right] \exp\left[i \frac{\pi}{2}\sigma_x^1\right] \nonumber\\&\cdot \exp\left[-i\Omega t\frac{1}{2}\sigma_z^1\sigma_z^2\right] \exp\left[-i \frac{\pi}{2}\sigma_x^1\right].\label{4.92}
\end{align}
According to the above equation, in order to realize the simulation of $U_\text{q}$ with a time length of $t$, the steps are as follows: First, apply a $\pi$ pulse in x direction on the first qubit and apply a free evolution gate with a duration of $\Omega t/\pi J$ ($J$ refers to the coupling strength of the two qubits), and then apply a $\pi$ pulse in -x direction on the first qubit and a pulse with the rotation angle $2\Omega t$ in -z direction on the second qubit. This z rotation can be further decomposed to x and y rotations. If the initial state is the superposition state of some of the states in $\left| \uparrow\uparrow \right\rangle $, $\left| \downarrow\uparrow \right\rangle$, $\left| \downarrow\downarrow \right\rangle$ and $\left| \uparrow\downarrow \right\rangle$, namely to simulate the superposition state of the basis states of the harmonic oscillator, then after the evolution $U_\text{qs}$ which simulates $U_\text{q}$, the phase changes determined by the difference in the energy levels of the quantum harmonic oscillator can be found in the coherence terms in the density matrix of the final state.

\begin{figure*}
\centerline{
\includegraphics[width=5.5in]{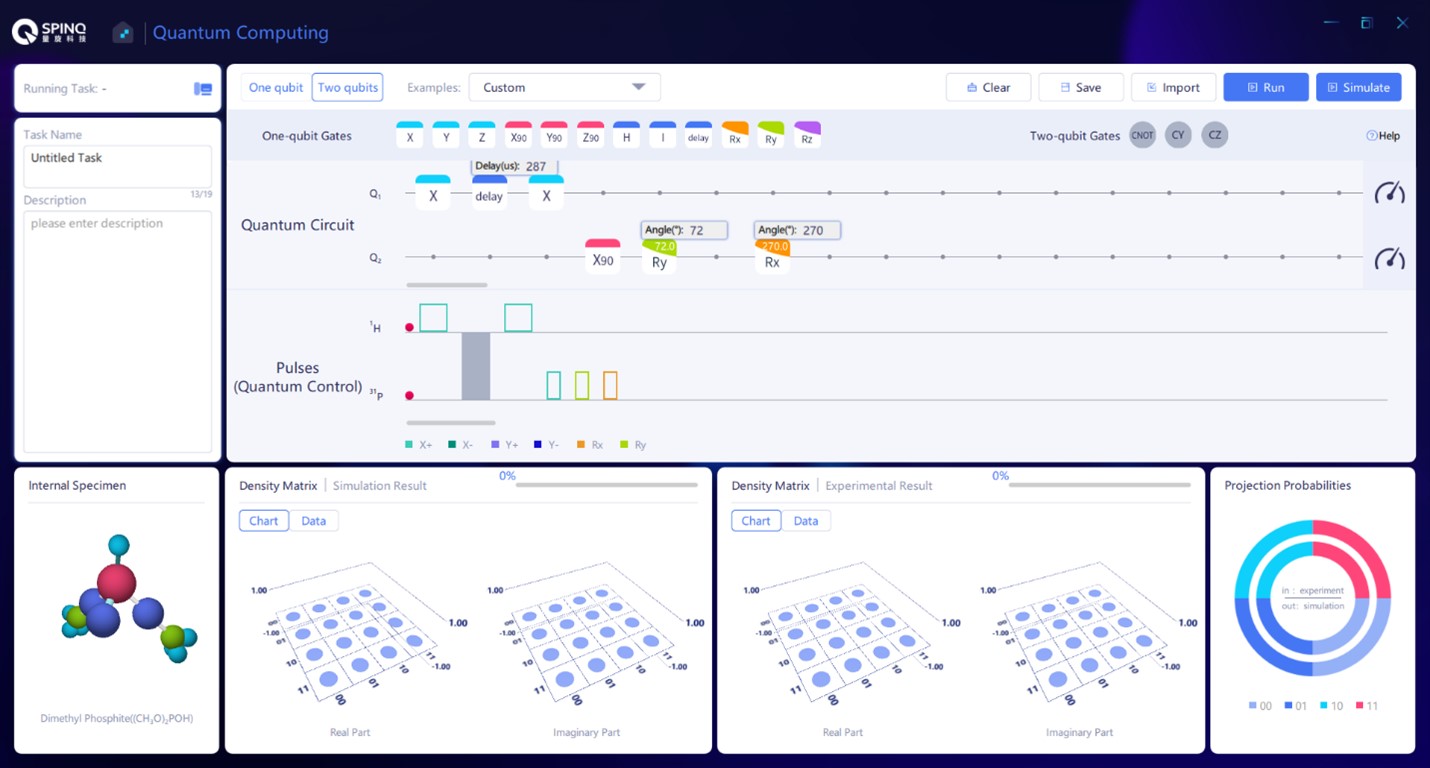}
}
\caption{Construct the quantum circuit to simulate the quatum harmonic osccilator starting from the initial state $|0\rangle$.}
\label{s4f53}
\end{figure*}

\begin{figure*}
\centerline{
\includegraphics[width=5.5in]{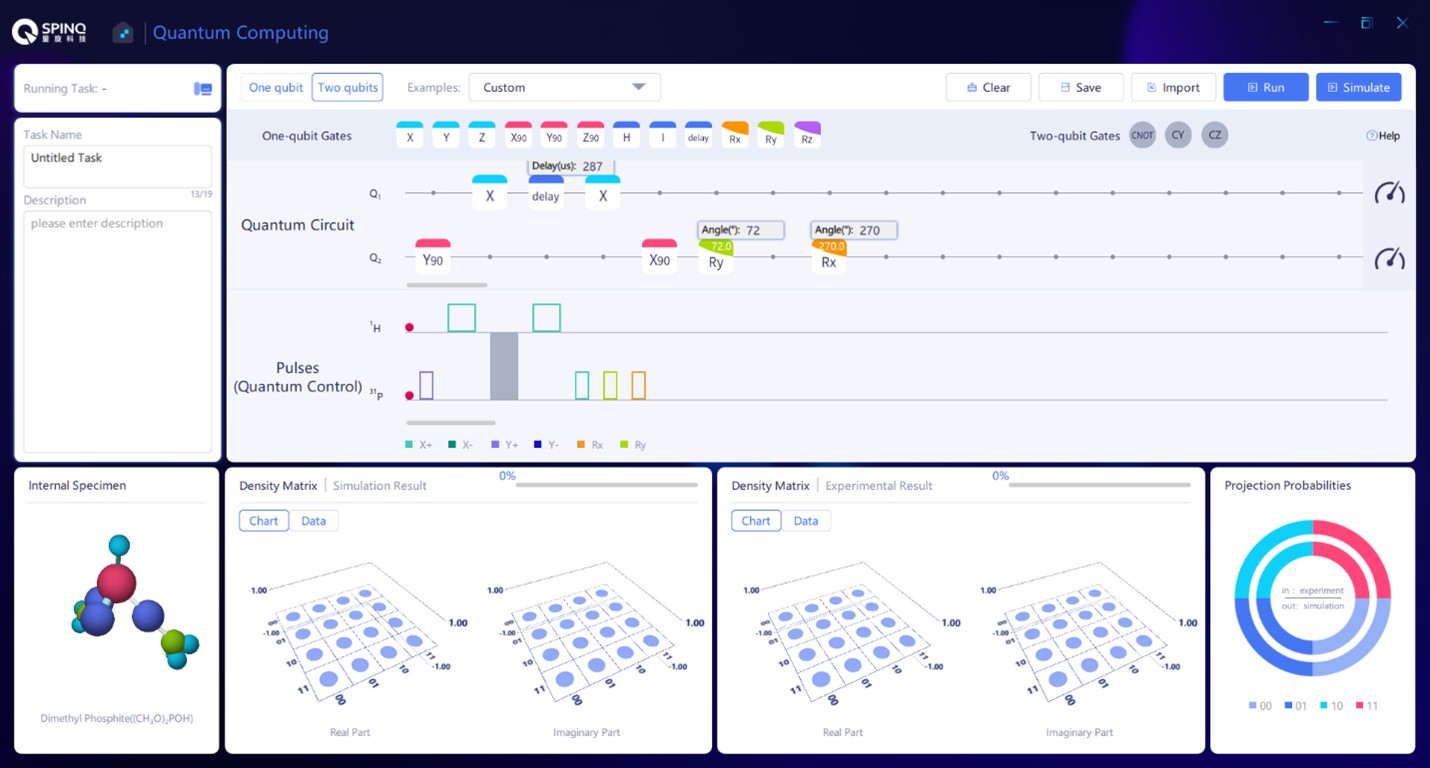}
}
\caption{Construct the quantum circuit to simulate the quatum harmonic osccilator starting from the initial state $|0\rangle+|3\rangle$.}
\label{s4f54}
\end{figure*}

\subsubsection{Experimental implementation}

In Fig. \ref{s4f53} we build the quantum circuit to simulate the evolution process of the quantum harmonic oscillator whose initial state is $|0\rangle$. “delay” gate refers to the free evolution gate, which requires an input parameter (in the unit of “us”) as the evolution duration. “Ry” and “Rx” refer to rotation gates around y axis and x axis, which require an input parameter as the rotation angle (in the unit of °). In order to simulate the evolution process of the quantum state under the Hamiltonian of quantum harmonic oscillator, it is necessary to change the value of $\Omega t$ to run multiple experiments, which is realized by changing the duration of “delay” gate and the angle of “Ry” gate in the gate sequence. Here we use a sequence of $\Omega t=[0.1 0.2 0.3 0.4 0.5 0.6 0.7 0.8 0.9 1]2\pi$ as an example. When $\Omega t$ is [0.1 0.2 0.3 0.4 0.5 0.6 0.7 0.8 0.9 1]$2\pi$, the duration of “delay” gate is [287 574 860 1147 1434 1721 2008 2294 2581 2868]us, and the angle of “Ry” gate is [72 144 216 288 360 432 504 576 648 720]°. It should be noted that we use the rotation by 270° around x axis to act in place of the rotation by 90° around -x axis.

We also build the quantum circuit to simulate the evolution process of the quantum harmonic oscillator whose initial state is $|0\rangle+|3\rangle$. As seen in Fig. \ref{s4f54}, the first quantum gate “Y90” is used to achieve the superposition state $|0\rangle+|3\rangle$. In this case, we shall observe the density matrix as $\Omega t$ increase, e.g., $\Omega t=[0.1 0.2 0.3 0.4 0.5 0.6 0.7 0.8 0.9 1]2\pi$, in the same way as the previous case. The duration of “delay” gate is [287 574 860 1147 1434 1721 2008 2294 2581 2868]us, and the angle of “Ry” gate is [72 144 216 288 360 432 504 576 648 720]°, respectively.

\subsubsection{Data analysis}

For the simulation when the initial state is $|0\rangle$, final state fidelities compared to $|0\rangle$ can be calculated using Eq. (\ref{1.27}) for each value of $\Omega t$. Here $|\varphi\rangle$ in Eq. (\ref{1.27}) is $|0\rangle$, $\rho$ is the experimental density matrix of the final state. Theoretically, the fidelity should be always 1, since no change will take place in $|0\rangle$ state under the effect of the Hamiltonian of the harmonic oscillator other than a global phase factor. Experimentally, decoherence and pulse errors all affect the results and can degrade the fidelities.

For the simulation when the initial state is $|0\rangle+|3\rangle$, the phase change can be observed in the entry $\rho_{12}$ of the density matrix. $\rho_{12}$ is the coherence term of $|0\rangle$ and $|3\rangle$. The energy difference of $|0\rangle$ and $|3\rangle$ is encoded in the phase of $\rho_{12}$ as its oscillation frequency.

\subsubsection{Discussion}

Readers can design quantum circuits to realize the quantum simulation of the quantum harmonic oscillator starting from the initial state $|0\rangle+|1\rangle+|2\rangle+|3\rangle$. It would be interesting to test how many oscillation frequencies can be observed in the coherence terms of the density matrix.

\section{Conclusions}
In this review, we introduced basic principles of quantum computing and NMR quantum computing. We also showed how to use a desktop NMR platform to demonstrate basic concepts of quantum computation and essential  quantum algorithms. Although NMR systems are considered not scalable  in the context of fault-tolerant quantum computing, they are very good systems for quantum computing education because of their clear physical images and well-developed quantum control techniques. Especially, with their maintenance-free and portable features, desktop NMR systems have great potentials in quantum computing education by facilitating hands-on experience for learners at all levels. We believe such portable quantum computing platforms will continue to contribute in equipping younger generations of students and researchers with the requisite knowledge for the future of quantum technologies.

\end{document}